\documentclass[11pt]{article}

\usepackage[letterpaper,margin=1in]{geometry}
\usepackage{authblk}
\usepackage[round,authoryear]{natbib}

\usepackage[normalem]{ulem}
\usepackage[none]{hyphenat}
\usepackage{breakcites}
\usepackage[utf8]{inputenc}
\usepackage[T1]{fontenc}
\usepackage{diagbox}
\usepackage{nicefrac}
\usepackage{microtype}
\usepackage{xcolor}
\usepackage{graphicx}
\graphicspath{{Arxiv/}{../}{../PNAS_submission/}}
\usepackage{caption}
\usepackage{subcaption}
\usepackage{booktabs}
\usepackage{amsmath}
\usepackage{amssymb}
\usepackage{mathtools}
\usepackage{amsthm}
\usepackage{amsfonts}
\usepackage{bm}
\usepackage{enumitem}
\usepackage{multirow}
\usepackage{wrapfig}
\usepackage{scalerel}
\usepackage{dsfont}
\usepackage{xspace}
\usepackage{thmtools,thm-restate}
\usepackage{algorithm}
\usepackage[noend]{algorithmic}
\usepackage{etoc}
\usepackage{hyperref}
\usepackage[capitalize,noabbrev]{cleveref}
\usepackage{autonum}

\hypersetup{
  colorlinks=true,
  linkcolor=blue,
  citecolor=blue,
  urlcolor=blue,
  pdftitle={Autotune: fast, accurate, and automatic tuning parameter selection for LASSO},
  pdfauthor={Tathagata Sadhukhan, Ines Wilms, Stephan Smeekes, and Sumanta Basu}
}
\allowdisplaybreaks
\AtBeginDocument{\etocdepthtag.toc{main}}

\theoremstyle{plain}
\newtheorem{theorem}{Theorem}[section]
\newtheorem{proposition}[theorem]{Proposition}
\newtheorem{lemma}[theorem]{Lemma}
\newtheorem{corollary}[theorem]{Corollary}
\theoremstyle{definition}
\newtheorem{definition}[theorem]{Definition}
\newtheorem{assumption}[theorem]{Assumption}
\theoremstyle{remark}
\newtheorem{remark}[theorem]{Remark}

\IfFileExists{Arxiv/tathagata-macros.tex}{%

\newcommand{\autotune}{\texttt{autotune}}

\newcommand{\Autotune}{\mathsf{Autotune}}
\newcommand{\glmnet}{\texttt{glmnet}}
\newcommand{\partialres}[1][k]{r^{(#1)}}
\newcommand{\softthres}[1][\lambda]{\mathcal{S}_{#1}}

\newcommand{\rss}{\text{RSS}}

\newcommand{\red}[1]{\textcolor{red}{{#1}}}

\newcommand{\loss}[1][\lambda]{\mathcal{L}_{#1}}

\newcommand{\solution}[1][\lambda]{\widehat\beta \parenth{#1}}
\newcommand{\snr}{\mathrm{SNR}}

\newcommand{\SUlink}{\hyperref[algo:sigma_update]{\textsc{Sigma-Update}}\xspace}
\newcommand{\Activelassolink}{\hyperref[algo:Autotune_lasso_active]{\textsc{Active-Autotune-Lasso}}\xspace}
\newcommand{\Activelink}{\hyperref[algo:active_set_selector]{\textsc{Active-Set-Selector}}\xspace}
\newcommand{\Active}{\mathsf{Predictor.Ranking}}
\newcommand{\Activeselect}{\mathsf{Active.Selection}}
\newcommand{\support}{\mathsf{Support.Set}}
\newcommand{\supportold}{\mathsf{Support.Set}^{\mathsf{(old)}}}
\newcommand{\betaold}{\what\beta^{\mathsf{(old)}}}
\newcommand{\ytemp}{\text{Y}^{\mathsf{(temp)}}}
\newcommand{\mrm}{\mathrm}
\newcommand{\partres}[1][i]{r^{(#1)}}

\newcommand{\hypo}[1][j]{\mathcal{H}_{0,#1}}
\newcommand{\seqp}[1][p]{\braces{1,\dots,#1}}

\newcommand{\sumn}[1][i]{\sum_{#1=1}^n}

\newcommand{\x}{x}

\newcommand{\axi}[1][i]{\x_{#1}}

\newcommand{\eps}{\epsilon}

\newcommand{\pseqxn}[1][n]{(\axi[i])_{i\geq 1}} 
\newcommand{\pseqxnn}[1][n]{(\axi[i])_{i=1}^n} 

\newcommand{\brackets}[1]{\left[ #1 \right]}

\newcommand{\parenth}[1]{\left( #1 \right)}

\newcommand{\braces}[1]{\left\{ #1 \right \}}

\newcommand{\abss}[1]{\left| #1 \right |}

\newcommand{\real}{\ensuremath{\mathbb{R}}}

\def\balign#1\ealign{\begin{align}#1\end{align}}
\def\baligns#1\ealigns{\begin{align*}#1\end{align*}}
\def\balignat#1\ealign{\begin{alignat}#1\end{alignat}}
\def\balignats#1\ealigns{\begin{alignat*}#1\end{alignat*}}
\def\bitemize#1\eitemize{\begin{itemize}#1\end{itemize}}
\def\benumerate#1\eenumerate{\begin{enumerate}#1\end{enumerate}}

\newenvironment{talign*}
 {\csname align*\endcsname}
 {\endalign}
\newenvironment{talign}
 {\csname align\endcsname}
 {\endalign}

\def\balignst#1\ealignst{\begin{talign*}#1\end{talign*}}
\def\balignt#1\ealignt{\begin{talign}#1\end{talign}}

\newcommand{\qtext}[1]{\quad\text{#1}\quad}

\let\originalleft\left
\let\originalright\right
\renewcommand{\left}{\mathopen{}\mathclose\bgroup\originalleft}
\renewcommand{\right}{\aftergroup\egroup\originalright}

\def\tinycitep*#1{{\tiny\citep*{#1}}}
\def\tinycitealt*#1{{\tiny\citealt*{#1}}}
\def\tinycite*#1{{\tiny\cite*{#1}}}
\def\smallcitep*#1{{\scriptsize\citep*{#1}}}
\def\smallcitealt*#1{{\scriptsize\citealt*{#1}}}
\def\smallcite*#1{{\scriptsize\cite*{#1}}}

\def\red#1{\textcolor{red}{{#1}}}

\def\mbb#1{\mathbb{#1}}
\def\mc#1{\mathcal{#1}}
\def\msf#1{\mathsf{#1}}
\def\mrm#1{\mathrm{#1}}

\def\<{\left\langle} 
\def\>{\right\rangle}

\def\norm#1{\left\|{#1}\right\|} 
\newcommand{\onenorm}[1]{\norm{#1}_1} 
\newcommand{\twonorm}[1]{\norm{#1}_2} 
\newcommand{\infnorm}[1]{\norm{#1}_{\infty}} 

\newcommand{\inner}[2]{\langle{#1},{#2}\rangle} 
\def\what#1{\widehat{#1}}

\def\E{\mbb{E}} 

\def\Var{\mrm{Var}} 

\newcommand{\Gsn}{\mathcal{N}}

\newcommand{\iid}{\textrm{i.i.d.}\xspace}

\providecommand{\argmin}{\mathop\mathrm{arg min}}

\providecommand{\diag}{\mathop\mathrm{diag}}

\ifdefined\nonewproofenvironments\else
\ifdefined\ispres\else

\newenvironment{proof-sketch}{\noindent\textbf{Proof Sketch}
  \hspace*{1em}}{\qed\bigskip\\}
\newenvironment{proof-idea}{\noindent\textbf{Proof Idea}
  \hspace*{1em}}{\qed\bigskip\\}
\newenvironment{proof-of-lemma}[1][{}]{\noindent\textbf{Proof of Lemma {#1}}
  \hspace*{1em}}{\qed\\}
\newenvironment{proof-of-theorem}[1][{}]{\noindent\textbf{Proof of Theorem {#1}}
  \hspace*{1em}}{\qed\\}
\newenvironment{proof-attempt}{\noindent\textbf{Proof Attempt}
  \hspace*{1em}}{\qed\bigskip\\}

\fi
\fi

\newtheorem{myproposition}{Proposition}

\newtheorem{mycorollary}{Corollary}

\newtheorem{mylemma}{Lemma}

\newtheorem{mydefinition}{Definition}

\newtheorem{myremark}{Remark}

\newtheorem{myassumption}{Assumption}

 \crefname{appendix}{App.}{App.}
\crefname{equation}{}{}
\crefname{lemma}{Lem.}{Lem.}
\crefname{theorem}{Thm.}{Thm.}
\crefname{Corollary}{Cor.}{Cors.}
\crefname{algorithm}{Alg.}{Algs.}

\crefname{section}{Sec.}{Sec.}
\crefname{table}{Tab.}{Tab.}
\crefname{remark}{Rem.}{Rem.}
\crefname{definition}{Def.}{Def.}
\crefname{Proposition}{Prop.}{Prop.}
\crefname{myremark}{Rem.}{Rem.}
\crefname{mylemma}{Lem.}{Lem.}
\crefname{mydefinition}{Def.}{Defs.}
\crefname{myproposition}{Prop.}{Prop.}
\crefname{mycorollary}{Cor.}{Cors.}
\crefname{myassumption}{Assum.}{Assum.}
\crefname{figure}{Fig.}{Fig.}
\crefname{enumi}{}{}
\crefname{name}{}{} 

}{%

}

\title{Autotune: fast, accurate, and automatic tuning parameter selection for LASSO}

\author[1]{Tathagata Sadhukhan}
\author[2]{Ines Wilms}
\author[2]{Stephan Smeekes}
\author[1]{Sumanta Basu\thanks{Corresponding author: \href{mailto:sumbose@cornell.edu}{sumbose@cornell.edu}}}
\affil[1]{Department of Statistics and Data Science, Cornell University}
\affil[2]{Department of Quantitative Economics, Maastricht University}
\date{}

\begin{document}

\maketitle

\begin{abstract}
Least absolute shrinkage and selection operator (Lasso), the popular variable selection engine for high-dimensional regression, is commonly tuned using cross-validation (CV). This is known to be slow and loses accuracy in high-dimension, low signal-to-noise ratio (SNR) settings. These issues are exacerbated for high-dimensional time series models such as the vector autoregression (VAR), where time series cross-validaiton (TSCV) is used for tuning.
We propose \texttt{autotune}, a strategy for the Lasso to tune itself automatically by exploiting information contained in partial residuals computed during a single Lasso fit.
The strategy can also be viewed as alternately estimating noise standard deviation and column space of relevant predictors.
Numerical experiments on regression and VAR models show that \texttt{autotune} is faster than existing alternatives, and more accurate when the SNR is low. It also provides an accurate estimator of noise scale and diagnostic plots akin to screeplots for checking model sparsity. We demonstrate the benefit of \texttt{autotune} on several real data sets and develop an R package available on CRAN.
\end{abstract}

\begin{center}
\small\textbf{Keywords:} LASSO; vector autoregression; automatic tuning
\end{center}

\section{Introduction}
\IfFileExists{Arxiv/01-intro.tex}{%
For a response vector $\text{Y} = (\text{y}_i)_{i=1}^n\in \real^n$ and $p$ predictors in a design matrix $\text{X} = \left((\text{x}_{ij})\right)_{1 \le i \le n, 1 \le j \le p}$, a Lasso estimator $\widehat{\beta}(\lambda)$ solves the convex problem 
\begin{align}
\label{eq: original lasso}
    \min_{\beta\in \real^p} \, \loss(\beta):= \,  (1/2n) \, \|\text{Y}-\text{X}\beta\|^2 + \lambda\|\beta\|_1.
\end{align}
Over the last two decades, Lasso has been adopted widely in high-dimensional time series models such as the VAR \cite{basu2015regularized,kock2025data,shojaie2022granger,masini2023machine,duker2025vector}. Fast and accurate selection of its  regularization parameter $\lambda$ remains an active area of research \cite{wu2020survey}, especially for time series data.


Tuning based on Akaike and Bayesian information criteria (AIC \cite{akaike2003new} and BIC \cite{schwarz1978estimating}) are fast, but known to perform poorly in high-dimension \cite{ chen2008extended, zhao2015mixture,wu2020survey}.  Cross-validation (CV), the most popular tuning strategy, is slow and also known to lose accuracy \cite{wang2007tuning,lei2020cross, bates2024cross} in high-dimension. 
These issues are exacerbated for time series data, where time series CV (TSCV), a version of CV constrained to obey temporal ordering of data when creating training and test folds, is used. Time series models that require computing regression coefficients recursively or iteratively \cite{Moulines2005OnRecursive,Kapetanios2003} have not yet adopted Lasso widely, partly due to the computational bottleneck of tuning.

More broadly, CV-based  practice of tuning Lasso deviates from the traditional practice of variable selection in two ways. In classical, low-dimensional regression with a handful of predictors, one estimates the noise scale along with $\beta$, and selects predictors in the final regression model based on a scale-free tuning parameter $\alpha \in (0, 1)$, the significance level, which different practitioners can choose differently based on their perceived cost of false positives in a given domain problem. In contrast, CV-based tuning advocates for a $L_2$-risk optimal model universally for all domain problems, irrespective of the relative costs of false positives and negatives. Moreover, in this ``fit first, tune last'' workflow, the overall computational cost of choosing a single scalar $\lambda$ is often much higher than solving a single high-dimensional problem \eqref{eq: original lasso}, which is somewhat counterintuitive.



A popular alternative to CV is the Scaled Lasso \cite{sun2012scaled}. Assuming a linear model $\text{Y} = \text{X} \beta + \varepsilon$, $\Var(\varepsilon) = \sigma^2 I$,  high-dimensional asymptotic theory suggests $\lambda \asymp \sigma \sqrt{\log p / n}$ is an ideal choice. Leveraging this connection between $\lambda$ and $\sigma$,  Scaled Lasso automates tuning by jointly estimating the noise scale $\sigma$ along with $\beta$. Bilevel optimization algorithms such as the \texttt{sparseho} \cite{bertrand2020implicit} also automate tuning by optimizing over $\lambda$ along with $\beta$. To date, neither is uniformly faster than CV Lasso implemented in \texttt{glmnet} \cite{glmnet}, and their time series asymptotics are not yet fully understood.  
For VAR models, \cite{kock2025data} has recently developed an automatic tuning strategy similar to Scaled Lasso, by 
estimating noise scales along with the autoregression coefficients. 

Our main premise is that estimating the noise scale $\widehat{\sigma}^2 = \Var(\text{Y} - \widehat{\text{Y}})$ only requires good prediction $\widehat{\text{Y}} = \widehat{\text{X}\beta}$, a problem that is known to be easier than estimating the high-dimensional vector $\beta$ \cite{chatterjee2013assumptionless,yu2019estimating}. So, an automatic tuning algorithm that primarily relies on good prediction may be able to first find a good $\sigma$ ( hence a good $\lambda$) quickly, so that we need to solve \eqref{eq: original lasso} only once in the end for the chosen $\lambda$. Such a ``tune first, fit last'' workflow can potentially offer a substantial speed gain over CV-style tuning that requires solving the high-dimensional problem \eqref{eq: original lasso} multiple times.

In an attempt to operationalize this premise, in this paper we 
ask: \\
\textit{$~~$ Is there a $\lambda$ on the original penalty grid of Lasso which is at least as accurate as CV, but can be found quickly?}\\
and 
develop $\autotune$, a strategy for the Lasso to tune itself by relying only on a pre-specified, scale-free parameter $\alpha \in (0, 1)$.

\begin{figure*}[!t]
    \centering
    \begin{subfigure}[h]{0.24\linewidth}
        \centering
    {\fontsize{6}{1}\selectfont \textbf{[A]}}
        \includegraphics[page = 1, trim=0in 0in 0in 0.4in, clip, width=\textwidth]{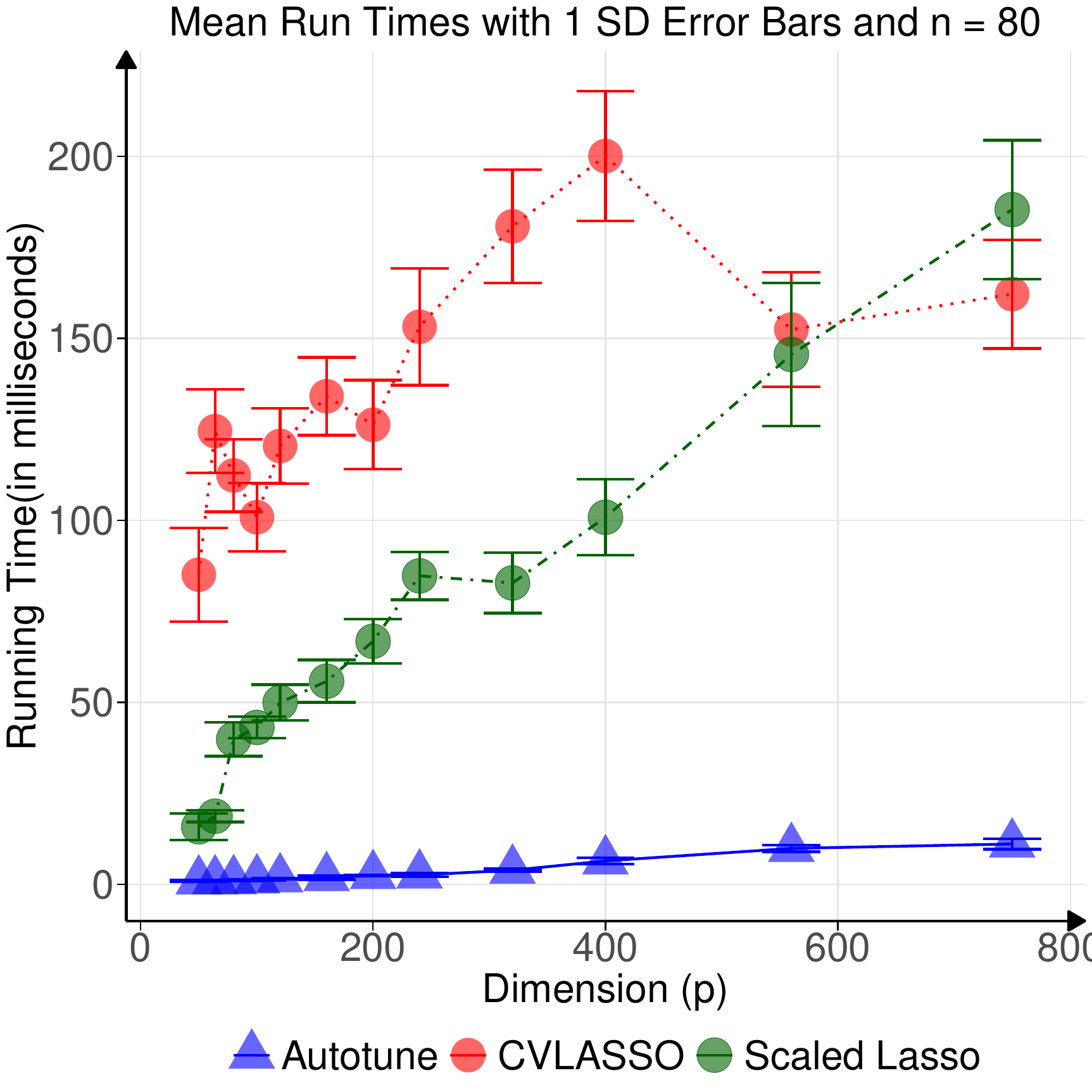}
        \label{fig: intro runtime plot}
    \end{subfigure}
    \begin{subfigure}[h]{0.24\linewidth}
        \centering
        {\fontsize{6}{1}\selectfont \textbf{[B]}}
        \includegraphics[page = 1,trim=0in 0in 0in 0.33in, clip, width=\textwidth]{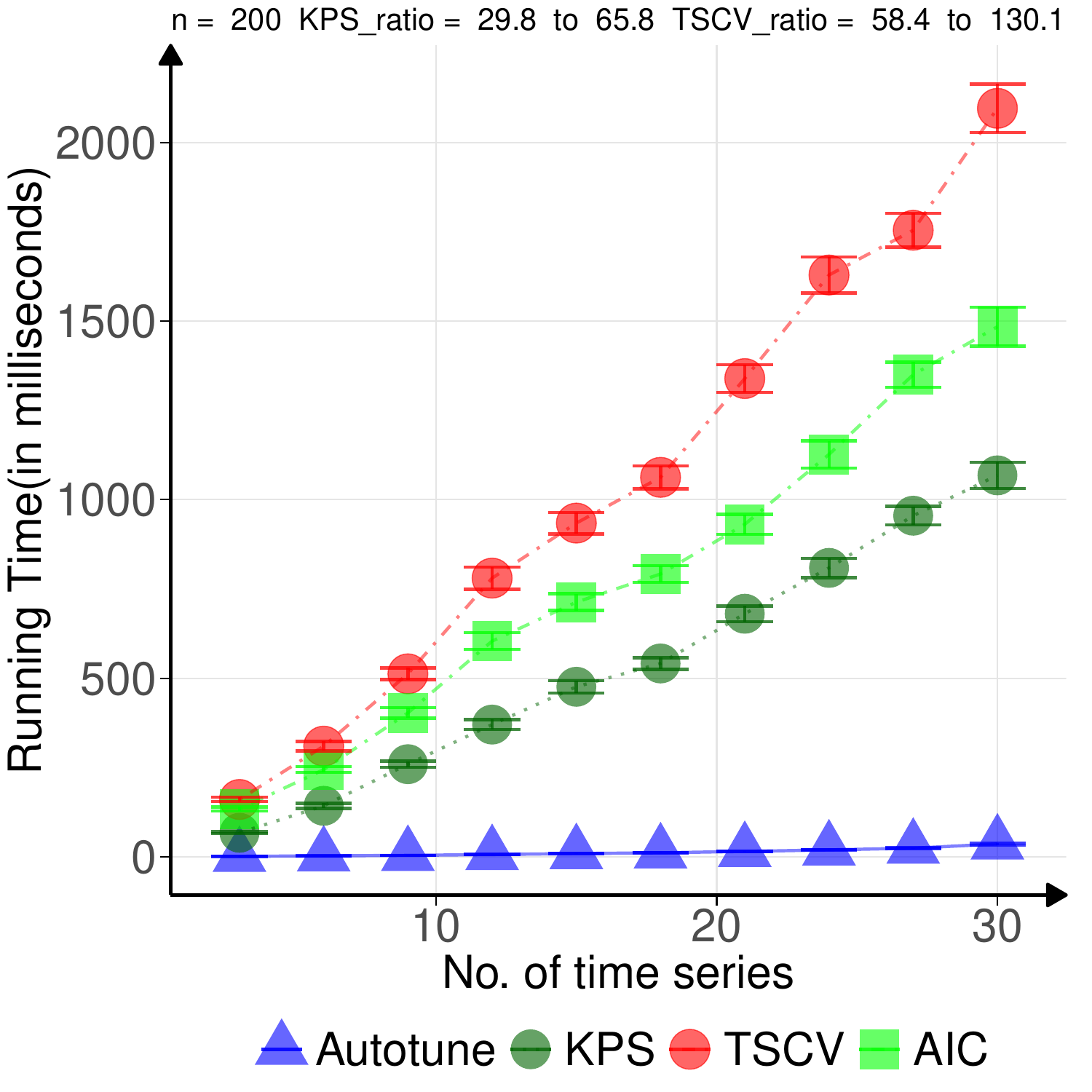}
        \label{fig: intro ts runtime plot} 
    \end{subfigure}
    \begin{subfigure}[h]{0.24\textwidth}
        \centering
        {\fontsize{6}{7}\selectfont \textbf{[C]}}
        \includegraphics[page = 1, trim=0in 0in 0in 0in, clip, width=\textwidth]{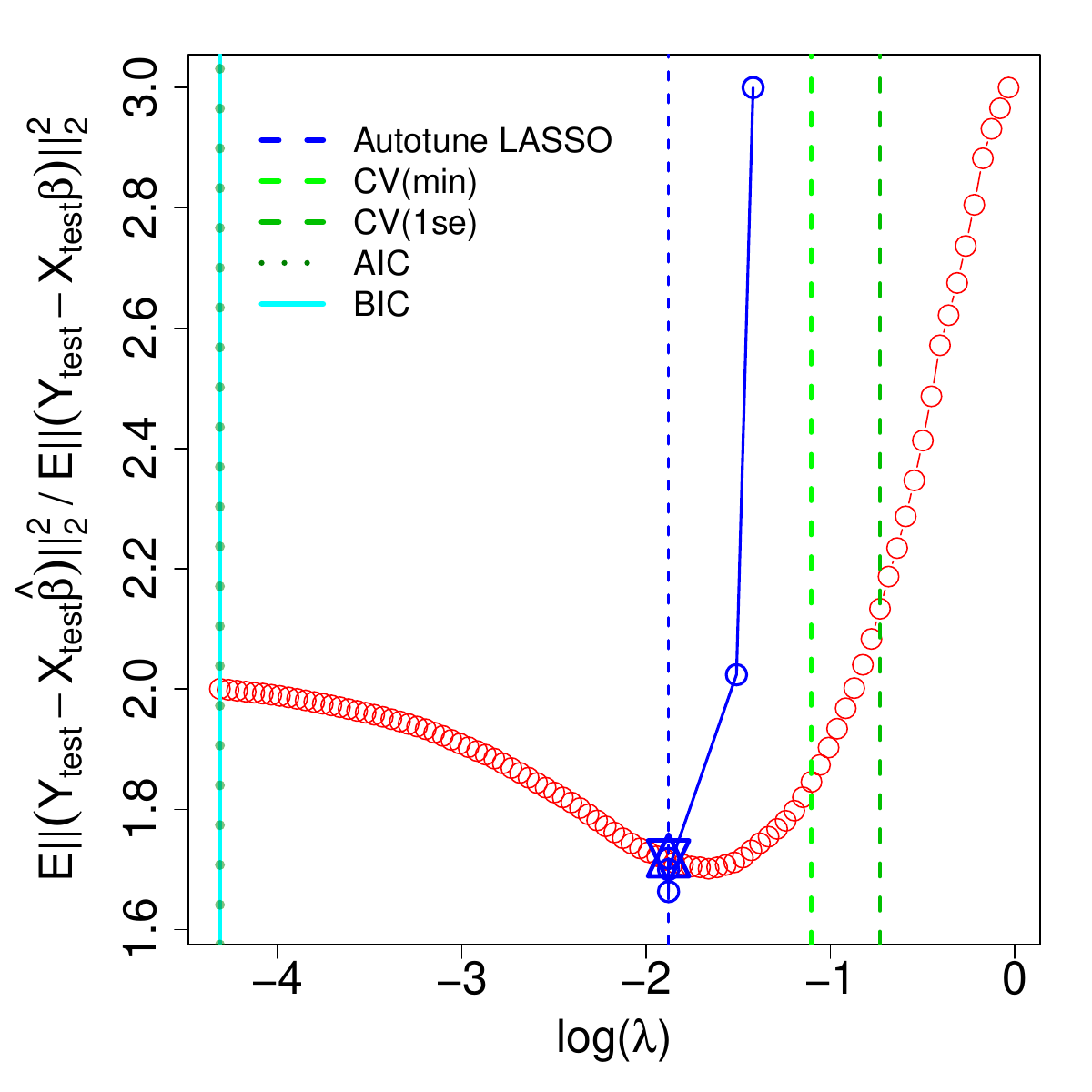}
    \end{subfigure}
    \begin{subfigure}[h]{0.25\textwidth}
        \centering
        {\fontsize{6}{7}\selectfont \textbf{[D]}}
        \includegraphics[width=\textwidth]{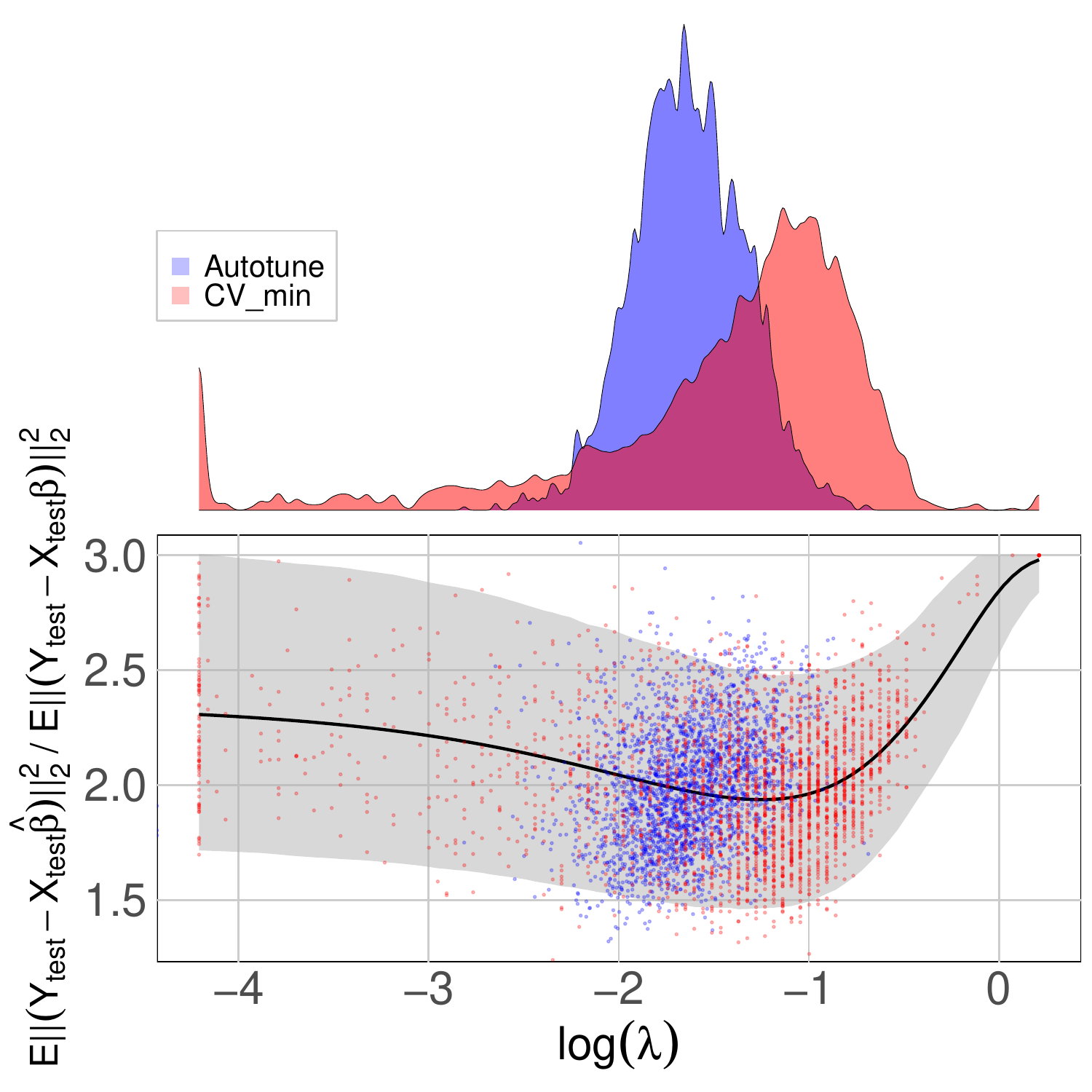}
    \end{subfigure}
  
  \caption{ Mean runtimes (with 1 SE bars) of $\autotune$ and its benchmarks in simulated \textbf{[A]} Lasso regression and \textbf{[B]} VAR Lasso.  
    {$\autotune$ shows 15x-200x speedup over CV and scaled Lasso style alternatives.} \textbf{[C]} Relative test error (RTE) of  $\autotune$ (blue) and Lasso (red) solution paths against $\log(\lambda)$ in a simulated Lasso regression ($n = 80,$ $ p=750$ and $\norm{\beta}_0 = 5$). {$\autotune$ ($\alpha = 0.01$) finds a $\lambda$ with lower RTE than AIC, BIC and CV in only $3$ attempts.} \textbf{[D]} Distribution of $\lambda$ selected by $\autotune$ and CV over $2500$ simulations from the same model shows the same pattern, {$\autotune$ tends to select $\lambda$  with significantly lower RTE than CV.}}
  \label{fig: teaser into autotune}
\end{figure*}

In Figure \ref{fig: teaser into autotune}, we demonstrate the speed of $\autotune$. For Lasso regression, $\autotune$ ($\alpha = 1\%$) is 15-140 times faster than CV Lasso of \texttt{glmnet}, and 15-55 times faster than Scaled Lasso over a number of simulation scenarios (details in Table \ref{tab: runtime table}). For VAR Lasso, it is 60-200 times faster TSCV and 30-80 times faster than  \cite{kock2025data}.

In Figure \ref{fig: teaser into autotune}, we illustrate on a single Lasso regression ($n=80, p=750$, $\norm{\beta}_0 = 5$) why it is faster. Instead of computing $\widehat{\beta}(\lambda)$ on a dense grid of $\lambda$ (red dots), $\autotune$ (blue dots) searches over only 3 different data-adaptively chosen $\lambda$ values before stopping (blue star).

Interestingly, $\autotune$ selects a $\lambda$ with lower relative test error (RTE, see Section \ref{sec: simulation}) than CV Lasso, whereas AIC and BIC both overfit. To ensure this pattern is not anecdotal, we simulate 2500 times from the same model, and compare the $\lambda$ selected by  $\autotune$ and CV Lasso. We find that $\autotune$ indeed produces statistically significantly smaller RTE than CV Lasso (paired t-test p-value $= 1.06\times 10^{-14})$.

Under the hood, $\autotune$ simply estimates $\beta$ and $\sigma^2$ alternately inside a full negative Gaussian log-likelihood with an $\ell_1$ penalty
\begin{align}
\label{eq: Autotune lasso objective}
\min_{\beta, \sigma^2}\quad \frac{1}{2} \log \sigma^{2} + \frac{1}{2n\sigma^2} \|\text{Y}-\text{X}\beta\|^2 + \lambda_0 \|\beta\|_1,
\end{align}
where $\lambda_0 = \|\text{X}^\top \text{Y}/2n\|_\infty / Var(\text{Y})$.  Since for any given $\widehat{\sigma}^2$, solving for $\beta$ in \eqref{eq: Autotune lasso objective} is essentially solving Lasso with tuning parameter $\lambda:= \lambda_0 \widehat{\sigma}^2$, sequentially guessing different values of $\sigma^2$ results in fitting Lasso on an automatic, data-driven sequence of $\lambda$ values. Such an alternating estimation is also at the heart of Scaled Lasso, where $\sigma^{2}$ is estimated using the current $\widehat{\beta}$ as $\widehat{\sigma}^2 = \|\text{Y}-\text{X} \widehat{\beta}\|^2/n$. The automatic tuning of VAR Lasso in \cite{kock2025data} also uses similar updates.

In $\autotune$, we replace this $\widehat{\sigma}^2$ by a new estimator. 
For a given $\lambda$, coordinate descent (CD) for Lasso iteratively computes $\widehat{\beta}_k = \mc{S}_\lambda(\sum_{i=1}^n \text{x}_{ij} r^{(k)}_i/n)$, $k = 1, \ldots, p, 1, \ldots, p, \ldots$, where $\mc{S}_{\lambda}(a) = \mathrm{sign}(a)(|a|-\lambda)_+$ is the soft thresholding operator and the partial residual (PR) vector $\partialres$ of the $k^{th}$ predictor is 
\begin{align}
\label{eq: partial residuals df}
    \partialres_i := \text{y}_i - \sum_{j \neq k}\text{x}_{i, j}\what\beta_j, \text{  for  $i\in\braces{1,\dots,n}$  }.
\end{align}
During the course of this iterations, we track the sizes ($\ell_2$ norm $\|r^{(k)}\|_2$) of these PR, and rank the predictors in their decreasing order. 
If the true model is sparse and only a few predictors contribute the most towards predicting $Y$, we expect these predictors' PR to be larger compared to others. Betting on this sparsity assumption, after every iteration of CD (one round of $k = 1, \ldots, p$),
we build a small linear model of $Y$ by sequentially including these predictors one at a time, using F-tests with a significance level $\alpha$ to decide when to stop. The mean squared error (MSE) of this linear model is then used as the new $\widehat{\sigma}^2$, which guides $\autotune$ to move on and try a new $\lambda = \lambda_0 \widehat{\sigma}^2$. We find that this $\widehat{\sigma}^2$ tends to converge during the first few iterations of CD, allowing us to devise an early stopping rule and avoid computing the full Lasso estimator $\widehat{\beta}(\lambda)$ at each of these candidate $\lambda$ values.

In Section \ref{sec:app-theory}, we show that this heuristic algorithm is equivalent to alternately estimating $\sigma$ and the projection matrix $P_{S}$ on the column space spanned by the support variables in $\text{X}$. We argue that this new target $P_{S}$ is related to prediction ($\widehat{\text{X}\beta}$) and not estimation ($\widehat{\beta}$), offering us a way to avoid expensive computation of $\beta$ at every stage of tuning. Taken together, our  ``tune first, fit last'' approach makes $\autotune$ much faster than the traditional approach of first fitting Lasso over a grid of $\lambda$ values, and then selecting a $\lambda$.

In a wide range of simulation experiments using all data generating processes from \cite{hastie2020best}, we show  that (Section \ref{subsec: regg setup sim comparison}) $\autotune$ yields the same or better accuracy than CV Lasso and Scaled Lasso  in terms of estimation (Relative Mean Squared Error), prediction (RTE), and variable selection (Area under Receiver Operating Characteristic curve), while also offering substantial speedup. For Lasso VAR simulations (Section \ref{subsec: VAR}), $\autotune$ yields even larger speed and accuracy gains over TSCV and \cite{kock2025data}. In Appendix \ref{app: comparison vs BLO skglm}, we benchmark against Bilevel Optimization (\texttt{sparse-ho}) and a python implementation of Lasso \cite{skglm_neurips2022} with qualitatively similar findings.

Interestingly, $\autotune$ only uses $10-20$\% of its total runtime to tune $\lambda$ and the rest to find $\widehat{\beta}(\lambda)$ (Table \ref{tab: finding lambda vs finding beta}). This suggests that the task of tuning may not be inherently hard and does not need to be the main computational bottleneck in fitting Lasso for high-dimensional problems.

In addition to automatic tuning, in Section \ref{subsubsec: simulation results for noise estimation and sparsity diagnostics} we show that the $\widehat{\sigma}^2$ from $\autotune$ can accurately estimate noise scale, leading to better high-dimensional inference with the Debiased Lasso \cite{dezeure2015high}. We also provide a promising visual diagnostic, similar to scree plot for principal component analysis, for checking the sparsity assumption of Lasso. 

Finally, on a real financial data analysis (Section \ref{sec: realdata}),  
we show that $\autotune$ generalizes better while fitting a sparser model as compared to CV Lasso and Scaled Lasso. In Appendix \ref{app: more real life datasets}, we show that the same performance gains of $\autotune$ over the Lasso benchmarks hold in three more real life datasets. An R package \href{https://cran.r-project.org/web/packages/autotune/index.html}{autotune} is made publicly available on CRAN.

\section{Background}


For completeness, here we briefly describe our benchmark methods and some technical ingredients needed to present $\autotune$. More details are in Appendix \ref{sec: appen background}.

\paragraph*{Tuning and noise variance estimation for Lasso regression} The R package $\mathsf{glmnet}$ \cite{glmnet} recommends selecting $\lambda$ over a grid by either CV(min): the $\lambda$ minimizing K-fold CV error over the grid, or CV(1se): the largest $\lambda$ with a K-fold CV error within 1 SD of the minimum CV error. We use these as benchmarks.

Asymptotic choice of $\lambda \asymp \sigma\sqrt{\log(p)/n}$ links tuning to estimating $\sigma$ \cite{bickel2009simultaneous}, leading to a number of new methods. Scaled Lasso \cite{sun2012scaled}, our other major benchmark \footnote{Other works \cite{belloni2011square, wang2020tuning,JMLR:v22:20-539,lederer2023} that changed the original objective function \eqref{eq: original lasso} of Lasso are not included in our benchmarking analysis since we focus only on tuning the Lasso.}, solves
\begin{align}
\label{eq: Scaled lasso objective}
\min_{\beta, \sigma^2}\quad \sigma/2 + \frac{1}{2n\sigma} \|\text{Y}-\text{X}\beta\|^2 + \lambda_0 \|\beta\|_1
\end{align}
by alternately minimizing over $\beta$ and $\sigma$. Also, \cite{yu2019estimating} proposed Natural and Organic Lasso for estimating $\sigma$ by modifying the Lasso objective. 
All three estimators $\what{\sigma}^2$ suffer from the shrinkage bias in $\what{\beta}$, an issue that $\autotune$ attempts to resolve explicitly.

\paragraph*{Tuning Lasso for VAR models} 
A VAR($d$) model on a $p$-dimensional stationary time series $\braces{Z^t}_{t \geq 1}$ with uncorrelated errors takes the form 
    $Z^t = \sum_{i=1}^d A_i Z^{t-i}+\varepsilon^t, \quad \varepsilon^t\stackrel{\iid}{\sim} \parenth{0, \Sigma_{\varepsilon}},$ 
where $A_1,\dots,A_d$ are $p\times p$ transition matrices of interest, and $\Sigma_{\varepsilon}= \diag\parenth{\sigma^2_1,\dots,\sigma^2_p}$, with  noise variances $\sigma^2_1,\dots,\sigma^2_p$.  Given $n + d$ consecutive observations from  $\braces{Z^t}_{t \geq 1}$, all the $d$ VAR transition matrices can be jointly estimated in the autoregressive design Y $ = \text{X} \Phi + \text{E}$, where Y = $\brackets{Z^{n+d},\dots,Z^{d+1}}^\top$, X is a $n\times dp$ matrix with $i^{th}$ row being $\brackets{ \left(Z^{i + d - 1}\right)^{\top},  \left(Z^{i + d - 2}\right)^{\top},  \dots,  \left(Z^{i}\right)^{\top} }$, $\Phi = \brackets{A_1, A_2,\dots,A_d}^\top$ and $E = \brackets{\varepsilon^{n+d},\dots,\varepsilon^{d+1}}^\top$. For any $i = 1, \ldots, p$, Lasso VAR estimates the $i^{th}$ row of $\Phi$ as   
\begin{equation}
    \label{eq: Var columnwise regg}\what{\Phi}_{\cdot,i}=\underset{\beta \in \mathbb{R}^{dp}}{\operatorname{argmin}} \frac{1}{2n}\left\|\text{Y}_{\cdot, i}-\text{X} \beta\right\|^2 + \lambda_i\onenorm{\beta}, 
\end{equation}
where $\lambda_1, \ldots, \lambda_p$ are $p$ different tuning parameters. They are often tuned using time series CV (TSCV) \cite{hyndman2018forecasting}. Instead of randomly selecting K folds as in CV, one divides the time series data into $2K$ consecutive folds. Then, for every choice of $\{ \lambda_i \}_{i=1}^p$, one fits K different models by taking folds $1:(K + h - 1)$ as training set and fold $K+h$ as test set, for $h = 1, \ldots, K$. Rest of the tuning proceeds as in regular CV. 

Recently \cite{kock2025data} proposed an automated tuning procedure built upon  \cite{belloni2012sparse}. For $t \geq 1$, the $t^{th}$ iteration computes $\what\Phi^{(t)}$  using an weighted Lasso:
\begin{align}
    \label{eq: kps columnwise regg}&\what{\Phi}^{(t)}_{\cdot,i}=\underset{\beta \in \mathbb{R}^{dp}}{\operatorname{argmin}}\hspace{0.1cm} \frac{1}{n}\left\|\text{Y}_{\cdot, i}-\text{X} \beta\right\|^2 + \frac{\lambda}{n}\onenorm{\Gamma^{(t)}_i\beta},
\end{align}
where $\Gamma^{(t)}_i = \diag\parenth{\what v^{(t)}_{i,1}, \dots, \what v^{(t)}_{i,pq}}, \quad \text { for } i=1, \ldots, p$, and  $\what{v}^{(t)}_{i,j} = \frac{1}{\sqrt{n}}\sqrt{\sum_{r = 1}^n\text{X}_{r,j}^2\parenth{\text{Y}_{r, i}-\text{X}_{r:}\what\Phi^{(t-1)}_{\cdot, i}}^2}$.
Except for the use of weights $X^2$ to adjust for heteroskedasticity, this algorithm is very similar to Scaled Lasso in the sense of sequentially estimating the noise scale using $\what{v}^{(t)}$. It also suffers from the shrinkage bias in $\what{\Phi}$ that affects $\what{v}$. 

\paragraph*{Coordinate descent (CD), partial residuals (PR)} 
The fastest implementation of Lasso in \texttt{glmnet} is based on CD (Algorithm \ref{algo: CD for lasso}). A pathwise version involving warm start, active set selection (see Appendix \ref{sec: appen background}) computes $\widehat{\beta}(\lambda)$ quickly over a grid of $\lambda$ values. After each iteration, CD produces a vector of intermediate residuals ($r$) and $p$ vectors of PR:  $\partialres[j] = r + \text{X}_{.,j}^{\top}\what\beta_j$, $j = 1, \ldots, p$.
We design  $\autotune$ to explicitly look for partial information on $\sigma^2$ in these objects, and use that to guide the tuning process.

\begin{algorithm}[tb]
  \caption{Pseudocode of Coordinate Descent for Lasso}
  \label{algo: CD for lasso}
  \begin{algorithmic}
    \STATE {\bfseries Input:} Y, Normalized predictors X (i.e. $\twonorm{\text{X}_{\cdot,j}}^2=n$), tuning parameter $\lambda$
    \STATE Initialize $\what{\beta}^{(0)} = 0_p$, $r =$  Y$ - \text{X}\what{\beta}^{(0)} = $Y.
    \FOR{$t = 1,\ldots,\mathsf{max.iterations}$}
    \FOR{$j = 1,\ldots,p$}
    \STATE $\what{\beta}^{(t)}_j 
    = \softthres\parenth{\frac{1}{n}\text{X}_{\cdot,j}^{\top}\parenth{r + \text{X}_{\cdot,j}\what{\beta}^{(t-1)}_j}}$ 
    \STATE $r \gets r - \text{X}_{\cdot,j}\parenth{\what{\beta}^{(t)}_j -\what{\beta}^{(t - 1)}_j}$
    \ENDFOR
    \STATE \textbf{if} $\norm{\what{\beta}^{(t)} - \what{\beta}^{(t-1)}}$ is small, \textbf{then} break
    \ENDFOR
    \STATE {\bfseries Return:} $\what{\beta}^{(t)}$
  \end{algorithmic}
\end{algorithm}

\paragraph*{Sequential F-tests}
%
Given a response Y and $p~(<n)$ predictors X$_j$, $j\in \seqp$, a classic model selection procedure in Statistics \cite{montgomery2020introduction} builds a sequence of nested linear models $\{\mc{M}_k\}_{k \ge 1}$: 
\begin{align}
    \mc{M}_k:  \text{ Y} &= \text{ X}_{1}\beta_1 + \text{ X}_{2}\beta_2 + \dots + \text{X}_{k}\beta_k + \varepsilon,\\
    \text{RSS}(\mc{M}_k) &:= \sumn \parenth{\text{y}_i - \what{\text{y}}_i}^2, \qquad  k \in \braces{1,\dots,p}, 
\end{align}
and tests $\hypo[k]:\braces{\beta_k=0}|\parenth{\beta_1, \dots, \beta_{k - 1}}$ with F-statistic
\begin{align}
    F = \frac{\rss(\mc{M}_{k - 1}) - \rss(\mc{M}_{k})}{\rss(\mc{M}_k)/(n - k)}\underset{\hypo[k]}{\sim} F_{1, n - k}
\end{align}
to decide whether to include $X_k$ in the linear model with predictors $\{X_j\}_{j=1}^{k-1}$. The significance level $(\alpha)$ of the F-tests allows practitioners to control the desired level of false positives in a given domain problem.
%
The rank order of predictors in which the nested models $\mc{M}_k$ are constructed is important for selecting the final model.

\section{Method: tuning with \texorpdfstring{$\autotune$}{autotune}}\label{sec: method}

Consider the negative penalized Gaussian log likelihood (Equation \ref{eq: Autotune lasso objective}). 
Note that any estimator that alternates between (M1) and (M2) below can lead to an automatic tuning procedure for Lasso with a data-driven sequence  $\lambda_n = \lambda_0 \, \widehat{\sigma}^2_n$, for $n \ge 1$.

    \begin{enumerate}[label = (M\arabic*)]
        \item Given $\sigma^{-2} = \what\sigma^{-2}$, update $\widehat{\beta}$ by running Algorithm \ref{algo: CD for lasso} with $\lambda = \lambda_0 \widehat{\sigma}^2$, \label{item: beta update step}
        \item Given $\beta = \what\beta$, update $\what\sigma^{-2}$ using the outputs of  Algorithm \ref{algo: CD for lasso}. 
        \label{item: sigma update step}
    \end{enumerate}

Scaled Lasso \cite{sun2010comments} falls in this general framework, with $\widehat{\sigma}^2$ in \cref{item: sigma update step} being estimated by the conditional maximum likelihood estimator $ \|\text{Y} - \text{X} \widehat{\beta}\|^2/n$. Next, we describe our estimator $\widehat{\sigma}^2$ within this framework.

%

\paragraph*{A basic  $\autotune$ strategy} 

If a small number ($s \ll p$) of predictors contribute the most towards predicting Y, we expect the PR of some $s$ predictors to be somewhat larger than the standard deviation of the other $(p-s)$ ones. So we can rank the $\ell_2$-norm of these $p$ PR ($\|r^{(k)}\|_2$, $k = 1, \ldots, p$) at the end of Algorithm \ref{algo: CD for lasso} from largest to smallest, and use this rank list to sequentially add predictors to build a nested sequence of linear models using F-tests at the significance level $\alpha$ ({see Background section and Appendix \ref{sec: appen background}} for details). The MSE of this final linear model is  the new estimator of $\widehat{\sigma}^2$ in (M2).

Since for conducting these F-tests, we only require the RSS and not the individual regression coefficients from these linear models, we can reduce the computation substantially by performing a single Gram-Schmidt (GS) orthogonalization of $\braces{\text{X}_{\cdot,1}, \text{X}_{\cdot,2}, \dots}$. Without loss of generality, let the rank list of predictors be $\{1, 2, \ldots, p\}$. Then we define $\braces{u_{\cdot,1}, \dots, u_{\cdot,k - 1}}$ to be the set of first $k - 1$ predictors in the rank list orthogonalized via GS, and $\eps^{(k-1)}$ to be the vector of residuals in the linear model $\text{ Y} \sim u_{\cdot,1} + u_{\cdot,2} + \dots + u_{\cdot,k - 1}$. Formally, $u_{\cdot,k} = \text{X}_{\cdot,k} - \sum_{i = 1}^{k-1}\frac{\langle \text{X}_{\cdot,k},u_{\cdot,i} \rangle}{\langle u_{\cdot,i},u_{\cdot,i} \rangle}u_{\cdot,i}$ and $\eps^{(k)} = \eps^{(k-1)} - \frac{\inner{\eps^{(k-1)}}{u_{\cdot,k}}}{\inner{u_{\cdot, k}}{u_{\cdot,k}}}u_{\cdot,k}.$

Note that $\eps^{(k)}$ is essentially the vector of residuals from the linear model $\mc{M}_{k}$ of Y on $\braces{\text{X}_{\cdot,1}, \text{X}_{\cdot,2}, \dots, \text{X}_{\cdot,k}}$. 
Then, we can perform F-test of $\mc{M}_{k -1}$ against $\mc{M}_{k}$ (the linear model we get after adding $u_{\cdot,k}$ to $\mc{M}_{k - 1}$) by using the formula in \cref{eq: seq f test} with $\mrm{RSS}(\mc{M}_k) = \twonorm{\eps^{(k)}}^2$ and $\mrm{RSS}(\mc{M}_{k - 1}) = \twonorm{\eps^{(k - 1)}}^2$.
A complete description of this $\sigma$ update is given in Algorithm \ref{algo:sigma_update}. So, this basic version of $\autotune$ is effectively using Algorithms \ref{algo: CD for lasso} and \ref{algo:sigma_update} alternately to perform steps \cref{item: beta update step} and \cref{item: sigma update step} above.

\paragraph*{Stopping CD early} We observed that $\what\sigma^2$, which estimates a scalar nuisance parameter, converged way faster as compared to the high-dimensional $\what\beta$ in Algorithm \ref{algo: CD for lasso} (demonstrated in Figure \ref{fig: partial residuals}). So we set an additional criterion for flagging the quick convergence of $\what\sigma^2$, and $\autotune$ stops updating $\what\sigma^2$ at the iteration when the estimated support set is a subset of that in the previous iteration. Then, fixing the final $\what\sigma^2$ (hence fixing $\lambda = \lambda_0\what\sigma^2$), $\autotune$ keeps on repeating \cref{item: beta update step} till the $\what\beta$'s converge. This final version is described in Algorithm \ref{algo:Autotune_lasso}.

\begin{algorithm}[h]
\caption{Sigma Update\ --\ Updates the estimate of noise variance $\what\sigma^2$} 
  \label{algo:sigma_update}
  \begin{algorithmic}
  \STATE {\bfseries Input:} Y, X, $\alpha$, $r$, $\what\beta$
  \small
    {
    \STATE
    Calculate partial residuals $\partres[j] \gets r + \text{X}_{\cdot,j}\what\beta_j$ $\forall j\in \seqp$ 
    \STATE
    $\bigg\{\twonorm{\partres[\pi(i)]}\bigg\}_{i=1}^n \gets \mrm{Sort}\bigg(\twonorm{\partres[1]}, \ldots, \twonorm{\partres[p]}\bigg)$ in decreasing order\\[2pt]
    \STATE
    $\Active\gets \braces{\pi(1),\ldots,\pi(p)}, ~~\ytemp \gets \text{Y}$\\[2pt]
    \STATE
    $\support\gets \phi$, $~~u \gets \text{X}[\hspace{0.1cm},\Active]$ \\[2pt]
    \FOR{$i = 1,\ldots,p$}
        \STATE $j\gets \Active[i]$\\[2pt]
        \IF{$i>1$}
            \STATE $u_{\cdot,i} = \text{X}_{\cdot,j} - \sum_{k = 1}^{i-1}\frac{\langle \text{X}_{\cdot,j},u_{\cdot,k} \rangle}{\langle u_{\cdot,k},u_{\cdot,k} \rangle}u_{\cdot,k} $ 
        \ENDIF
            
        \STATE $\mc{M}_i: \ytemp \sim  u_{\cdot,i}$, leading to the prediction\\ $\what{\text{Y}} \gets \frac{\langle\ytemp, u_{\cdot,i}\rangle}{\langle u_{\cdot,i}, u_{\cdot,i}\rangle}u_{\cdot,i}$, and RSS$(\mc{M}_i) = \twonorm{\ytemp - \what{\text{Y}}}^2$ \\[2pt]
        \STATE Perform Sequential F-test \cref{eq: seq f test} using F-test statistic = $\frac{\parenth{\twonorm{\ytemp}^2 - \mrm{RSS}(\mc{M}_i)}}{\mrm{RSS}(\mc{M}_i)/ (n-i)}$, and cutoff = $F_{\alpha;1, n - i}$
        \\[2pt]
        \STATE \textbf{if} $\mrm{F}$-$\mrm{test}$ $\mrm{statistic} \leq \mrm{cutoff}$ \textbf{then} break the for loop
        \STATE $\support \gets \support \cup \{j\}$\\[2pt]
        \STATE $\ytemp\gets\ytemp - \what{\text{Y}}$
    \ENDFOR
    \STATE $\what\sigma^2\gets\frac{1}{n - \abss{\support}}\sumn \parenth{\text{y}_i^{\mathsf{(temp)}}}^2$
    }
\STATE \textbf{Return:} $\what\sigma^2$, 
$\support$, $\Active$
  \end{algorithmic}
\end{algorithm}

\begin{algorithm}[h]
\caption{Autotune Lasso\ --\ Return estimates of regression coefficients $\beta$ } 
  \label{algo:Autotune_lasso}
  \begin{algorithmic}
  \STATE \textbf{Input:} Y, Normalized predictors \text{X} (i.e. $\twonorm{\text{X}_{\cdot,j}}^2=n$), Sequential F-test significance level $\alpha $
  \small
  {\STATE
    Initialize $\what\beta \gets 0$, $r \gets \text{Y}$, $\widehat{\sigma}^2\gets \mathrm{Var}(\text{Y})$, $\lambda_0\gets\frac{1}{\mathrm{Var}(\text{Y})}\|\frac{\text{X}^{\top}\text{Y}}{2n}\|_{\infty}$, $\Active\gets \seqp$, $\support\gets\phi$, sigma.update.flag $\gets\mathsf{TRUE}$ 
    \WHILE{$\mrm{error}\geq\msf{error.tolerance}$} 
        {
        \STATE Set $\lambda = \lambda_0\what\sigma^2$\\
        \STATE Set $\supportold\gets\support$, $\betaold \gets \what\beta$\\[2pt]
        \FOR{$i = 1,\ldots,p$}
        
        \STATE $j\gets \Active[i]$\\[2pt]
        \STATE $\what\beta_j \gets \mathsf{Soft.Threshold}_{\lambda}\parenth{\frac{1}{n}\text{X}_{\cdot,j}^{\top}r + \what\beta_j}$\\[2pt]
        \STATE $r \gets r + \text{X}_{\cdot,j}\parenth{\betaold_j-\what\beta_j}$
        
        \ENDFOR
        
    \IF{$\mrm{sigma.update.flag} == \mathsf{TRUE}$} 
        
        \STATE $\what\sigma^2, \support,\Active\gets$ \\{\normalsize\SUlink}\,$\parenth{\text{Y, X}, \alpha, r, \what\beta}$\\
        \IF{$\support\subseteq\supportold$} 
        \STATE sigma.update.flag $\gets\mathsf{FALSE}$
        \ENDIF
        \ENDIF
        \STATE $\mrm{error} \gets \|\what{\beta}- \betaold\|_1 \big/ \| \betaold\|_1$
        }
    \ENDWHILE
    }
    \STATE \textbf{Return:} $\what\beta$,  $\what\sigma^2$, $\lambda$
  \end{algorithmic}
\end{algorithm}

\paragraph*{Warm starts and initialization of $\lambda_0$} We use a warm start strategy, as is common in \texttt{glmnet}, which helps CD converge faster.

Asymptotic theory suggests  $\lambda \asymp \left\|\text{X}^\top \varepsilon /n \right\|_\infty \asymp \sigma \sqrt{\log p / n}$ for Lasso. Adjusting for the loss function in $\autotune$, $\lambda_0 \asymp \left\|\text{X}^\top \varepsilon /n \right\|_\infty / \text{Var}(\varepsilon) \asymp \sigma \sqrt{\log p / n}/\sigma^2$ is a natural choice. Since $\varepsilon$ is unknown in practice, a reasonable starting point for our iterative algorithm is to use $Y$ instead of $\varepsilon$ in the above expression and set $\lambda_0 = \tau \infnorm{\text{X}^{\top} \text{Y}/n}/Var(Y)$ for some constant $\tau$. We set $\tau = 1/2$ in $\autotune$ and report the sensitivity of $\autotune$ to $\tau$ in Appendix \ref{app: sensitivity to tau plots}.


\paragraph*{Speeding up predictor ranking} Switching from  $\ell_2$ norm of PR, which requires $p$ multiplications, to $\ell_1$ norm of PR helps improve speed without compromising accuracy. In other words, important predictors are still ranked high in terms of both norms but we can reduce the number of expensive multiplication operations.

\paragraph*{Active set selection} Following \cite{10.1111/j.1467-9868.2011.01004.x}, we designed an active set version of $\autotune$, \texttt{autotune\_lasso(
  x,
  y,
  active = TRUE)} for further speedup (pseudocode in \\ \Activelassolink, Appendix \ref{sec: app methods}).

\begin{figure}[tbp]
    \centering
    \begin{subfigure}[t]{0.4\textwidth}
        \centering
        {\fontsize{6}{6}\selectfont \textbf{CD Iteration 1 out of 30}}
        \includegraphics[page = 1, width=\textwidth]{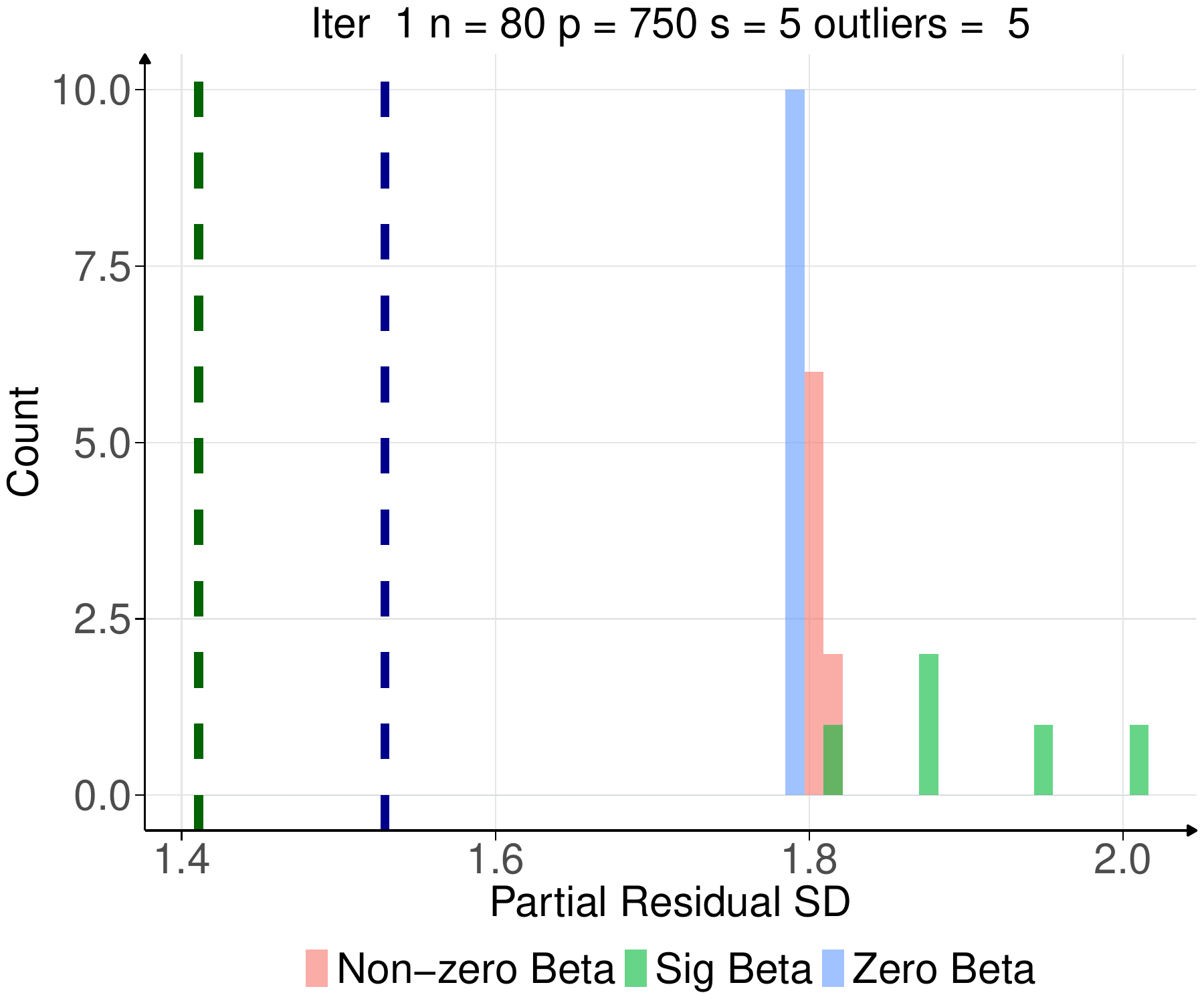}
    \label{fig: partial residuals iter 1}
    \end{subfigure}
    \begin{subfigure}[t]{0.4\textwidth}
        \centering
        {\fontsize{6}{6}\selectfont \textbf{CD Iteration 2 out of 30}}
        \includegraphics[page = 2, width=\textwidth]{Pics/Partial_Res_SD_plots/Par_SD_Repli_2_plot_rho_0_type_2.pdf}
        \label{fig: partial residuals iter 2}
    \end{subfigure}
    \caption{Standard Deviation (SD) of PR of different predictors after the first 2 iterations of CD in $\autotune$ Lasso in a simulated high-dimensional data set ($n =80, p = 750$). It  required 30 iterations of CD for $\beta$ to converge. However, PR of support variables (``Sig Beta'', green bars) tend to be much larger than others in as early as the first 2 iterations, while the $\what\beta$ still contain many false positives (``Non-zero Beta'', pink bars). These SD values are quite different from the true noise scale (SD($\varepsilon$), green dashed line) and its current estimate (SD($r$), blue dashed line). This suggests a good ranking of predictors can be obtained from early iterations of CD even when estimates of $\sigma$ or $\beta$ are not very accurate.
     }
  \label{fig: partial residuals}
\end{figure}

\subsection{Using \texorpdfstring{$\autotune$}{autotune} for VAR Lasso}
\label{subsec: autotune VAR}

Estimating a VAR($d$) model on a $p$-dimensional time series under $\ell_1$ penalty is equivalent to solving $p$ separate Lasso regression problems (\cref{eq: Var columnwise regg}, Appendix \ref{sec: appen background}). For a $p$-dimensional stationary time series $\braces{Z^t}_{t \geq 1} = \braces{\parenth{Z^t_1,\dots,Z^t_p}}_{t \geq 1}$, eqn. \cref{eq: VAR regression design} shows how we construct the response and predictor matrix Y and X from $\braces{Z^t}_{t \geq 1}$ for estimating $\Phi$, the matrix formed by stacking $d$ VAR transition matrices. VAR Lasso with $\autotune$  performs $p$ separate $\autotune$ Lasso regression of each column of Y on X to yield $\what\Phi$. 
\begin{equation}
    \label{eq: autotune var columnwise regg}\what{\Phi}_{\cdot,i}=\underset{\beta \in \mathbb{R}^{dp}}{\operatorname{argmin}} \frac{1}{2n}\left\|\text{Y}_{\cdot, i}-\text{X} \beta\right\|^2 + \lambda_i\onenorm{\beta}, ~ \text { for } i=1, \ldots, p.
\end{equation}
Separate regularization for each Lasso regression allows $\autotune$ to adapt to heterogeneity in noise variances across different time series components.



\subsection{An illustration of PR behavior along CD iterations}
\label{app: partial res in action}
We empirically illustrate the behavior of PR of different predictors across different iterations of CD, since this enables us to stop $\autotune$ early. With a slight abuse of terminology, we use ``standard deviation (SD)'' instead of the $\ell_2$-norms of PR $r^{(k)}$. We simulate from the following linear model
\begin{align}
    \text{y}_i = \text{X}_{i,\cdot}^{\top}\beta_0+\epsilon_i,&\quad \epsilon_i \sim\mathcal{N}(0,\sigma^2),\qtext{for all} i\in \braces{1,\dots,n}\label{eq: Linear DGP with gaussian errors}.
\end{align}
with $n = 80$ observations, $p = 750$ covariates, and $\beta_0 = (1,1,1,1,1,0,0,\dots)\in \real^p$, i.e. each y$_i$ depend on only the first 5 covariates. Each $\text{X}_{i,\cdot}$ is independently generated from $p$-dimensional standard normal distribution. Fixing $\text{X}_{i,\cdot}$, we generate  y$_i$ by drawing errors $\epsilon_i$ from $\mathcal{N}(0,\sigma^2)$ with $\sigma^2$ chosen so that signal-to-noise ratio is 1.

In Figure \ref{fig: partial residuals}, we show histograms of $SD(r^{(k)})$, $k=1, \ldots, p$,  at the end of the first two iterations of CD in Algorithm \ref{algo:Autotune_lasso}. The predictors are categorized into three groups. Significant predictors i.e. predictors whose true coefficients are non-zero, are depicted in green. Spurious predictors whose estimated coefficients are zero and nonzero are shown in blue and maroon bars respectively.

As expected, since the predictors $\text{X}_{\cdot, j}$ are uncorrelated, $SD(r^{(k)})$ of significant predictors separate from the those of insignificant predictors right from the first iteration. In the second iteration, the separation between significant and insignificant predictors becomes wider.

These plots emphasize two interesting benefits of tracking SD of PR instead of simply tracking the support set of $\what\beta$. First, in both iterations, there are a large number of false positives in $\what\beta$ making it difficult to separate relevant predictors. Second, at this $\lambda$, Lasso needed \textit{30 iterations of CD to converge to a $\what\beta$}, whereas the predictor ranking needed for estimating $\sigma^2$ stabilized only after the first 2 CD iterations, avoiding the need to compute the high-dimensional vector  $\what\beta(\lambda)$. We observe this phenomenon in most of the simulation setups we consider. 

In general, support set recovery in Lasso is known to be hard, especially in presence of correlated predictors \cite{aos_support_set_recovery} where $\autotune$ can rank some of the spurious  predictors at the top. But we only require good prediction accuracy (not support set recovery) for good $\sigma$ estimation, and even if the spurious predictors which are strongly correlated with significant predictors are ranked high by $\autotune$, we find that $\what\sigma$ of $\autotune$ remains  reasonably accurate. We empirically validate this in Section \ref{sec:app-theory}.

\subsection{A statistical explanation of autotune}
\label{sec:app-theory}

We offer an intuitive and statistical explanation of $\autotune$ here. At a high level, when we reformulate the problem of searching for $\lambda$ to the problem of learning a statistical nuisance parameter $\sigma$, we can leverage the following insight from theoretical statistics: in high-dimension ($n \ll p$), the large vector $\beta \in \mathbb{R}^p$ is very hard to find, but the scalar $\sigma \in \mathbb{R}$ does not have to be. Since $\widehat{\sigma}^2 = Var(\widehat{\epsilon}) = Var(Y - \widehat{X\beta})$, to find $\sigma$ one only needs good prediction $\widehat{X\beta} = \widehat{Y} \in \mathbb{R}^n$, which can be much simpler than finding a good estimator of $\widehat{\beta} \in \mathbb{R}^p$ \cite{yu2019estimating}. 

Consider a simple example with two predictors $X_1$ and $X_2$ which are nearly perfectly correlated, with $\beta_1 \neq 0$ and $\beta_2 = 0$. It would be very hard to find the right regression coefficients. But because the predictors are highly correlated, one can build nearly equally good predictive model $\widehat{Y}$ using either $X_1$ or $X_2$. And that is all we need to find a good $\sigma^2$, and hence a good $\lambda$.

All existing approaches such as cross-validation (CV), scaled Lasso, and bilevel optimization (BLO) are at odds with this insight. They all solve (at least partially) the hard problem of finding $\beta$ over and over, before finding $\lambda$. CV uses grid search over $\lambda$, and optimizes a loss over $\beta$ at each value of $\lambda$. Popular BLO methods such as \texttt{sparseho} \cite{bertrand2020implicit} in turn replaces the grid search over $\lambda$ by a gradient descent to speed up, but iteratively uses current values of $\beta$ to compute that gradient.

In contrast, we can show that instead of $\beta$ itself, $\autotune$ searches over an easier target: projection matrix on the column space of $X_S$, where $S = support(\beta)$. Estimating $S$ is closely related to the top few predictors of \texttt{Predictor.Ranking} in Algorithm \ref{algo:sigma_update} which are selected by the sequential F-test. 
In the example above, even if we select the highly correlated spurious predictor $X_2$ instead of the true predictor $X_1$ in our F-test, the estimator of $\sigma^2$ is not affected very much since the column spaces of $X_1$ and $X_2$ are very similar by virtue of their correlation.

Formally, consider the sparse linear model $Y = X \beta + \varepsilon$, where $S = support(\beta)$ with $|S| = s$. Then the oracle predictions are given by  $\widehat{Y} = X_S \widehat{\beta}_{OLS} = P_S Y$, where $P_S = X_S (X_S^\top X_S)^{-1} X_S^\top$ is the projection matrix on the column space of $X_S$. For fixed design, $P_S$ can be viewed as a transformation of $\beta$ and a reparameterization which contains just enough information to construct a good prediction $\widehat{Y}$. As shown in the example above, it may be possible to estimate $P_S$ well without estimating $S$ accurately, which is known to be a difficult problem   \cite{vandeGeer2009OnThe}.

Next, we show that $\autotune$ without early stopping can be viewed as an alternate search over ($P_S$, $\sigma$), instead of an alternate search over ($\beta$, $\sigma$) as in Scaled Lasso. To that end, note that 
\begin{eqnarray*}
\|Y - X \beta \|^2 &=& \|(Y - P_S Y) + (P_S Y - X_S \beta_S) \|^2
\\
&=& \| (I-P_S) Y + P_S (Y - X_S \beta_S) \|^2  
\\
&=& \|(I-P_S) Y \|^2 + \|P_S \varepsilon \|^2. 
\end{eqnarray*}
So, we can re-write the negative penalized Gaussian log-likelihood over the new parameters $(P_S,\sigma^2)$ for a fixed $\lambda_0$ as follows:
\begin{align}
\label{eq: obj over P_S and sigma} \frac{n}{2}\log(\sigma^2) +\frac{\|Y-P_S Y\|_2^2}{2\sigma^2} +\frac{\|P_S\varepsilon\|_2^2}{2\sigma^2} + \lambda_0 \| \beta \|_1 
\end{align}
 and design alternating estimation algorithms of the form

\begin{enumerate}[label = (T\Roman*)]
    \item given $P_S$, estimate $\sigma^2$; \label{item: m1}
    \item given $\sigma^2$, estimate $P_S$. \label{item: m2}
\end{enumerate} 


For solving \cref{item: m1}, assuming $|S| < n$, our $\autotune$ estimator $\widehat{\sigma}^2:= \|(I-P_S)Y\|^2/(n-|S|)$ is a natural choice as the MSE of ordinary least squares (OLS) regression. Interestingly, it can also be viewed as the solution to an Expectation-Maximization (EM) style iterative algorithm which treats $X$ as observable and $\varepsilon$ as a latent random vector.

\begin{lemma} 
Consider the problem \cref{item: m1}. Given $P_S$ with $|S| < n$, there exists an EM-style iterative algorithm to solve for $\sigma^2$ in \eqref{eq: obj over P_S and sigma} that converges to 
\begin{align}
    \what\sigma^2 = \frac{\|Y-P_S Y\|_2^2}{n - |S|}.
\end{align}
\end{lemma}
   
\begin{proof} 
We drop the last term in \eqref{eq: obj over P_S and sigma} as it does not involve $\sigma^2$. For $t \ge 0$, let $\sigma^2_{(t)}$ be the $t^{th}$ iterate of our EM-style algorithm. Since $P_S$ is idempotent and  $\mathbb{E}_{\varepsilon|X \sim (0, \sigma^2_{(t)})}[\varepsilon^\top P_S \varepsilon] = s\sigma^2_{(t)}$, the E-step takes the form
\begin{align}
Q(\sigma^2 | \sigma^2_{(t)}) := -\frac{n}{2}\log(\sigma^2)
    -\frac{s\,\sigma^2_{(t)}}{2\sigma^2}
    -\frac{\|Y-P_S Y\|_2^2}{2\sigma^2},
\end{align}
and the M-step takes the form
\begin{align}
    {\sigma}^2_{(t+1)}& = \arg\max_{\sigma^2>0} Q(\sigma^2 | \sigma^2_{(t)}) \\
&= \frac{s\,\sigma^2_{(t)} + \|Y-P_S Y\|_2^2}{n} \\
&= \frac{\|Y-P_S Y\|_2^2}{n}\left(1 + \frac{s}{n}+ \dots + \frac{s^t}{n^t}\right) + \frac{s^{t + 1}}{n^{t+1}}{\sigma}^2_{(0)}.
\end{align}
In the limit $t \rightarrow \infty$, this iterative algorithm converges to
\begin{align}
    \lim_{t\to\infty} \sigma^2_{(t)} = \frac{\|Y-P_S Y\|_2^2}{n} \left(1- \frac{s}{n}\right)^{-1}, 
\end{align}
which matches the $\autotune$ estimator $\what\sigma^2$.
\end{proof}


\cref{item: m2} can be solved in many ways. The most natural estimator is $P_{\widehat{S}}$, using the support set $\widehat{S}$ of the current Lasso estimate $\widehat{\beta}$. However, it is known that $\widehat{S}$ tends to select many false positives. 
To avoid this, existing works such as the post-Lasso \cite{lee2016exact} recommends fitting an ordinary least squares (OLS) model of $Y$ on $X_{\widehat{S}}$, and select variables based on a significance level $\alpha$. Instead of building a single OLS on a potentially large set of predictors $\hat{S}$, a faster alternative is to  incrementally build a linear model by including predictors $X_j$ with large $|\what\beta_j|$ one at a time, use sequential F-tests at level $\alpha$ to stop, and use those predictors to construct a new $\widehat{S}$ and $P_{\widehat{S}}$.

Indeed, we can show that in \texttt{autotune} without early stopping, the predictor ranking obtained by ranking $\ell_2$-norms of PR is identical to the ranking obtained from large  Lasso coefficients as described above.


\begin{lemma}
    Let $\what \beta$ be the Lasso coefficients at $\lambda$, i.e. a minimizer of \cref{eq: original lasso} at $\lambda$. Let $r^{(j)}$ be the PR vector of the $j^{th}$ predictor. Then
    \begin{align}
        \|r^{(j)}\|^2 = \|r\|^2 + 2 n \lambda |\widehat{\beta}_j| + n|\widehat{\beta}_j|^2.
    \end{align}
    Hence, $\|r^{(j)}\|^2$ is a monotone increasing function of the magnitude of the LASSO coefficients $|\widehat{\beta}_j|$ and the \texttt{Predictor.Ranking} in $\autotune$ is identical to the one given by ranking $|\what\beta_j|$ in decreasing order.
\end{lemma}

\begin{proof}
    \begin{align}
        \|r^{(j)}\|_2^2 =& \twonorm{r + X_j\widehat{\beta}_j}^2
        =\twonorm{r}^2 + 2\inner{r}{X_j\widehat{\beta}_j} + \twonorm{X_j\widehat{\beta}_j}^2\\
        =& \twonorm{r}^2 + 2\widehat{\beta}_j\inner{r}{X_j} + \widehat{\beta}_j^2\twonorm{X_j}^2
        = \twonorm{r}^2 + 2\widehat{\beta}_j\inner{Y - X\widehat{\beta}}{X_j} + n|\widehat{\beta}_j|^2 \label{eq: Predictors are normalized}\\
        =& \twonorm{r}^2 + 2\widehat{\beta}_jn\lambda \mathrm{sign}(\what\beta_j) + n|\widehat{\beta}_j|^2 \label{eq: kkt condition of lasso residuals}\\
        =& \|r\|_2^2 + 2 n \lambda |\widehat{\beta}_j| + n|\widehat{\beta}_j|^2.
    \end{align}
Here, \cref{eq: Predictors are normalized} holds since the predictors are standardized, and \cref{eq: kkt condition of lasso residuals} follows from the KKT condition of Lasso.
\end{proof}



\section{Numerical experiments}
\label{sec: simulation}

We demonstrate $\autotune$ in three settings: 1) in high-dimensional regression, we compare its runtime and accuracy (prediction, variable selection and estimation) of $\what{\beta}$ against Scaled Lasso and CV Lasso, 2) in large VAR models, we compare its runtime and accuracy  (estimation and variable selection) of $\what{A}$ against TSCV and \cite{kock2025data}, 3) we compare the $\widehat{\sigma}^2$ of \texttt{autotune} against Scaled Lasso, Organic and Natural Lasso, and describe new diagnostic plots to check model sparsity. All the experiments were run in Linux (version 5.19.17), and more implementation details are in Appendix \ref{subsec: benchmark R packages}.

In the interest of space, we defer the comparison of $\autotune$ with Bilevel Optimization and $\texttt{skglm}$, a state-of-the-art python implementation of Lasso to Appendix \ref{app: comparison vs BLO skglm}.


\begin{figure}[btp]
    \centering
    \begin{tabular}{cc}
        \multicolumn{2}{c}{
        \textbf{High Dimensional setup:} 
        $n = 80, \hspace{0.1cm} p=750, \hspace{0.1cm} s=5$,
        $\rho = 0.35$, 
        \textbf{[A]:} Beta-type 1 and 
        \textbf{[B]:} Beta-type 2
        }\\

        {\fontsize{6}{7}\selectfont \textbf{[A]}} &
        {\fontsize{6}{7}\selectfont \textbf{[B]}} \\[0mm]

        \includegraphics[
            trim=0in 0in 0in 0in, clip,
            width=0.49\textwidth
        ]{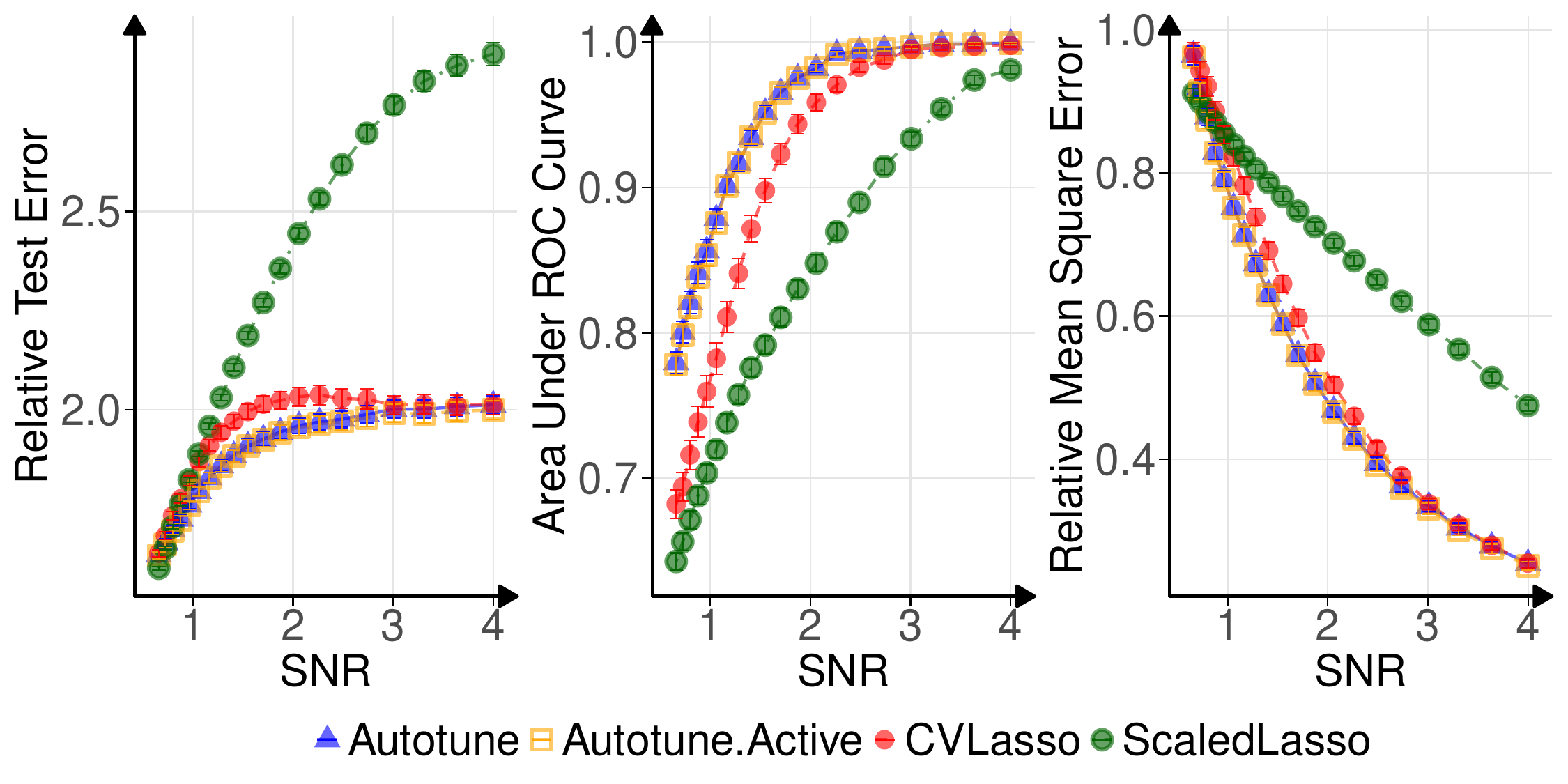}
        &
        \includegraphics[
            trim=0in 0in 0in 0in, clip,
            width=0.49\textwidth
        ]{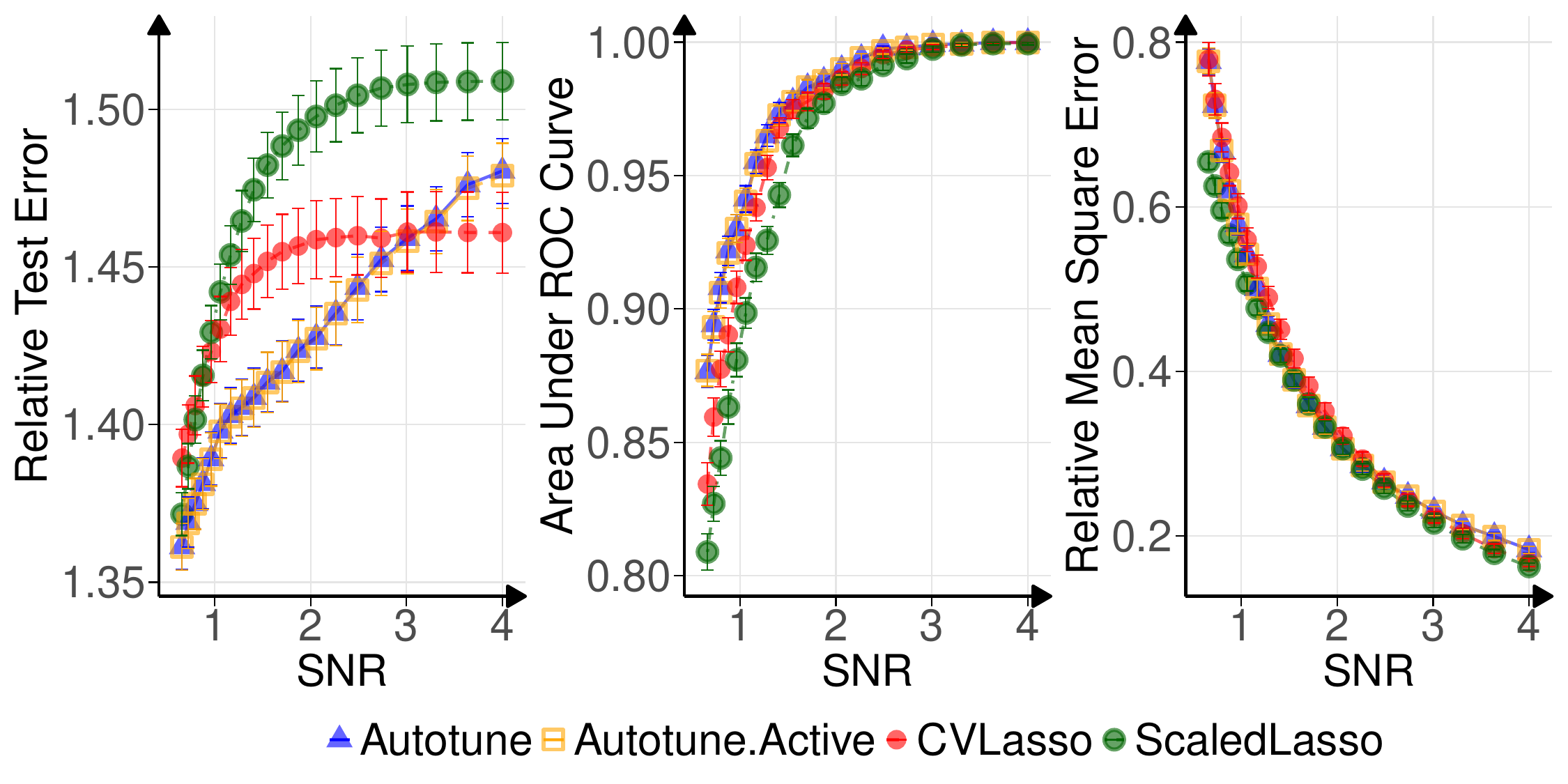}
    \end{tabular}
    \\[2mm]

    \caption{RMSE, RTE, and AUROC (with 1SE bars) of $\autotune$ and benchmark methods for Lasso regression, plotted against SNR.}
    \label{fig: high dim accuracy plot}
\end{figure}

\subsection{High-dimensional regression}
\label{subsec: regg setup sim comparison}
We closely follow the simulation setup of \cite{hastie2020best}, where the data generating process (DGP) is governed by 6 parameters: $n$, $p$, $s$ (level of sparsity), $\rho$ (correlation between the predictors), $\snr$ (signal-to-noise ratio), and Beta-type (pattern of sparsity). We use the linear model Y $=\text{X}\beta + \epsilon$, where $\text{Y}\in \real^n$, $\text{X}\in \real^{n\times p}$, and  $\epsilon\sim\Gsn_n\parenth{0, \sigma^2 \mrm{I}}$.
In this section, we focus on two types of sparse $\beta\in \real^p$. In Beta-type 1, $\beta$ has $s$ entries equal to 1, placed equidistantly between indices 1 and $p$, with all other entries set to 0. In Beta-type 2, the first $s$ entries of $\beta$ are equal to 1, and the remaining entries are 0. We then generate the predictors X$_{i, \cdot}\sim \Gsn_p\parenth{0, \Sigma}$ for all $i\in \braces{1,\dots,n}$, where $\Sigma_{kl} = \rho^{|k - l|}$, and $\eps_i$ has variance $\sigma^2 = \beta^\top \Sigma \beta/\snr$.
We run $\autotune$ and its benchmarks on the data (Y,X), record relevant performance metrics, and  repeat these steps (X fixed, $\epsilon$ random) to report the average and SE of these  metrics. 
We describe the results for three more sparse Beta-types from \cite{hastie2020best} in Appendix \ref{subsec: All Beta-types}.

Let $\what\beta = \parenth{\what\beta_1,\dots, \what\beta_p}$ be the estimated coefficients of an algorithm, and $\text{Y}_{\mrm{test}}|\text{X}_{\mrm{test}} \sim \Gsn_n\parenth{\text{X}_{\mrm{test}}\beta, \sigma^2\mathrm{I}}$ where the rows of $\text{X}_{\mrm{test}}$ are independently drawn from $ \Gsn_p(0, \Sigma)$. For prediction error, we use the  Relative Test Error, $\mrm{RTE} = \frac{\parenth{\what\beta - \beta}^{\top}\Sigma\parenth{\what\beta - \beta} + \sigma^2}{\sigma^2} = \frac{\E\twonorm{\text{Y}_{\mrm{test}} - \text{X}_{\mrm{test}}\what\beta}^2}{\E\twonorm{\text{Y}_{\mrm{test}} - \text{X}_{\mrm{test}}\beta}^2}$ which measures the ratio of expected prediction error using estimates of the algorithm $\what\beta$ to that of the oracle, which knows the true $\beta$. For estimation error, we use the Relative Mean Square Error, $\mrm{RMSE} = {\twonorm{\what\beta - \beta}^2}/{\twonorm{\beta}^2} 
$. 
Note that if $\what\beta = 0$, then RMSE $=1$ and RTE $= \mrm{SNR} + 1$, whereas for perfect fit, i.e. $\what\beta = \beta$, RMSE $=0$ and RTE $= 1$. 
So, a good frame of reference for RMSE is $(0, 1)$, while RTE varies in the range $(1, 1+\snr)$. We quantify how well an algorithm recovers the sparsity pattern of $\beta$ by measuring the area under receiver operating characteristic (AUROC) curve.


\paragraph*{Dimension specifications} We consider three different dimension regimes: low ($p=50$, $n=80$, $s = 5$), moderate ($p=150$, $n=80$, $s = 5$), and high ($p=750$, $n=80$, $s = 5$).

\begin{wrapfigure}{r}{0.58\textwidth}
    \centering
    \begin{subfigure}[t]{0.48\linewidth}
        \centering
        \includegraphics[page=1,trim=0in 0in 0in 0.4in,clip,width=\linewidth]
        {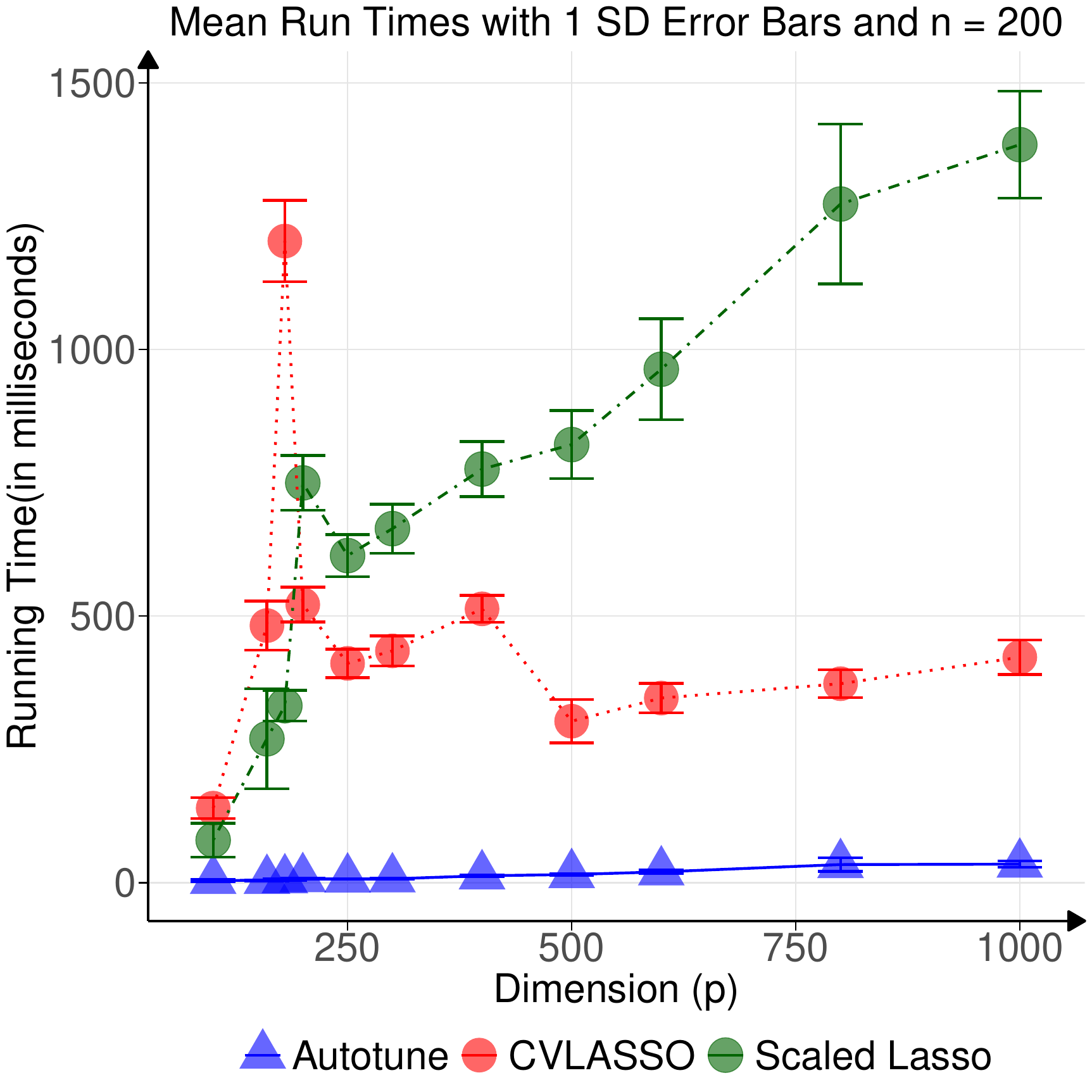}
        \label{fig: runtime_b}
    \end{subfigure}
    \hfill
    \begin{subfigure}[t]{0.48\linewidth}
        \centering
        \includegraphics[page=1,trim=0in 0in 0in 0.4in,clip,width=\linewidth]
        {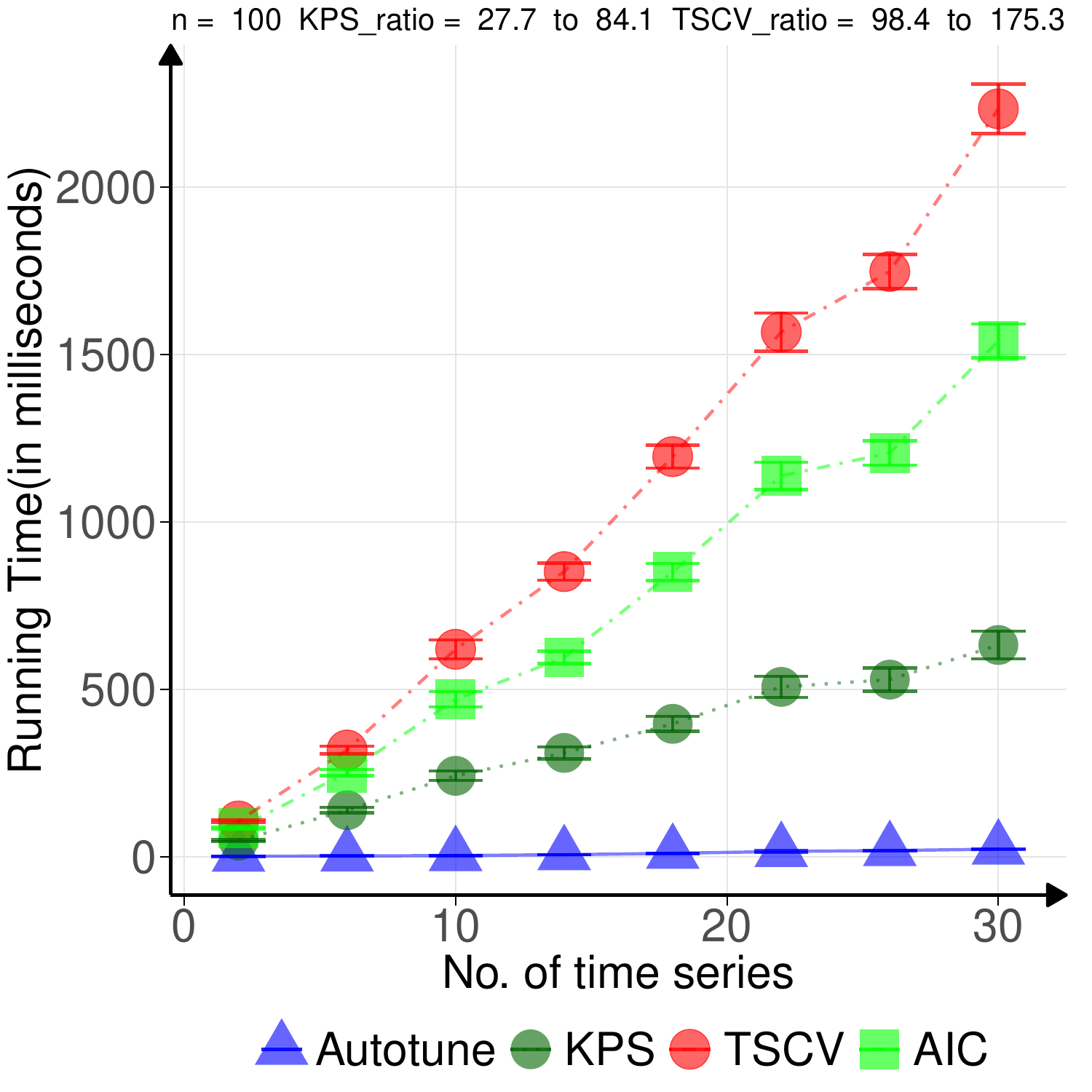}
        \label{fig: ts runtime3}
    \end{subfigure}
    \caption{Runtimes (with 1SE bars) of $\autotune$ and benchmark
    methods in \textbf{[left]} Lasso regression \&
    \textbf{[right]} block diagonal VAR,
    \textit{showing up to 210$\times$ speedup.}}
    \label{fig: runtime plots}
\end{wrapfigure}
\paragraph*{Estimation, prediction and variable selection accuracy} We plot the RMSE, RTE, and AUROC of 
$\autotune$, CV Lasso, and Scaled Lasso against an SNR grid (0.66 to 4 in log scale) in high-dimension setup (details of grid choice in Appendix \ref{subsec: interpretation of SNR appen} and \ref{subsec: why this range of SNR}).
For brevity, we only present a slice of the numerical comparisons in Figure \ref{fig: high dim accuracy plot} and Figure \ref{fig: moderate dim accuracy plot} where $\rho =0.35$. 
A full suite of results, qualitatively similar, are in Appendix \ref{subsec: all accuracy plots}. There  we consider the dimension regimes, the sparsity patterns (5 Beta-types) and  $\rho\in \braces{0, 0.35, 0.7}$ (same as in \cite{hastie2020best}).  

In Figure \ref{fig: high dim accuracy plot} and \ref{fig: moderate dim accuracy plot} (high and moderate dimensional setup respectively), we see that $\autotune$ has significantly smaller prediction error than its benchmarks across most of the setups considered.  It also shows uniformly higher AUROC across all setups. Lastly, $\autotune$ either matches or improves upon the estimation performance of the benchmarks in terms of RMSE. 
We also observe that the performance gains of $\autotune$ across all metrics become wider in  low to moderate SNR and as the dimensionality of the problem increases (compare Figure \ref{fig: high dim accuracy plot} and \ref{fig: moderate dim accuracy plot}).

\begin{wraptable}{R}{0.52\textwidth}
\centering
\scriptsize
\resizebox{\linewidth}{!}{%
\begin{tabular}{lrrr}
\toprule
Setting & Tuning $\lambda$ & Finding $\solution$ & Tuning runtime \% \\
\midrule
Beta-type 1 & & & \\
$\rho=0$    & 2.79 (0.2) & 14.90 (0.6) & 15.77\% \\
$\rho=0.35$ & 3.89 (0.2) & 27.20 (1.0) & 12.51\% \\
$\rho=0.7$  & 4.01 (0.2) & 29.10 (1.1) & 12.11\% \\
Beta-type 2 & & & \\
$\rho=0$    & 2.56 (0.1) & 21.60 (1.0) & 10.60\% \\
$\rho=0.35$ & 4.30 (0.2) & 21.00 (0.9) & 17.00\% \\
$\rho=0.7$  & 3.64 (0.2) & 27.00 (0.8) & 11.88\% \\
\bottomrule
\end{tabular}%
}
\caption{$\autotune$ spends only 10-17\% of the total runtime for
finding an ideal $\lambda$. Rest of the time is used to compute
$\solution$ at that $\lambda$.}
\label{tab: finding lambda vs finding beta}
\end{wraptable}


Finally, we notice that where a benchmark method shows significant improvement in estimation or prediction accuracy over $\autotune$, the SNR is very high and both methods there have an AUROC close to 1 (see Beta-type 2 in Figure \ref{fig: high dim accuracy plot}). 
Additionally, introducing correlation among true predictors in Beta-type 2 makes Scaled Lasso competitive to $\autotune$ in terms of estimation error, but still inferior in terms of variable selection and prediction.

\begin{figure*}[!tb]
    \centering
    \begin{tabular}{c|c}
        \textbf{2x2 Block VAR(1) setup:} SNR = 2.5 & \textbf{ diagonal VAR(1) setup:} SNR = 2.5\\
        Time Series length 100 & Time Series length 200 \\
         \includegraphics[trim=0in 0in 0in 0in, clip, width = 0.485\textwidth]{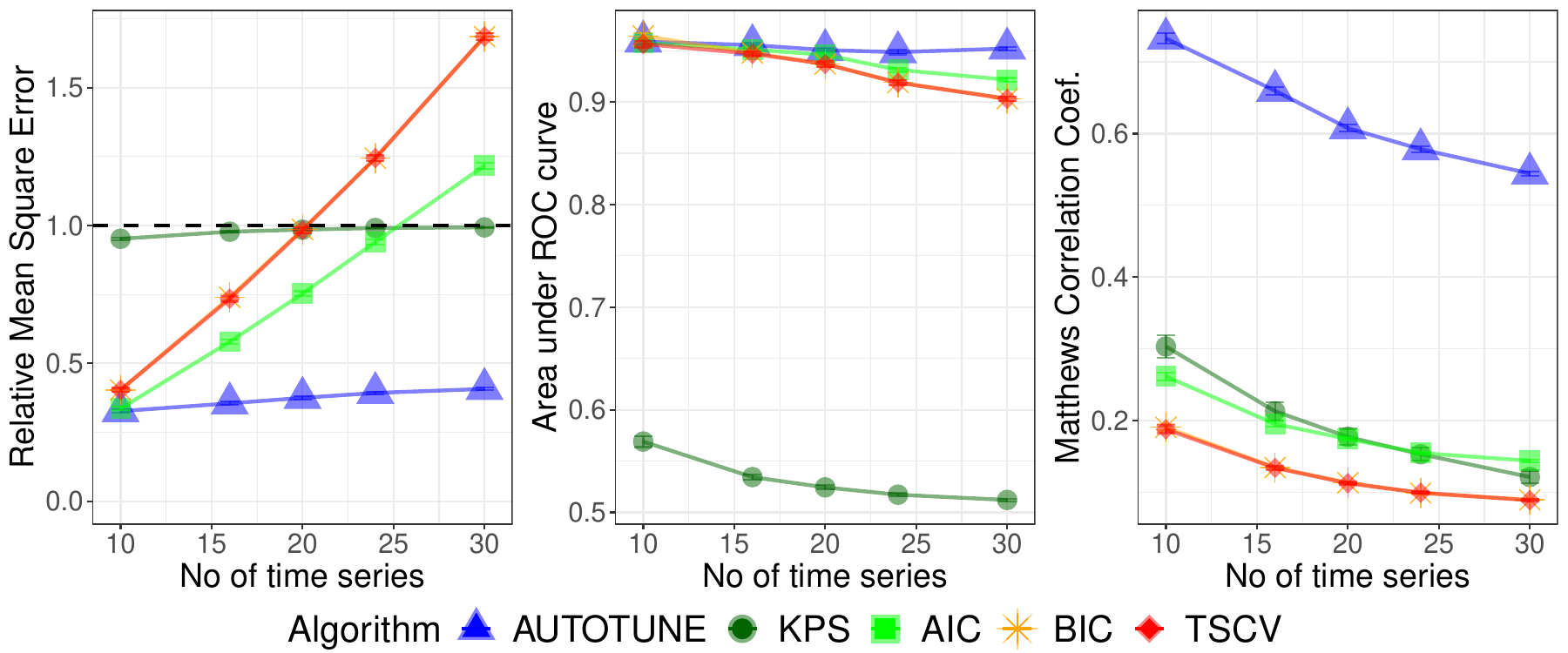} &
        \includegraphics[trim=0in 0in 0in 0in, clip, width = 0.485\textwidth]{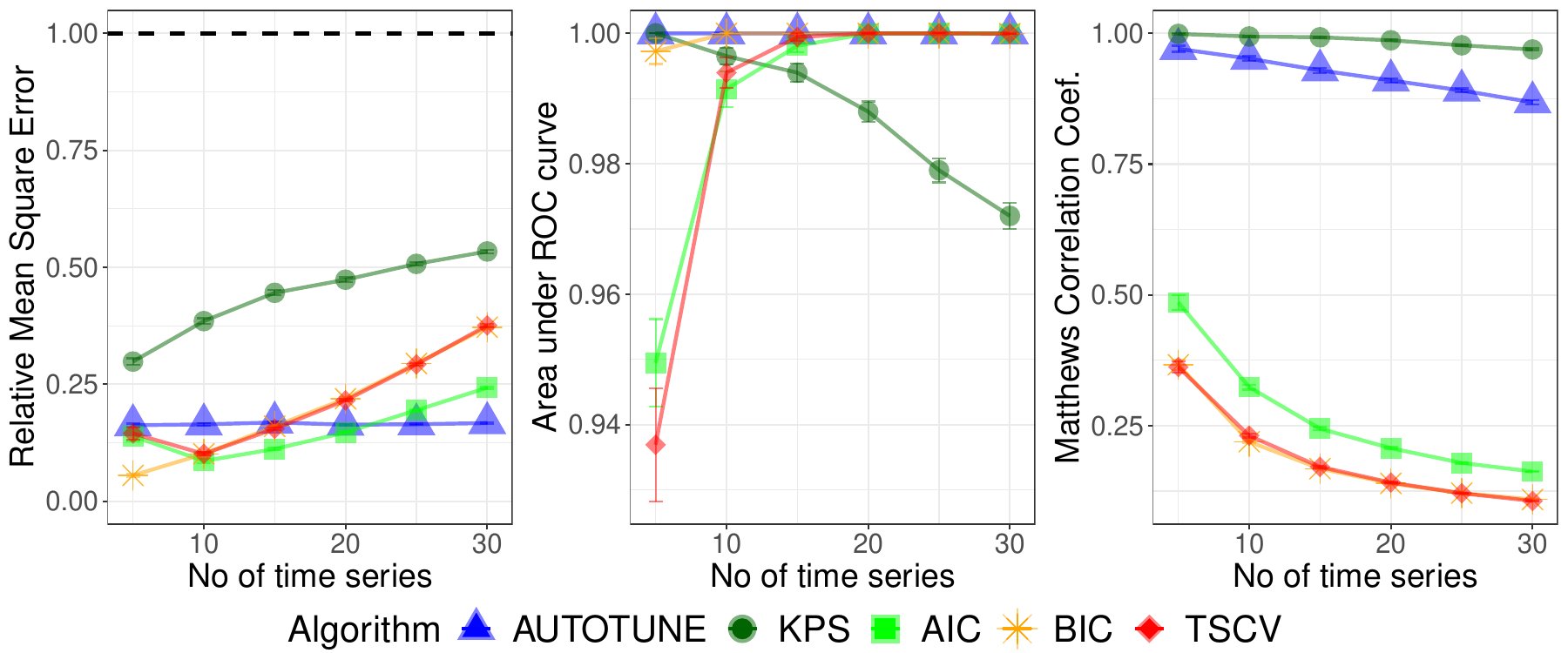}
    \end{tabular}
    \\[-1mm]
    \caption{RMSE, AUROC, and MCC (with 1SE bars) of $\autotune$ and benchmark methods for VAR, plotted against $p$.
    }
    \label{fig: var comparison main text}
\end{figure*}

\paragraph*{Runtime comparsion} 
In Figure \ref{fig: runtime plots}, we plot mean runtimes (in milliseconds) and SE (over 100 replications) of different methods against increasing number of predictors $(p)$ in Lasso regression tasks. We fix $\rho= 0.35,$ $\mrm{SNR} = 1$, Beta-type 2, $n = 200$ and $s = 5$. 
%
We see that $\autotune$ is 15-210 times faster than CV Lasso and 40-100 times faster than Scaled Lasso, details are given in Table \ref{tab: runtimes for fig b}.
Since $\autotune$ finds an ideal $\lambda$ first, and then computes $\solution$, we can ask: \textit{how much time it takes to tune, and how much to fit the final LASSO model?}. So we monitor the times (in seconds) taken by $\autotune$ to (i) converge to a $\lambda$,  and (ii) compute $\solution$ after fixing the $\lambda$. We report the results for high-dimensional setup $\{p=750$, $n=80$, $s = 5\}$ in Table \ref{tab: finding lambda vs finding beta}. Interestingly, $\autotune$ only uses about 10-17\% of its total runtime to tune and the rest to fit a single Lasso. This finding suggests that the task of finding a good tuning parameter may not be inherently hard, and does not need to be the main computational bottleneck in high-dimensional regression and VAR problems.

\paragraph*{$\autotune$ vs. $\autotune$ with active set selection} Additionally, we see that $\autotune$ with active set selection has lower runtime (Appendix \ref{subsec: autotune_vs_active_runtime}) and achieves the same performance in terms of RTE, AUROC and RMSE compared to $\autotune$ with no active set selection (Figure \ref{fig: high dim accuracy plot}, \ref{fig: moderate dim accuracy plot} and Appendix \ref{subsec: all accuracy plots}).

\subsection{Using \texorpdfstring{$\autotune$}{autotune} for VAR Lasso}
\label{subsec: VAR}

We compare VAR Lasso tuned with $\autotune$ against VAR Lasso tuned with  TSCV, AIC, BIC, and  \cite{kock2025data} in terms of estimation, variable selection and runtime. From here on, we refer to VAR Lasso tuned via \cite{kock2025data} as ``KPS Lasso''.

Instead of out-of-sample prediction, we focus on estimation and variable selection accuracy since the primary interest in large VAR models is often learning Granger causality (lead-lag) relationships among the component time series \cite{shojaie2022granger}. In variable selection, we also investigate another criterion, Matthews Correlation Coefficient (MCC), since $\autotune$ produced an AUROC of nearly 1 in all our simulation setups. 

We conduct our experiments on two DGP: (i) $p$-dimensional diagonal VAR(1) process X$_t = A\text{X}_{t-1} + \eps_t$ with diagonal $A=0.5\mrm{I}_p$ and $\eps_t\stackrel{\iid}{\sim}\Gsn(0,\Sigma)$,  where $\Sigma = \diag(\sigma_1^2,\dots,\sigma_p^2)$ with $\sigma_j^2= \frac{0.5^2}{\nu}$ and $\nu$ is the SNR level,  (ii) same DGP as in (i) except $A$ has a block diagonal structure. The blocks of $A$ are $2\times 2$ matrices with all entries equal to 0.3, making the maximum eigenvalue of the companion matrix 0.6. DGP(ii) presents relatively harder problem for estimating $A$ as compared to DGP(i). Dimension of parameter space for both DGP is $p^2$. We simulate time series initializing at $\text{X}_0 = 0_{1\times p}$ and with a burn-in period of 1000 to ensure stationarity. We vary $p\in\braces{5,10,\dots,30}$ for DGP(i) and $p\in\braces{10,16,24,30}$ for DGP(ii). For experiments, we fix $n\in\braces{100, 200}$ and SNR $\nu$ to 2.5. 


In Figure \ref{fig: var comparison main text} and \ref{fig: var comparison appendix}, the estimation accuracy of $\autotune$ VAR is significantly better in high dimensions and matches in moderate dimensions to the estimation accuracy of the benchmarks. Moreover, the estimation error, RMSE of  $\autotune$ remains robust to the increasing dimensionality of the problem, even when the benchmarks provides trivial estimate $\widehat{A}\approx 0_{p\times p}$ or even worse, i.e. RMSE $\geq$1. $\autotune$ VAR shows best variable selection among the VAR methods across all the setups except the easiest setup of DGP(i) with $n=200, p \leq 10$, where DGP is low dimensional.

In Figure \ref{fig: teaser into autotune}, \ref{fig: runtime plots} and \ref{fig: time series runtime}, we plot the mean runtimes of different methods against $p$, BIC was dropped as it gave identical timings as AIC, which is always in between KPS and TSCV. $\autotune$ is 60 - 175 times faster than TSCV and 30 - 80 times faster than KPS Lasso.

\subsection{Noise estimation and sparsity diagnostics}

We conduct two different experiments to assess $\sigma^2$ estimator of $\autotune$, and illustrate a visual diagnostic like scree plot to check the sparsity assumption.


\label{subsubsec: simulation results for noise estimation and sparsity diagnostics}


\begin{wrapfigure}{R}{0.55\textwidth}
    \centering
    \begin{subfigure}[t]{0.48\linewidth}
        \centering
        \includegraphics[page=1,trim=0in 0in 0in 0.4in,clip,width=\linewidth]
        {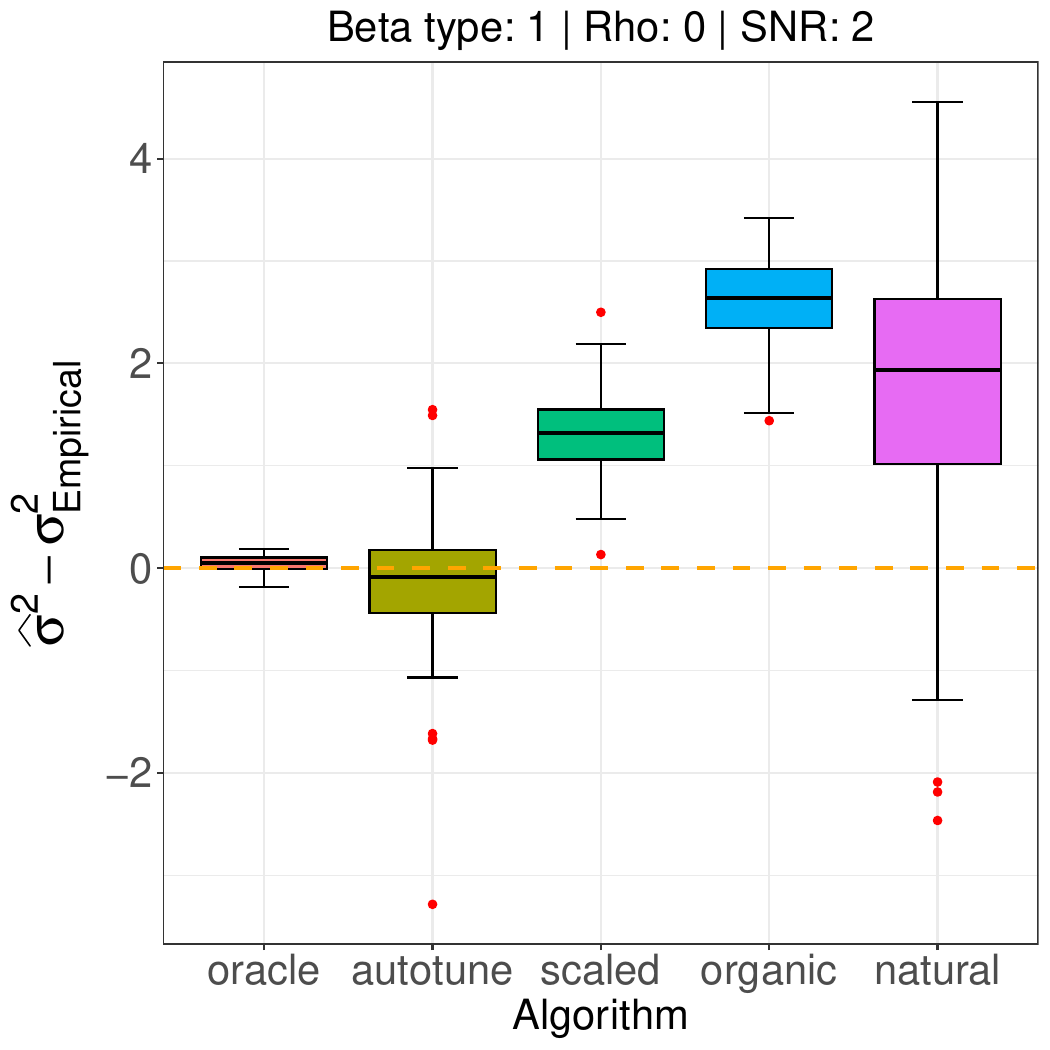}
        \label{fig: sigma1}
    \end{subfigure}
    \begin{subfigure}[t]{0.48\linewidth}
        \centering
        \includegraphics[page=1,trim=0in 0in 0in 0.4in,clip,width=\linewidth]
        {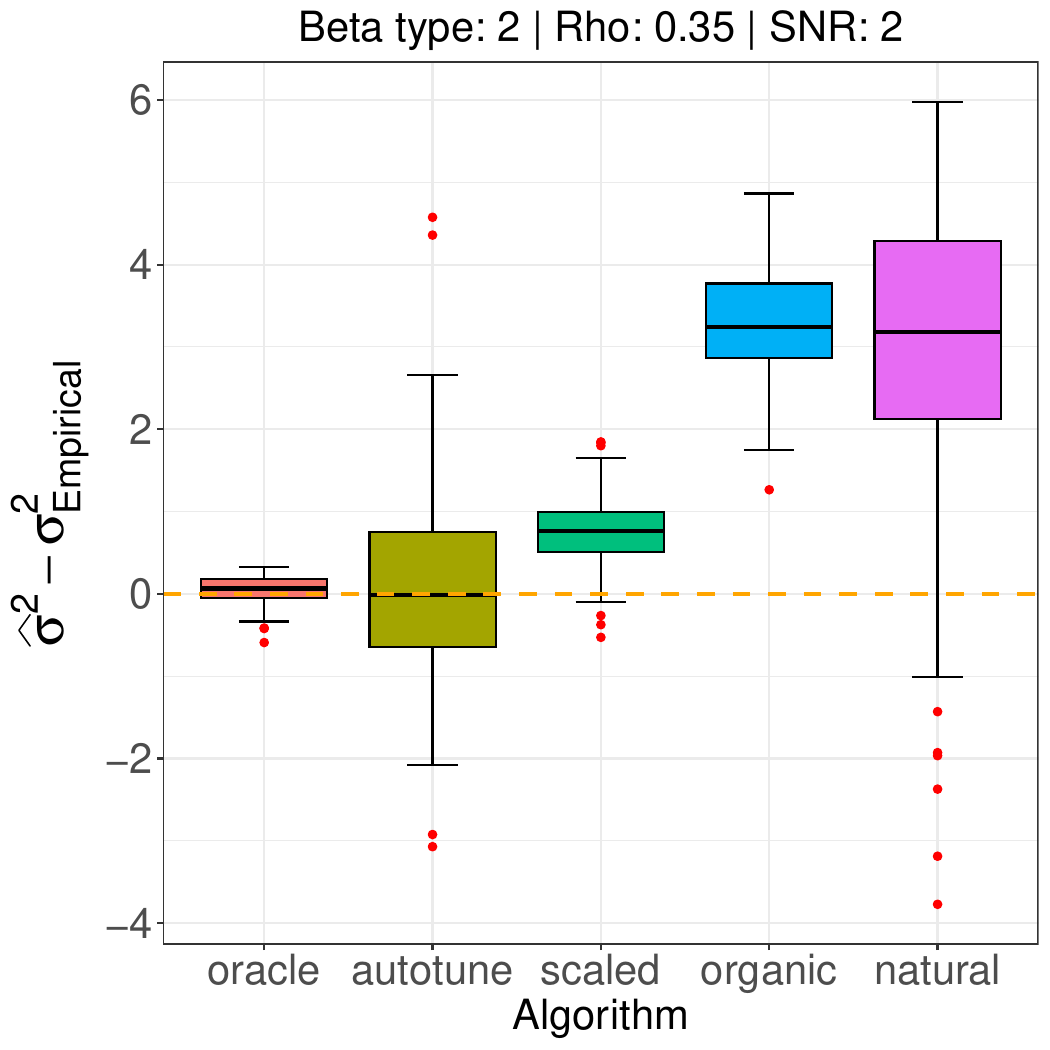}
        \label{fig: sigma4}
    \end{subfigure}
    \caption{Distribution of $\what\sigma^2$ from $\autotune$ and
    benchmark methods across $100$ simulated high-dimensional regressions
    with \textbf{[Left]} uncorrelated design ($\rho=0$) and
    \textbf{[Right]} correlated design ($\rho=0.35$).
    $\autotune$ estimates $\sigma$ better than all non-oracle benchmarks,
    except for Scaled Lasso in correlated design, where the improvement is
    not statistically significant.}
    \label{fig: sigma est comparison plots}
\end{wrapfigure}

\paragraph*{Noise scale ($\sigma^2$) estimation} We compare the $\sigma^2$ estimation accuracy of $\autotune$ against Scaled Lasso, Organic Lasso and Natural Lasso. We add an oracle method which performs linear regression of Y on the true support set and reports the MSE as the $\widehat{\sigma}^2$.
We also skip CV Lasso based $\what\sigma^2$ as \cite{yu2019estimating} found Organic Lasso to be uniformly better in $\sigma^2$ estimation than CV Lasso.
\begin{wrapfigure}{R}{0.52\textwidth}
    \centering
    \begin{subfigure}[t]{\linewidth}
        \includegraphics[
            page=1,
            trim=0in 0.55in 0in 0.4in,
            clip,
            width=\linewidth
        ]{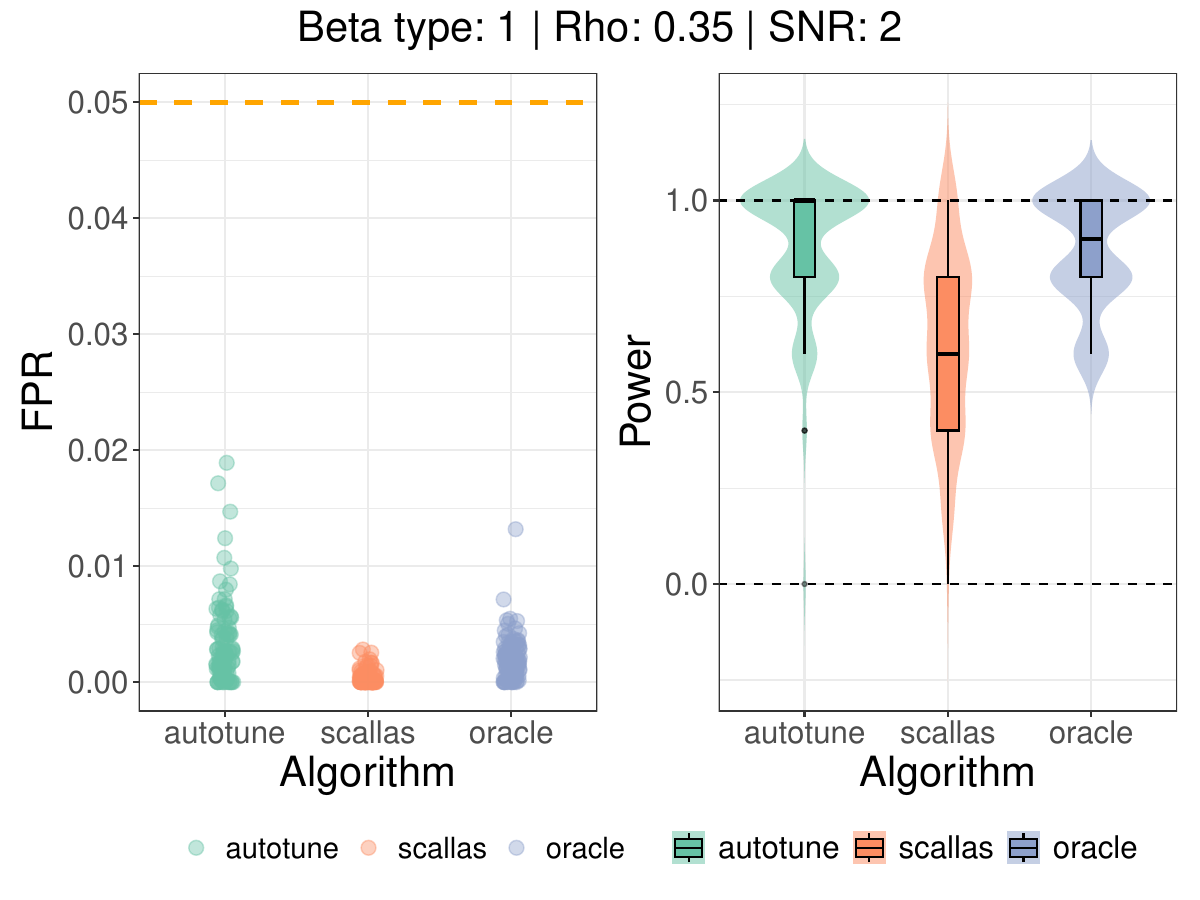}
        \label{fig: inference1}
    \end{subfigure}
    \caption{Comparing the quality of statistical inference with debiased
    Lasso when using $\what{\sigma}^2$ from $\autotune$ and Scaled Lasso.
    Distributions of \textbf{[Left]} false positive rate (FPR) and
    \textbf{[Right]} statistical power over 100 simulations.
    $\autotune$ shows improved power similar to oracle while keeping FPR
    below the nominal level (0.05), while Scaled Lasso tends to be more
    conservative.}
    \label{fig: sigma est diff plots}
\end{wrapfigure}

We contrast the $\what\sigma^2$ of different algorithms against the  $\sigma^2_{\mrm{empirical}} = \mrm{Var}(\text{Y}- \text{X}\beta)$, i.e. variance of the residuals obtained via regression of Y on X using true coefficients $\beta$. Note that  $\sigma^2_{\mrm{empirical}}$ knows the true $\beta$ whereas $\what\sigma^2$ of the oracle method only knows the true  support set of $\beta$. 

We plot the distribution of $\what\sigma^2 - \sigma^2_{\mrm{empirical}}$ across 100 replications in Figure \ref{fig: sigma est comparison plots} and \ref{fig: sigma est diff plots appendix}. For Figure \ref{fig: sigma est comparison plots}, we fix $n$ = 80, $p$ = 750, $s$ = 5, SNR = 2, (left plot) $\rho = 0$, Beta-type 1 and (right plot) $\rho = 0.35$, Beta-type 2.

In Figures \ref{fig: sigma est comparison plots} and \ref{fig: sigma est diff plots appendix}, $\autotune$ estimates the $\sigma$ significantly better than all the non-oracle benchmarks as per the paired t-test comparisons. Only exception is 2$^{nd}$ plot of Figure \ref{fig: sigma est comparison plots}, when the true predictors are correlated with one another; 
we see Scaled Lasso matching the accuracy of $\autotune$ and paired t-test between them revealed no significant difference.

Additionally, we report the sensitivity of $\sigma^2$ estimation to chosen $\alpha$ in the sequential F-test of $\autotune$ in the Section \ref{subsec: sigma alpha comp}. We find that $\lambda$ selected by $\autotune$ (effectively the $\widehat{\sigma}^2$) monotonically decreases as $\alpha$ increases.

\paragraph*{Using $\widehat{\sigma}^2$ for high-dimensional inference}
We compare the quality of estimator of $\what\sigma^2$ of $\autotune$, Scaled Lasso and the oracle for downstream inference of coefficients in high-dimensional regression. Results of Organic and Natural Lasso were not competitive, so we drop them from our benchmarks. 

In this inference experiment, we use debiased or desparsified Lasso \cite{dezeure2015high}. It first computes the debiased Lasso coefficients, and then it needs an estimate of $\sigma^2$ to compute the p-values for two-sided testing of $\hypo:\beta_j=0$ for $j\in \seqp$. For each competing method, we plugin their $\widehat \sigma^2$ in debiased Lasso, and for each $j\in\seqp$, we reject $\hypo$ if the p-value of $j^{th}$ covariate is less than 0.05. 
Then, we measure false positive rate, $\mrm{FPR} = \sum_{j\in\mc{S}^C}\mbb{I}\brackets{\hypo \text{ was rejected } }/\parenth{p-s}$ and statistical power  $ = \sum_{j\in\mc{S}}\mbb{I}\brackets{\hypo \text{ was rejected } }/s$, where $\mc{S}$ is the true support set and $s = |\mc{S}|$. In Figure \ref{fig: sigma est diff plots} and \ref{fig: inference plots appendix}, we employ the same DGP used in the previous experiment (which analyzed $\what\sigma^2 - \sigma^2_{\mrm{empirical}}$) and plot the FPR and power across 100 replications. 

In Figure \ref{fig: sigma est diff plots} and  \ref{fig: inference plots appendix}, the gains of $\autotune$ are more pronounced as $\autotune$ and the oracle nearly shows the same power across all the setups considered while always keeping FPR below 0.05 level. In contrast, Scaled Lasso numerically shows suboptimal power in all the setups we consider.

\begin{figure}[!b]
    \centering
    
    \begin{subfigure}[t]{0.45\textwidth}
        \centering
        \includegraphics[page = 4, trim=0in 0in 0in 0.4in, clip, width=\textwidth]{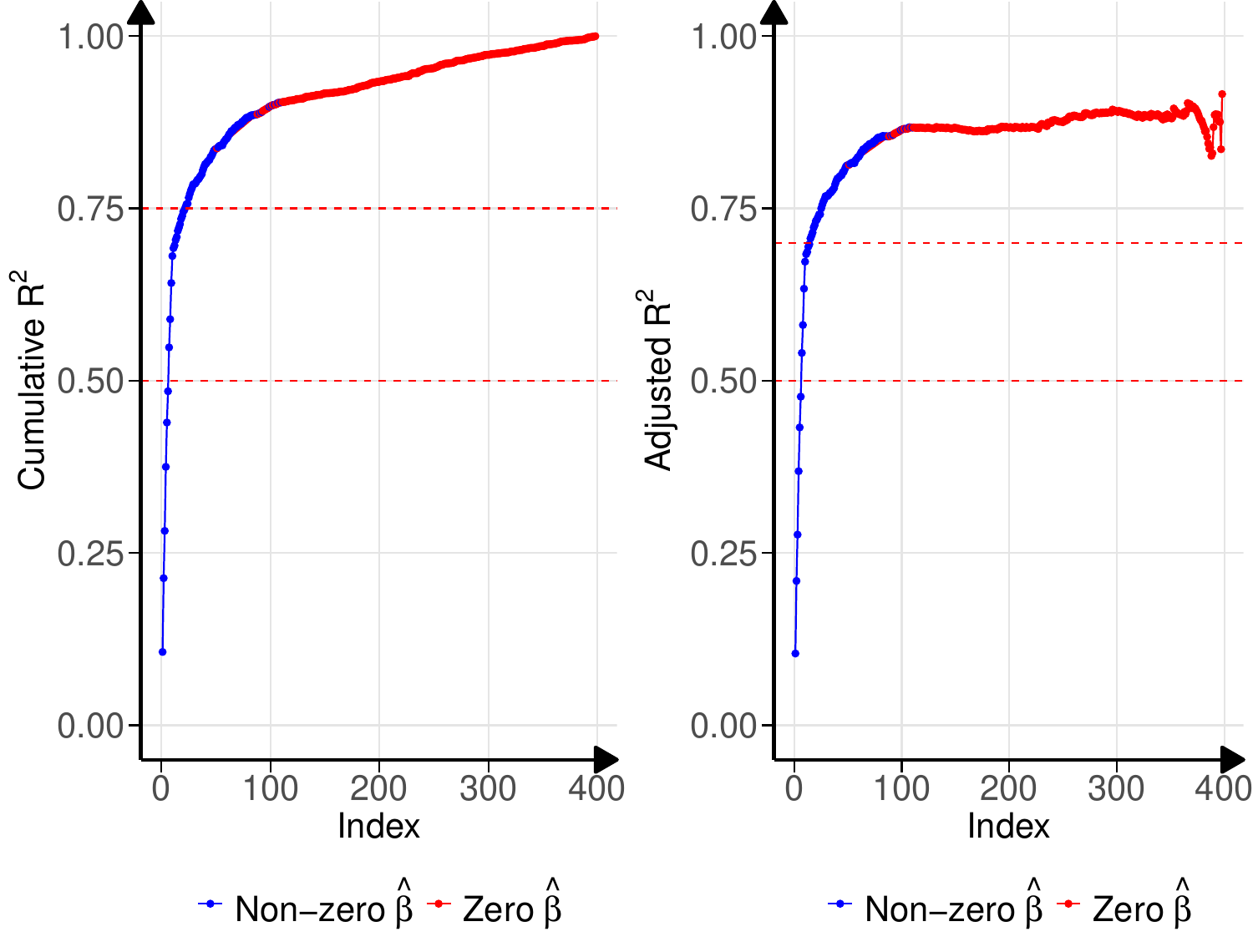}
        \label{fig: good diagnostics2}
    \end{subfigure}
    \begin{subfigure}[t]{0.45\textwidth}
        \centering
        \includegraphics[page = 9, trim=0in 0in 0in 0.4in, clip, width=\textwidth]{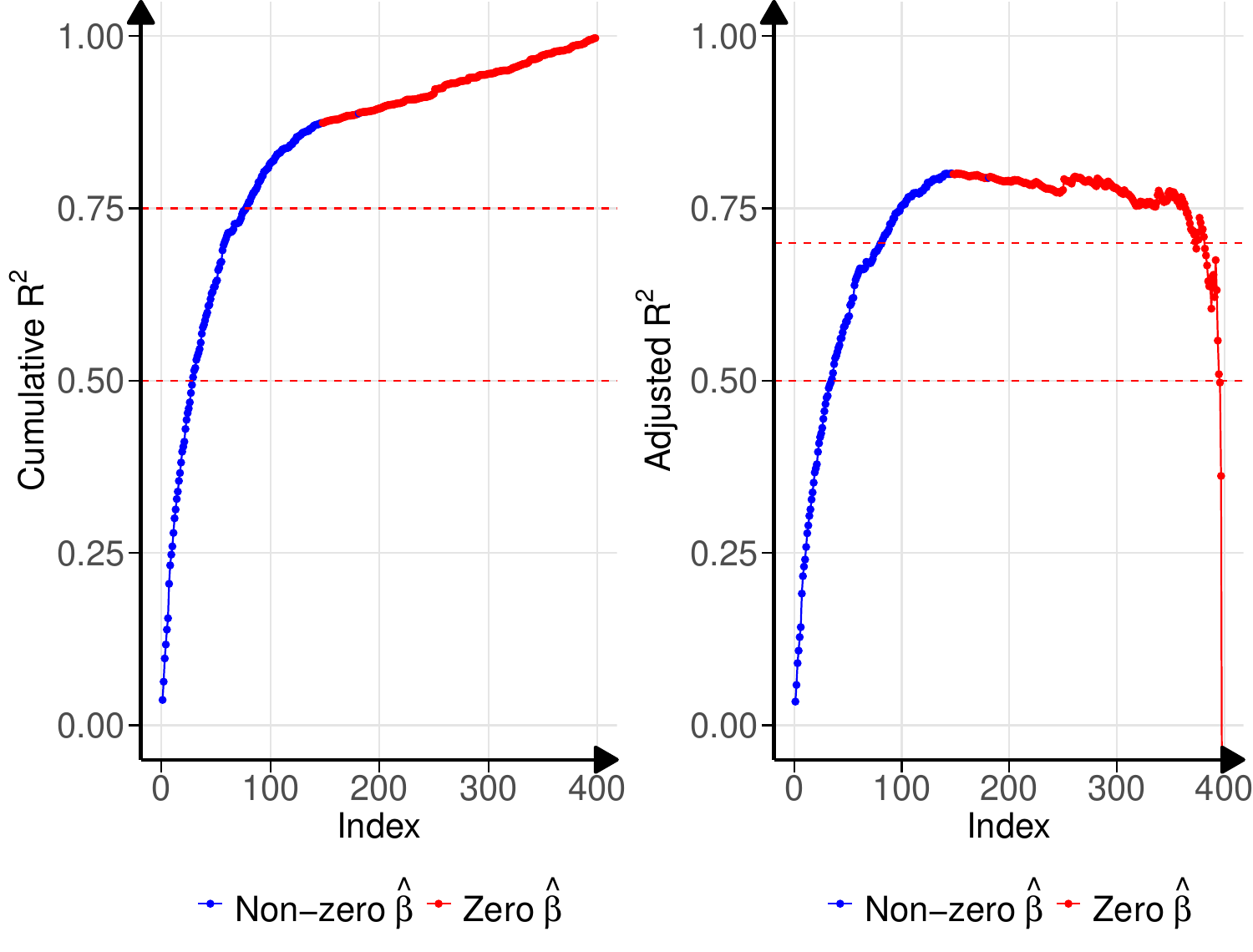}
        \label{fig: bad diagnostics}
    \end{subfigure}
    \caption{Diagnostic plots of cumulative and adjusted R$^2$ for\textbf{ [Left]} sparse ($s=10$) and \textbf{[Right]} non-sparse ($s=350$) Lasso regression. An elbow pattern is (not) discernible for (non) sparse DGP.}
    \label{fig: full diagnostics plots}
\end{figure}
\paragraph*{Visual diagnostics for checking sparsity} Our intuition for the diagnostics stems from principal component analysis where scientists look for an ``elbow'' in the scree plot and hope that the principal components appearing before the elbow will give an accurate, sparse representation of their high-dimensional data. We plot the cumulative and adjusted R$^2$ of the least squares model formed by incrementally adding predictors to the linear model in the rank order given by PR of $\autotune$ Lasso. This means in Figure \ref{fig: full diagnostics plots}, a $(s,t)$ point denotes the linear model formed by first $s$ ranked predictors has a cumulative or adjusted R$^2$ value of $t$. We fix $n$ = 400, $p$ = 2000, SNR = 2, $\rho$ = 0, Beta-type 2 and $s\in\braces{10, 350}$ in Figure \ref{fig: full diagnostics plots} and $s\in\braces{5, 30}$ in Figure \ref{fig: diagnostics appendix}. We use blue and red dots to refer to the predictors with nonzero and zero $\what\beta$ respectively. We try to locate an elbow, a point which is preceded by blue dots with visibly big contributions to the cumulative or adjusted R$^2$ and followed by a tight cluster of blue and red dots with marginal contribution to the R$^2$, i.e. a drastic change in the slope of the trend of the points around it. All these indicators are present (absent) in the sparse case of $s = 10$ (non-sparse case of $s = 350$) of Figure \ref{fig: full diagnostics plots}. We acknowledge that this is a very preliminary  attempt at sparsity diagnostics, a task that is inherently subjective. While we found this strategy promising in many cases, we also encountered some scenarios where our diagnostics do not give a clear indication of sparsity (see Figure \ref{fig: diagnostics appendix}). 

\section{Empirical illustration on a financial dataset}\label{sec: realdata}


We benchmark $\autotune$ against several other tuning strategies (CV, AIC, BIC and Scaled Lasso) on a real-life financial dataset, $\msf{sp500}$ from $\texttt{scalreg}$\footnote{$\texttt{scalreg}$ R package was archived by CRAN in July 2026. So for ease of reproducility, we also provide $\msf{sp500}$ dataset in our R package \href{https://cran.r-project.org/web/packages/autotune/index.html}{$\autotune$}.} \cite{scalreg}, the R package of Scaled Lasso. Due to space constraints, we defer the results of benchmarking $\autotune$ on three more real-world datasets $(\msf{ames}\_\msf{housing}, \msf{Scheetz2006}, \msf{e2006-tfidf})$ to Appendix \ref{app: more real life datasets}.
\begin{wrapfigure}{R}{0.7\textwidth}
    \centering
    \begin{subfigure}[b]{0.48\linewidth}
        \centering
        \includegraphics[width=\linewidth]
        {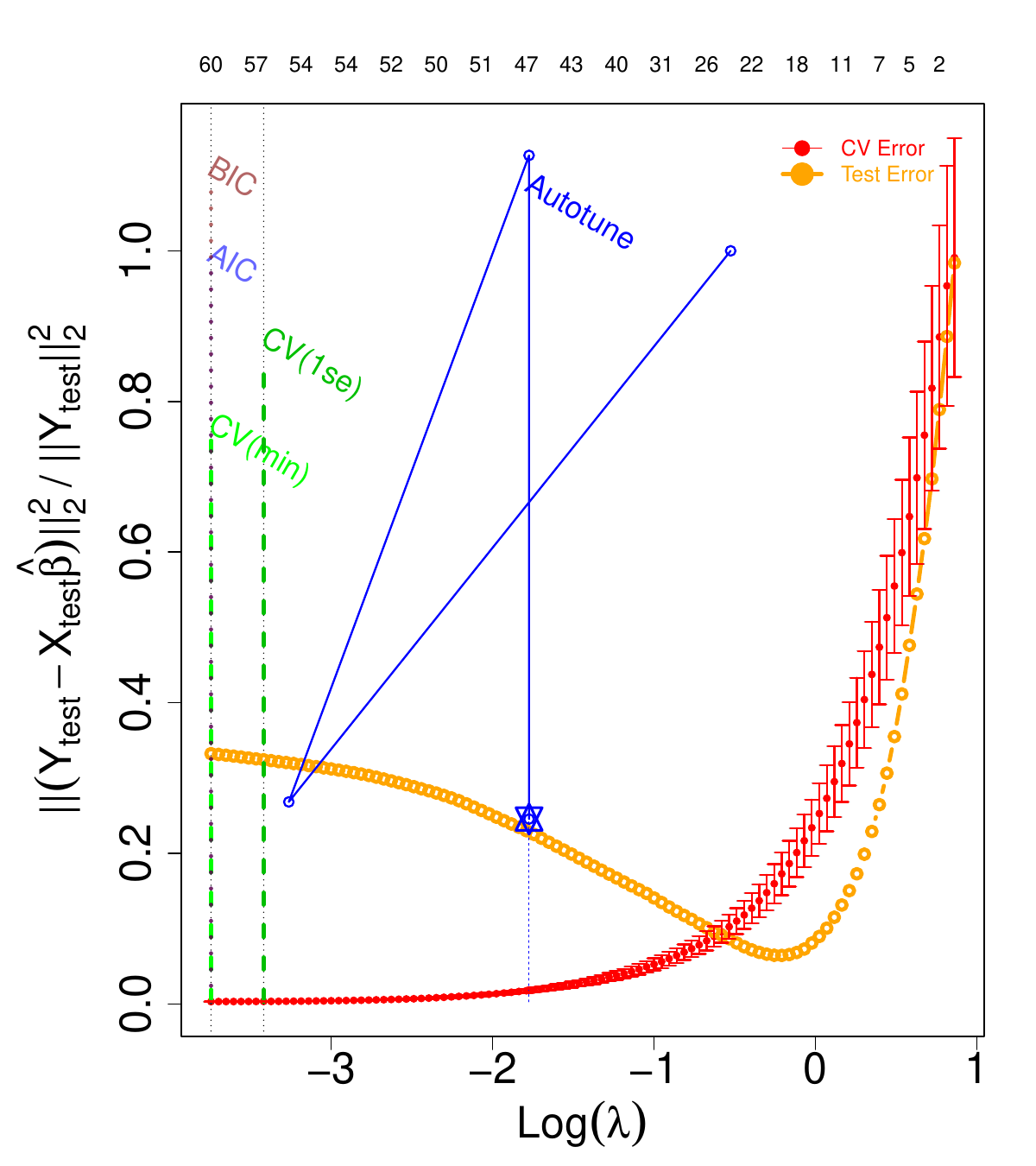}
        \label{fig: sp500 n190}
    \end{subfigure}
    \begin{subfigure}[b]{0.48\linewidth}
        \centering
        \includegraphics[width=\linewidth]
        {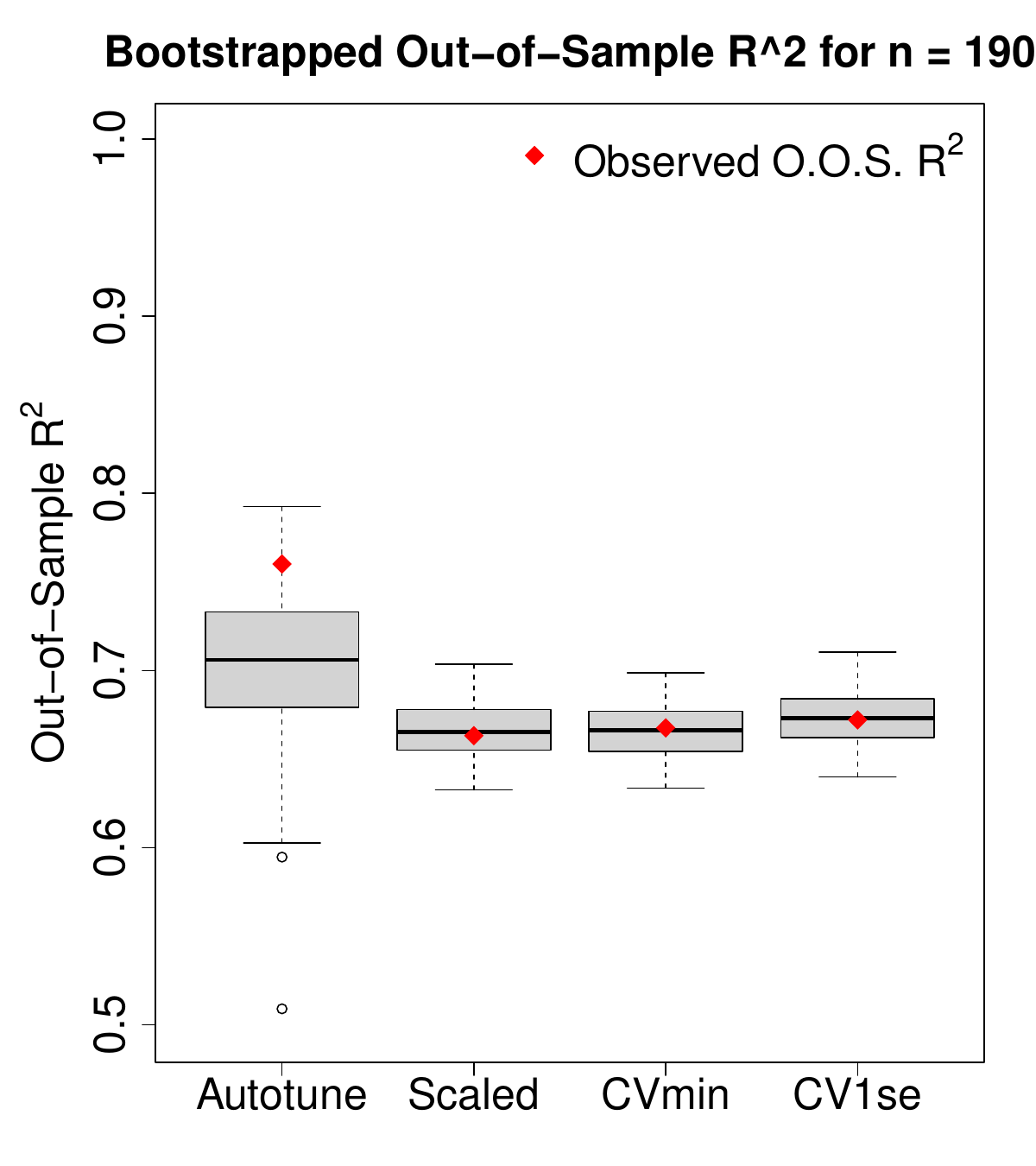}
        \label{fig: sp500 n190 boot}
    \end{subfigure}
    \caption{\textbf{[left]} Test error curve of the Lasso solution path
    [orange] and $\autotune$ [blue]. Vertical dotted bars indicate
    $\lambda$ selected by different tuning strategies.
    \textbf{[right]} O.O.S. R$^2$ of models fitted on 100 bootstrapped
    training datasets.}
    \label{fig: sp500}
\end{wrapfigure}

The dataset $\msf{sp500}$ has been used in prior literature \cite{wu2020survey} for comparing different Lasso algorithms. It contains close-of-day percentage change data for $p = 492$ out of Standard and Poor's 500 stocks and Dow  Jones Industrial Average (DJIA) index across $252$ days of the year 2008. So, one observation in $\msf{sp500}$ will be the percentage change on a particular day for 492 stocks and the DJIA index. Following the vignette of $\msf{scalreg}$, our goal is modeling the DJIA using those 492 stocks. We assign the $1:n$ observations to the training set, and other observations $(n+1):252$ to the test set, for $n = 150,\dots, 210$. This splitting mechanism ensures test error curve of CV Lasso has
a ``U-shape''. 
We compare the methods in terms of RTE, Out-of-sample $\text{R}^2$ (O.O.S. R$^2$) and size of the final model selected, details are given in Appendix \ref{app: experimental details of data analysis}. Here, RTE $=\twonorm{\text{Y}_{\mrm{test}} - \text{X}_{\mrm{test}}\what\beta}^2 \big/\twonorm{\text{Y}_{\mrm{test}}}^2$ and O.O.S. R$^2$ is the multiple R$^2$ of the trained model on the test data, as a result, O.O.S. R$^2$ $= 1 -$ RTE.

In Figure \ref{fig: sp500} and \ref{fig: appen sp500 plots}, we see that $\autotune$ quickly reaches its final $\lambda$ within 3-6 iterations, yielding the lowest RTE among all benchmarks (Table \ref{tab: sp500 test mse}). Bootstrapped O.O.S. R$^2$  validates the prediction gains of $\autotune$ (see Figure \ref{fig: sp500} and \ref{fig: boot sp500 appen}). Lastly, we find that $\autotune$ generally selects the sparsest model among all the benchmarks (Table \ref{tab: sp500 model selection}).

  \section{Conclusion}\label{sec:conclusion}

Traditional approaches to tune Lasso, such as CV, requires repeatedly solving $\hat{\beta}(\lambda)$ at many different $\lambda$ values. In high-dimensional problems, this comes at the cost of speed and sometimes loss of accuracy. For dependent data problems such as in time series where CV is restricted to obey temporal ordering, this issue is even more prominent. Recent attempts to automate tuning, such as the Scaled Lasso or bilevel optimization, \textit{jointly optimize} over $\beta$ and $\lambda$ (or noise standard deviation $\sigma$) but remain slower compared to state-of-the-art implementation of CV-Lasso.

Building upon a core statistical premise that prediction is easier than estimation in high-dimension, we develop $\autotune$, an approach where Lasso first tunes itself automatically and selects a good $\lambda$, and then solves for $\what\beta(\lambda)$ only once in the end. In simulation and real data experiments, it exhibits improved speed and accuracy over its benchmarks. As byproducts, it also offers a new estimator $\what\sigma^2$, useful for noise scale estimation and high-dimensional inference, as well as new diagnostic plots to check model sparsity, a first attempt in the literature to the best of our knowledge. Finally, we show that $\autotune$ can be viewed as alternately estimating noise standard deviation $\sigma$ and a new object $P_S$, projection matrix on the column space of $X_S$, where $S = support(\beta)$. We argue that this new target $P_S$ is related to \textit{prediction} as opposed to \textit{estimation} $\beta$, opening up new ways to design algorithms that alternately estimates these two objects.

A number of questions remain open regarding our current $\autotune$ strategy. The algorithm remains somewhat reliant on initialization $\lambda_0$. In high SNR settings, it starts to exhibit a loss in accuracy. Our choice of estimating $P_S$ via sequential F-test is a reasonable start, but it may be possible to design more accurate estimators of $P_S$ by directly constraining its rank. A complete convergence analysis of the iterative algorithm is also needed to understand its strengths and limitations. We leave these questions for future research.





\section*{Acknowledgements}

SB acknowledges partial support from NSF awards DMS-1812128, DMS-2210675, CAREER  DMS-2239102, and
 NIH awards R01GM135926, R21NS120227.

}{%

}

\section*{Author contributions}
T.S., I.W., S.S. and S.B. designed the problem statement, performed research,
contributed analytic tools, analyzed data and wrote the paper.

\section*{Competing interests}
The authors declare no conflict of interest.

\bibliographystyle{plainnat}
\IfFileExists{Arxiv/pnas-sample.bib}{%
  \bibliography{Arxiv/pnas-sample}
}{%
  \bibliography{pnas-sample}

@inproceedings{bertrand2020implicit,
  author    = {Bertrand, Quentin and Klopfenstein, Quentin and Blondel, Mathieu and Vaiter, Samuel and Gramfort, Alexandre and Salmon, Joseph},
  title     = {Implicit differentiation of {Lasso}-type models for hyperparameter optimization},
  booktitle = {Proceedings of the 37th International Conference on Machine Learning ({ICML})},
  year      = {2020},
  series    = {Proceedings of Machine Learning Research},
  publisher = {PMLR},
  volume    = {119},
  pages     = {832--842},
  url       = {https://proceedings.mlr.press/v119/bertrand20a.html}
}

@Manual{AmesHousingR,
  title  = {AmesHousing: The Ames Iowa Housing Data},
  author = {Max Kuhn and Dmytro Perepolkin},
  year   = {2020},
  note   = {R package version 0.0.4},
  url    = {https://CRAN.R-project.org/package=AmesHousing}
}

@inproceedings{kogan-etal-2009-predicting,
    title = "Predicting Risk from Financial Reports with Regression",
    author = "Kogan, Shimon  and
      Levin, Dimitry  and
      Routledge, Bryan R.  and
      Sagi, Jacob S.  and
      Smith, Noah A.",
    booktitle = "Proceedings of Human Language Technologies: The 2009 Annual Conference of the North American Chapter of the Association for Computational Linguistics",
    month = jun,
    year = "2009",
    address = "Boulder, Colorado",
    publisher = "Association for Computational Linguistics",
    pages = "272--280",
    url = "https://aclanthology.org/N09-1031",
}

@inproceedings{skglm_neurips2022,
  title     = {Beyond L1: Faster and better sparse models with skglm},
  author    = {Bertrand, Quentin and Klopfenstein, Quentin and Bannier, Pierre-Antoine and Gidel, Gauthier and Massias, Mathurin},
  booktitle = {Advances in Neural Information Processing Systems},
  volume    = {35},
  pages     = {29337--29348},
  year      = {2022},
  url       = {https://proceedings.neurips.cc/paper_files/paper/2022/hash/fe5c31e525e9a26a1426ab0b589f42fe-Abstract-Conference.html}
}

@book{nesterov2013introductory,
  title={Introductory lectures on convex optimization: A basic course},
  author={Nesterov, Yurii},
  volume={87},
  year={2013},
  publisher={Springer Science \& Business Media}
}

@Manual{scalreg,
    title = {scalreg: Scaled Sparse Linear Regression},
    author = {Tingni Sun},
    year = {2019},
    note = {R package version 1.0.1},
    url = {https://CRAN.R-project.org/package=scalreg},
  }

@Article{glmnet,
    title = {Regularization Paths for Generalized Linear Models via
      Coordinate Descent},
    author = {Jerome Friedman and Trevor Hastie and Robert Tibshirani},
    journal = {Journal of Statistical Software},
    year = {2010},
    volume = {33},
    number = {1},
    pages = {1--22},
    doi = {10.18637/jss.v033.i01},
  }

@article{aos_support_set_recovery,
author = {Nicolai Meinshausen and Bin Yu},
title = {{Lasso-type recovery of sparse representations for high-dimensional data}},
volume = {37},
journal = {The Annals of Statistics},
number = {1},
publisher = {Institute of Mathematical Statistics},
pages = {246 -- 270},
year = {2009},
doi = {10.1214/07-AOS582},
URL = {https://doi.org/10.1214/07-AOS582}
}

@article{belloni2012sparse,
  title={Sparse models and methods for optimal instruments with an application to eminent domain},
  author={Belloni, Alexandre and Chen, Daniel and Chernozhukov, Victor and Hansen, Christian},
  journal={Econometrica},
  volume={80},
  number={6},
  pages={2369--2429},
  year={2012},
  publisher={Wiley Online Library}
}

@article{postlasso,
 ISSN = {00905364},
 URL = {http://www.jstor.org/stable/43818915},
 author = {Jason D. Lee and Dennis L. Sun and Yuekai Sun and Jonathan E. Taylor},
 journal = {The Annals of Statistics},
 number = {3},
 pages = {907--927},
 publisher = {Institute of Mathematical Statistics},
 title = {EXACT POST-SELECTION INFERENCE, WITH APPLICATION TO THE LASSO},
 urldate = {2025-12-14},
 volume = {44},
 year = {2016}
}

@article{MEINSHAUSEN2007374,
title = {Relaxed Lasso},
journal = {Computational Statistics \& Data Analysis},
volume = {52},
number = {1},
pages = {374-393},
year = {2007},
issn = {0167-9473},
doi = {https://doi.org/10.1016/j.csda.2006.12.019},
url = {https://www.sciencedirect.com/science/article/pii/S0167947306004956},
author = {Nicolai Meinshausen}
}

@article{efron2004least,
  title={LEAST ANGLE REGRESSION},
  author={Efron, Bradley and Hastie, Trevor and Johnstone, Iain and Tibshirani, Robert},
  journal={The Annals of Statistics},
  volume={32},
  number={2},
  pages={407--499},
  year={2004}
}

@article{roberts2014stabilizing,
  title={Stabilizing the lasso against cross-validation variability},
  author={Roberts, Steven and Nowak, Gen},
  journal={Computational Statistics \& Data Analysis},
  volume={70},
  pages={198--211},
  year={2014},
  publisher={Elsevier}
}

@article{bovelstad2007predicting,
  title={Predicting survival from microarray data—a comparative study},
  author={B{\o}velstad, Hege M and Nyg{\aa}rd, St{\aa}le and St{\o}rvold, Hege L and Aldrin, Magne and Borgan, {\O}rnulf and Frigessi, Arnoldo and Lingj{\ae}rde, Ole Christian},
  journal={Bioinformatics},
  volume={23},
  number={16},
  pages={2080--2087},
  year={2007},
  publisher={Oxford University Press}
}

@article{chen2008extended,
  title={Extended Bayesian information criteria for model selection with large model spaces},
  author={Chen, Jiahua and Chen, Zehua},
  journal={Biometrika},
  volume={95},
  number={3},
  pages={759--771},
  year={2008},
  publisher={Oxford University Press}
}

@article{zhao2015mixture,
  title={Mixture model selection via hierarchical BIC},
  author={Zhao, Jianhua and Jin, Libin and Shi, Lei},
  journal={Computational Statistics \& Data Analysis},
  volume={88},
  pages={139--153},
  year={2015},
  publisher={Elsevier}
}

@article{lei2020cross,
  title={Cross-validation with confidence},
  author={Lei, Jing},
  journal={Journal of the American Statistical Association},
  volume={115},
  number={532},
  pages={1978--1997},
  year={2020},
  publisher={Taylor \& Francis}
}

@article{bickel2009simultaneous,
  title={SIMULTANEOUS ANALYSIS OF LASSO AND DANTZIG SELECTOR},
  author={Bickel, Peter J and Ritov, Ya'acov and Tsybakov, Alexandre B},
  journal={The Annals of Statistics},
  volume={37},
  number={4},
  pages={1705--1732},
  year={2009},
  publisher={Citeseer}
}

@article{bates2024cross,
  title={Cross-validation: what does it estimate and how well does it do it?},
  author={Bates, Stephen and Hastie, Trevor and Tibshirani, Robert},
  journal={Journal of the American Statistical Association},
  volume={119},
  number={546},
  pages={1434--1445},
  year={2024},
  publisher={Taylor \& Francis}
}

@article{basu2015regularized,
  title={Regularized estimation in sparse high-dimensional time series models},
  author={Basu, Sumanta and Michailidis, George},
  year={2015},
journal = {Annals of Statistics}
}

@article{smith2012future,
  title={The future of FMRI connectivity},
  author={Smith, Stephen M},
  journal={Neuroimage},
  volume={62},
  number={2},
  pages={1257--1266},
  year={2012},
  publisher={Elsevier}
}

@article{wang2007tuning,
  title={Tuning parameter selectors for the smoothly clipped absolute deviation method},
  author={Wang, Hansheng and Li, Runze and Tsai, Chih-Ling},
  journal={Biometrika},
  volume={94},
  number={3},
  pages={553--568},
  year={2007},
  publisher={Oxford University Press}
}

@article{hastie2015statistical,
  title={Statistical learning with sparsity},
  author={Hastie, Trevor and Tibshirani, Robert and Wainwright, Martin},
  journal={Monographs on statistics and applied probability},
  volume={143},
  number={143},
  pages={8},
  year={2015}
}

@article{10.1111/j.1467-9868.2011.01004.x,
    author = {Tibshirani, Robert and Bien, Jacob and Friedman, Jerome and Hastie, Trevor and Simon, Noah and Taylor, Jonathan and Tibshirani, Ryan J.},
    title = {Strong Rules for Discarding Predictors in Lasso-Type Problems},
    journal = {Journal of the Royal Statistical Society Series B: Statistical Methodology},
    volume = {74},
    number = {2},
    pages = {245-266},
    year = {2011},
    month = {11},
    issn = {1369-7412},
    doi = {10.1111/j.1467-9868.2011.01004.x},
    url = {https://doi.org/10.1111/j.1467-9868.2011.01004.x},
    eprint = {https://academic.oup.com/jrsssb/article-pdf/74/2/245/49513900/jrsssb\_74\_2\_245.pdf},
}

@article{10.1214/07-AOAS131,
author = {Jerome Friedman and Trevor Hastie and Holger H{\"o}fling and Robert Tibshirani},
title = {{Pathwise coordinate optimization}},
volume = {1},
journal = {The Annals of Applied Statistics},
number = {2},
publisher = {Institute of Mathematical Statistics},
pages = {302 -- 332},
year = {2007},
doi = {10.1214/07-AOAS131},
URL = {https://doi.org/10.1214/07-AOAS131}
}

@article{stadler2010,
  title={l1-penalization for mixture regression models},
  author={St{\"a}dler, Nicolas and B{\"u}hlmann, Peter and Van De Geer, Sara},
  journal={Test},
  volume={19},
  pages={209--256},
  year={2010},
  publisher={Springer}
}

@article{antoniadis2010comments,
  title={Comments on: l1-penalization for mixture regression models},
  author={Antoniadis, Anestis},
  journal={Test},
  volume={19},
  number={2},
  pages={257--258},
  year={2010},
  publisher={Springer Nature BV}
}

@article{sun2010comments,
  title={Comments on: 1-penalization for mixture regression models},
  author={Sun, Tingni and Zhang, Cun-Hui},
  journal={Test},
  volume={19},
  number={2},
  pages={270},
  year={2010},
  publisher={Springer Nature BV}
}

@article{sun2012scaled,
  title={Scaled sparse linear regression},
  author={Sun, Tingni and Zhang, Cun-Hui},
  journal={Biometrika},
  volume={99},
  number={4},
  pages={879--898},
  year={2012},
  publisher={Oxford University Press}
}

@article{belloni2011square,
  title={Square-root lasso: pivotal recovery of sparse signals via conic programming},
  author={Belloni, Alexandre and Chernozhukov, Victor and Wang, Lie},
  journal={Biometrika},
  volume={98},
  number={4},
  pages={791--806},
  year={2011},
  publisher={Oxford University Press}
}

@article{hastie2020best,
  title={Best subset, forward stepwise or lasso? Analysis and recommendations based on extensive comparisons},
  author={Hastie, Trevor and Tibshirani, Robert and Tibshirani, Ryan},
  journal={Statistical Science},
  volume={35},
  number={4},
  pages={579--592},
  year={2020},
  publisher={JSTOR}
}

@article{wu2020survey,
  title={A survey of tuning parameter selection for high-dimensional regression},
  author={Wu, Yunan and Wang, Lan},
  journal={Annual review of statistics and its application},
  volume={7},
  pages={209--226},
  year={2020},
  publisher={Annual Reviews}
}

@article{wang2020tuning,
  title={A tuning-free robust and efficient approach to high-dimensional regression},
  author={Wang, Lan and Peng, Bo and Bradic, Jelena and Li, Runze and Wu, Yunan},
  journal={Journal of the American Statistical Association},
  volume={115},
  number={532},
  pages={1700--1714},
  year={2020},
  publisher={Taylor \& Francis}
}

@inproceedings{lederer2015don,
  title={Don't fall for tuning parameters: tuning-free variable selection in high dimensions with the TREX},
  author={Lederer, Johannes and M{\"u}ller, Christian},
  booktitle={Proceedings of the AAAI conference on artificial intelligence},
  volume={29},
  number={1},
  year={2015}
}

@article{yu2019estimating,
  title={Estimating the error variance in a high-dimensional linear model},
  author={Yu, Guo and Bien, Jacob},
  journal={Biometrika},
  volume={106},
  number={3},
  pages={533--546},
  year={2019},
  publisher={Oxford University Press}
}

@article{dezeure2015high,
  title={High-dimensional inference: confidence intervals, p-values and R-software hdi},
  author={Dezeure, Ruben and B{\"u}hlmann, Peter and Meier, Lukas and Meinshausen, Nicolai},
  journal={Statistical science},
  pages={533--558},
  year={2015},
  publisher={JSTOR}
}

@book{hyndman2018forecasting,
  title={Forecasting: principles and practice},
  author={Hyndman, Rob J and Athanasopoulos, George},
  year={2018},
  publisher={OTexts}
}

@article{bergmeir2018note,
  title={A note on the validity of cross-validation for evaluating autoregressive time series prediction},
  author={Bergmeir, Christoph and Hyndman, Rob J and Koo, Bonsoo},
  journal={Computational Statistics \& Data Analysis},
  volume={120},
  pages={70--83},
  year={2018},
  publisher={Elsevier}
}

@article{bergmeir2012use,
  title={On the use of cross-validation for time series predictor evaluation},
  author={Bergmeir, Christoph and Ben{\'\i}tez, Jos{\'e} M},
  journal={Information Sciences},
  volume={191},
  pages={192--213},
  year={2012},
  publisher={Elsevier}
}

@Manual{bigtimeRpackage,

    title = {bigtime: Sparse Estimation of Large Time Series Models},

    author = {Ines Wilms and David S. Matteson and Jacob Bien and Sumanta Basu and Will Nicholson and Enrico Wegner},

    year = {2023},

    note = {R package version 0.2.3},

    url = {https://CRAN.R-project.org/package=bigtime},

    doi = {10.32614/CRAN.package.bigtime}
}

@article{michailidis2013autoregressive,
  title={Autoregressive models for gene regulatory network inference: Sparsity, stability and causality issues},
  author={Michailidis, George and d’Alch{\'e}-Buc, Florence},
  journal={Mathematical biosciences},
  volume={246},
  number={2},
  pages={326--334},
  year={2013},
  publisher={Elsevier}
}

@article{kock2025data,
  title={Data-Driven Tuning Parameter Selection for High-Dimensional Vector Autoregressions},
  author={Kock, Anders B and Pedersen, Rasmus S and S{\o}rensen, Jesper R-V},
  journal={Journal of the American Statistical Association},
  number={just-accepted},
  pages={1--19},
  year={2025},
  publisher={Taylor \& Francis}
}

@article{akaike2003new,
  title={A new look at the statistical model identification},
  author={Akaike, Hirotugu},
  journal={IEEE transactions on automatic control},
  volume={19},
  number={6},
  pages={716--723},
  year={2003},
  publisher={Ieee}
}

@article{schwarz1978estimating,
  title={Estimating the dimension of a model},
  author={Schwarz, Gideon},
  journal={The annals of statistics},
  pages={461--464},
  year={1978},
  publisher={JSTOR}
}

@article{chatterjee2013assumptionless,
  title={Assumptionless consistency of the lasso},
  author={Chatterjee, Sourav},
  journal={arXiv preprint arXiv:1303.5817},
  year={2013}
}

@book{montgomery2020introduction,
  title={Introduction to statistical quality control},
  author={Montgomery, Douglas C},
  year={2020},
  publisher={John wiley \& sons}
}

@Book{carpackage,
    title = {An {R} Companion to Applied Regression},
    edition = {Third},
    author = {John Fox and Sanford Weisberg},
    year = {2019},
    publisher = {Sage},
    address = {Thousand Oaks {CA}},
    url = {https://www.john-fox.ca/Companion/},
  }

@article{10.1093/jjfinec/nbab023,
    author = {Hecq, Alain and Margaritella, Luca and Smeekes, Stephan},
    title = {Granger Causality Testing in High-Dimensional VARs: A Post-Double-Selection Procedure*},
    journal = {Journal of Financial Econometrics},
    volume = {21},
    number = {3},
    pages = {915-958},
    year = {2021},
    month = {11},
    issn = {1479-8409},
    doi = {10.1093/jjfinec/nbab023},
    url = {https://doi.org/10.1093/jjfinec/nbab023},
    eprint = {https://academic.oup.com/jfec/article-pdf/21/3/915/50621867/nbab023.pdf},
}

@article{banbura2010large,
  title={Large Bayesian vector auto regressions},
  author={Ba{\'n}bura, Marta and Giannone, Domenico and Reichlin, Lucrezia},
  journal={Journal of applied Econometrics},
  volume={25},
  number={1},
  pages={71--92},
  year={2010},
  publisher={Wiley Online Library}
}

@article{bernanke2005measuring,
  title={Measuring the effects of monetary policy: a factor-augmented vector autoregressive (FAVAR) approach},
  author={Bernanke, Ben S and Boivin, Jean and Eliasz, Piotr},
  journal={The Quarterly journal of economics},
  volume={120},
  number={1},
  pages={387--422},
  year={2005},
  publisher={MIT Press}
}

@article{fan2011sparse,
  title={Sparse high-dimensional models in economics},
  author={Fan, Jianqing and Lv, Jinchi and Qi, Lei},
  journal={Annu. Rev. Econ.},
  volume={3},
  number={1},
  pages={291--317},
  year={2011},
  publisher={Annual Reviews}
}

@article{broll2025prolong,
  title={PROLONG: penalized regression for outcome guided longitudinal omics analysis with network and group constraints},
  author={Broll, Steven and Basu, Sumanta and Lee, Myung Hee and Wells, Martin T},
  journal={Bioinformatics},
  volume={41},
  number={4},
  pages={btaf099},
  year={2025},
  publisher={Oxford University Press}
}

@article{seth2013granger,
  title={Granger causality analysis of fMRI BOLD signals is invariant to hemodynamic convolution but not downsampling},
  author={Seth, Anil K and Chorley, Paul and Barnett, Lionel C},
  journal={Neuroimage},
  volume={65},
  pages={540--555},
  year={2013},
  publisher={Elsevier}
}

@article{shojaie2022granger,
  author  = {Shojaie, Ali and Fox, Emily B.},
  title   = {Granger Causality: A Review and Recent Advances},
  journal = {Annual Review of Statistics and Its Application},
  volume  = {9},
  number  = {1},
  pages   = {289--319},
  year    = {2022},
  doi     = {10.1146/annurev-statistics-040120-010930},
  url     = {https://www.annualreviews.org/doi/10.1146/annurev-statistics-040120-010930}
}

@article{duker2025vector,
  author  = {Düker, Marie-Christine and Matteson, David S. and Tsay, Ruey S. and Wilms, Ines},
  title   = {Vector AutoRegressive Moving Average Models: A Review},
  journal = {WIREs Data Science and Statistics},
  volume  = {7},
  number  = {1},
  pages   = {e70009},
  year    = {2025},
  doi     = {10.1002/wics.70009}
}

@article{masini2023machine,
  title={Machine learning advances for time series forecasting},
  author={Masini, Ricardo P and Medeiros, Marcelo C and Mendes, Eduardo F},
  journal={Journal of Economic Surveys},
  volume={37},
  number={1},
  pages={76--111},
  year={2023},
  publisher={Wiley Online Library}
}

@article{Moulines2005OnRecursive,
  author = {Moulines, Eric and Priouret, Pierre and Roueff, Fran\c{c}ois},
  title = {On recursive estimation for time varying autoregressive processes},
  journal = {The Annals of Statistics},
  volume = {33},
  number = {6},
  pages = {2610--2654},
  year = {2005},
  publisher = {Institute of Mathematical Statistics},
  doi = {10.1214/009053605000000624},
  url = {https://projecteuclid.org/journals/annals-of-statistics/volume-33/issue-6/On-recursive-estimation-for-time-varying-autoregressive-processes/10.1214/009053605000000624.full}
}

@article{Kapetanios2003,
  title = {A note on an iterative least-squares estimation method for ARMA and VARMA models},
  author = {Kapetanios, George},
  journal = {Economics Letters},
  volume = {79},
  number = {3},
  pages = {305--312},
  year = {2003},
  publisher = {Elsevier},
  doi = {10.1016/S0165-1765(03)00005-3}
}

@article{JMLR:v22:20-539,
  author  = {Johannes Lederer and Michael Vogt},
  title   = {Estimating the Lasso's Effective Noise},
  journal = {Journal of Machine Learning Research},
  year    = {2021},
  volume  = {22},
  number  = {276},
  pages   = {1--32},
  url     = {http://jmlr.org/papers/v22/20-539.html}
}

@ARTICLE{lederer2023,
  author={Taheri, Mahsa and Lim, Néhémy and Lederer, Johannes},
  journal={IEEE Transactions on Information Theory}, 
  title={Balancing Statistical and Computational Precision: A General Theory and Applications to Sparse Regression}, 
  year={2023},
  volume={69},
  number={1},
  pages={316-333},
  doi={10.1109/TIT.2022.3203857}}

@article{vandeGeer2009OnThe,
  author    = {Sara A. van de Geer and Peter B{\"u}hlmann},
  title     = {On the conditions used to prove oracle results for the Lasso},
  journal   = {Electronic Journal of Statistics},
  volume    = {3},
  pages     = {1360--1392},
  year      = {2009},
  doi       = {10.1214/09-EJS506},
  url       = {https://doi.org}
}

@article{lee2016exact,
  author = {Lee, Jason D. and Sun, Dennis L. and Sun, Yuekai and Taylor, Jonathan E.},
  title = {Exact post-selection inference, with application to the lasso},
  journal = {The Annals of Statistics},
  volume = {44},
  number = {3},
  pages = {907--927},
  year = {2016},
  publisher = {Institute of Mathematical Statistics},
  doi = {10.1214/15-AOS1371},
  url = {https://projecteuclid.org/journals/annals-of-statistics/volume-44/issue-3/Exact-post-selection-inference-with-application-to-the-lasso/10.1214/15-AOS1371.short}
}
}

\clearpage
\appendix
\newgeometry{left=1in,right=1in,top=0.6in,bottom=0.6in}

\etocdepthtag.toc{appendix}
\begin{center}\bfseries\large Table of Contents for the Appendix\end{center}
\etocsettocstyle{}{}
{\etocsettagdepth{appendix}{subsection}
\etocsettagdepth{main}{none}
\tableofcontents}

\IfFileExists{Arxiv/07-appendix.tex}{%




\section{Supplementary material for background}

\label{sec: appen background}


In this section, we describe existing tuning strategies for Lasso in regression and VAR modeling, and discuss their limitations. Then we briefly describe PR in CD algorithms for Lasso, and sequential F-tests for model selection in classical statistics - two key technical ingredients used in $\autotune$.

\subsection{Tuning with cross-validation (CV)} Tuning $\lambda$ via K-fold cross-validation~\cite{hastie2015statistical} remains the most common tuning strategy among practitioners.  
Denote $\solution = \argmin_{\beta\in \real^p}  (1/2n) \, \|\text{Y}-\text{X}\beta\|^2 + \lambda\|\beta\|_1$. Once we have $\what{\beta}(\lambda)$ on a grid of $\lambda$ values, there are two choices in $\mathsf{glmnet}$ \cite{glmnet}:
\begin{enumerate}[label = ($\lambda$\arabic*)]
    \item CV(min): $\lambda$ which minimized the K-fold CV error among all the $\lambda$ values in the $\lambda$-grid,
    \label{item: cv lambda min}
    \item CV(1se): largest $\lambda$ such that its K-fold CV error was within 1 SD of the minimum CV error.
    \label{item: lambda 1se}
\end{enumerate}
Despite its popularity, the number of folds in CV is set to 5 or 10 in an ad hoc manner. The final chosen $\lambda$ \cref{item: cv lambda min} is also known to be  susceptible to the fold assignment used in CV  \cite{bovelstad2007predicting, roberts2014stabilizing}. 
%
%
%
%
Recent works have also shown that CV overfits the data \cite{wang2007tuning,lei2020cross} and loses accuracy \cite{bates2024cross}. 
To the best of our knowledge, there has been no systematic study to assess the quality of CV(1se) \cref{item: lambda 1se} solutions.


\subsection{Information criterion (IC) based tuning} 
AIC \cite{akaike2003new} and BIC \cite{schwarz1978estimating} introduced the idea of penalizing complexity of the model fitted for inducing model selection. 
The general formula is $n\log\twonorm{\text{Y} - \text{X}\what\beta}^2 + kp$ where $k=2$ for AIC  and $k = \log(n)$ for BIC. For tuning $\lambda$ in Lasso, their ease of computation, and avoiding repeated data-splitting calculations made them a simpler alternative to CV. But these information criteria were formulated under the assumption that $p$ is fixed and sample size $n\to\infty$, making them unsuitable for high-dimensional regime $n<p$ \cite{chen2008extended, zhao2015mixture}. Other information criterion like eBIC~\cite{chen2008extended}, HBIC \cite{zhao2015mixture} are also proposed for scenarios where $p$ grows with $n$. But these are not yet adopted widely and recent research \cite{10.1093/jjfinec/nbab023} suggests that they do not improve upon the performance of BIC in time series analysis. 


\subsection{\texorpdfstring{Linking $\lambda$ with error variance}{Linking lambda with error variance}} High-dimensional asymptotics of Lasso for linear model with Gaussian errors suggests choosing $\lambda$ in the order of $\sigma\sqrt{\log(p)/n}$ \cite{bickel2009simultaneous}. 
Since $\sigma$ is unknown in practice, a line of work reformulates the loss function \cref{eq: original lasso} or \cref{eq: Autotune lasso objective} to make the tuning of $\lambda$ independent of $\sigma$ \cite{belloni2011square, lederer2015don, wang2020tuning}. Statistical properties of these methods in high-dimensional time series are still topics ongoing research. In this work, we instead focus on tuning Lasso whose statistical properties are by now well-understood for time series data.

Another line of work, related more to our method, focuses on first estimating $\sigma$ from data and then using it for tuning $\lambda$. 
For example, Scaled Lasso \cite{sun2012scaled} introduced a $\sigma$-based tuning procedure for $\lambda$ by modifying the penalized Gaussian
log-likelihood objective \cref{eq: Autotune lasso objective}. 
\begin{align}
\min_{\beta, \sigma^2}\quad \sigma/2 + \frac{1}{2n\sigma} \|\text{Y}-\text{X}\beta\|^2 + \lambda_0 \|\beta\|_1.
\end{align}
In fact, \cref{eq: Scaled lasso objective} was motivated by the discussions \cite{sun2010comments, antoniadis2010comments} on the paper \cite{stadler2010}. Scaled Lasso \cite{sun2012scaled} proposes an iterative minimization algorithm alternating over $\beta$ and $\sigma^2$ for tuning $\lambda$ in Lasso: use the $\widehat{\beta}$ values from the Lasso solution path to calculate the Mean Squared Error (MSE), then do a degrees of freedom correction on MSE to estimate the $\sigma$, and then the $\what{\sigma}$ tunes the $\lambda$ for the next iteration in Scaled Lasso. Moreover, \cite{sun2012scaled} derived asymptotic guarantees of $\what\sigma$ of Scaled Lasso.

Additionally, \cite{yu2019estimating} proposed Natural and Organic Lasso which optimized the following modified loss functions respectively:
\begin{align}
\min_{\frac{1}{\sigma^2},\frac{\beta}{\sigma^2}\in \real_+ \times \real^p}& \, \, -\frac{1}{2}\log\parenth{\frac{1}{\sigma^2}} + \frac{1}{2n\sigma^2} \twonorm{\text{Y}-\text{X}\frac{\beta/\sigma^2}{1/\sigma^2}}^2 + \lambda\|\beta/\sigma^2\|_1,\qtext{and}
    \min_{\beta\in \real^p} \, \, \frac{1}{2n} \|\text{Y}-\text{X}\beta\|^2 + \lambda\|\beta\|_1^2
\label{eq: organic lasso}
\end{align}
\cite{yu2019estimating} arrived at Natural Lasso by making the original Lasso objective \cref{eq: original lasso} scale invariant. Natural Lasso still needs CV to tune its $\lambda$. For Organic Lasso, authors (Thm 12 of \cite{yu2019estimating}) gave a theoretical choice of $\lambda$ using the asymptotic theory of Lasso.

But it is known that $\widehat{\beta}$ from Lasso has shrinkage bias \cite{hastie2015statistical}, and using its corresponding MSE may introduce that bias in estimating $\what{\sigma}$, leading to inaccurate tuning. While such bias correction is usually done in the literature after the final Lasso fit using ordinary least squares \cite{MEINSHAUSEN2007374, postlasso}, an attempt to correct or even reduce such shrinkage bias in $\what{\sigma}$ during the Lasso fit may offer some advantage in tuning. We attempt to achieve this in $\autotune$.







\subsection{\texorpdfstring{Tuning $\lambda$ in VAR models}{Tuning lambda in VAR models}} 
Vector autoregression (VAR) is a workhorse model in multivariate time series data, widely used in macroeconomics \cite{banbura2010large, bernanke2005measuring}, portfolio selection in finance \cite{fan2011sparse}, identifying gene regulatory networks in genomics \cite{michailidis2013autoregressive, broll2025prolong} and for understanding functional connectivity between different brain regions in neuroscience  \cite{smith2012future, seth2013granger}. Formally, a VAR($d$) model on a $p$-dimensional stationary time series $\braces{Z^t}_{t \geq 1}$
with no contemporaneous association takes the form 
\begin{align}
    \mrm{VAR}\parenth{d}: Z^t = \sum_{i=1}^d A_i Z^{t-i}+\varepsilon^t, \quad \varepsilon^t\stackrel{\iid}{\sim} \parenth{0, \Sigma_{\varepsilon}},\\ \quad \Sigma_{\varepsilon}= \diag\parenth{\sigma^2_1,\dots,\sigma^2_p},
\end{align}
where $A_1,\dots,A_d$ are $p\times p$ transition matrices which need to be estimated and $\sigma^2_1,\dots,\sigma^2_p$ are the noise variances.  Given $n + d$ stationary observations of $\braces{Z^t}_{t \geq 1}$, all the $d$ VAR transition matrices can be estimated in the autoregressive design \cite{hyndman2018forecasting}
\begin{align}
    \label{eq: VAR regression design}\underbrace{\left[\begin{array}{c}
\left(Z^{n+d}\right)^{\top} \\
\vdots \\
\left(Z^{d+1}\right)^{\top}
\end{array}\right]}_{\text{Y}_{n\times p}}=\underbrace{\left[\begin{array}{cccc}
\left(Z^{n + d - 1}\right)^{\top} & \left(Z^{n + d - 2}\right)^{\top} & \dots & \left(Z^{n}\right)^{\top} \\
\vdots & \vdots & \vdots& \vdots \\
\left(Z^{d}\right)^{\top} & \left(Z^{d - 1}\right)^{\top} & \dots & \left(Z^{1}\right)^{\top}
\end{array}\right]}_{\text{X}_{n\times dp}}\underbrace{\brackets{ \begin{array}{c}
     A_1^{\top}  \\
      A_2^{\top} \\
      \vdots \\
      A_d^{\top}
\end{array}}}_{\Phi_{dp\times p}} +\underbrace{\left[\begin{array}{c}
\left(\varepsilon_{n+d}\right)^{\top} \\
\vdots \\
\left(\varepsilon^{d+1}\right)^{\top}
\end{array}\right]}_{E_{n \times p}},
\end{align}
where $\Phi$ is the $dp\times p$ matrix constructed by stacking $A_1^{\top}, A_2^{\top},\dots, A_d^{\top}$. The parameter space dimension ($dp^2$) grows quadratically with the number of time series $p$, and the OLS estimate of $\Phi$ will be consistent only if $n>dp^2$, which is uncommon in modern VAR applications. So, it is common to assume $\Phi$ is sparse and use $\ell_1$-penalized least squares regressions \cite{basu2015regularized}. We obtain the VAR Lasso estimator of $\Phi$ by conducting $p$ separate, columnwise Lasso regressions. Mathematically,

\begin{equation}
    \what{\Phi}_{\cdot,i}=\underset{\beta \in \mathbb{R}^{dp}}{\operatorname{argmin}} \frac{1}{2n}\left\|\text{Y}_{\cdot, i}-\text{X} \beta\right\|^2 + \lambda_i\onenorm{\beta}, \quad \text { for } i=1, \ldots, p.
\end{equation}

\subsection{Time series cross-validation (TSCV)} A common approach to tune these $p$ different values of $\lambda_i$ is TSCV (Sec. 5.10 of \cite{hyndman2018forecasting}). It is more constrained because the temporal order of the data must be preserved when choosing the different folds. A ``rolling window'' is adopted, where the dataset is divided into $2K$ folds along the temporal dimension. Then the $(K+1)^{th}$ fold is assigned the role of a test set, and all the folds corresponding to previous time points are used as the training data to fit the VAR Lasso on the original grid of $\lambda_i$ separately for each $i\in \braces{1,\dots,p}$. We repeat this process $K$ times, where every fold in the latter half plays the role of test fold once, and in each repetition, all the previous folds are used as training data. Lastly for each $i\in \braces{1,\dots,p}$, TSCV averages the $K$ mean squared errors (MSE) for every value in the grid of $\lambda_i$ to get the mean TSCV error, and chooses a  $\lambda_i$ which minimizes this. 
As a result, TSCV has some statistical limitations over the usual CV \cite{bergmeir2012use, bergmeir2018note}. For example, constraints in data-splitting for TSCV reduces the effective sample size available for tuning of $\lambda$.

Moreover, running $p$ separate TSCV on each column can be computationally prohibitive for analysis of modern day large time series datasets. Hence, 
prior software \cite{bigtimeRpackage} have often focused on uniform penalization across all the separate column-wise lasso regressions, which is inadequate for handling heterogeneity across multiple time series.


\subsection{Automatic tuning for VAR Lasso}
Recently \cite{kock2025data} proposed an automated tuning mechanism for VAR Lasso, building upon the iterative algorithm proposed in \cite{belloni2012sparse}. Denote $\what\Phi^{(t)}$ as the estimator of $\Phi$ at the $t^{th}$ iteration. Then, \cite{kock2025data} iteratively computes the following columnwise weighted Lasso estimator for $t \geq 1$
\begin{align}
    &\what{\Phi}^{(t)}_{\cdot,i}=\underset{\beta \in \mathbb{R}^{dp}}{\operatorname{argmin}}\hspace{0.1cm} \frac{1}{n}\left\|\text{Y}_{\cdot, i}-\text{X} \beta\right\|^2 + \frac{\lambda}{n}\onenorm{\Gamma^{(t)}_i\beta},\\
    &\text{where }\Gamma^{(t)}_i = \diag\parenth{\what v^{(t)}_{i,1}, \dots, \what v^{(t)}_{i,pq}}, \quad \text { for } i=1, \ldots, p\\ & \text{ with } \what{v}^{(t)}_{i,j} = \frac{1}{\sqrt{n}}\sqrt{\sum_{r = 1}^n\text{X}_{r,j}^2\parenth{\text{Y}_{r, i}-\text{X}_{r:}\what\Phi^{(t-1)}_{\cdot, i}}^2}.
\end{align}
Here the weights $\Gamma_i$ are adjusted according to the residuals in the previous iteration. This strategy is similar to the iterative algorithm precursor to Scaled Lasso \cite{wu2020survey}, and shares its limitation of shrinkage bias in $\what{\Phi}$ that are used in noise estimator $\what{v}$. In \cref{subsec: VAR} we compare  different tuning strategies in VAR Lasso including $\autotune$, TSCV, AIC, BIC, and \cite{kock2025data} on simulated time series data.

\subsection{Coordinate descent (CD) and partial residuals} \label{app: cd and partial res} The ease of implementing coordinate descent, availability of coordinate-wise closed-form updates, augmented with all the computational tricks of warm starts, and active set selection makes it the default and the fastest algorithm~\cite{10.1214/07-AOAS131} for researchers using Lasso over other algorithms like proximal gradient descent \cite{nesterov2013introductory} and LARS \cite{efron2004least}. 
The idea of CD is first to fix the $\lambda$ in \cref{eq: original lasso} and then iteratively optimize over each parameter holding other parameters constant at their current values.

We provide the pseudocode of Lasso at a fixed $\lambda$ in Algorithm \ref{algo: CD for lasso}. A core component of CD updates in Lasso is the PR $\partialres[j]$, $\forall j\in\braces{1,\dots,p}$, which comes out to be $\partialres[j] = r + \text{X}_{.,j}^{\top}\what\beta_j$, where $r = \text{Y} - \sum_{k\in \braces{1,\dots,p}}\text{X}_{.,k}^{\top}\what{\beta}_k$ is the full residual. PR are continuously computed and updated inside every CD step, but there has been only a limited discussion about it in the prior literature. We take a deeper look into the partial residuals in the next section.

In Lasso, the standard practice \cite{hastie2015statistical} is to calculate $\solution$ over a decreasing $\lambda$ grid of equally spaced values (on log scale) from $\lambda_{\max}$ to (almost) 0. $\lambda_{\max} = \max_{j\in[p]}\frac{1}{n}\inner{\text{Y}}{\text{X}_{.,j}}$ is the smallest $\lambda$ which ensures $\solution[\lambda] = 0$. The Lasso solution for each successive $\lambda$ in the grid is then obtained by initializing the coordinate descent (CD) algorithm with the solution from the previous $\lambda$, a technique known as warm starts \cite{hastie2015statistical}. This process is repeated until the smallest $\lambda$ in the grid is reached, yielding $\solution[\lambda_{\min}]$. This entire procedure is often called pathwise coordinate descent, and the set of solutions $\braces{\solution[\lambda_{\max}],\dots,\solution[\lambda_{\min}]}$ is interchangeably called solution path or regularization path. Warm starts accelerates the solving of pathwise coordinate descent as compared to running CD separately across different values of $\lambda$ in the grid.

Sometimes, to further reduce computational costs, a subset of coefficients called ``active set'' is selected \cite{10.1111/j.1467-9868.2011.01004.x, hastie2015statistical} at each $\lambda$ in the grid.

\subsection{Inner algorithm for solving Scaled Lasso} Scaled Lasso internally uses least angle regression (LARS), which is slower than pathwise CD \cite{10.1214/07-AOAS131}. 

\subsection{Bilevel optimization (\textsf{sparse-ho})}
The \textsf{sparse-ho} method \cite{bertrand2020implicit} treats the
log-penalty $\eta=\log(\lambda)$ as a continuous hyperparameter and
selects it through bilevel optimization. Given training and validation
samples $(\text{X}_{\mathrm{tr}},\text{Y}_{\mathrm{tr}})$ and
$(\text{X}_{\mathrm{val}},\text{Y}_{\mathrm{val}})$, respectively, it
solves
\begin{align}
\widehat{\eta}
&\in
\underset{\eta\in\mathbb{R}}{\operatorname{argmin}}\;
\underbrace{
\mathcal{L}_{\mathrm{val}}
\left(\widehat{\beta}(\eta)\right)
}_{\text{outer problem}},
\qquad
\mathcal{L}_{\mathrm{val}}(\beta)
=
\frac{1}{2n_{\mathrm{val}}}
\left\|
\text{Y}_{\mathrm{val}}-\text{X}_{\mathrm{val}}\beta
\right\|_2^2,
\label{eq:sparseho-outer}
\\
\text{subject to}\qquad
\widehat{\beta}(\eta)
&\in
\underset{\beta\in\mathbb{R}^{p}}{\operatorname{argmin}}\;
\underbrace{
\left\{
\frac{1}{2n_{\mathrm{tr}}}
\left\|
\text{Y}_{\mathrm{tr}}-\text{X}_{\mathrm{tr}}\beta
\right\|_2^2
+
e^\eta\|\beta\|_1
\right\}
}_{\text{inner problem}} .
\label{eq:sparseho-inner}
\end{align}
Thus, the inner problem fits a Lasso model at penalty
$\lambda=e^\eta$, while the outer problem chooses the penalty by
minimizing its validation error.

Let
$\widehat{J}(\eta)=\partial\widehat{\beta}(\eta)/\partial\eta$ be the weak graident with respect to $\eta$.
The outer problem is optimized using the weak gradient
\begin{align}
\frac{\mathrm{d}}{\mathrm{d}\eta}
\mathcal{L}_{\mathrm{val}}
\left(\widehat{\beta}(\eta)\right)
=
\widehat{J}(\eta)^\top
\frac{\text{X}_{\mathrm{val}}^\top
\left\{
\text{X}_{\mathrm{val}}\widehat{\beta}(\eta)
-\text{Y}_{\mathrm{val}}
\right\}}
{n_{\mathrm{val}}}.
\label{eq:sparseho-hypergradient}
\end{align}
In particular, if
$\widehat{\mathcal{S}}(\eta)
=\operatorname{supp}\{\widehat{\beta}(\eta)\}$
is locally unchanged and the corresponding Gram matrix is nonsingular,
implicit differentiation of the inner optimality conditions gives
\begin{align}
\widehat{J}_{\widehat{\mathcal{S}}}(\eta)
&=
-e^\eta
\left(
\frac{
\text{X}_{\mathrm{tr},\widehat{\mathcal{S}}}^{\top}
\text{X}_{\mathrm{tr},\widehat{\mathcal{S}}}
}{n_{\mathrm{tr}}}
\right)^{-1}
\operatorname{sign}
\left(
\widehat{\beta}_{\widehat{\mathcal{S}}}(\eta)
\right),
&
\widehat{J}_{\widehat{\mathcal{S}}^c}(\eta)
&=0.
\label{eq:sparseho-jacobian}
\end{align}
Rather than explicitly forming this inverse, \textsf{sparse-ho}
computes the Jacobian iteratively on the active set and uses the
resulting hypergradient to update $\eta$. This exploits the sparsity of
the Lasso solution and avoids searching over a prespecified grid of
penalty values.

\subsection{Sequential F-tests}

Sequential F-tests (Chapter 4.6 of \cite{montgomery2020introduction}) is a classic model selection tool which is often employed to calculate ANOVA tables \cite{carpackage}. Given the response Y and $p$ predictors X$_j$, $j\in \seqp$, it looks into the following sequence of nested linear models and their corresponding residual sum of squares (RSS):
\begin{align}
    \mc{M}_k:  \text{ Y} &= \text{ X}_{1}\beta_1 + \text{ X}_{2}\beta_2 + \dots + \text{X}_{k}\beta_k + \varepsilon,\\
    \text{RSS}(\mc{M}_k) &= \sumn \parenth{\text{y}_i - \what{\text{y}}_i}^2, \qquad  k \in \braces{1,\dots,p} 
\end{align}
where the predictions $\what{\text{y}}_i$ comes from fitting $\mc{M}_k$ to data. Then, a sequential F-test iteratively  performs the hypothesis test $\hypo[k]:\braces{\beta_k=0}|\parenth{\beta_1, \dots, \beta_{k - 1}}$ using the statistic
\begin{align}
\label{eq: seq f test}
    F = \frac{\rss(\mc{M}_{k - 1}) - \rss(\mc{M}_{k})}{\rss(\mc{M}_k)/(n - k)}\underset{\hypo[k]}{\sim} F_{1, n - k}
\end{align}
until either we arrive at a $k_0\in\seqp$ where we fail to reject the $\hypo[k_0]$ or we reject $\hypo[j]$ for all $ j\in \seqp$. If the procedure stops, we conclude that the significant predictors for the regression of Y on X are $\braces{1,\dots,k_0 - 1}$, otherwise all the predictors are included in the model. Note that in $n<p$ regime, we will always arrive at a $k_0\leq n$ where the sequential F-test stops. 
The rank order of predictors in which the nested models $\mc{M}_k$ are constructed is important for selecting the final model.

\section{Supplementary methods section}
\subsection{\texorpdfstring{$\autotune$ lasso with active set selection}{autotune lasso with active set selection}}
\label{sec: app methods}

\begin{algorithm}[ptb]
\caption{Active Autotune Lasso\ --\ Return estimates of regression coefficients $\beta$. } 
  \begin{algorithmic}
      \label{algo:Autotune_lasso_active}
  \STATE \textbf{Input:} Predictors X (normalized), response Y, sequential F-test significance level $\alpha $, $\Activeselect$
  \STATE Initialize $\what\beta \gets 0$, $r \gets $Y, $\widehat{\sigma}^2\gets \mathrm{Var}$(Y), $\lambda_0\gets\frac{1}{\mathrm{Var}(\mrm{Y})}\|\frac{\mrm{X}^{\top}\mrm{Y}}{2n}\|_{\infty}$, $\Active\gets 1:p$, $\support\gets\phi$, sigma.update.flag $\gets\mathsf{TRUE}$
    \WHILE{$\mrm{error}\geq10^{-3}$} 
        {
        \STATE $\supportold\gets\support$, $\betaold \gets \what\beta$
        \FOR{$i = 1,\ldots,p$}
        \STATE
        $j\gets \Active[i]$
        \STATE
        $\what\beta_j \gets \mathsf{Soft.Threshold}_{\lambda_0\what\sigma^2}\parenth{\frac{1}{n}\mrm{X}_{\cdot,j}^{\top}r + \what\beta_j}$
        \STATE
        $r \gets r + \mrm{X}_{\cdot,j}\parenth{\betaold_j-\what\beta_j}$
        
        \ENDFOR
    \STATE Calculate partial residuals $\partres \gets r + \mrm{X}_{\cdot,i}\what\beta_i$  
    \STATE
    $\bigg\{\twonorm{\partres[\pi(i)]}\bigg\}_{i=1}^n \gets \mrm{Sort}\bigg(\twonorm{\partres[1]}, \ldots, \twonorm{\partres[p]}\bigg)$ in decreasing order.
    \STATE
    $\Active\gets \braces{\pi(1),\ldots,\pi(p)}$
    \IF{$\mrm{sigma.update.flag} == \mathsf{TRUE}$}
        \STATE$\what\sigma^2, \support, \Active \gets$ {\normalsize\SUlink}\,$\parenth{\mrm{X, Y}, \alpha, \Active}$
    \ENDIF
    \IF{$\support\subseteq\supportold$}
            \STATE sigma.update.flag $\gets\mathsf{FALSE}$
            \STATE$\supportold\gets\phi$\\
            \IF{$\Activeselect$ is $\msf{TRUE}$}
                \STATE $\what\beta, r$, $\Active \gets$ $\Activelink\parenth{\mrm{X},r,\what\beta, \Active, \support,\lambda_0\what\sigma^2}$ 
            \ENDIF
        \ENDIF
        \STATE
        $\mrm{error} \gets \|\what{\beta}- \betaold\|_1 / \| \betaold\|_1$
        }
    \ENDWHILE
    \STATE \textbf{Return:} $\what\beta$
  \end{algorithmic}
\end{algorithm}

\begin{algorithm}[ptb]
\caption{Active Set Selector\ --\ Performs active set screening of predictors.} 
  \label{algo:active_set_selector}
\begin{algorithmic}
\STATE \textbf{Input:} X, $r$, $\what\beta$, $\Active$, $\support$, thresholding parameter $\lambda$
  \STATE
    Initialize $\msf{Active.Set}\gets\support$\\
    \FOR{$i = \parenth{\abss{\support}+1},\ldots,p$}
    \STATE
        $j\gets \Active[i]$
        \IF{$\abss{\frac{1}{n}\inner{\mrm{X}_{\cdot,j}}{r}}\geq \lambda$}
        \STATE
            Add $\mrm{X}_{\cdot,j}$ to the back of $\msf{Active.Set}$
        \ELSE
        \STATE
            $r\gets r + \mrm{X}_{\cdot,j}\what\beta_j$\\
            $\what\beta_j\gets 0$
        \ENDIF
    \ENDFOR
    \STATE $\Active \gets \msf{Active.Set}$
\STATE \textbf{Return:} $\what\beta$,  $r$, $\Active$.
\end{algorithmic}
\end{algorithm} 

\newpage
\subsection{\texorpdfstring{Sensitivity of $\autotune$ to the choice of $\tau$ in $\lambda_0 = \tau \infnorm{\text{X}^{\top} \text{Y}/n}/Var(Y)$}{Sensitivity of autotune to tau in lambda0}}
\label{app: sensitivity to tau plots}
\leavevmode\par
\begin{figure}[H]
   \centering
   \includegraphics[page = 2, width=0.48\linewidth]{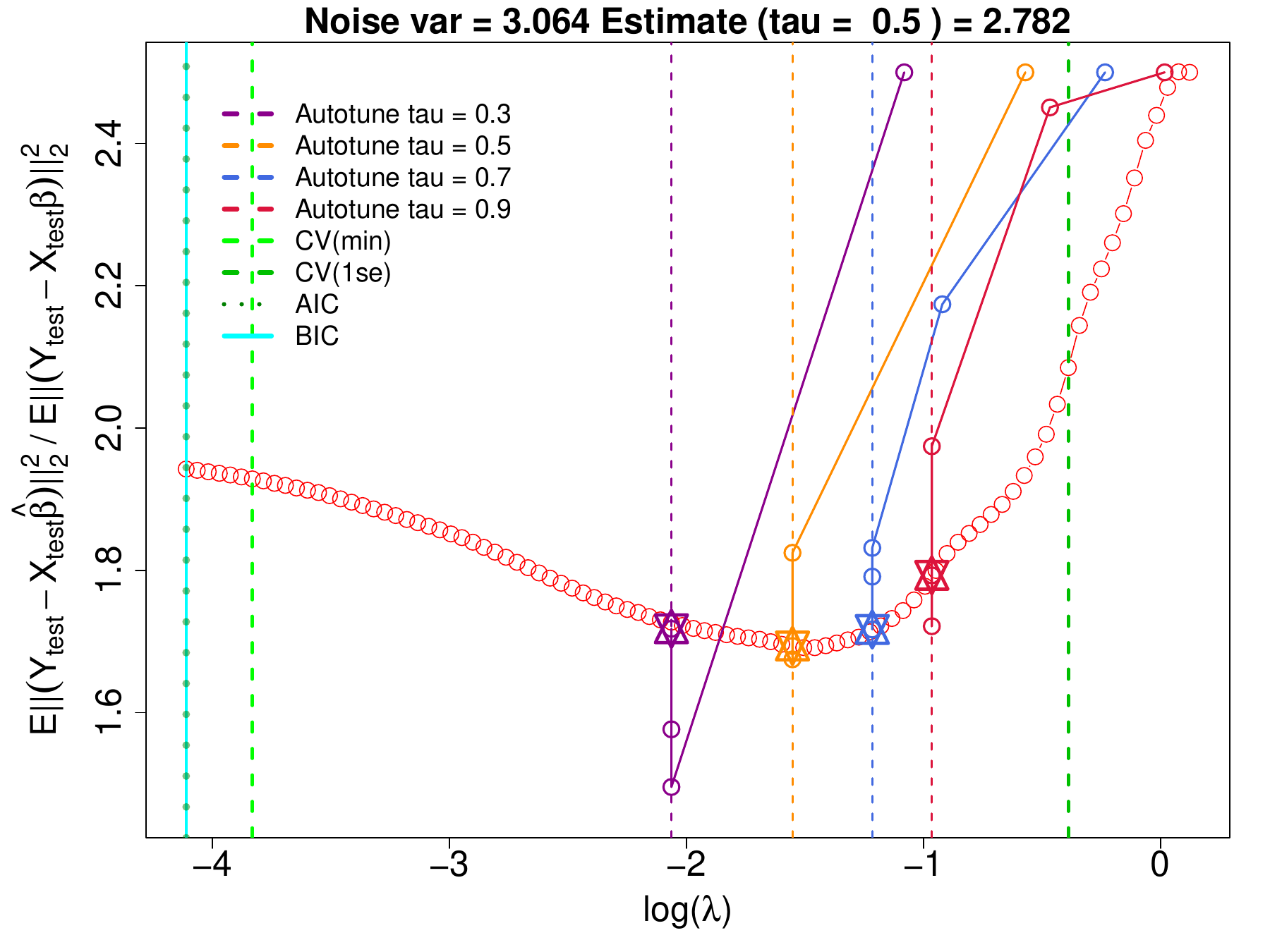}
   \includegraphics[page = 4, width=0.48\linewidth]{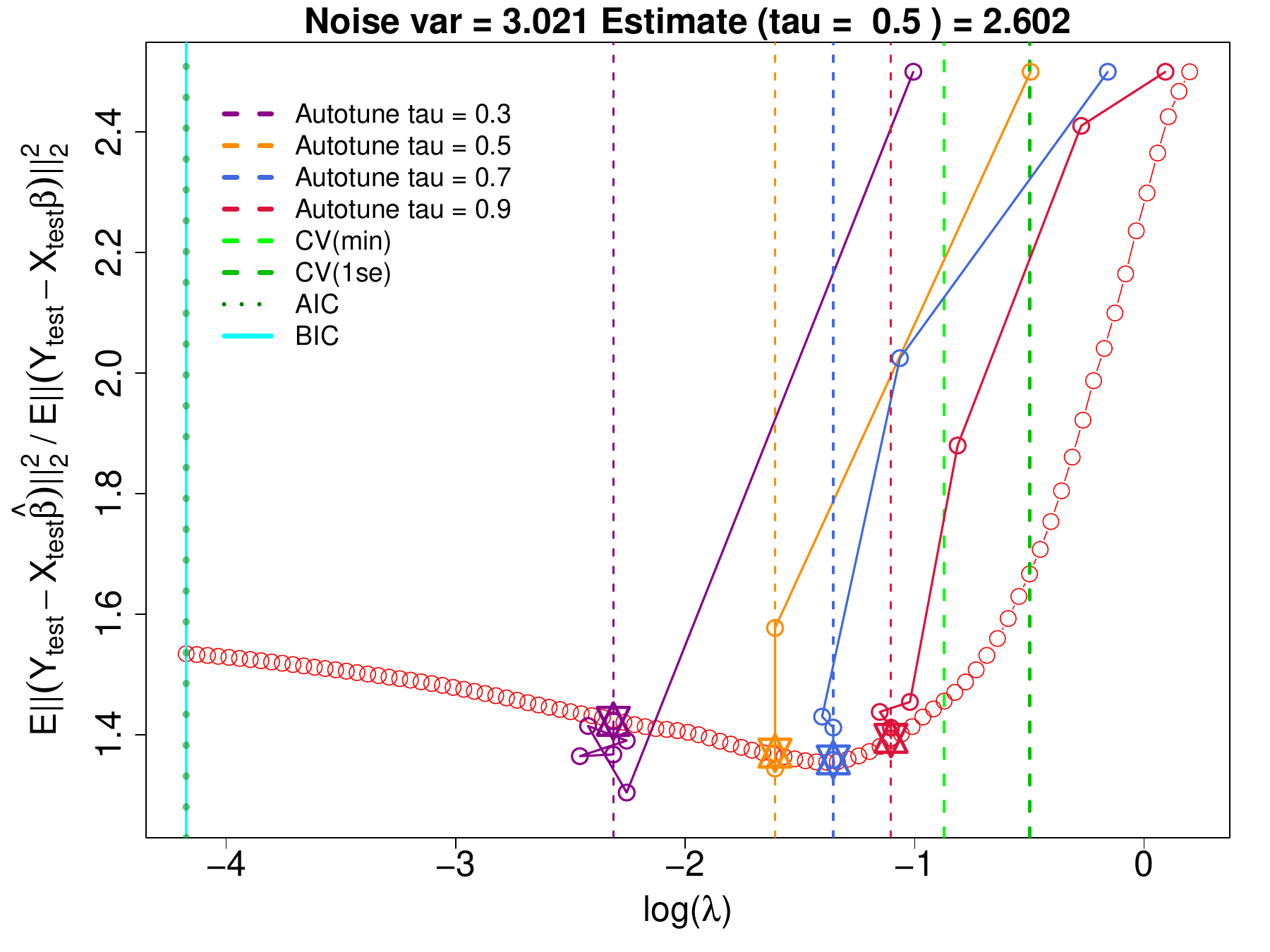}
   \includegraphics[page = 13, width=0.48\linewidth]{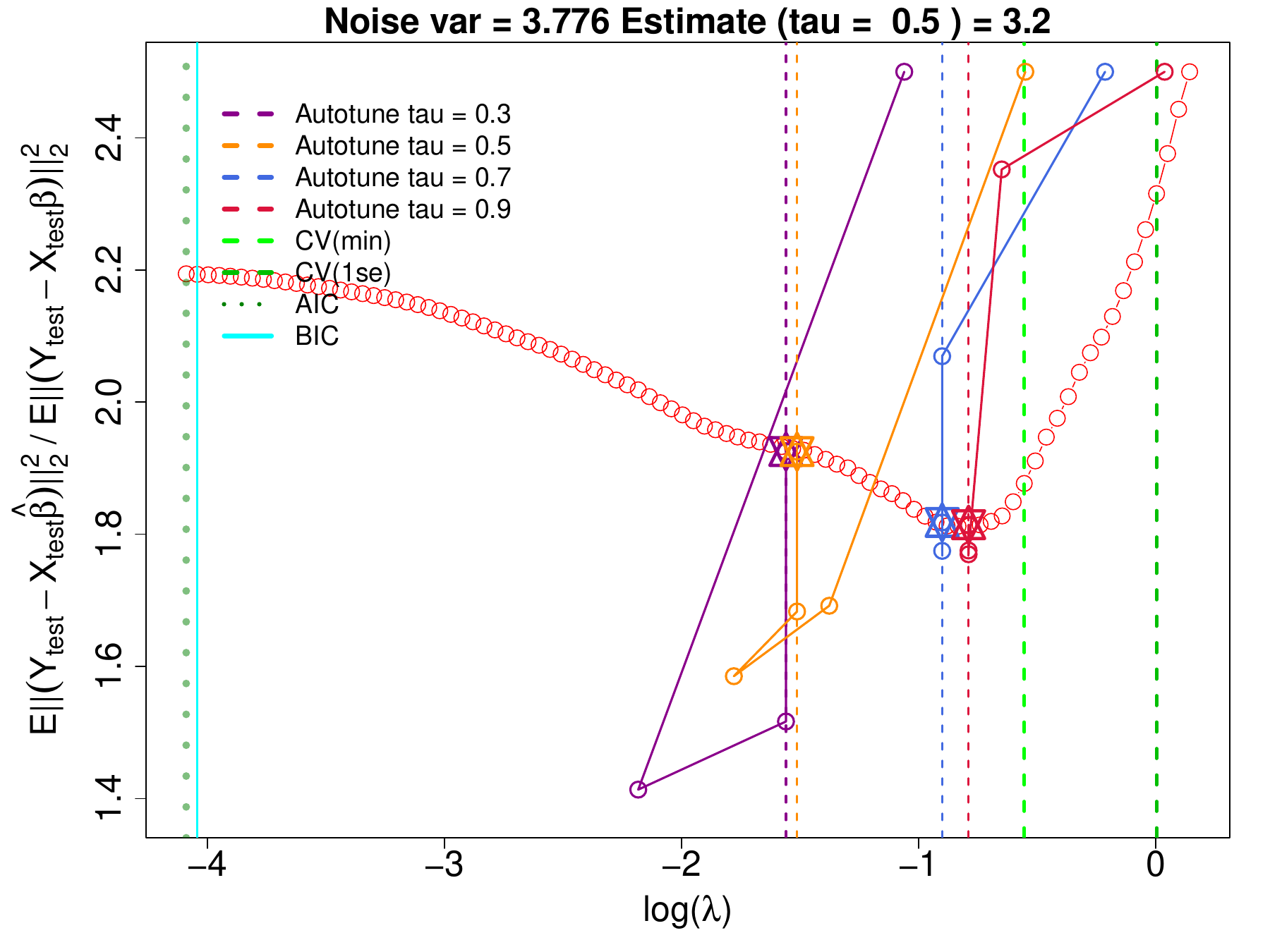}
   \includegraphics[page = 20, width=0.48\linewidth]{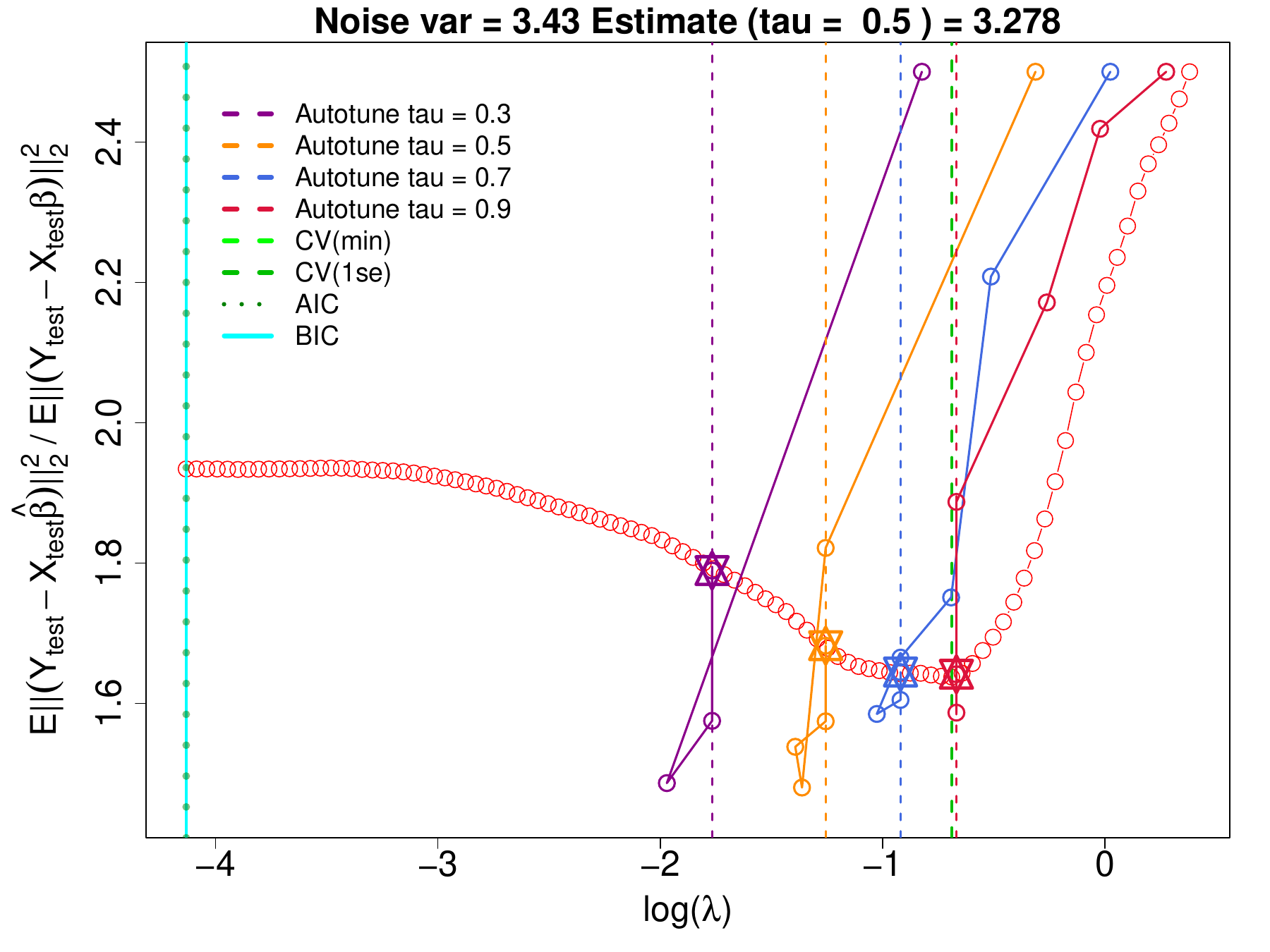}
   \caption{Relative test error (RTE) of  $\autotune$ with different values of $\tau$ and CV Lasso (red) solution paths against $\log(\lambda)$ in a simulated Lasso regression ($n = 80,$ $ p=750$, $s = 5$, SNR $=1.5,~\rho = 0$ and Beta-type 2). Color coded vertical bars show the $\lambda$ picked by different benchmarks and different values of $\tau$ inside $\autotune$. Simulation shows the final $\lambda$ picked by $\autotune$ is sensitive to the choice of $\tau$. 
   In \textbf{[Bottom- Right]} figure, CV(min) overfits and selects the smallest $\lambda$ in the grid, hence the vertical line depicting $\lambda$ selected by CV(min) gets hidden behind that of BIC. Empirical noise variance and estimate of noise variance given by $\autotune(\tau = 0.5)$ are shown on top of the figure.}
\end{figure}

\clearpage

\section{Deferred explanations and plots of simulation section}

\subsection{Complete record of Beta-types used in the simulation experiments}
\label{subsec: All Beta-types}

We consider five different schemes for generating the true coefficients $\beta\in\real^p$ in simulations:
\begin{enumerate}[label = (B\arabic*)]
    \item Beta-type 1: $\beta$ has $s$ entries equal to 1, placed equidistantly between indices 1 and $p$, with all other entries set to 0. \label{item: betatype1}
    \item Beta-type 2: The first $s$ entries of $\beta$ are equal to 1, and the remaining entries are 0. \label{item: betatype2}
    \item Beta-type 3: The first $s$ entries of $\beta$ decreases linearly from 10 to 0.5, while the rest are set to 0.
    \item Beta-type 4: (Approximate Sparsity) The first $s$ entries of $\beta$ are equal to 1, and the remaining entries decay exponentially, specifically $\beta_i = 0.5^{i-s}$ for $i = s + 1, \dots, p$.
    \item Beta-type 5: $\beta$ has $s$ nonzero entries decreasing linearly from 10 down to 0.5 and placed equally spaced across the index range $1$ to $p$, with all other entries set to 0.
    \label{item: betatype5}
\end{enumerate}

The first four setups were investigated in \cite{hastie2020best}, while we also bring in the last setup i.e. Beta-type 5 to study the effects when true predictors have unequal coefficients and get correlated with only the spurious predictors.

\subsection{Software used for implementing benchmarks}
\label{subsec: benchmark R packages}

For all our benchmarks, we used the default settings of their corresponding R packages.

The benchmarks of high-dimensional regression in \cref{subsec: regg setup sim comparison}, CV Lasso and Scaled Lasso are implemented via $\mathsf{cv.glmnet}$ function (from $\glmnet$ package \cite{glmnet}) and $\mathsf{scalreg}$ function (from $\mathsf{scalreg}$ package \cite{scalreg}) respectively.

For VAR experiments of \cref{subsec: VAR}, authors of \cite{kock2025data} had kindly shared their codebase of KPS Lasso with us for numerical comparison. For other benchmarks, we first fit the whole Lasso path using the $\glmnet$ R package for each of the $p$ columnwise regressions and then choose the final $\lambda_i$ via TSCV, AIC or BIC defined in \cref{sec: appen background} for $i \in \braces{1,\dots, p}$.

In \cref{subsubsec: simulation results for noise estimation and sparsity diagnostics}, for noise estimation experiments, we implement Natural and Organic Lasso \cite{yu2019estimating} via their R package, \texttt{natural}. For  inference experiments, we use the R package \texttt{hdi} for implementing  De-sparsified Lasso \cite{dezeure2015high} which provides us the option of manually plugging in the value of $\what\sigma$ of different algorithms used in \cref{subsubsec: simulation results for noise estimation and sparsity diagnostics}.

For comparison against BLO and a state-of-the-art python implementation in Appendix \ref{app: comparison vs BLO skglm}, we utilize the python packages \texttt{sparse-ho} and \texttt{skglm}.

\subsection{Interpretation of SNR} 
\label{subsec: interpretation of SNR appen}
Our definition of SNR is in line with the SNR in \cite{hastie2020best}. They gave a natural interpretation of the SNR in terms of a quantity called the Proportion of Variance Explained (PVE).
\begin{align}
    \mrm{PVE}(\what\beta) = 1 - \frac{\E\|\text{Y}_{\mrm{test}} - \text{X}_{\mrm{test}}\what\beta\|_2^2}{\mrm{Var}(\text{Y}_{\mrm{test}})}\\ = 1 - \frac{\parenth{\what\beta - \beta}^\top\Sigma\parenth{\what\beta - \beta} + \sigma^2}{\beta^\top\Sigma\beta + \sigma^2}\label{eq: pve}
\end{align}
Notice that the PVE of the optimal prediction $\what\beta = \beta$ is PVE($\beta$) $= \frac{\snr}{\snr + 1}$. So, SNR sets the best possible PVE that can be achieved by a linear model, hence it serves as a good indicator in the simulations of how hard is the regression problem in terms of prediction. For instance, SNR of 1 means that linear model can explain atmost 50\% of the total variation in the response variable Y.

\subsection{Choice of the range for SNR}
\label{subsec: why this range of SNR}
Our primary focus in the paper is the performance of Lasso in the high-dimensional configuration of $p = 750$. We noticed that for SNRs below 0.66 (max PVE drops to 40\%), all the methods, including  $\autotune$, are not doing any meaningful estimation (RMSE nearly 1), hence our SNR grid starts from 0.66. On the other side, we observe that all methods achieve the perfect AUROC of 1 in a large range of setups in high dimensions at SNR $=4$, so we set the maximum SNR to 4.

\subsection{\texorpdfstring{Deferred plots of \cref{subsec: regg setup sim comparison}}{Deferred plots of high-dimensional regression}}
\leavevmode\par

\begin{figure}[H]
    \centering
    \begin{tabular}{c}
         \quad\qquad \textbf{Moderate Dimensional setup:} $n = 80, \hspace{0.1cm} p=150, \hspace{0.1cm} s=5$, \textbf{correlation:} $\rho = 0.35$ \\
         \textbf{left half:} Beta-type 1 and \textbf{right half:} Beta-type 2\\
        \includegraphics[trim=0in 0in 0in 0in, clip, width = 0.49\textwidth]{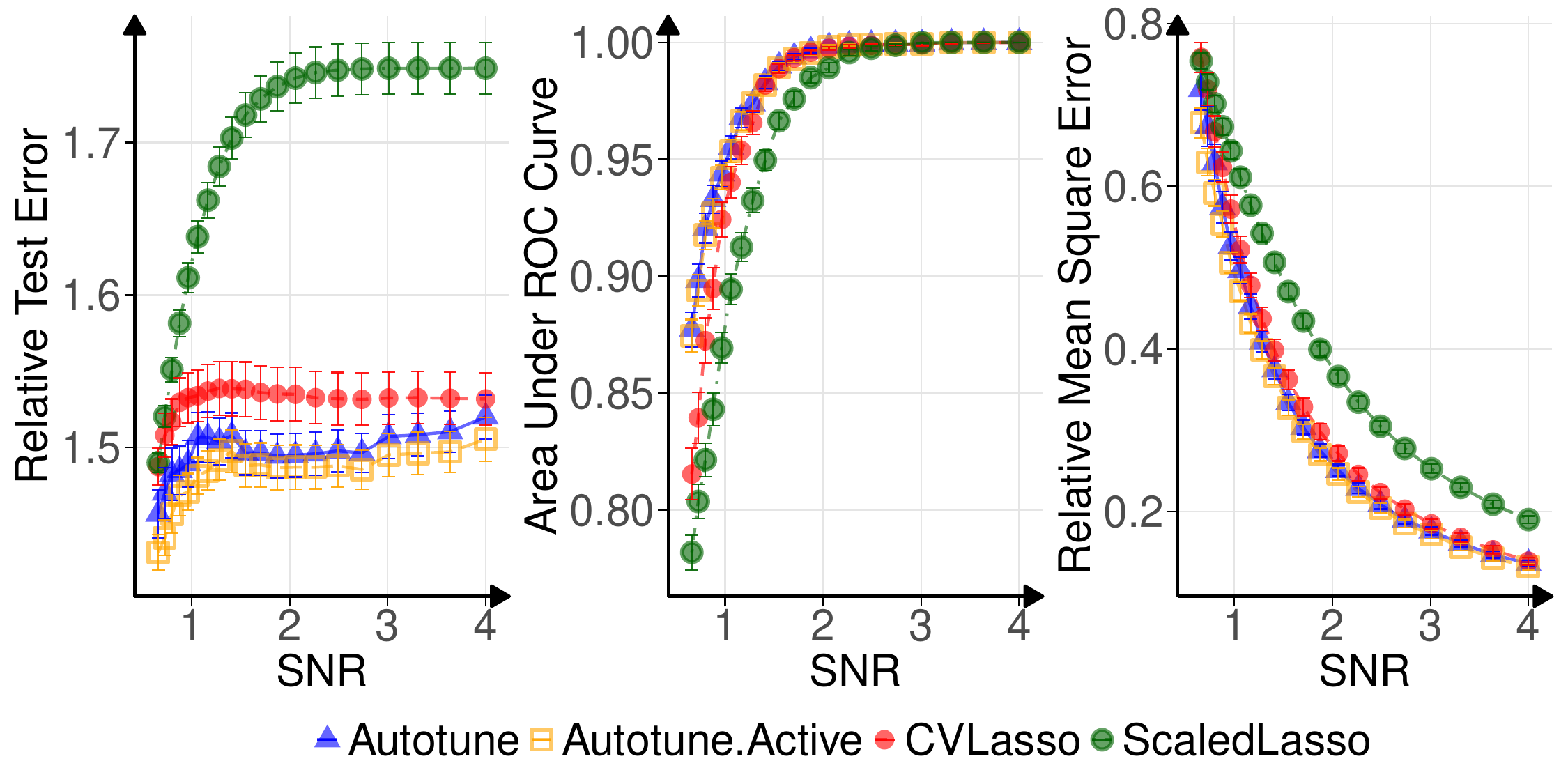}
        \includegraphics[trim=0in 0in 0in 0in, clip, width = 0.49\textwidth]{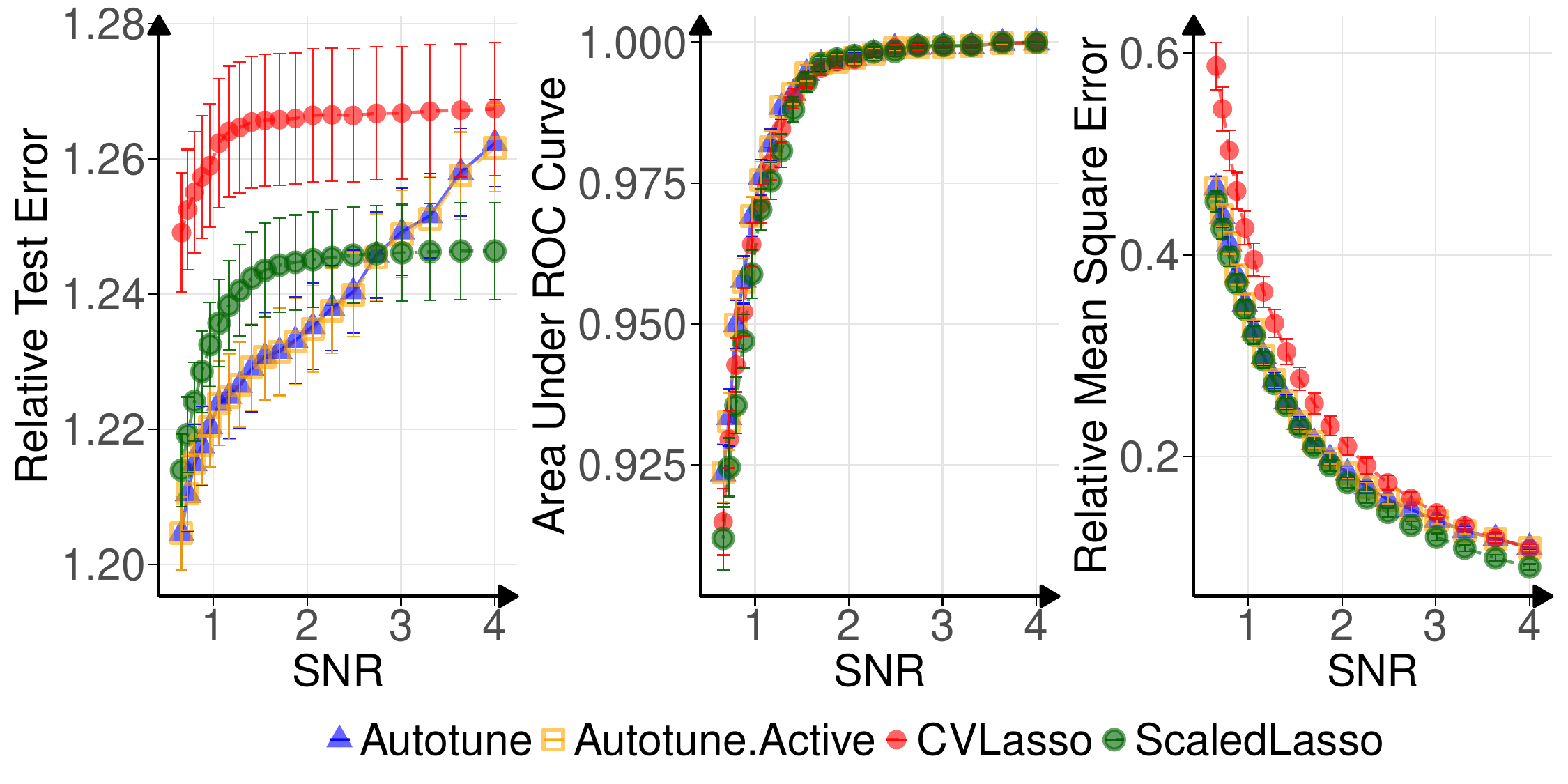}
    \end{tabular}
    \\[2mm]
    
    \caption{RTE, AUROC and RMSE (with 1SE bars) of $\autotune$ and benchmark methods for Lasso regression, plotted against SNR.
    }
    \label{fig: moderate dim accuracy plot}
\end{figure}

\subsection{\texorpdfstring{Deferred plots of \cref{subsec: VAR}}{Deferred plots of VAR Lasso}}
\leavevmode\par

\begin{figure}[H]
    \centering
    \begin{tabular}{c|c}
        \textbf{2x2 Block VAR(1) setup:} SNR = 2.5 & \textbf{ diagonal VAR(1) setup:} SNR = 2.5\\
        Time Series length 200 & Time Series length 100 \\
         \includegraphics[trim=0in 0in 0in 0in, clip, width = 0.49\textwidth]{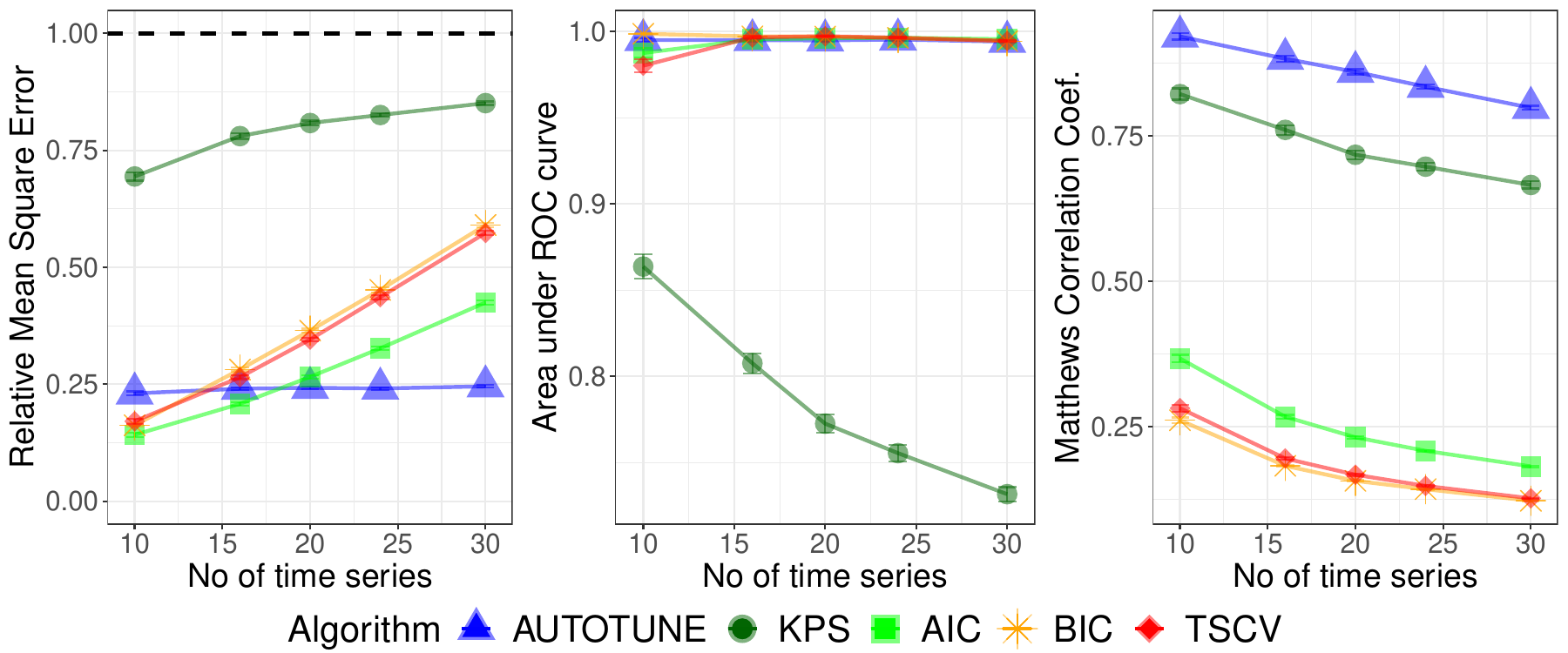} &
        \includegraphics[trim=0in 0in 0in 0in, clip, width = 0.49\textwidth]{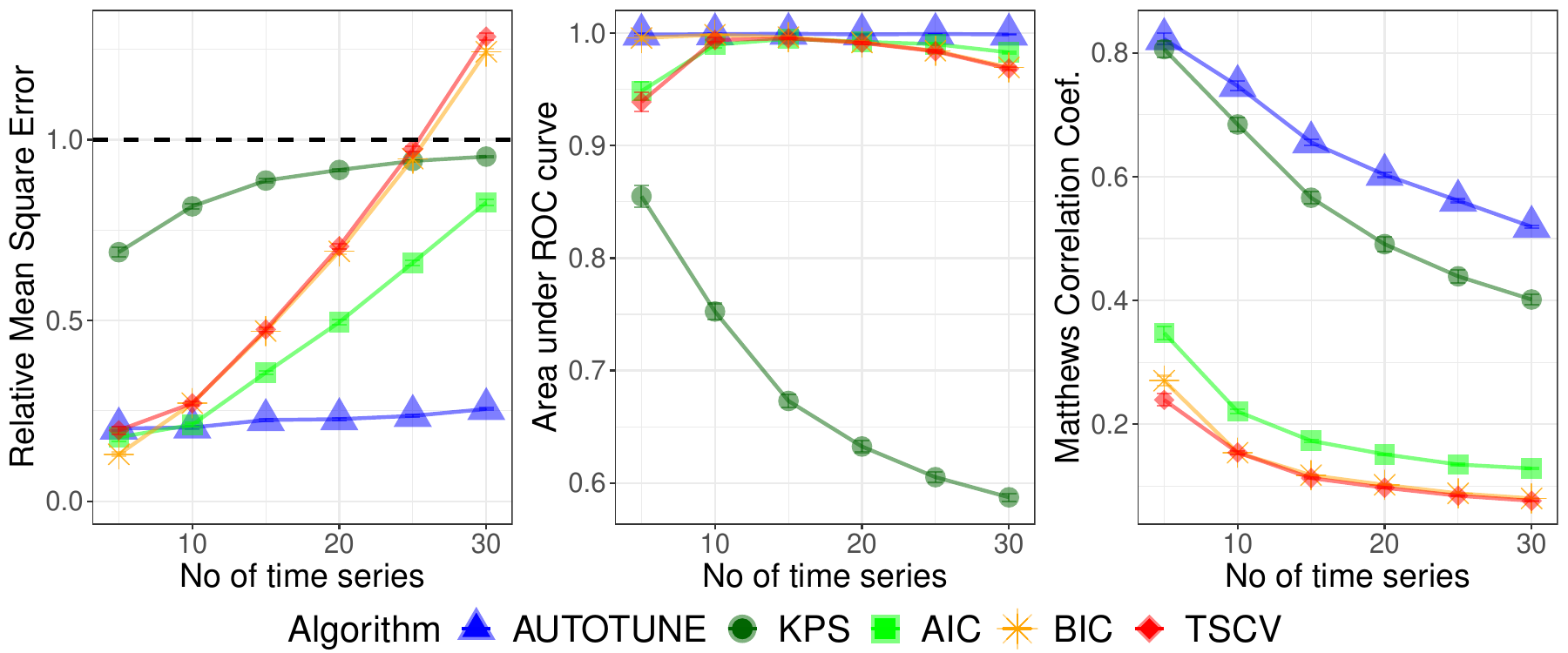}
    \end{tabular}
    \\[-1mm]
    \caption{RMSE, AUROC, and Matthews Correlation Coefficient plotted as functions of number of time series $(p)$ for (left) DGP(ii), i.e. 2x2 Block VAR(1) with $n = 200$ and (right) DGP(i), i.e. diagonal VAR(1) with $n = 100$.}
    \label{fig: var comparison appendix}
\end{figure}

\begin{figure}[htp]
    \centering
    \begin{subfigure}{0.4\textwidth}
        \centering
    \includegraphics[page = 1, trim=0in 0in 0in  0.4in, clip, width=\textwidth]{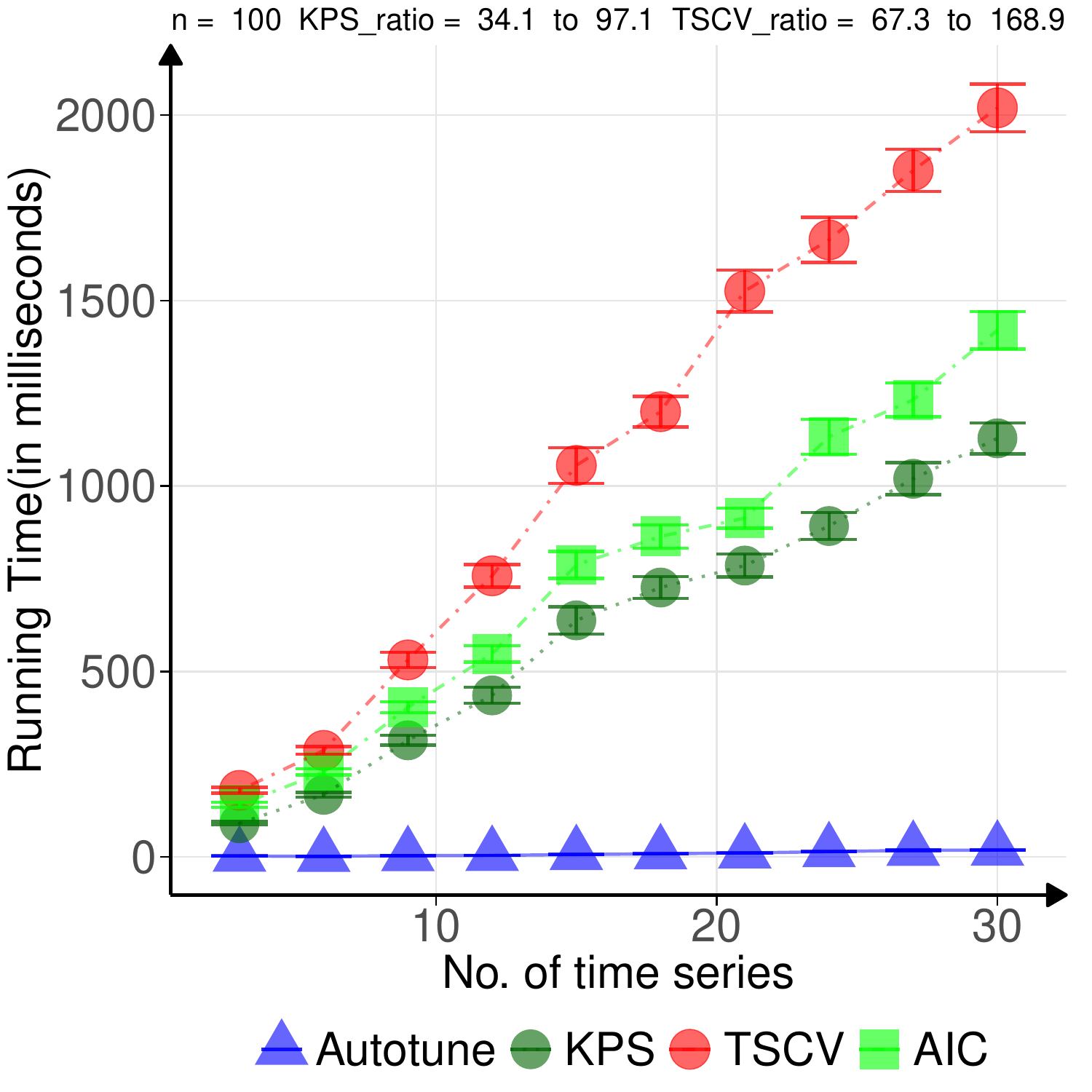}
    \caption{diagonal VAR, $n = 100$}
    \end{subfigure}
    \begin{subfigure}{0.4\textwidth}
        \centering
    \includegraphics[page = 1, trim=0in 0in 0in  0.4in, clip, width=\textwidth]{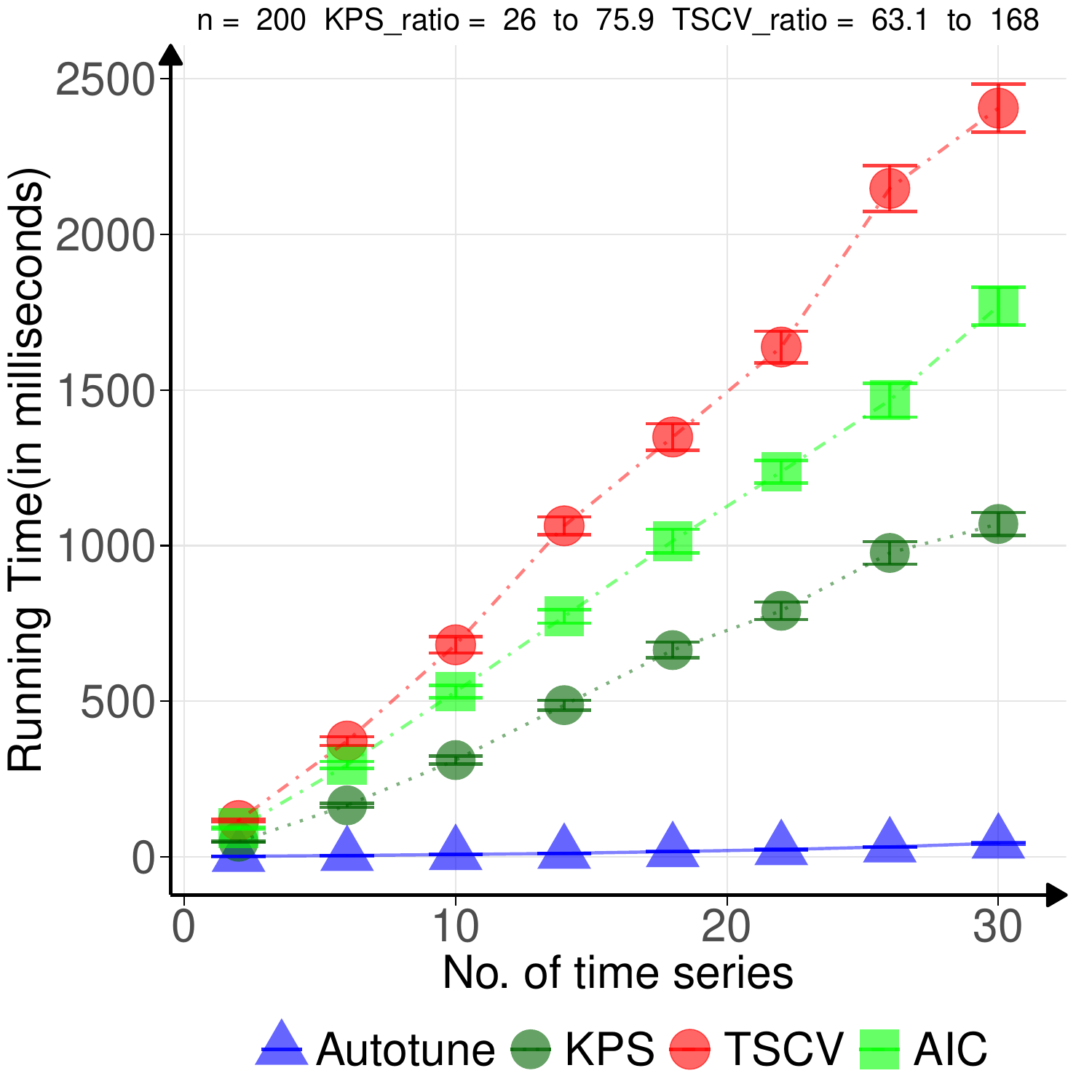}
    \caption{Block VAR, $n = 200$}
    \end{subfigure}
    \caption{Comparison of runtimes of VAR Lasso using $\autotune$ and benchmarks in two different VAR models with SNR = 2.5.}
    \label{fig: time series runtime}
\end{figure}

\clearpage

\subsection{\texorpdfstring{Deferred plots of \cref{subsubsec: simulation results for noise estimation and sparsity diagnostics}}{Deferred plots of noise estimation and sparsity diagnostics}}
\leavevmode\par
\begin{figure}[H]
    \centering
    \begin{subfigure}[t]{0.4\textwidth}
        \centering
        \includegraphics[page = 1, trim=0in 0in 0in 0.4in, clip, width=\textwidth]{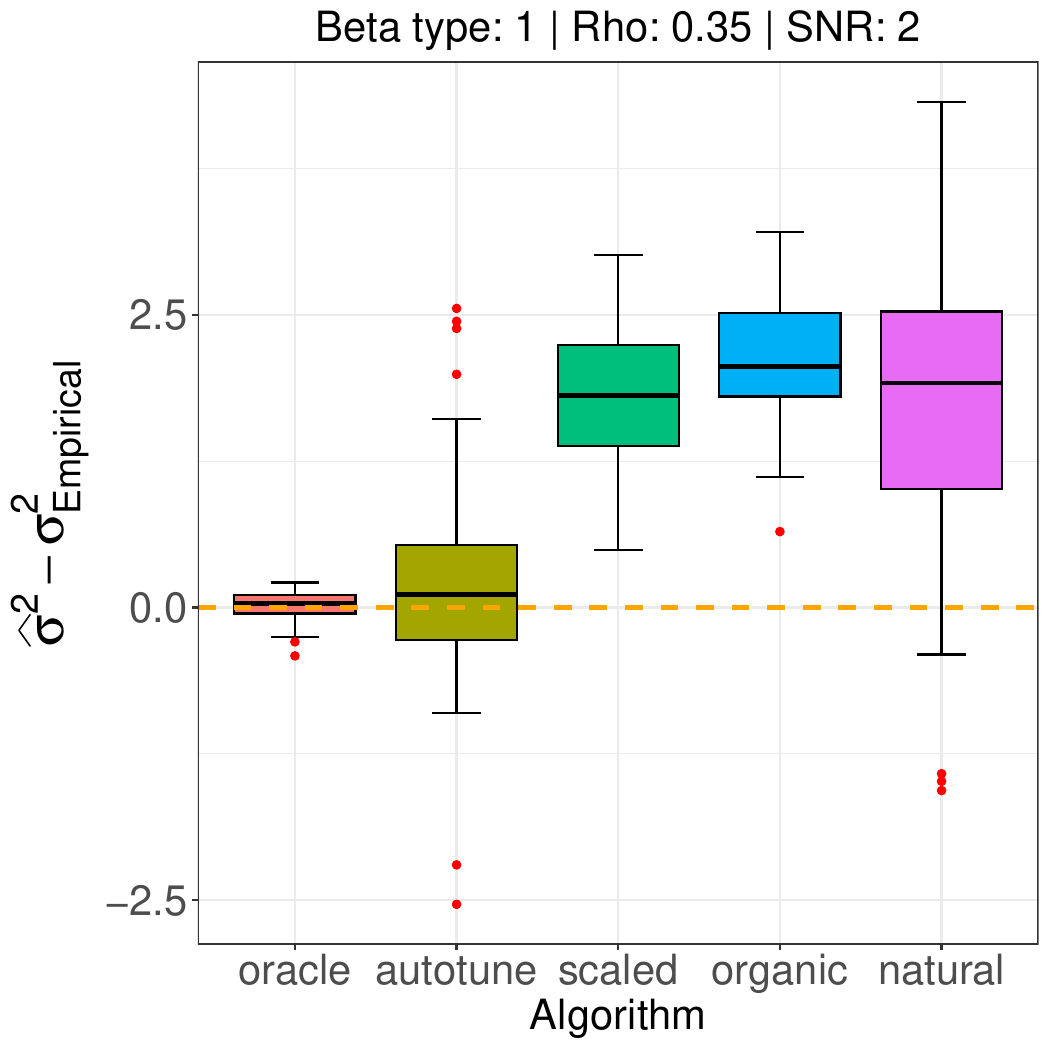}
        \caption{Beta-type 1, $\rho = 0.35$}
        \label{fig: sigma2}
    \end{subfigure}
    \begin{subfigure}[t]{0.4\textwidth}
        \centering
        \includegraphics[page = 1, trim=0in 0in 0in 0.4in, clip, width=\textwidth]{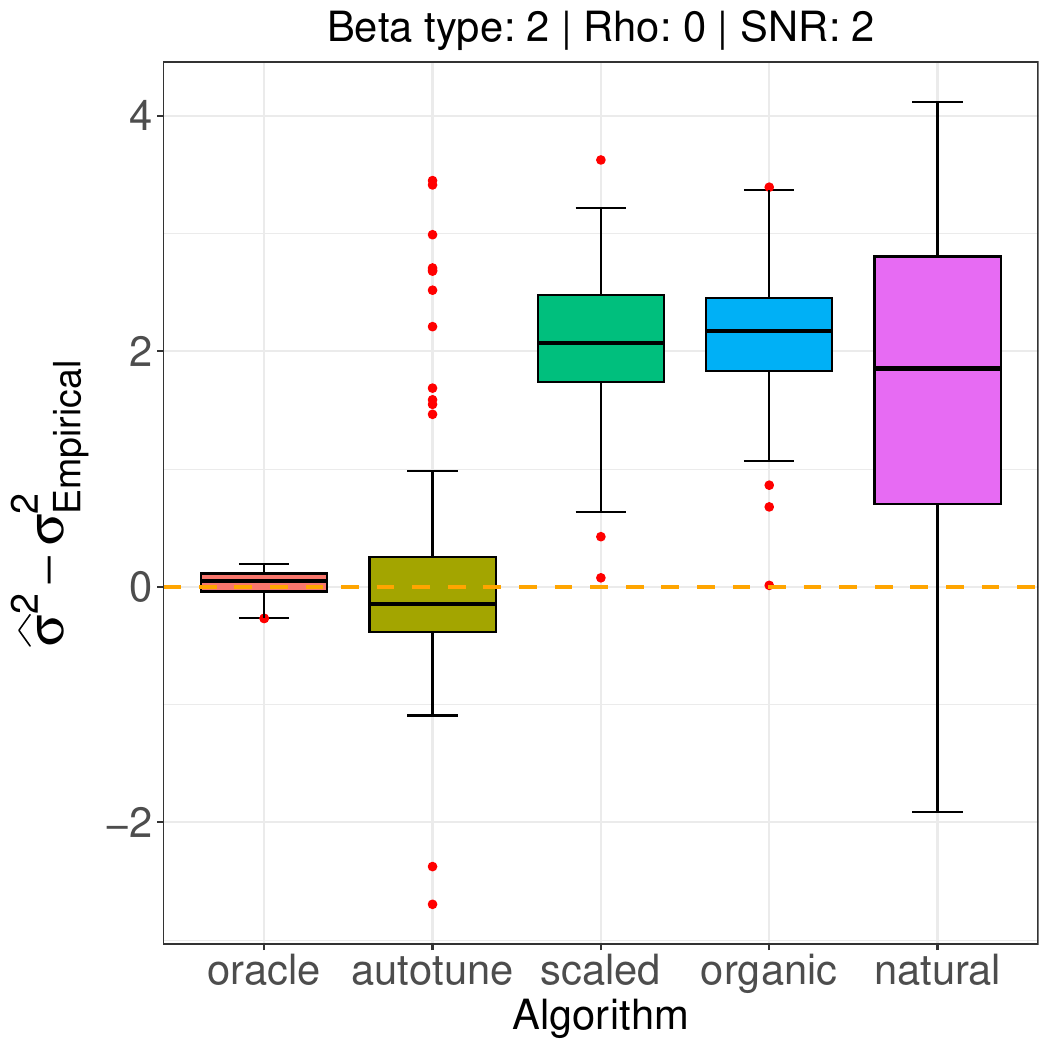}
        \caption{Beta-type 2, $\rho = 0$}
        \label{fig: sigma3}
    \end{subfigure}
    
    \caption{Comparison of noise variance estimation in the high-dimensional setup of $n$ = 80, $p$ = 750, $s$ = 5, and SNR = 2.}
    \label{fig: sigma est diff plots appendix}
\end{figure}

\begin{figure}[H]
    \centering

    \begin{subfigure}[t]{0.32\textwidth}
        \centering
        \includegraphics[page = 1, trim=0in 0in 0in 0.4in, clip, width=\textwidth]{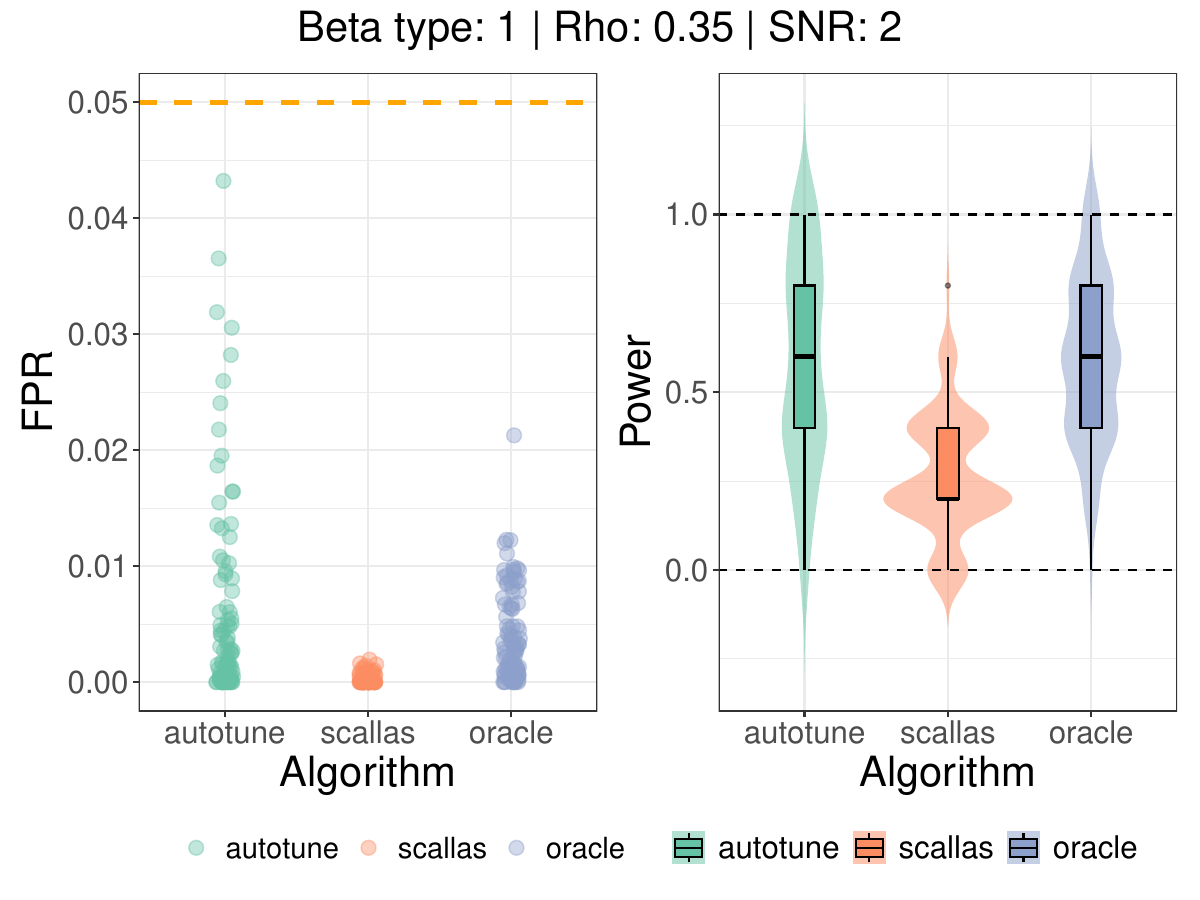}
        \caption{Beta-type 1, $\rho = 0.35$}
        \label{fig: inference2}
    \end{subfigure}
    \vspace{0.5em}  
    \centering
    \begin{subfigure}[t]{0.32\textwidth}
        \centering
        \includegraphics[page = 1, trim=0in 0in 0in 0.4in, clip, width=\textwidth]{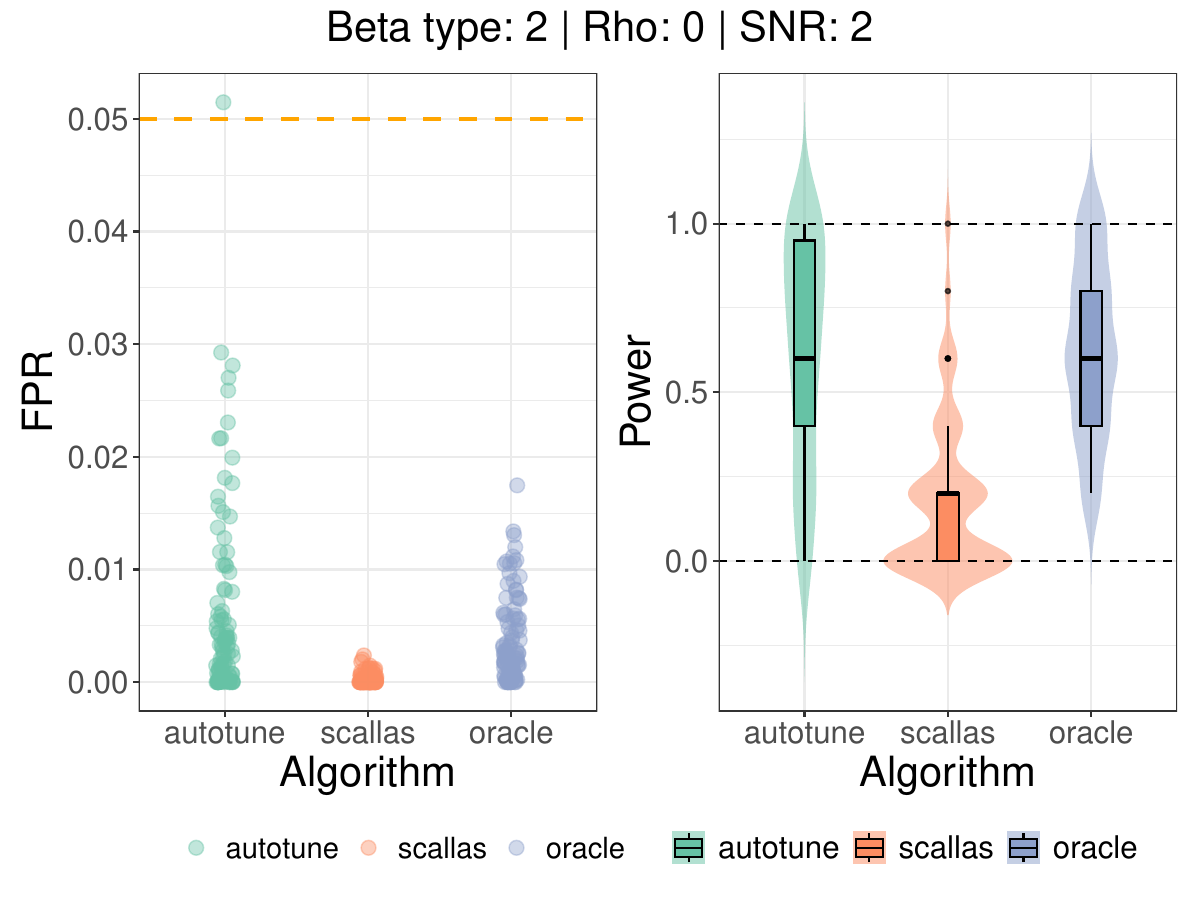}
        \caption{Beta-type 2, $\rho = 0$}
        \label{fig: inference3}
    \end{subfigure}
    \begin{subfigure}[t]{0.32\textwidth}
        \centering
        \includegraphics[page = 1, trim=0in 0in 0in 0.4in, clip, width=\textwidth]{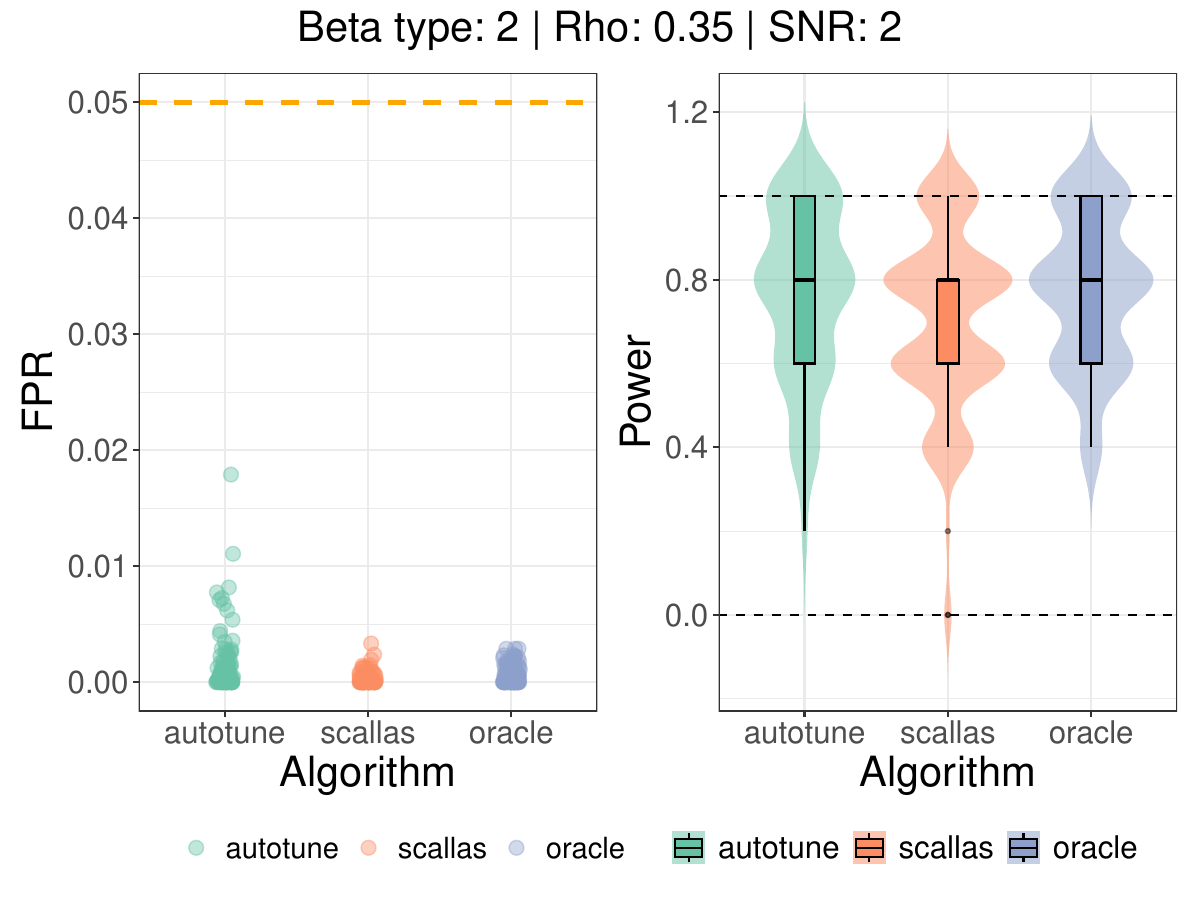}
        \caption{Beta-type 2, $\rho = 0.35$}
        \label{fig: inference4}
    \end{subfigure}
    \caption{Inference of regression coefficients using the estimated $\what{\sigma}^2$ of different algorithms in high-dimensional setup of $n$ = 80, $p$ = 750, $s$ = 5, and SNR = 2.}
    \label{fig: inference plots appendix}
\end{figure}

\begin{figure}[H]
    \centering
    \begin{subfigure}[h]{0.45\textwidth}
        \centering
        \includegraphics[page = 5, trim=0in 0in 0in 0in, clip, width=\textwidth]{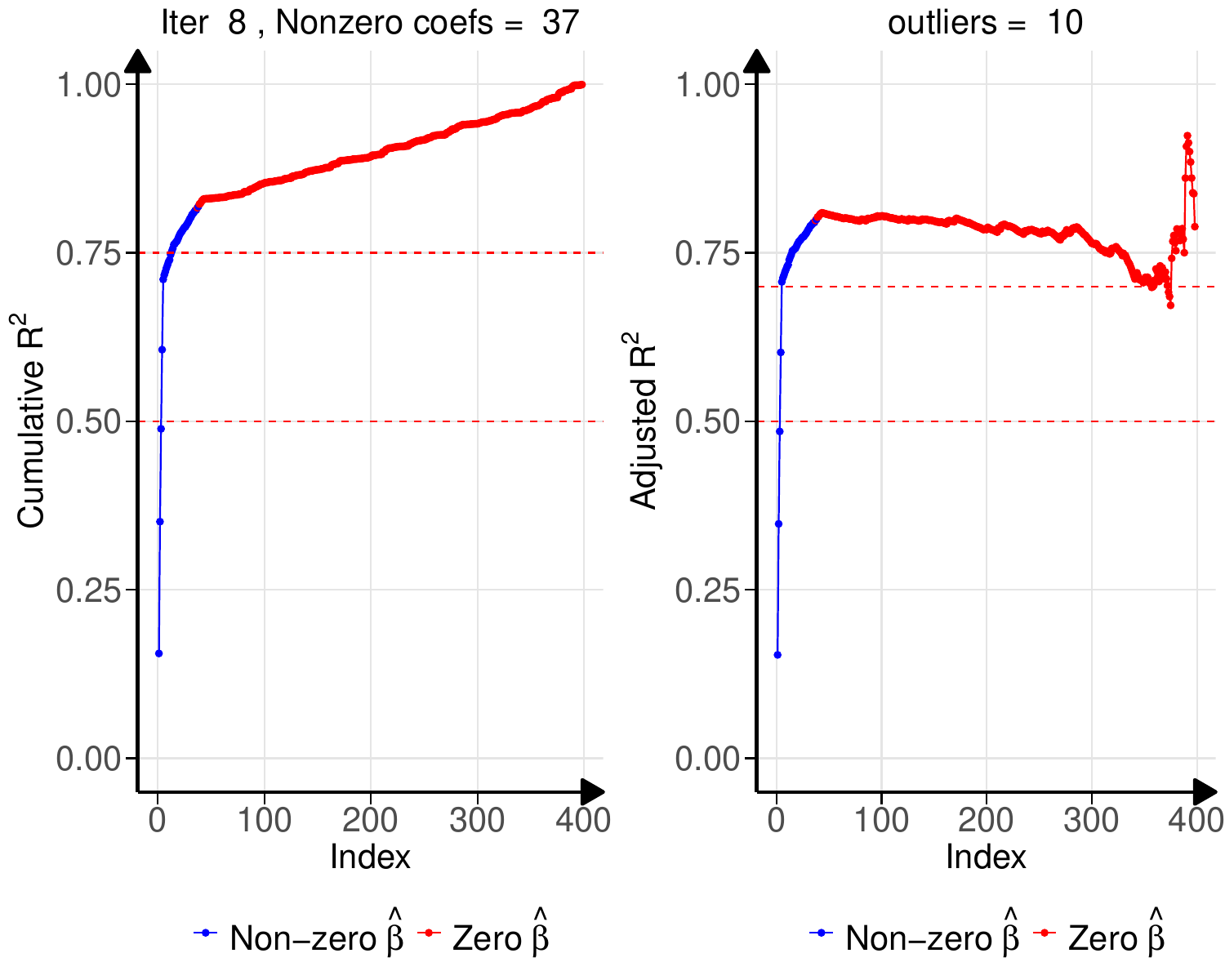}
        \caption{$s=5$}
        \label{fig: good diagnostics1}
    \end{subfigure}
    \begin{subfigure}[h]{0.45\textwidth}
        \centering
        \includegraphics[page = 5, trim=0in 0in 0in 0in, clip, width=\textwidth]{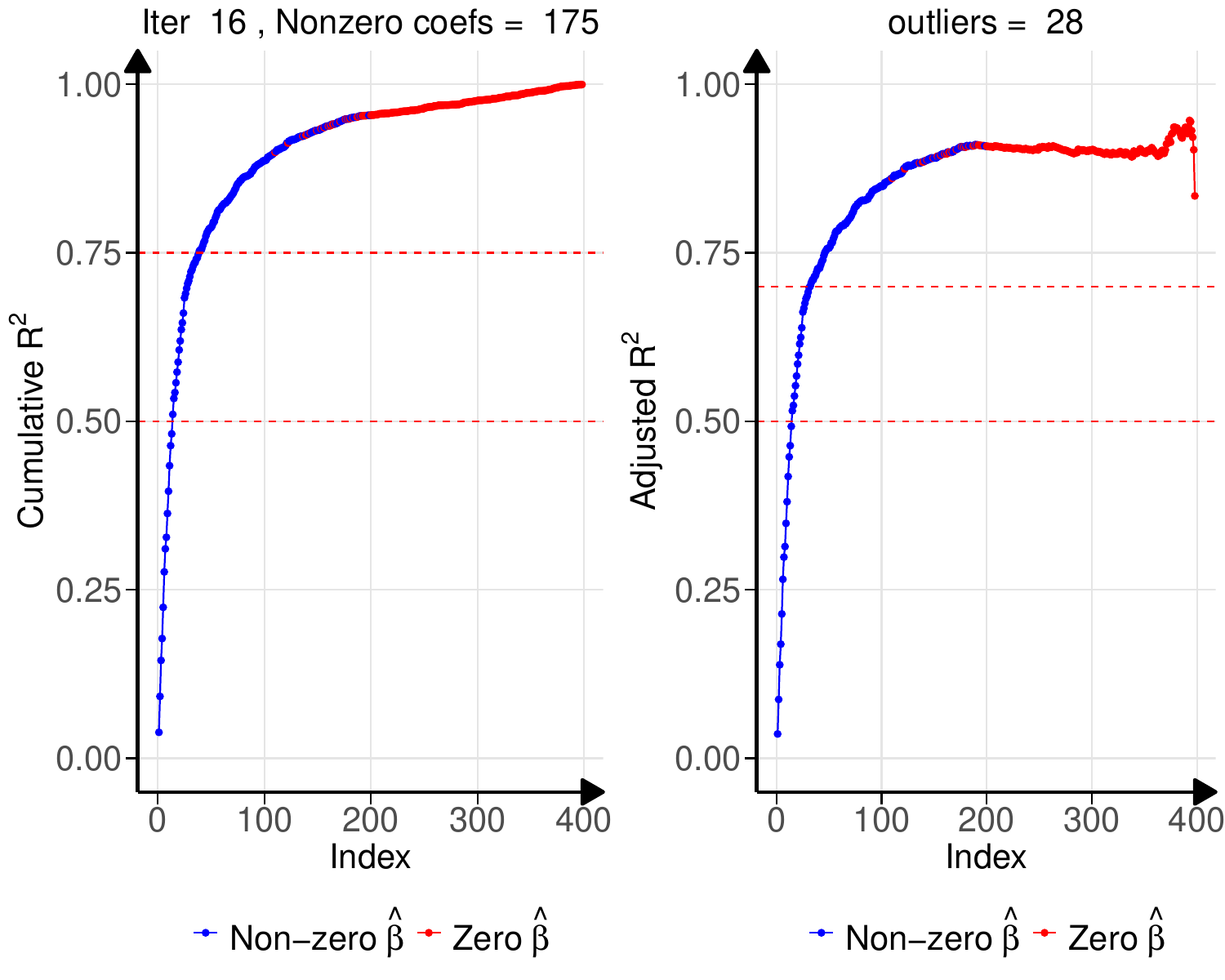}
        \caption{$s=30$}
        \label{fig: semi good diagnostics}
    \end{subfigure}
    \caption{Additional plots of behavior of our sparsity diagnostics for varying levels of sparsity (different values of $s$) when $n = 400$ and $p = 2000$}
    \label{fig: diagnostics appendix}
\end{figure}



\clearpage
\section{Comparison against Bi-level Optimization and skglm}

\label{app: comparison vs BLO skglm}








We augment the demonstration of performance of $\autotune$ in high dimensional regression with new benchmarks as per the constructive feedback of the reviewers. We compare the accuracy (prediction, variable selection and estimation) of $\autotune$, CV Lasso of $\glmnet$ against a Bilevel Optimization method \texttt{sparse-ho} \cite{bertrand2020implicit} and python based Lasso package \texttt{skglm} \cite{skglm_neurips2022}.

We mirror the performance metrics and DGP used in \cref{subsec: regg setup sim comparison}. In Figure \ref{fig: high dim accuracy plot vs blo sklgm}, we plot the RMSE, RTE and AUROC of autotune, CV Lasso (of $\glmnet$), sparse-ho and skglm against an SNR grid of 0.66 to 4 in the high dimensional regime of $p=750$, $n=80$, $s = 5$ and $\rho = 0.35$. In Figure \ref{fig: runtime plots vs blo skglm}, we plot the mean runtime of all the four methods for the DGP used in \cref{subsec: regg setup sim comparison}.

The Lasso method of skglm uses CV to tune the $\lambda$ and implements working set strategy and anderson acceleration inside the CD computations. We keep $\glmnet$ in the numerical comparison as skglm \cite{skglm_neurips2022} benchmarks against $\glmnet$ in terms of accuracy and runtime for Lasso objective at a prespecified single $\lambda$ and not for $\lambda$ chosen via CV from a $\lambda$ grid.

We get the same empirical results of \cref{subsec: regg setup sim comparison} with new benchmarks where $\autotune$ shows broader runtime improvements. In Figure \ref{fig: high dim accuracy plot vs blo sklgm}, both $\autotune$ and CV Lasso uniformly either matches or shows greater accuracy than sparse-ho and skglm in terms of RTE, AUROC and RMSE. Interestingly at low - moderate SNR, $\autotune$ depicts significantly better (nonoverlapping 1SE bars) accuracy across all metrics than the new benchmarks.

In terms of runtime, in Figure \ref{fig: runtime plots vs blo skglm}, $\autotune$ is 25 - 360 times faster than sparse-ho and 80-400 times faster than skglm. These experiments corroborate to our claim that the statistical insights used inside $\autotune$ leads to improvement in prediction accuracy, variable selection, retains estimation accuracy while delivering substantial speed improvements.


\begin{figure}[H]
    \centering
    \begin{tabular}{c}
         \textbf{High Dimensional setup:} $n = 80, \hspace{0.1cm} p=750, \hspace{0.1cm} s=5$, 
         $\rho = 0.35$, \\ \textbf{Top Row:} Beta-type 1 and \textbf{Bottom Row:} Beta-type 2\\
        \includegraphics[trim=0in 0.6in 0in 0in, clip, width=0.90\textwidth]{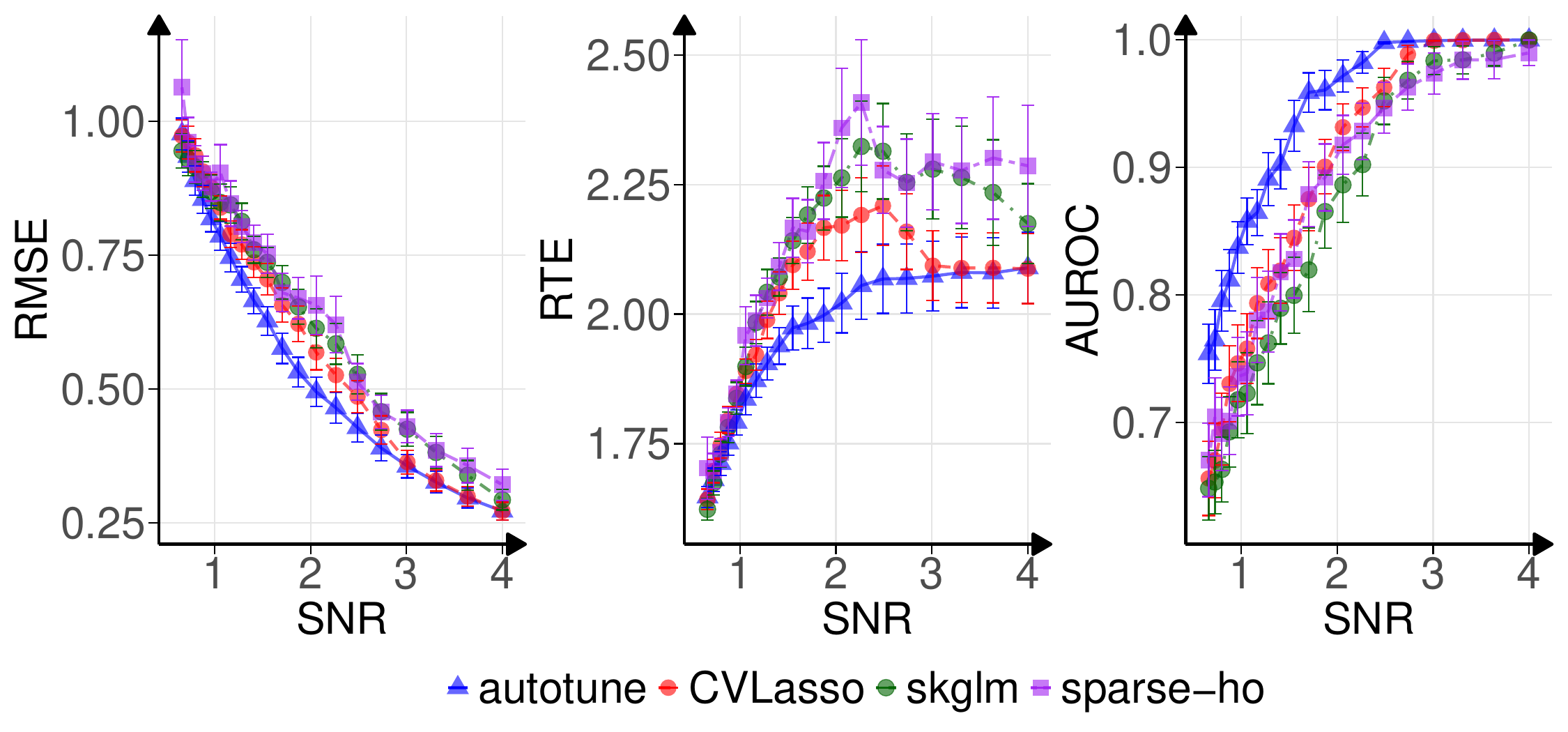} \\
        \includegraphics[trim=0in 0in 0in 0in, clip, width=0.90\textwidth]{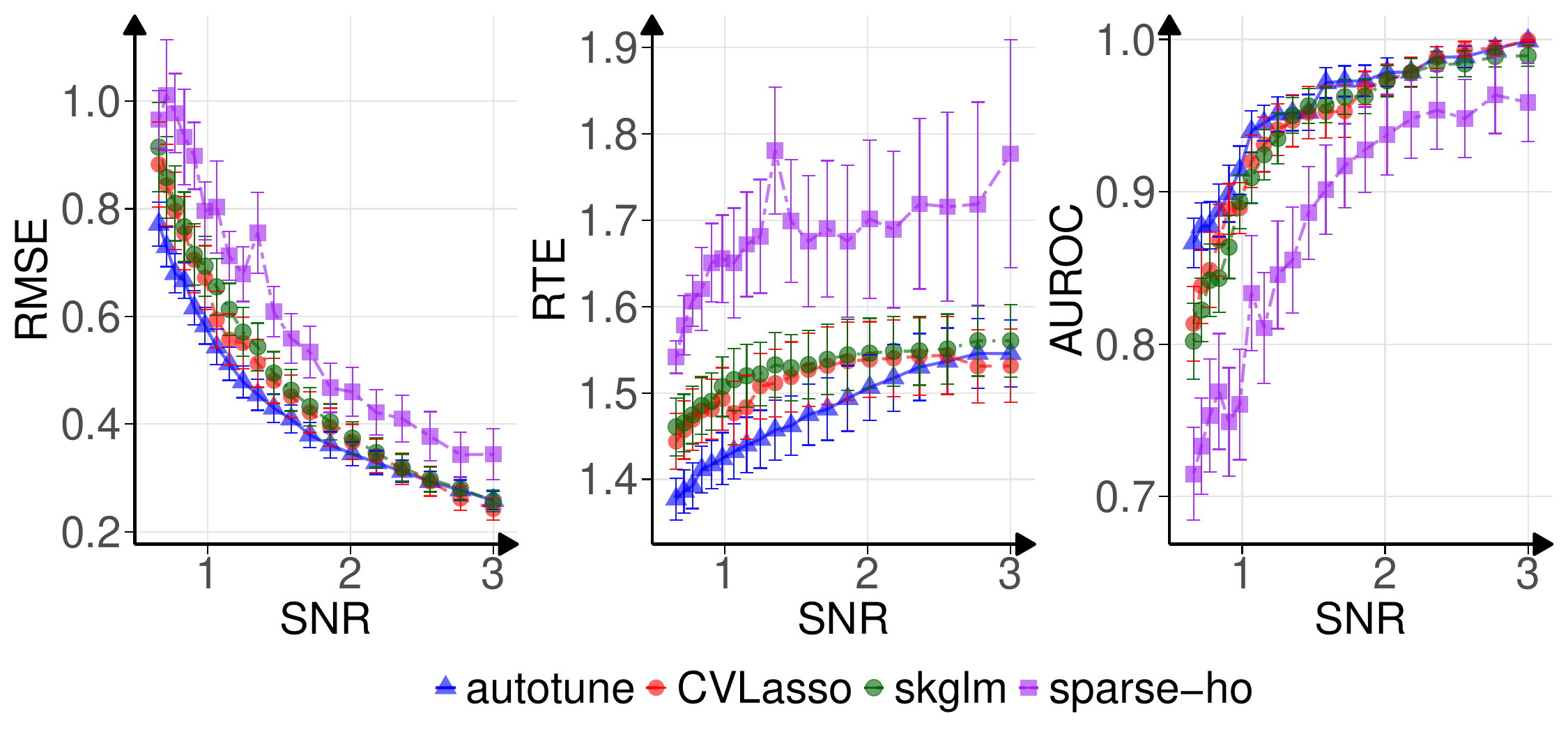}
    \end{tabular}
    \\[2mm]
    
    \caption{RTE, AUROC and RMSE (with 1SE bars of 100 replications) of $\autotune$ and new benchmark methods for Lasso regression, plotted against SNR.}
    \label{fig: high dim accuracy plot vs blo sklgm}
\end{figure}

\begin{figure}[H]
    \begin{tabular}{c}
         \textbf{Setup:} $n = 200, \hspace{0.1cm} s=5$, snr $=1$,
         $\rho = 0.35$, \textbf{Left:} Beta-type 1 and \textbf{Right:} Beta-type 2\\
          \begin{subfigure}[t]{0.48\textwidth}
        \centering
        \includegraphics[page = 1, trim=0in 0in 0in 0in, clip, width=\textwidth]{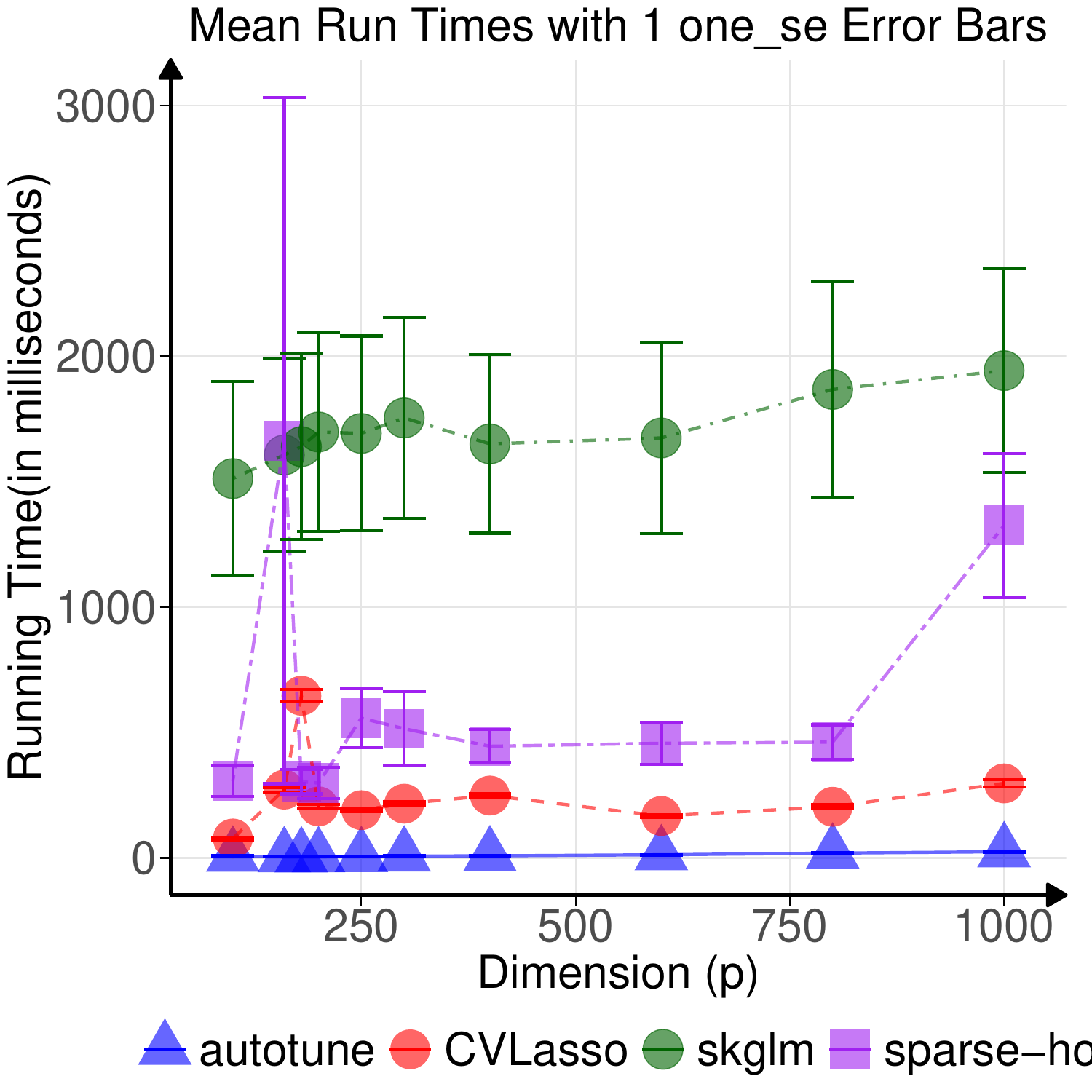}
    \end{subfigure}
    \hfill
    \begin{subfigure}[t]{0.48\textwidth}
        \centering
        \includegraphics[page = 1, trim=0in 0in 0in  0in, clip, width=\textwidth]{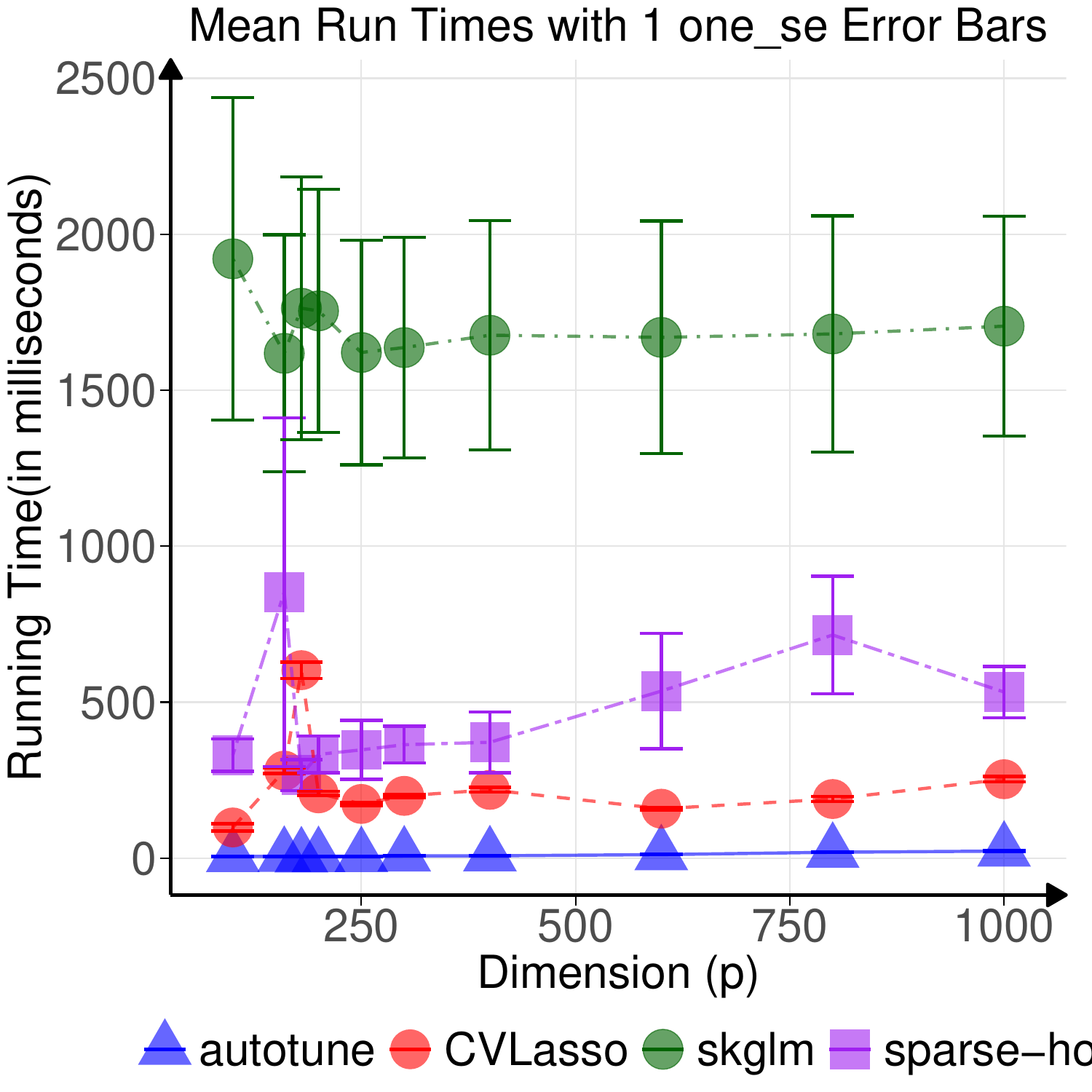}
    \end{subfigure}
    \end{tabular}
   
    \caption{Runtimes (with 1SE bars) of $\autotune$, CV Lasso (of $\glmnet$) and new benchmark methods, showing upto 400x speedup.
    }
    \label{fig: runtime plots vs blo skglm}
\end{figure}

\clearpage

\section{Complete suite of comparison plots in variable selection, estimation, and prediction accuracy}
\label{subsec: all accuracy plots}

\subsection{High Dimensional Setting: n = 80, p = 750, s = 5}
\leavevmode\par
\begin{figure}[H]
    \centering
    \begin{tabular}{c}
        \textbf{High Dimensional setup:} $n = 80, \hspace{0.1cm} p=750, \hspace{0.1cm} s=5$, \textbf{Beta-type: 1}, Varying levels  \\
        of correlation, \textbf{Top Row:} $\rho = 0$, \textbf{Middle Row:} $\rho = 0.35$ and \textbf{Bottom Row:} $\rho = 0.7$\\
        \includegraphics[trim=0in 0.8in 0in 0in, clip, width=0.85\textwidth]{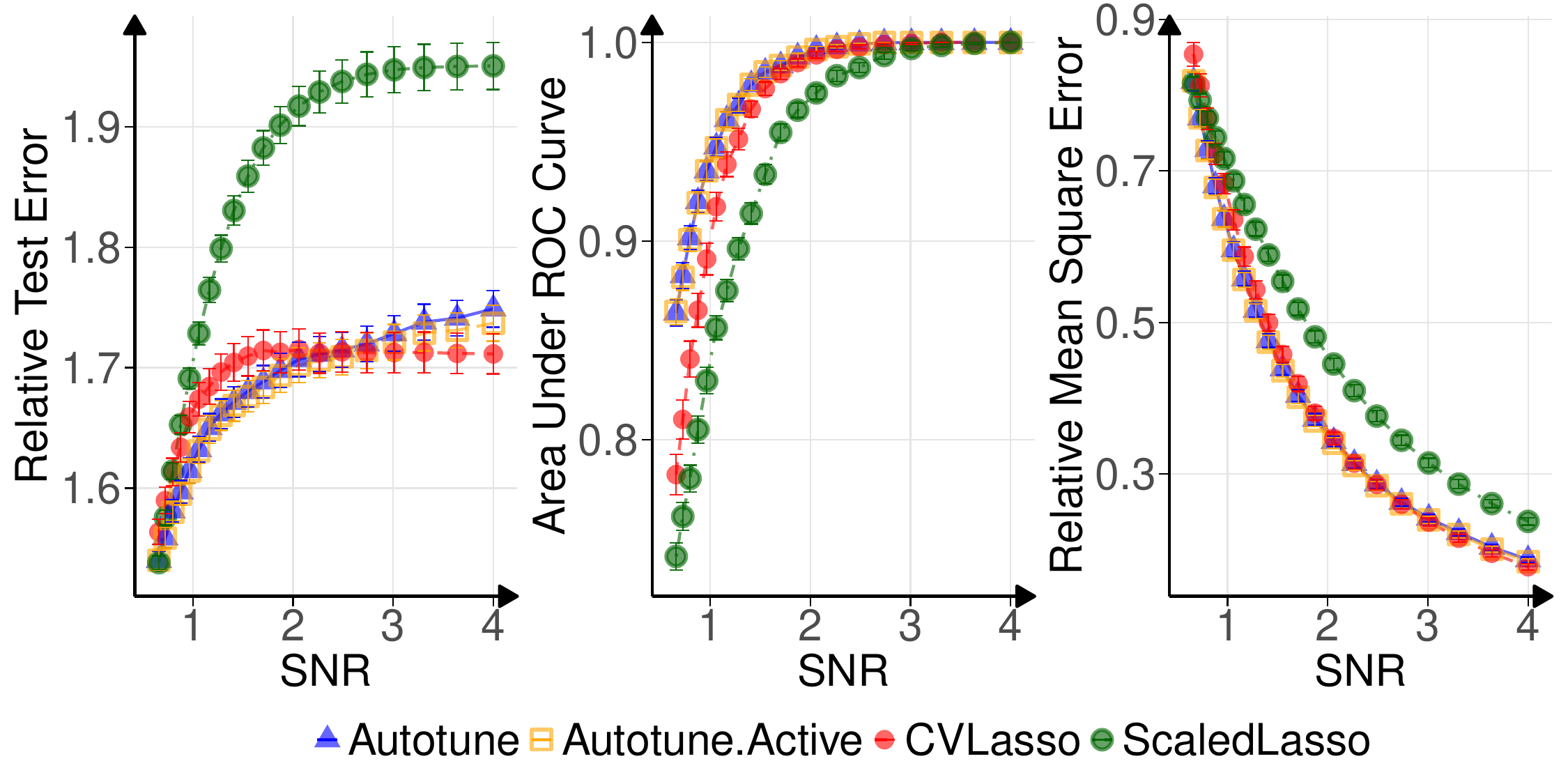}\\
        \includegraphics[trim=0in 0.8in 0in 0in, clip, width=0.85\textwidth]{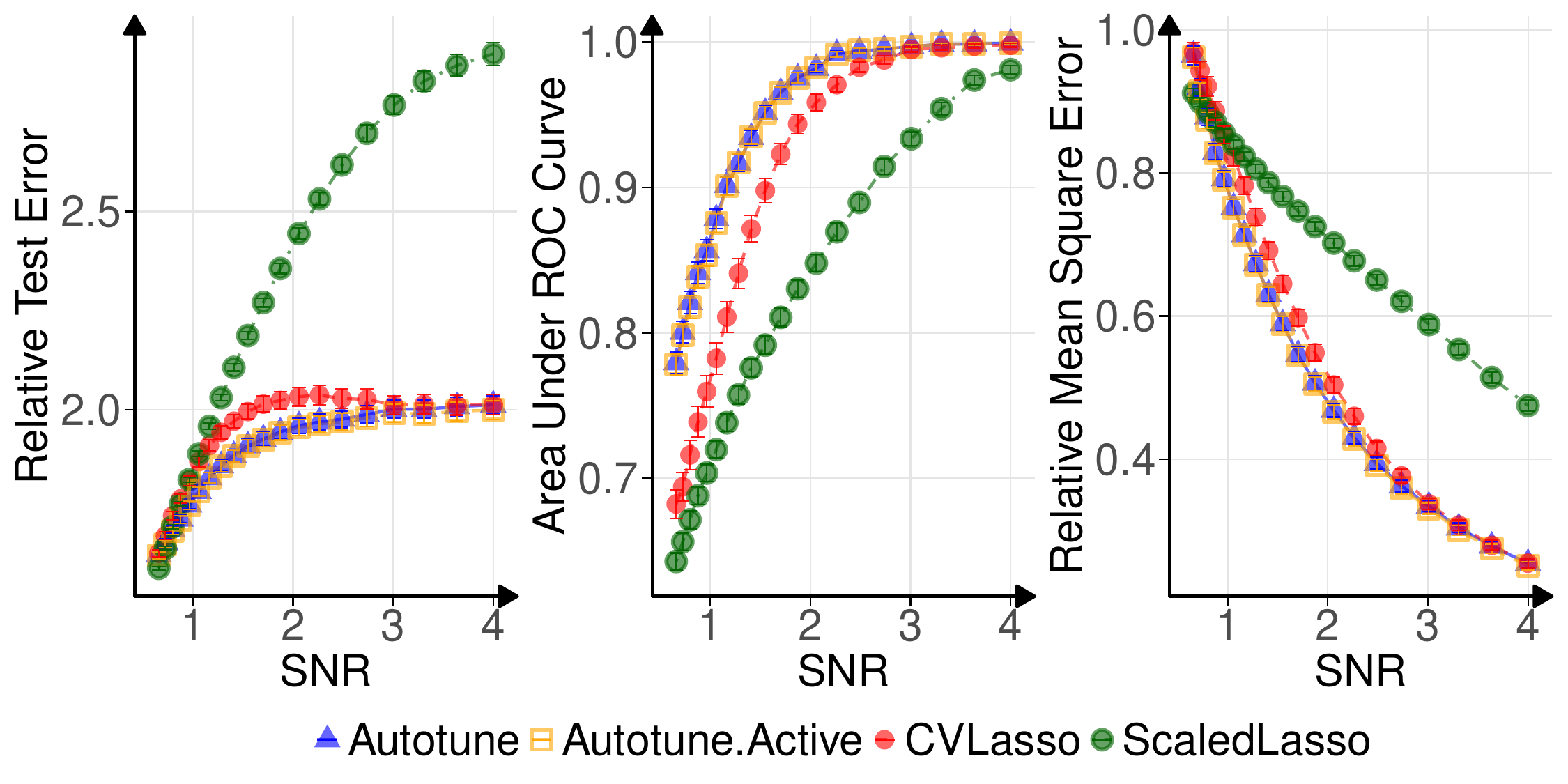}\\
        \includegraphics[trim=0in 0in 0in 0in, clip, width=0.85\textwidth]{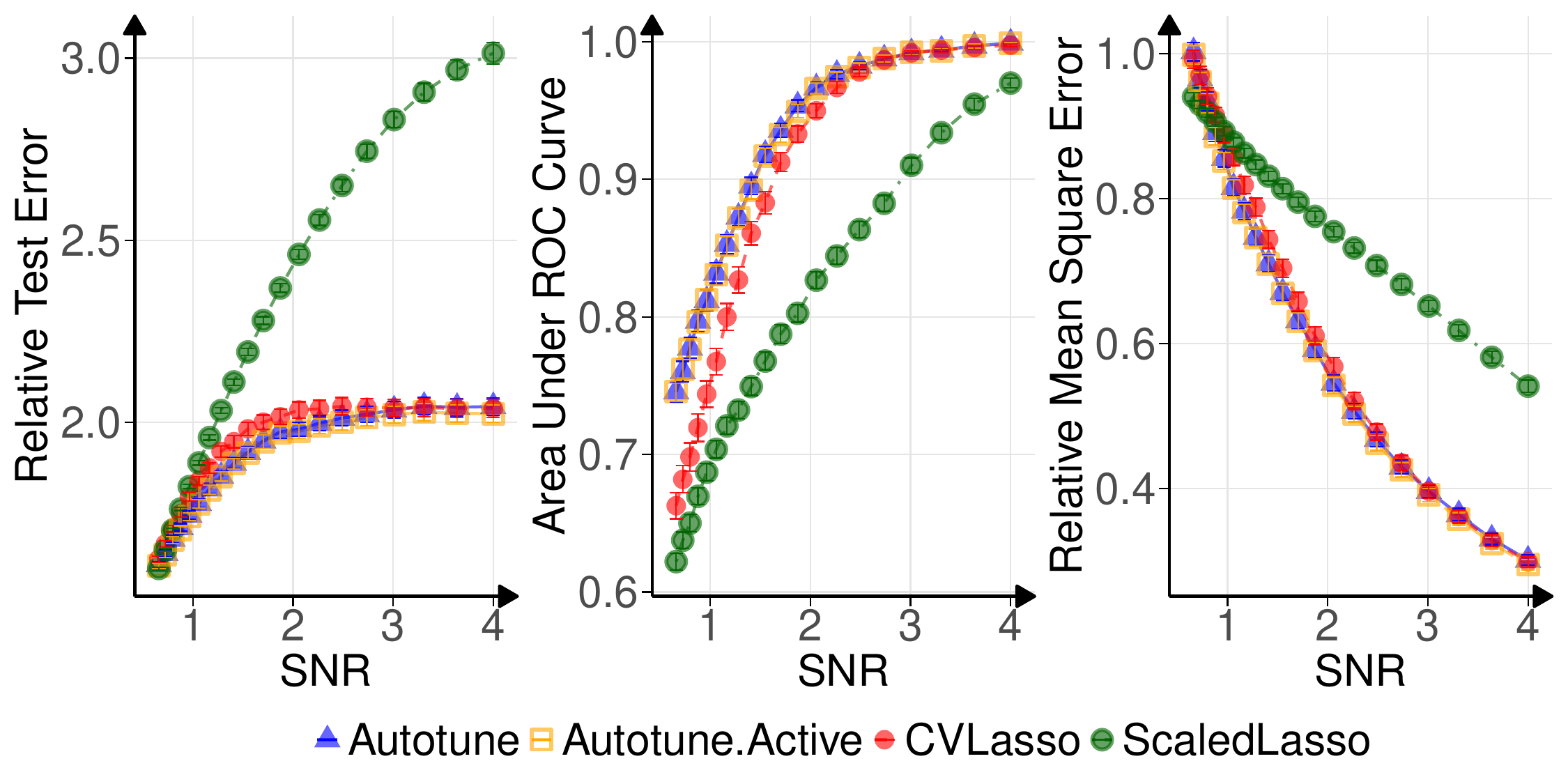}
    \end{tabular}
    \\
    \caption{RTE, AUROC and RMSE of $\autotune$, $\msf{autotune.active}$, CV, and Scaled Lasso plotted as a function of SNR for high-dimensional setup.}
    \label{fig: appen high dim accuracy plot btype 1}
\end{figure}

\begin{figure}[H]
    \centering
    \begin{tabular}{c}
         \textbf{High Dimensional setup:} $n = 80, \hspace{0.1cm} p=750, \hspace{0.1cm} s=5$, \textbf{Beta-type: 2}, Varying levels  \\
        of correlation, \textbf{Top Row:} $\rho = 0$, \textbf{Middle Row:} $\rho = 0.35$ and \textbf{Bottom Row:} $\rho = 0.7$\\
        \includegraphics[trim=0in 0.8in 0in 0in, clip, width=0.92\textwidth]{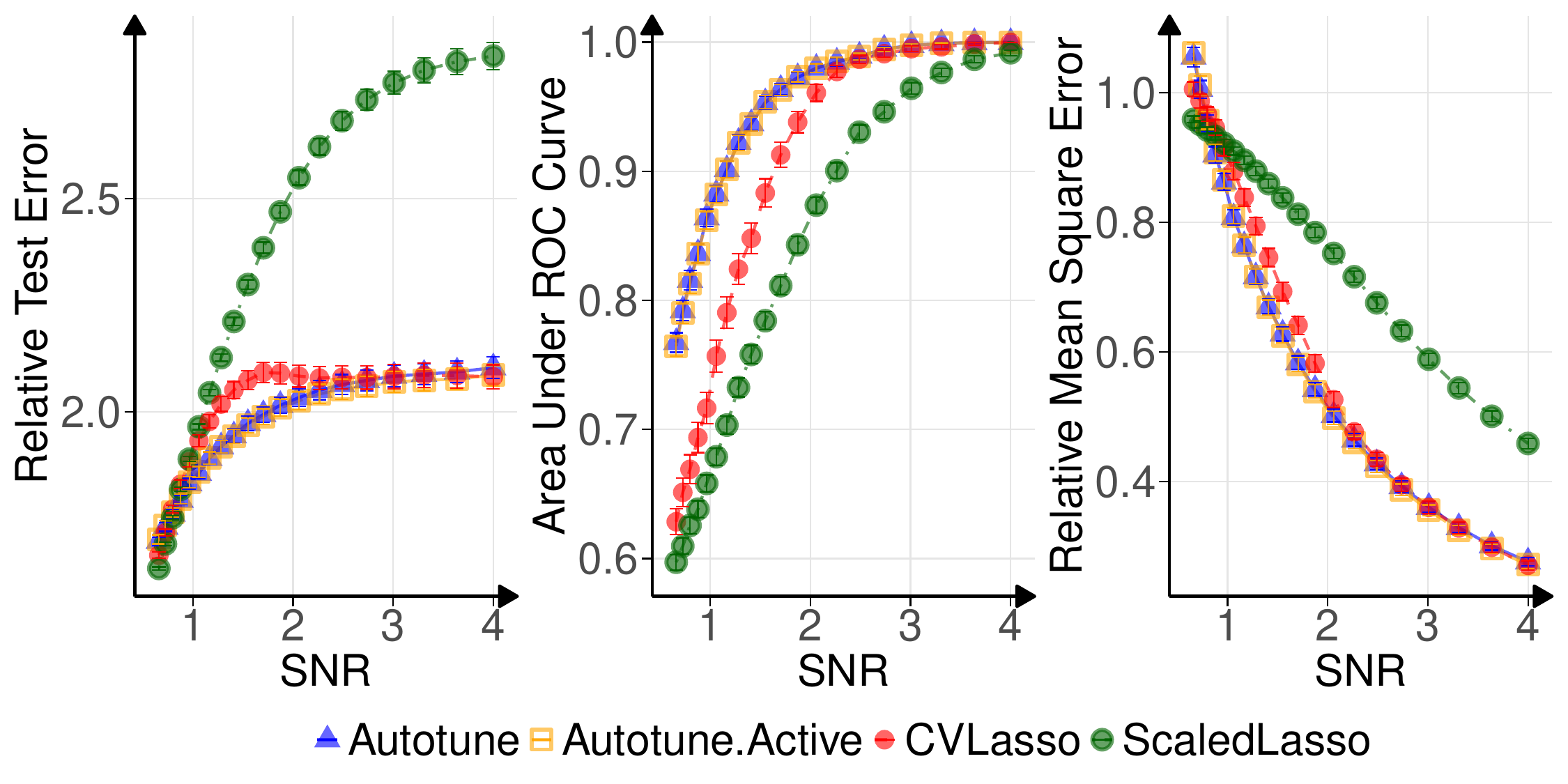}\\
        \includegraphics[trim=0in 0.8in 0in 0in, clip, width=0.92\textwidth]{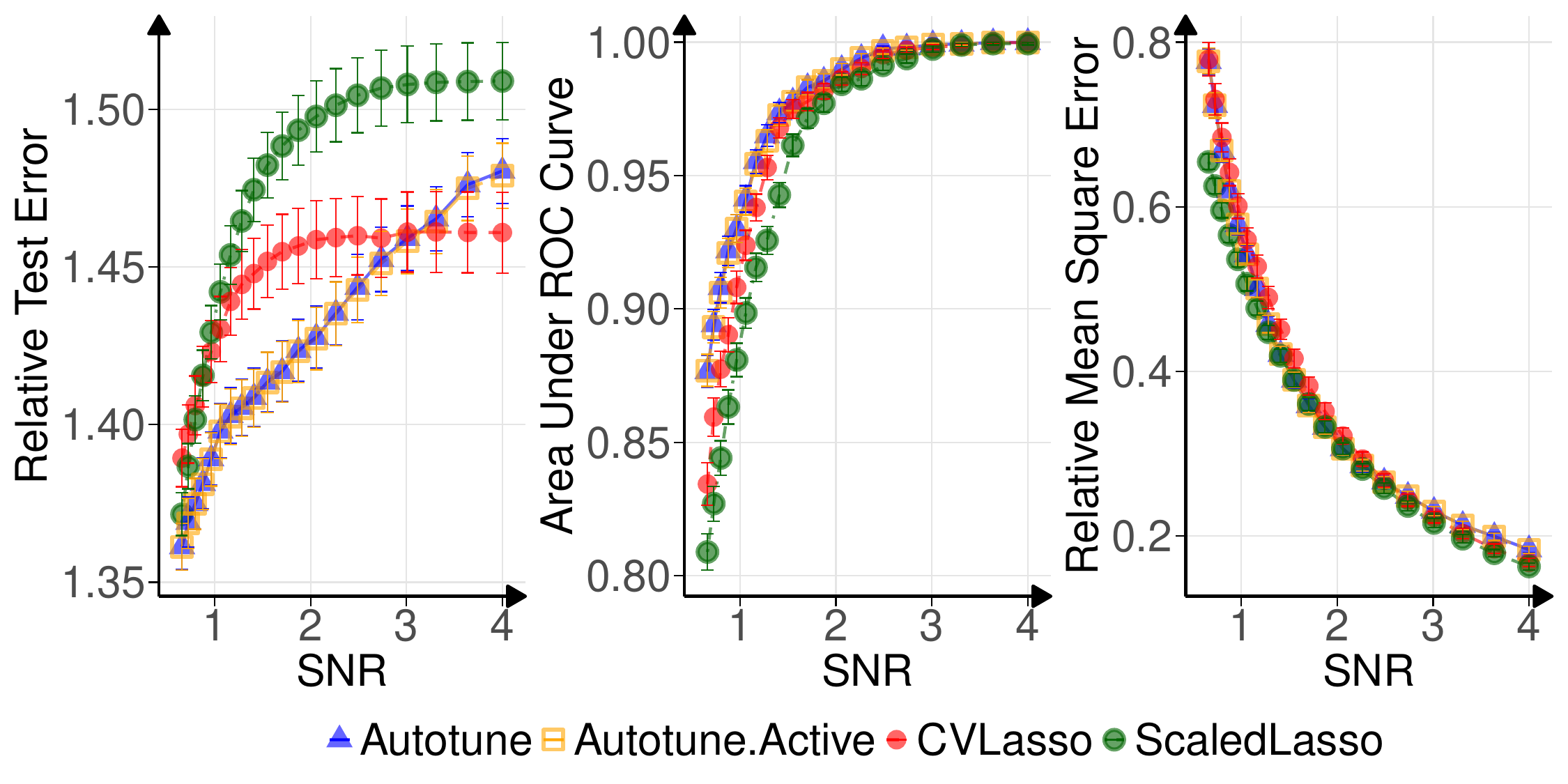}\\
        \includegraphics[trim=0in 0in 0in 0in, clip, width=0.92\textwidth]{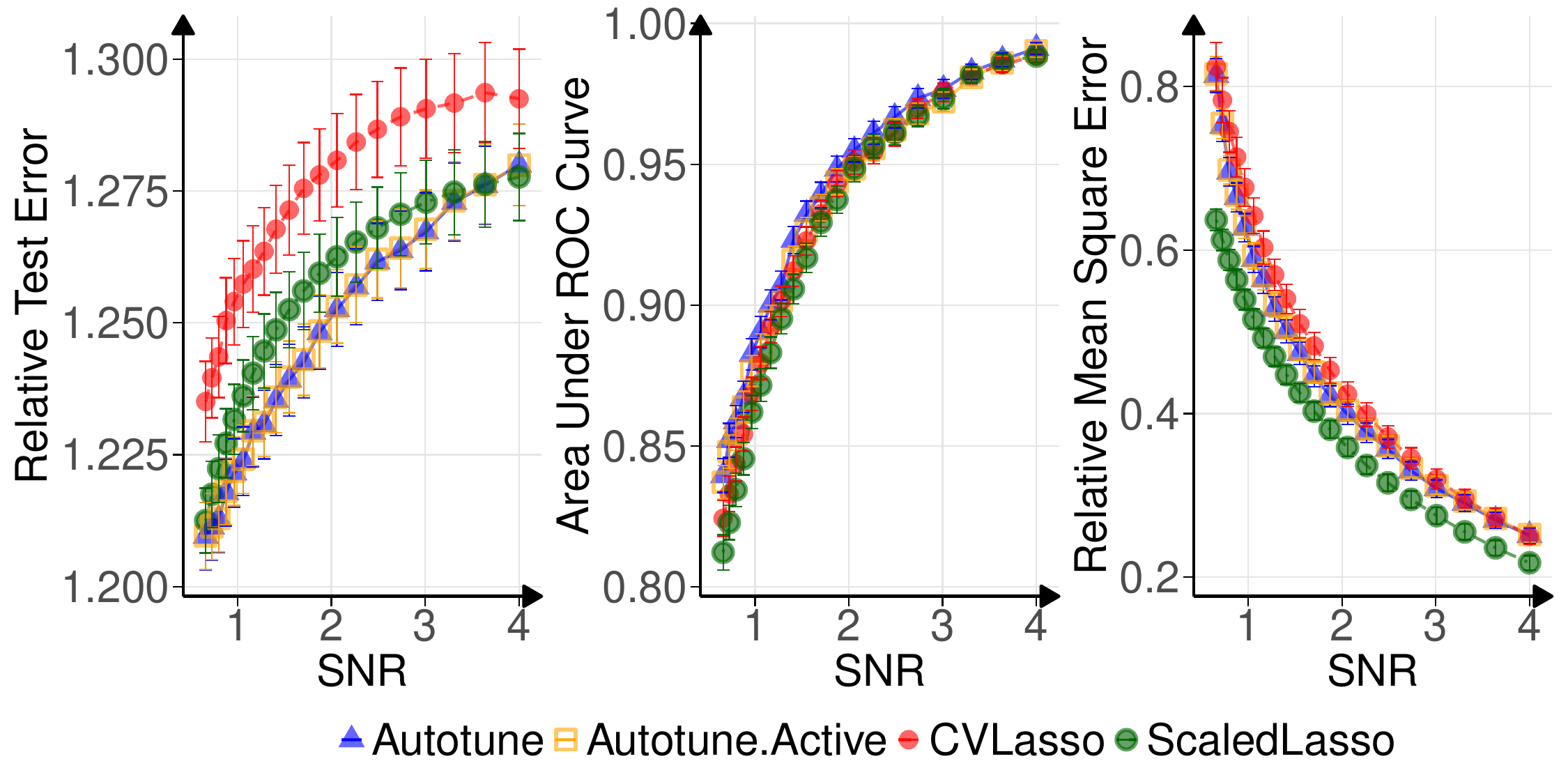}
    \end{tabular}
    \\
    \caption{RTE, AUROC and RMSE of $\autotune$, $\msf{autotune.active}$, CV, and Scaled Lasso plotted as a function of SNR for high-dimensional setup.}
    \label{fig: appen high dim accuracy plot btype 2}
\end{figure}

\begin{figure}[H]
    \centering
    \begin{tabular}{c}
\textbf{High Dimensional setup:} $n = 80, \hspace{0.1cm} p=750, \hspace{0.1cm} s=5$, \textbf{Beta-type: 3}, Varying levels  \\
        of correlation, \textbf{Top Row:} $\rho = 0$, \textbf{Middle Row:} $\rho = 0.35$ and \textbf{Bottom Row:} $\rho = 0.7$\\
        \includegraphics[trim=0in 0.8in 0in 0in, clip, width=0.92\textwidth]{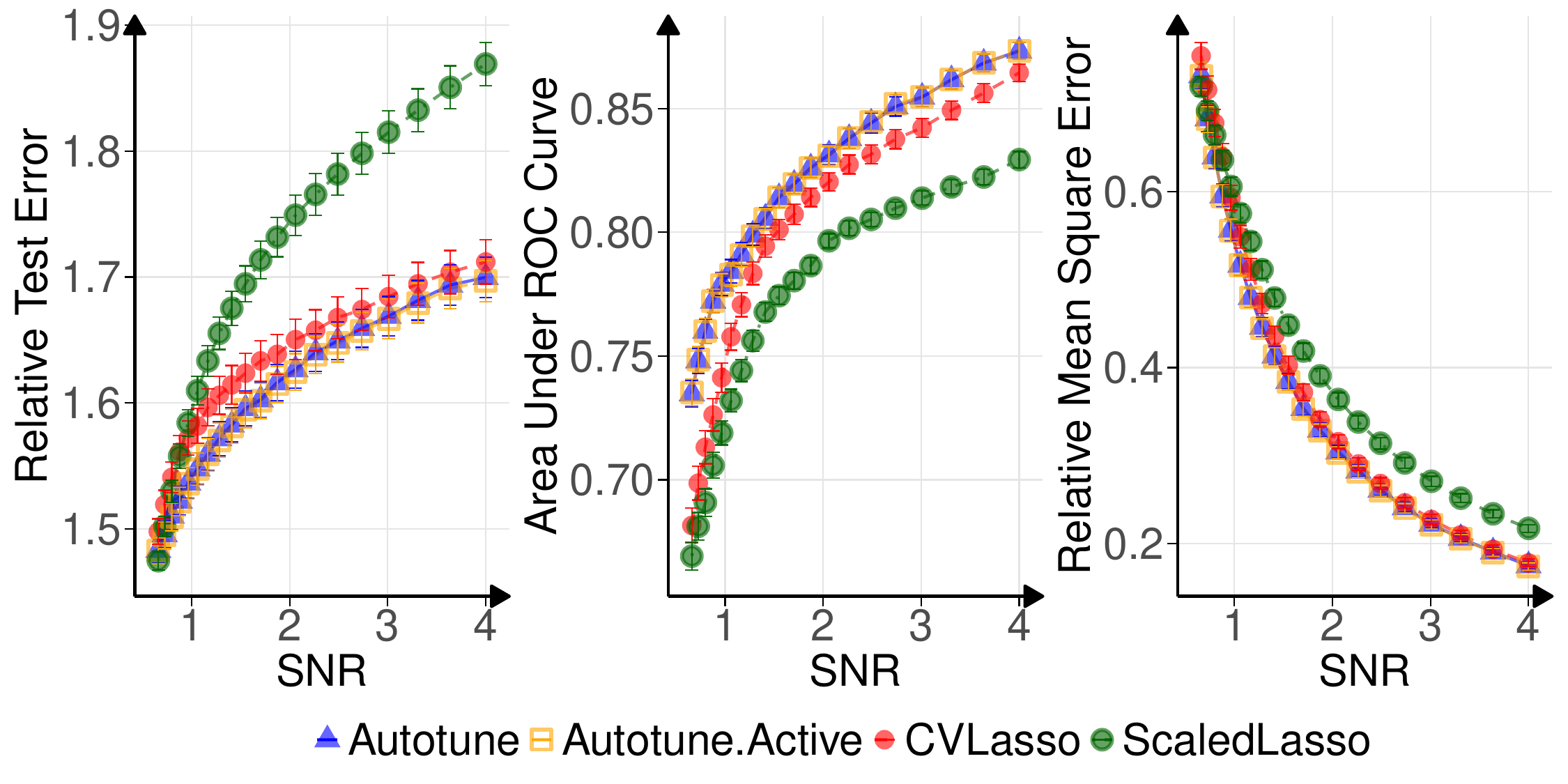}\\
        \includegraphics[trim=0in 0.8in 0in 0in, clip, width=0.92\textwidth]{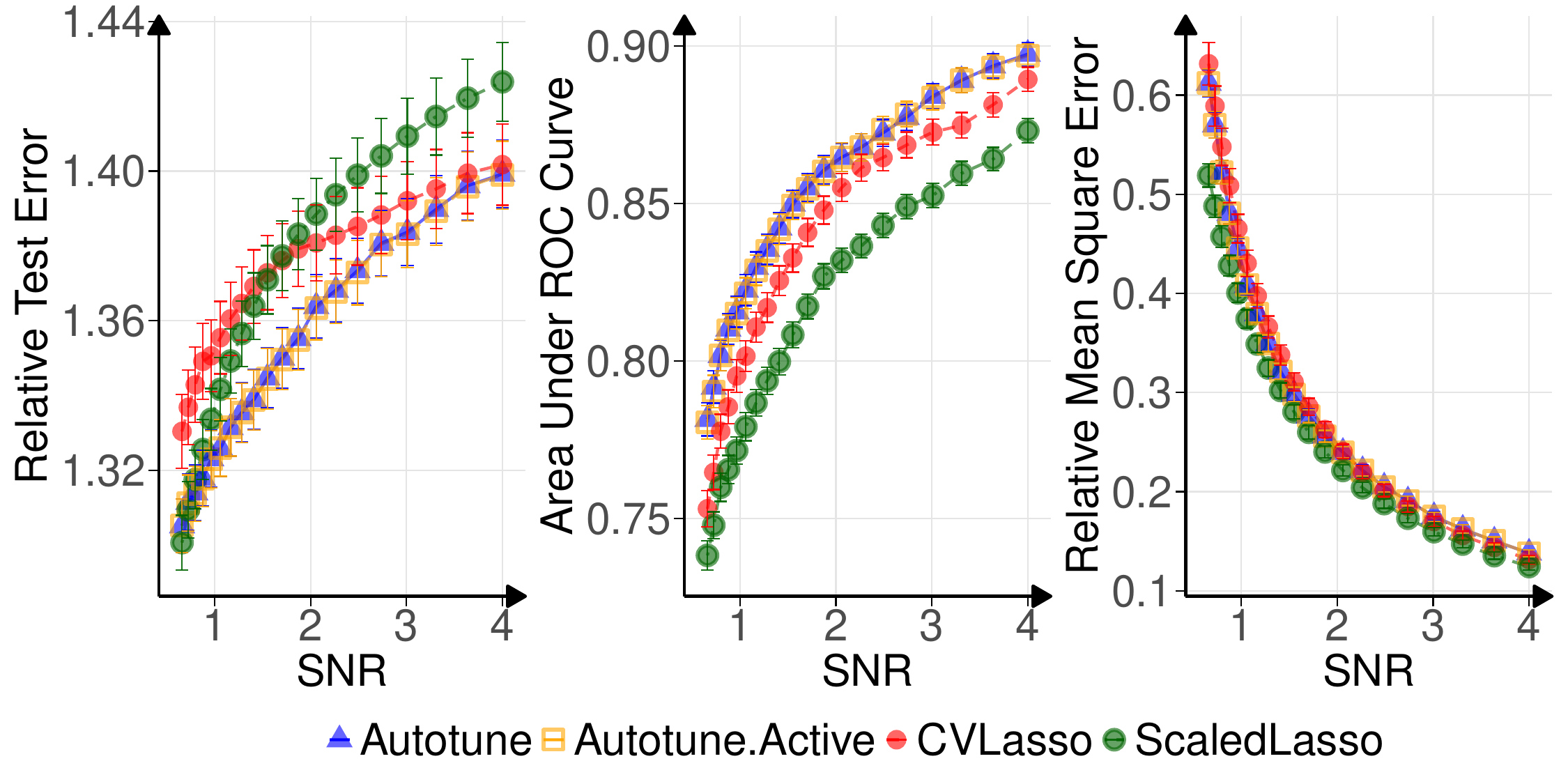}
    \end{tabular}
    \\
    \caption{RTE, AUROC and RMSE of $\autotune$, $\msf{autotune.active}$, CV, and Scaled Lasso plotted as a function of SNR for high-dimensional setup.}
    \label{fig: appen high dim accuracy plot btype 3}
\end{figure}
\clearpage
\begin{figure}[H]
    \ContinuedFloat
    \centering
    \begin{tabular}{c}
        \textbf{High Dimensional setup, Beta-type 3 (continued):} \textbf{Bottom Row:} $\rho = 0.7$\\
        \includegraphics[trim=0in 0in 0in 0in, clip, width=0.92\textwidth]{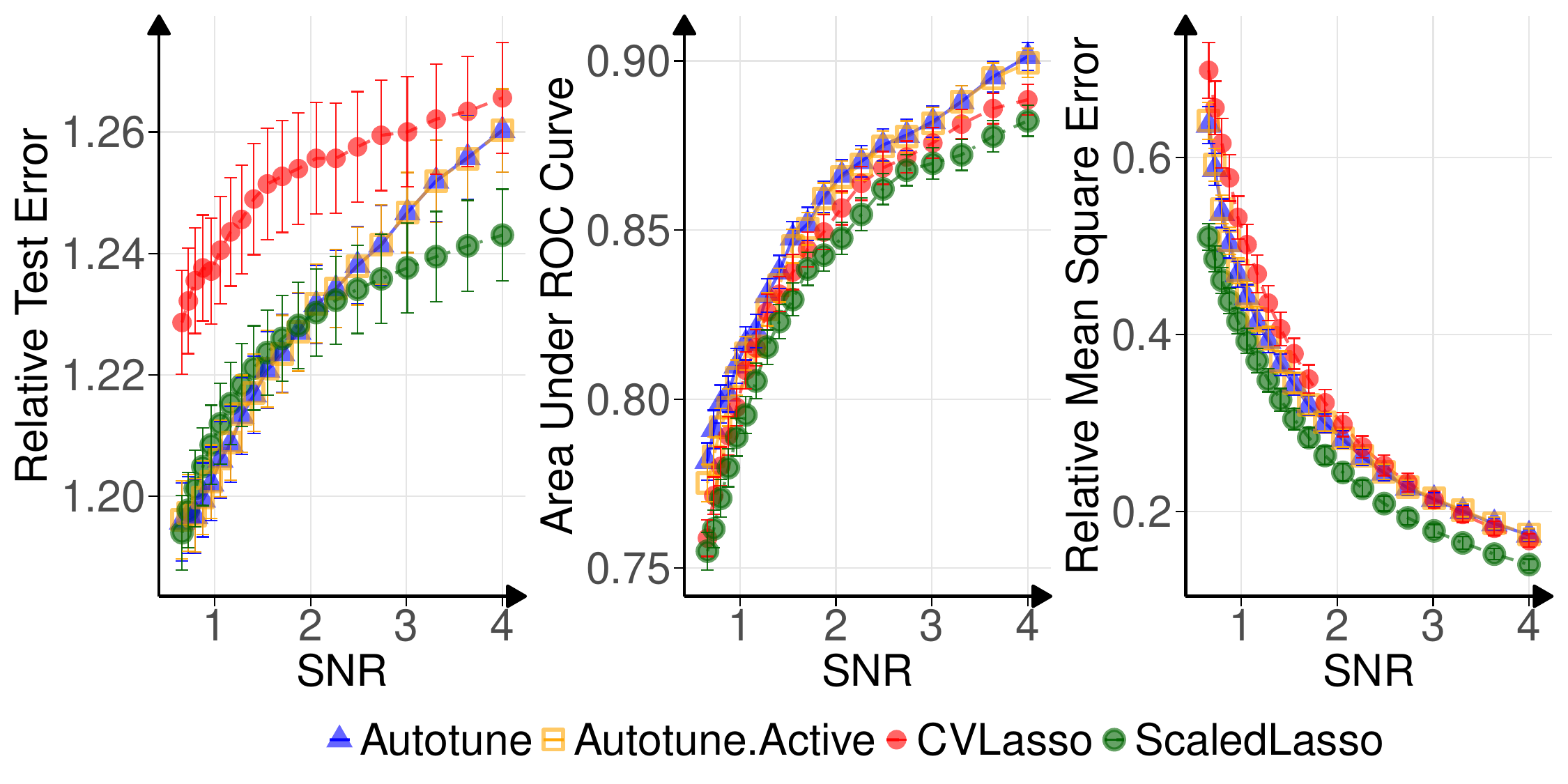}
    \end{tabular}
    \caption[]{Continued: bottom row with $\rho = 0.7$.}
\end{figure}

\begin{figure}[H]
    \centering
    \begin{tabular}{c}
\textbf{High Dimensional setup:} $n = 80, \hspace{0.1cm} p=750, \hspace{0.1cm} s=5$, \textbf{Beta-type: 5}, Varying levels  \\
        of correlation, \textbf{Top Row:} $\rho = 0$, \textbf{Middle Row:} $\rho = 0.35$ and \textbf{Bottom Row:} $\rho = 0.7$\\
        \includegraphics[trim=0in 0.8in 0in 0in, clip, width=0.92\textwidth]{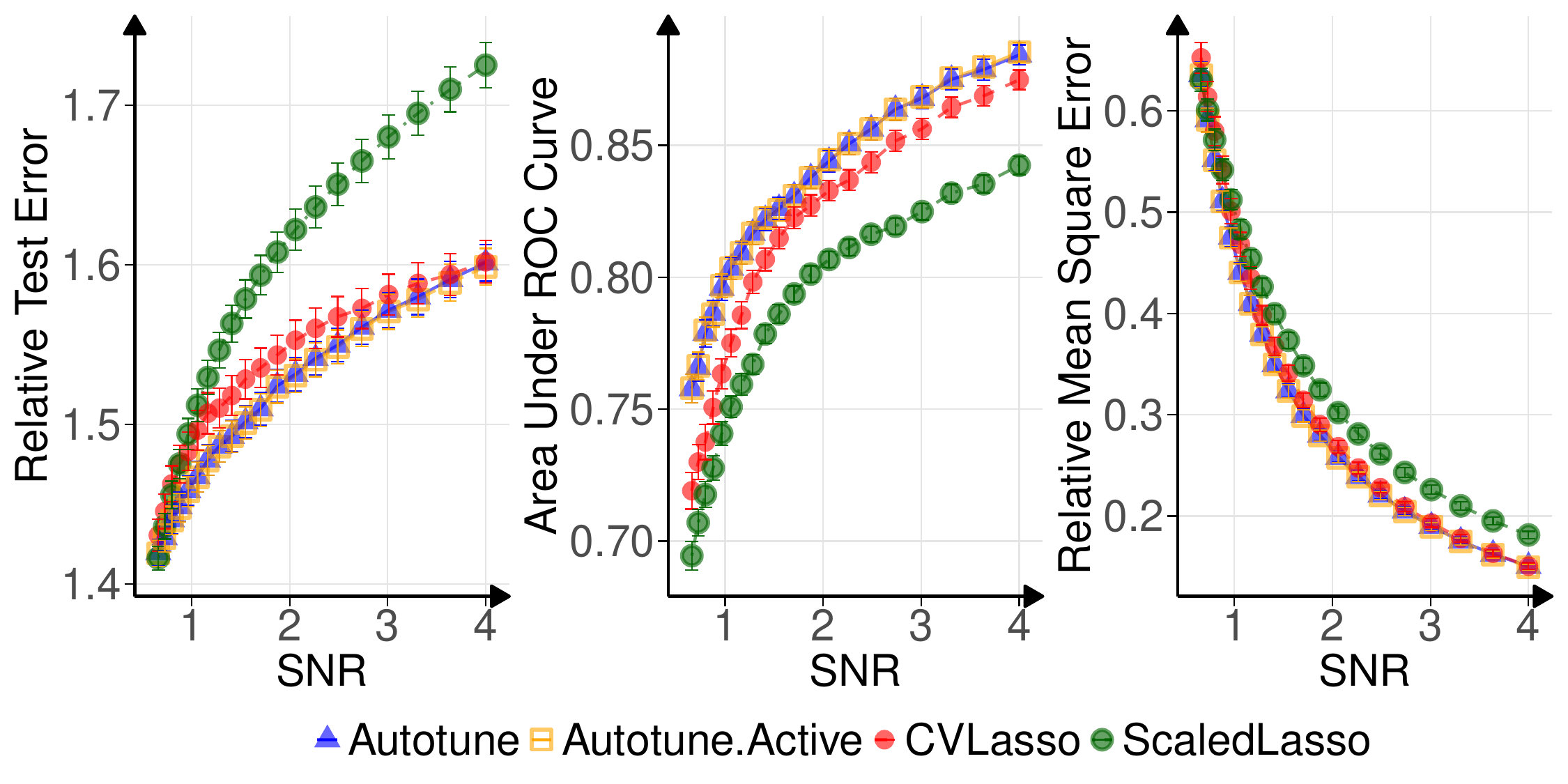}\\
        \includegraphics[trim=0in 0.8in 0in 0in, clip, width=0.92\textwidth]{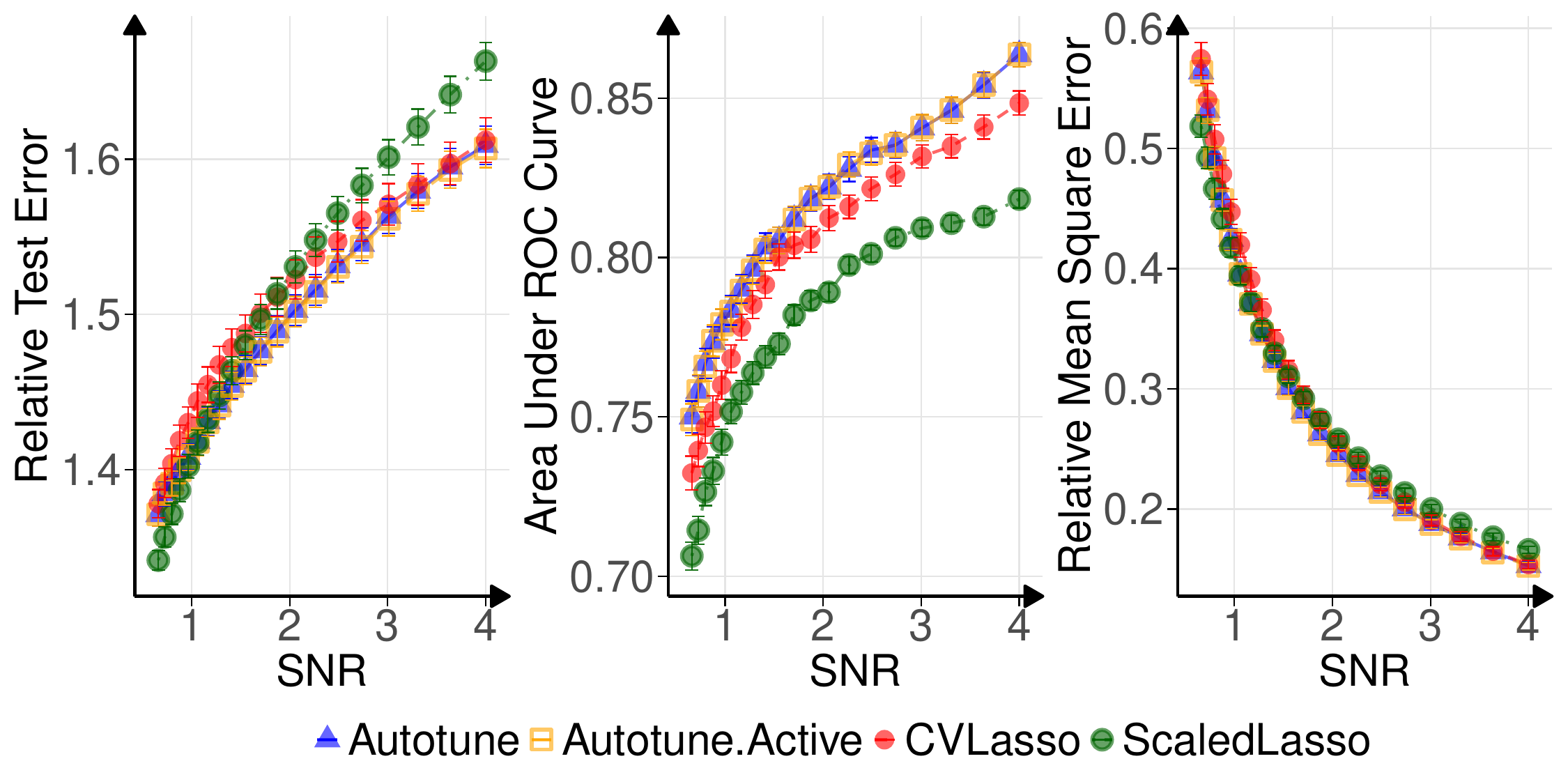}\\
        \includegraphics[trim=0in 0in 0in 0in, clip, width=0.92\textwidth]{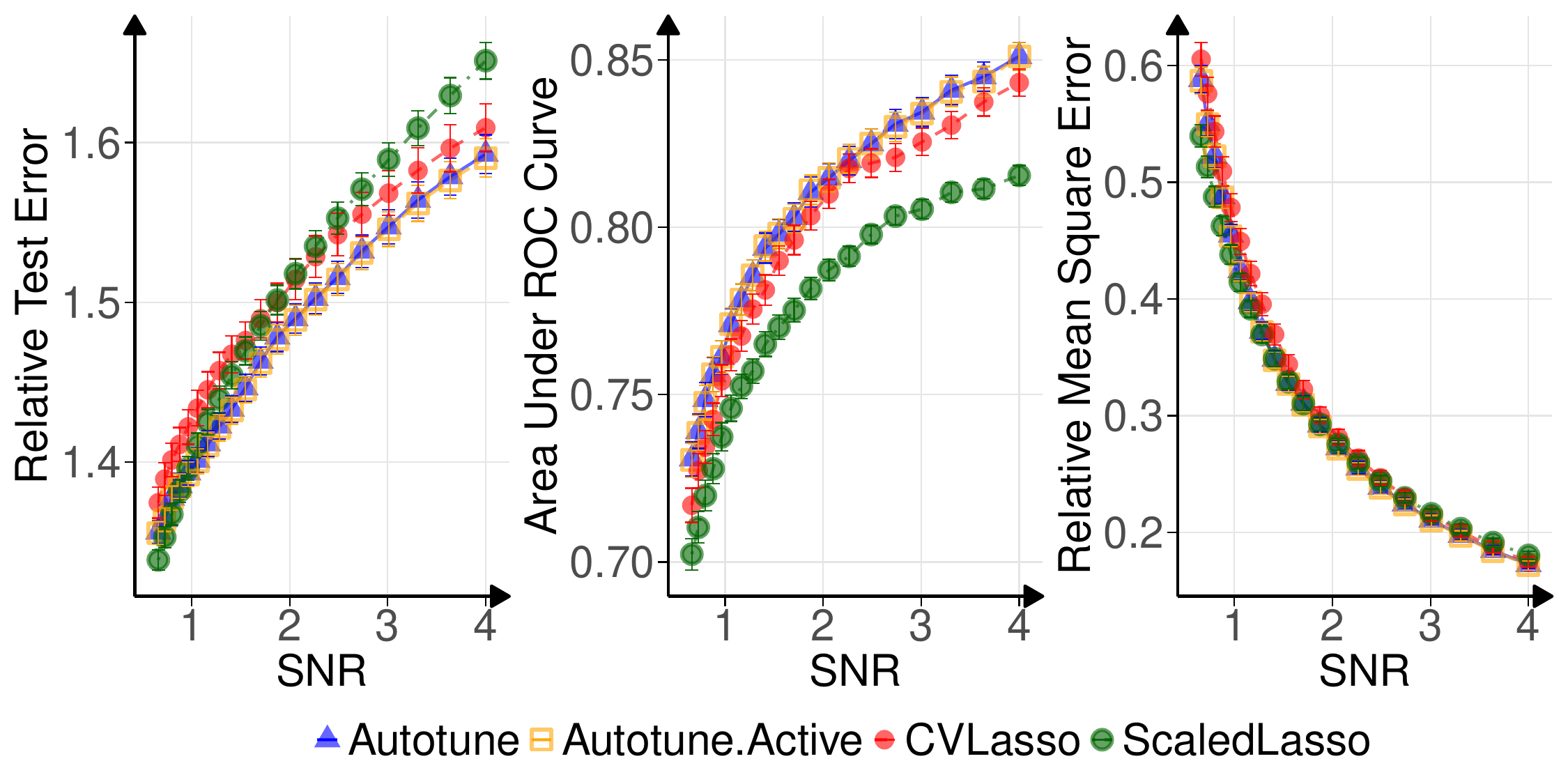}
    \end{tabular}
    \\
    \caption{RTE, AUROC and RMSE of $\autotune$, $\msf{autotune.active}$, CV, and Scaled Lasso plotted as a function of SNR for high-dimensional setup.}
    \label{fig: appen high dim accuracy plot btype 5}
\end{figure}

\subsection{Moderate Dimensional Setting: n = 80, p = 150, s = 5}
\leavevmode\par
\begin{figure}[H]
    \centering
    \begin{tabular}{c}
        \textbf{Moderate Dimensional setup:} $n = 80, \hspace{0.1cm} p=150, \hspace{0.1cm} s=5$, \textbf{Beta-type: 1}, Varying levels  \\
        of correlation, \textbf{Top Row:} $\rho = 0$, \textbf{Middle Row:} $\rho = 0.35$ and \textbf{Bottom Row:} $\rho = 0.7$\\
        \includegraphics[trim=0in 0.8in 0in 0in, clip, width=0.85\textwidth]{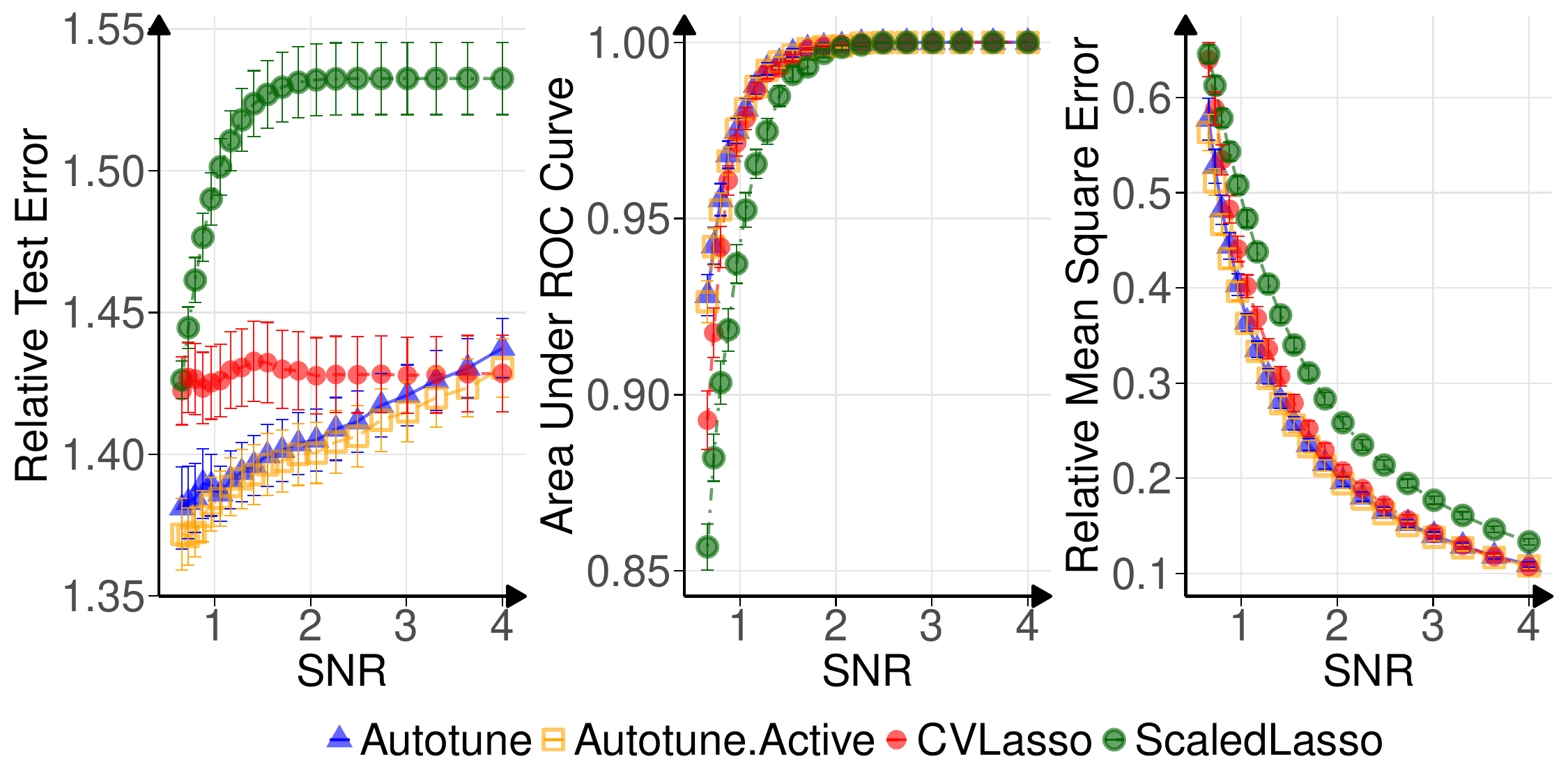}\\
        \includegraphics[trim=0in 0.8in 0in 0in, clip, width=0.85\textwidth]{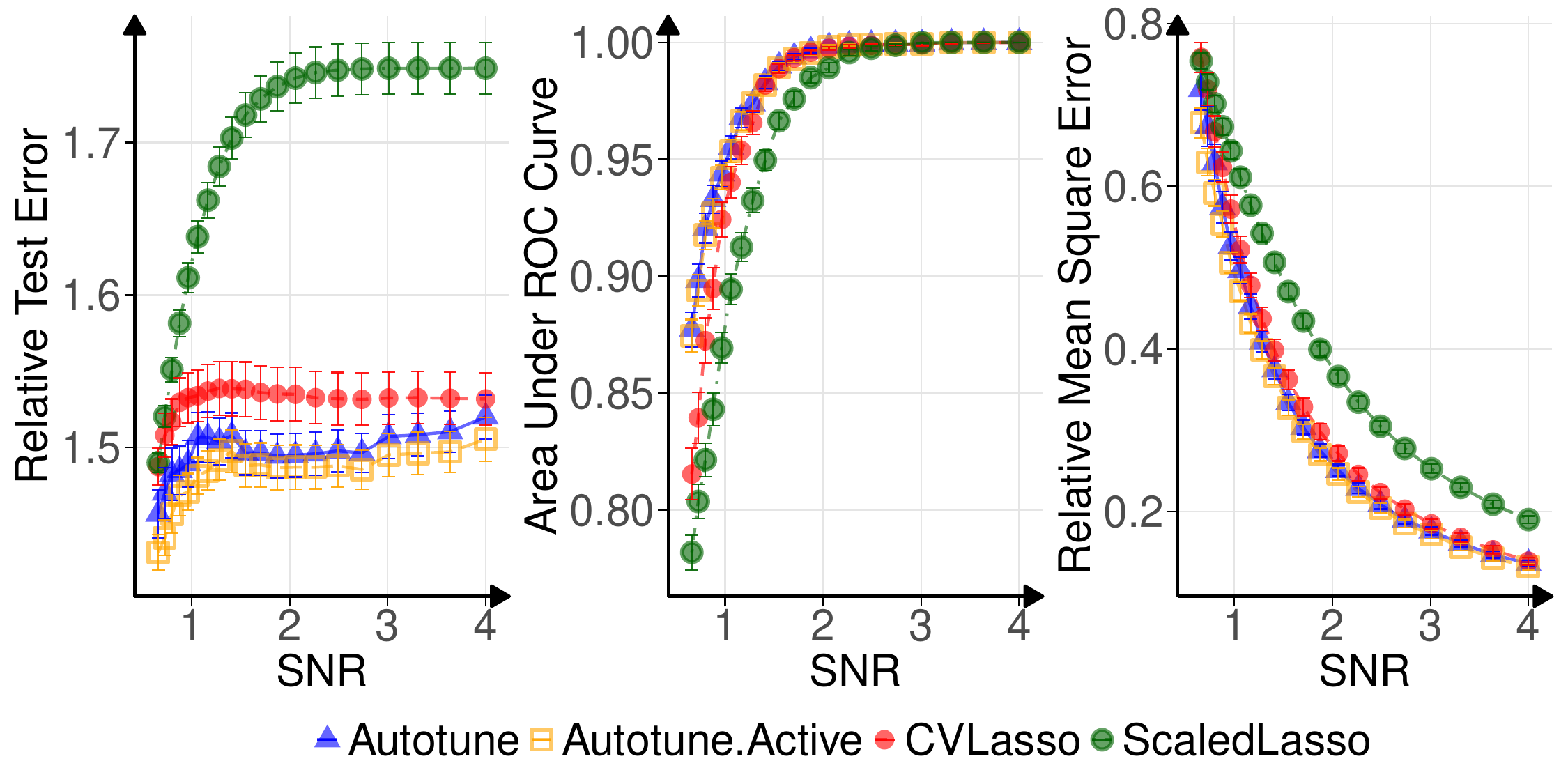}\\
        \includegraphics[trim=0in 0in 0in 0in, clip, width=0.85\textwidth]{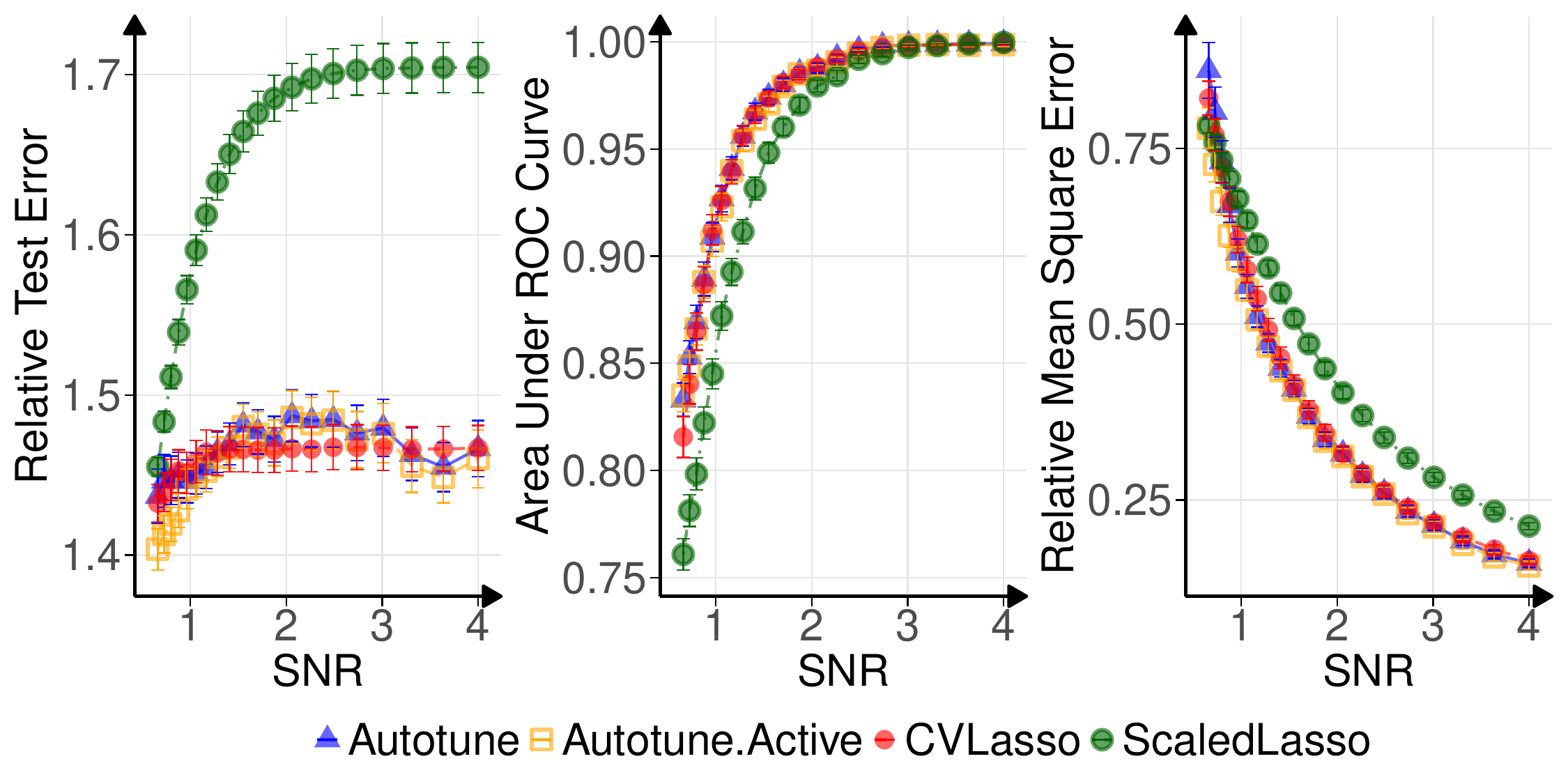}
    \end{tabular}
    \\
    \caption{RTE, AUROC and RMSE of $\autotune$, $\msf{autotune.active}$, CV, and Scaled Lasso plotted as a function of SNR for moderate-dimensional setup.}
    \label{fig: appen moderate dim accuracy plot btype 1}
\end{figure}

\begin{figure}[H]
    \centering
    \begin{tabular}{c}
        \textbf{Moderate Dimensional setup:} $n = 80, \hspace{0.1cm} p=150, \hspace{0.1cm} s=5$, \textbf{Beta-type: 2}, Varying levels  \\
        of correlation, \textbf{Top Row:} $\rho = 0$, \textbf{Middle Row:} $\rho = 0.35$ and \textbf{Bottom Row:} $\rho = 0.7$\\
        \includegraphics[trim=0in 0.8in 0in 0in, clip, width=0.92\textwidth]{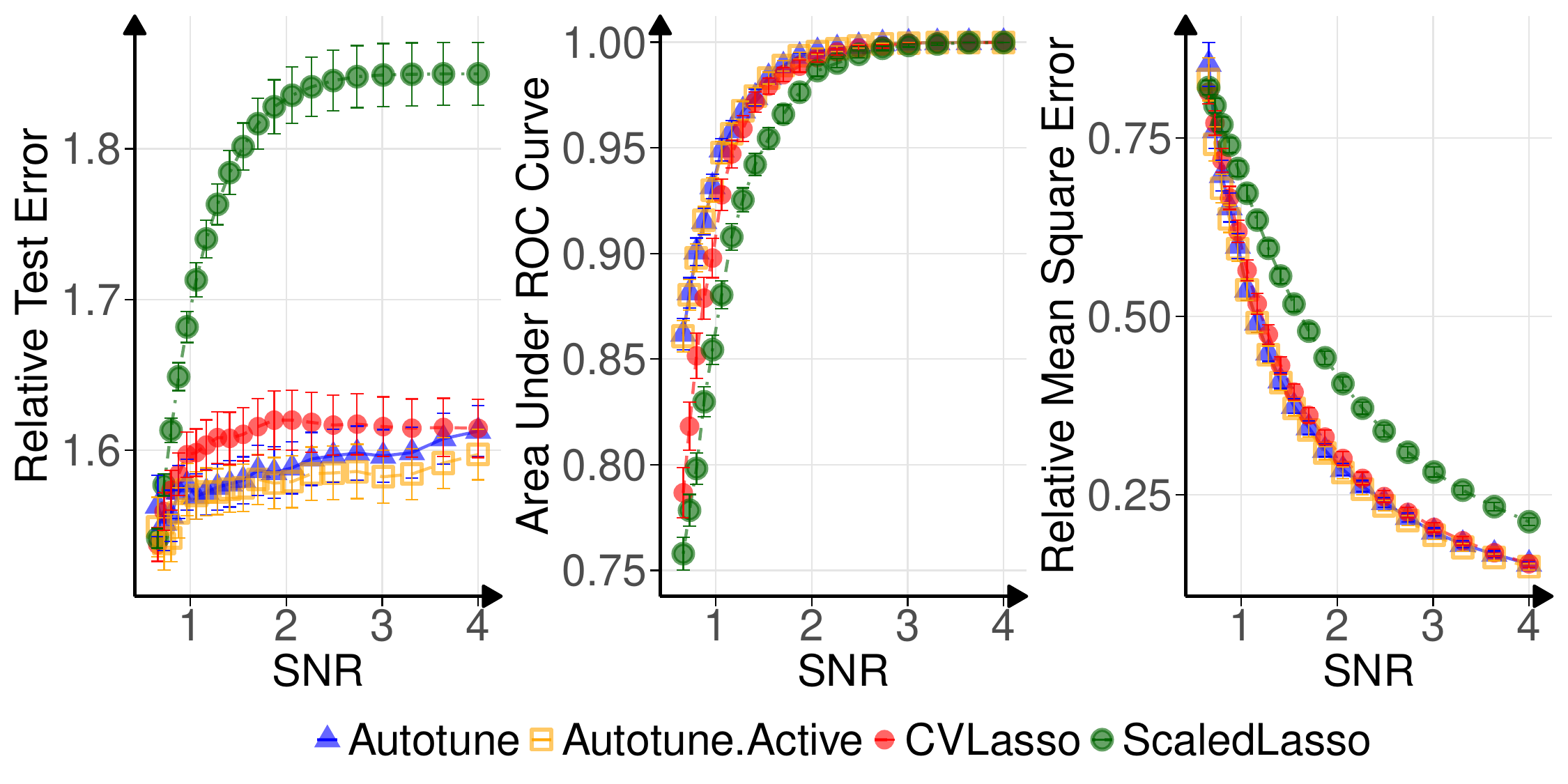}\\
        \includegraphics[trim=0in 0.8in 0in 0in, clip, width=0.92\textwidth]{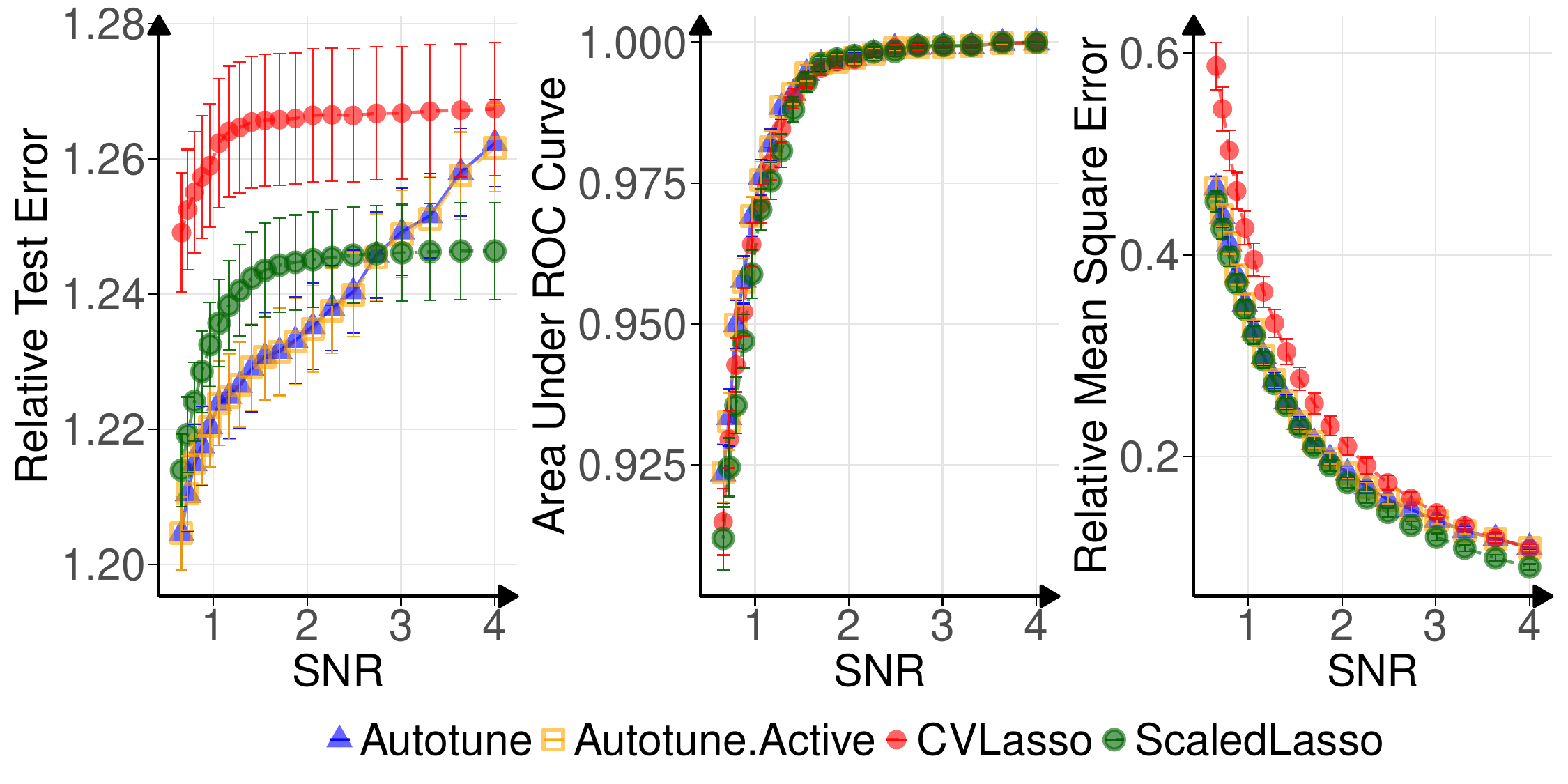}
    \end{tabular}
    \\
    \caption{RTE, AUROC and RMSE of $\autotune$, $\msf{autotune.active}$, CV, and Scaled Lasso plotted as a function of SNR for moderate-dimensional setup.}
    \label{fig: appen moderate dim accuracy plot btype 2}
\end{figure}
\clearpage
\begin{figure}[H]
    \ContinuedFloat
    \centering
    \begin{tabular}{c}
        \textbf{Moderate Dimensional setup, Beta-type 2 (continued):} \textbf{Bottom Row:} $\rho = 0.7$\\
        \includegraphics[trim=0in 0in 0in 0in, clip, width=0.92\textwidth]{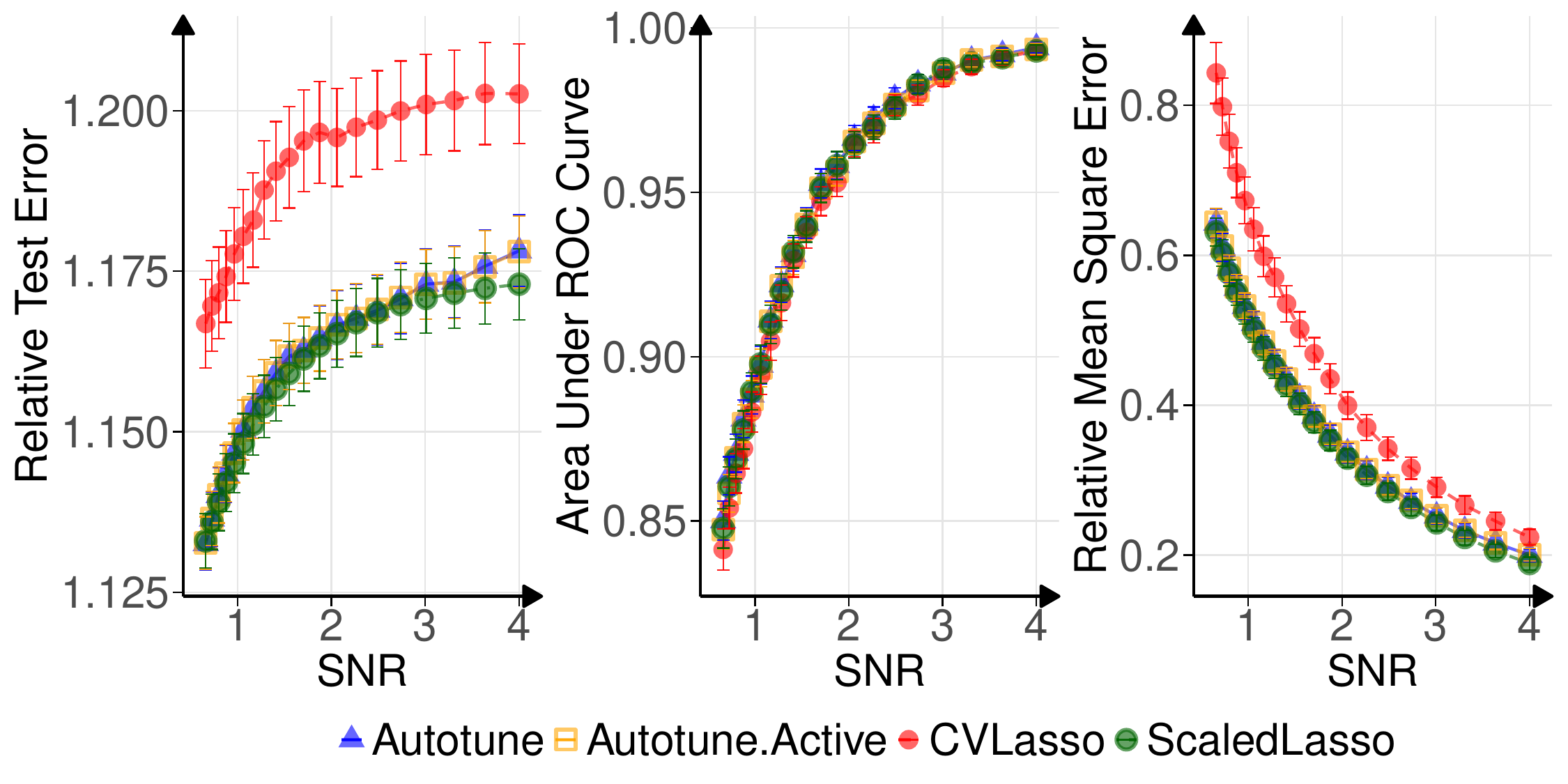}
    \end{tabular}
    \caption[]{Continued: bottom row with $\rho = 0.7$.}
\end{figure}

\begin{figure}[H]
    \centering
    \begin{tabular}{c}
        \textbf{Moderate Dimensional setup:} $n = 80, \hspace{0.1cm} p=150, \hspace{0.1cm} s=5$, \textbf{Beta-type: 3}, Varying levels  \\
        of correlation, \textbf{Top Row:} $\rho = 0$, \textbf{Middle Row:} $\rho = 0.35$ and \textbf{Bottom Row:} $\rho = 0.7$\\
        \includegraphics[trim=0in 0.8in 0in 0in, clip, width=0.92\textwidth]{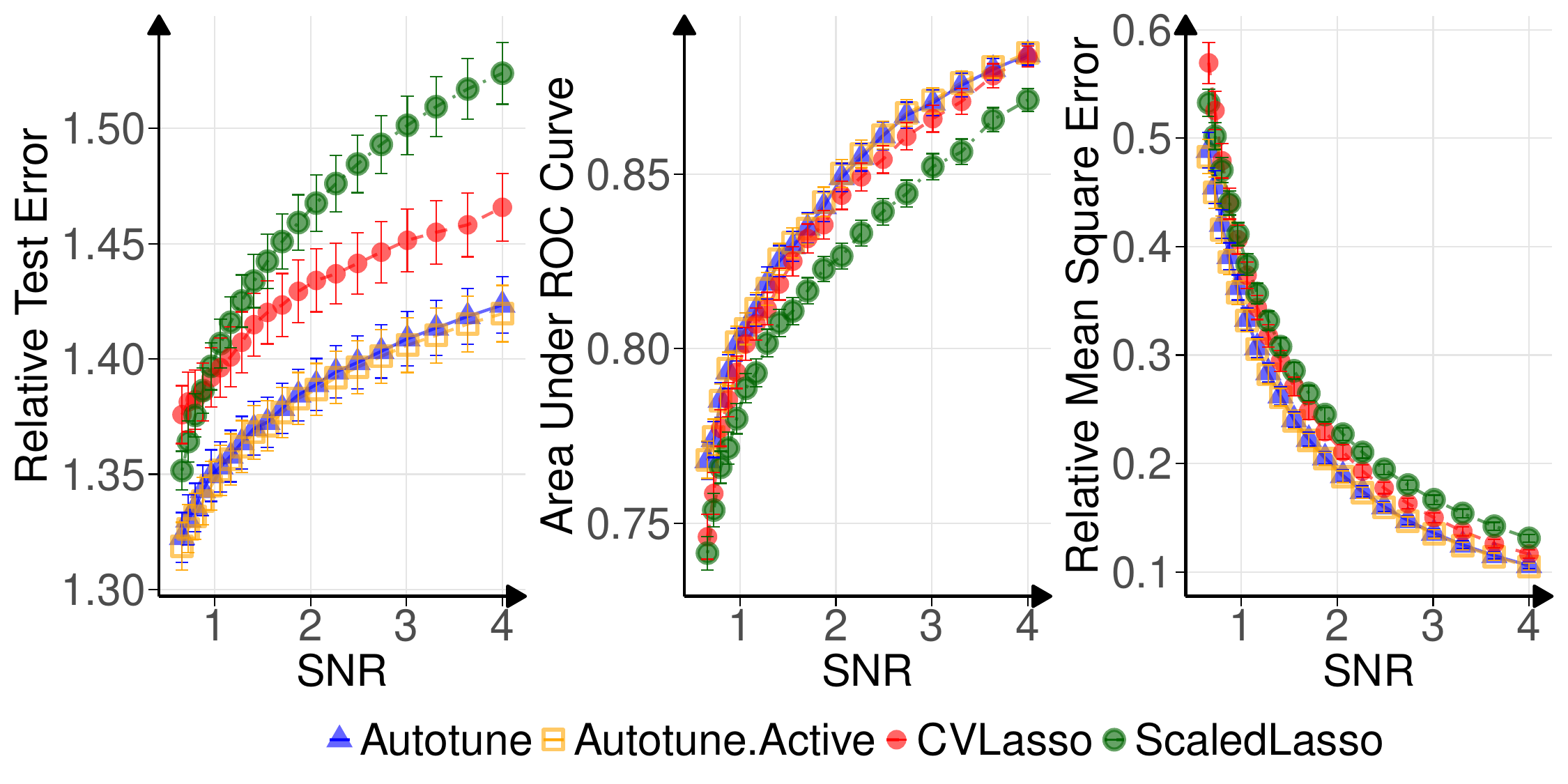}\\
        \includegraphics[trim=0in 0.8in 0in 0in, clip, width=0.92\textwidth]{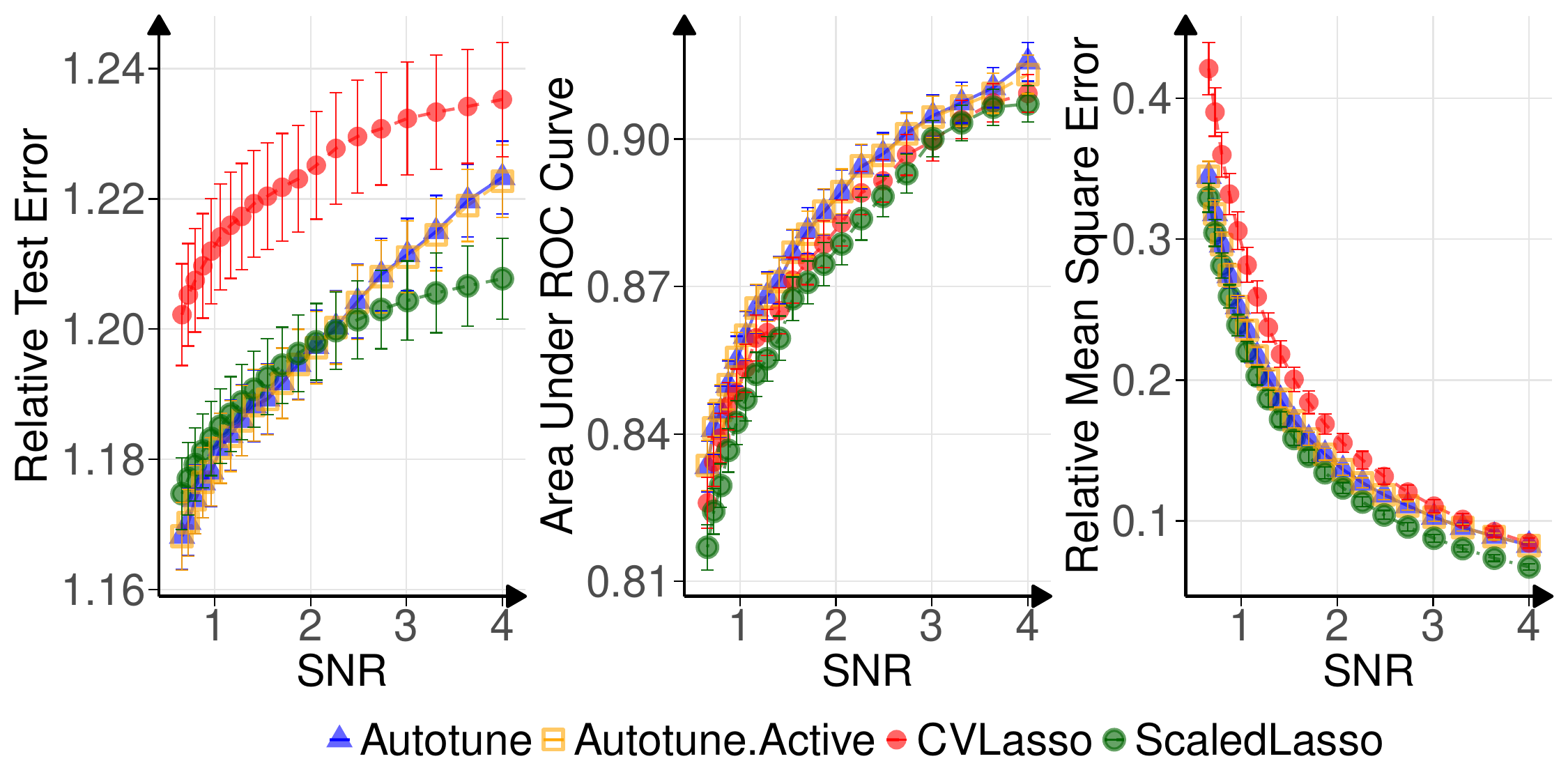}\\
        \includegraphics[trim=0in 0in 0in 0in, clip, width=0.92\textwidth]{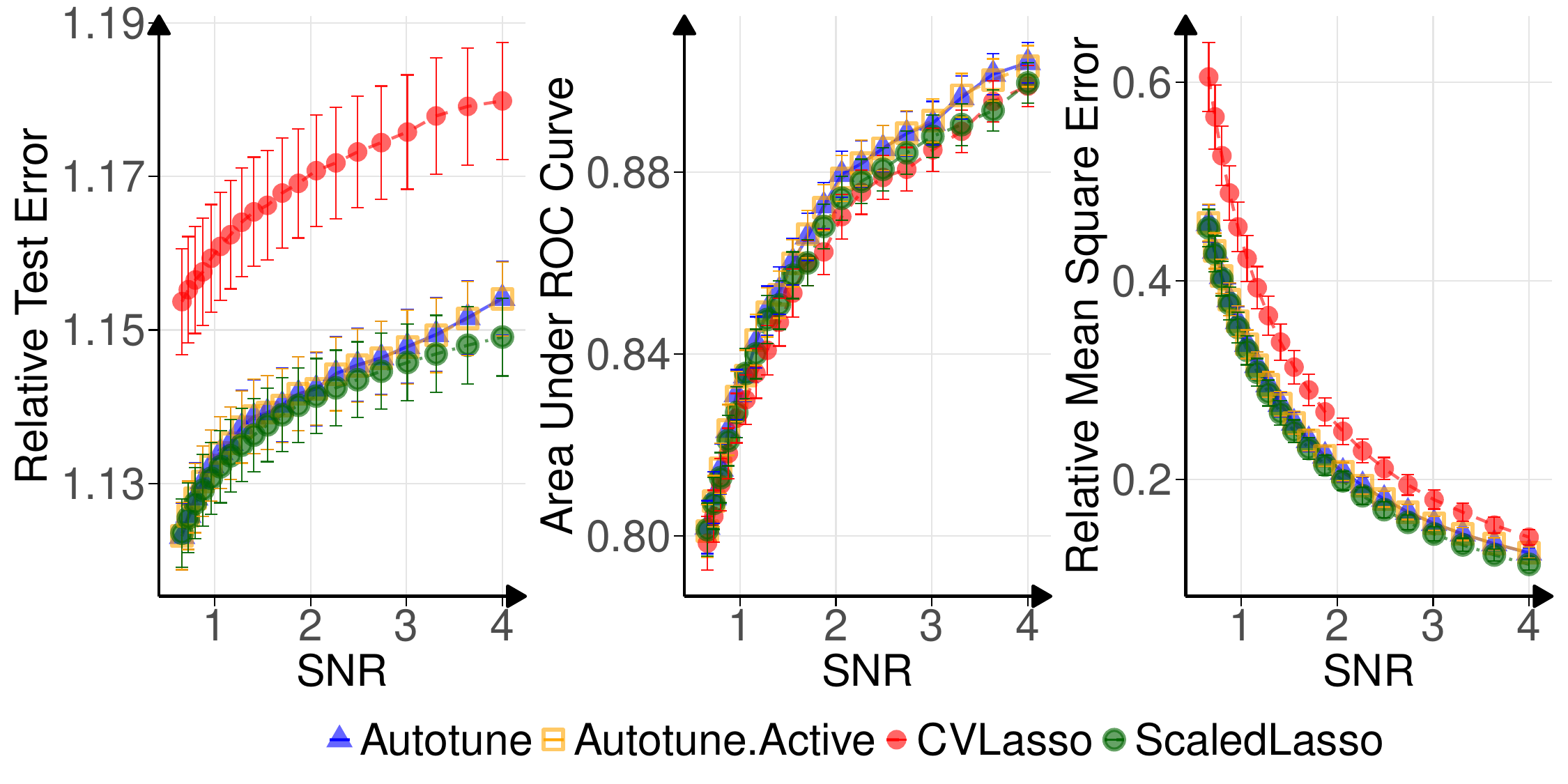}
    \end{tabular}
    \\
    \caption{RTE, AUROC and RMSE of $\autotune$, $\msf{autotune.active}$, CV, and Scaled Lasso plotted as a function of SNR for moderate-dimensional setup.}
    \label{fig: appen moderate dim accuracy plot btype 3}
\end{figure}

\begin{figure}[H]
    \centering
    \begin{tabular}{c}
        \textbf{Moderate Dimensional setup:} $n = 80, \hspace{0.1cm} p=150, \hspace{0.1cm} s=5$, \textbf{Beta-type: 5}, Varying levels  \\
        of correlation, \textbf{Top Row:} $\rho = 0$, \textbf{Middle Row:} $\rho = 0.35$ and \textbf{Bottom Row:} $\rho = 0.7$\\
        \includegraphics[trim=0in 0.8in 0in 0in, clip, width=0.92\textwidth]{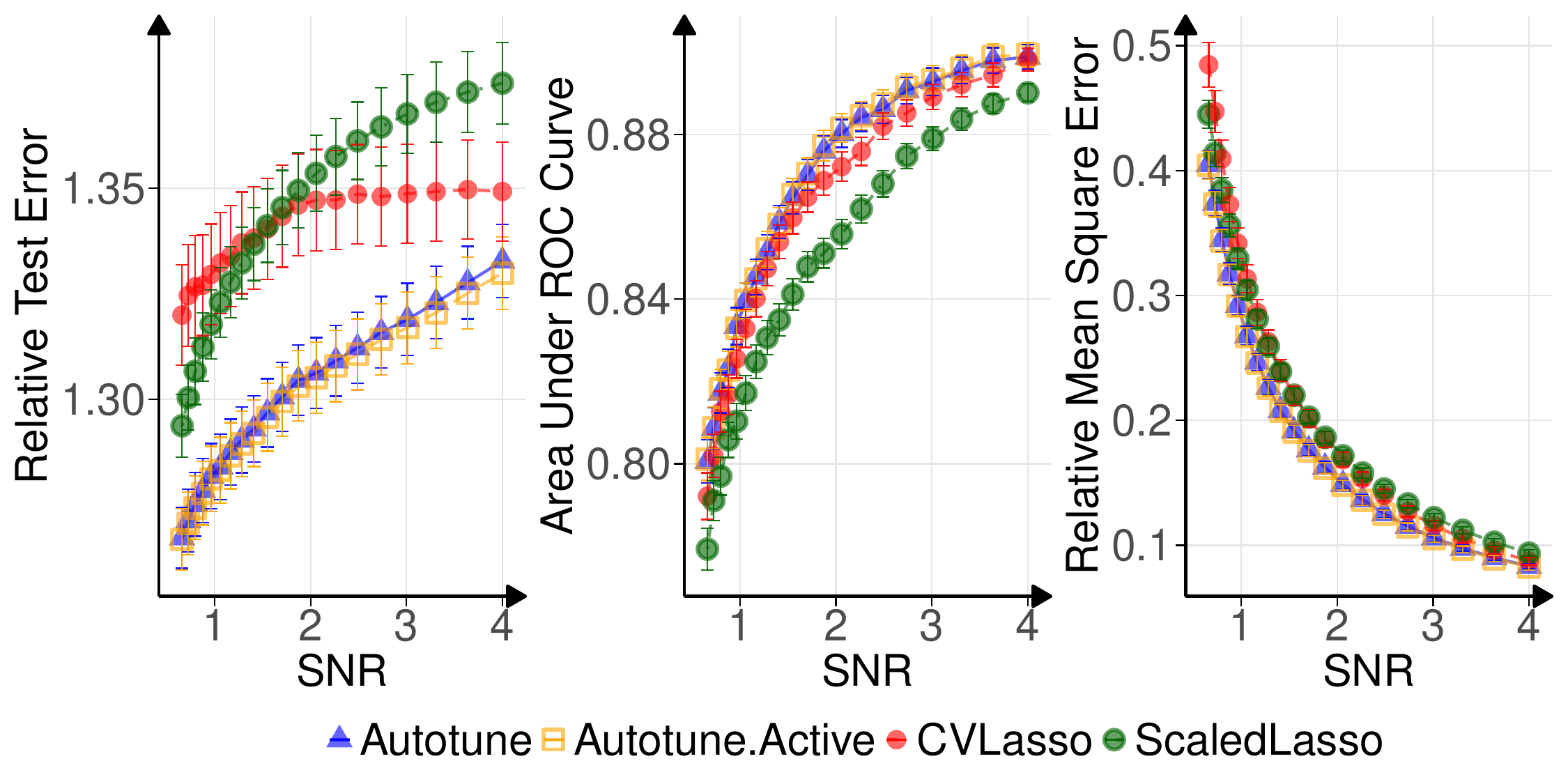}\\
        \includegraphics[trim=0in 0.8in 0in 0in, clip, width=0.92\textwidth]{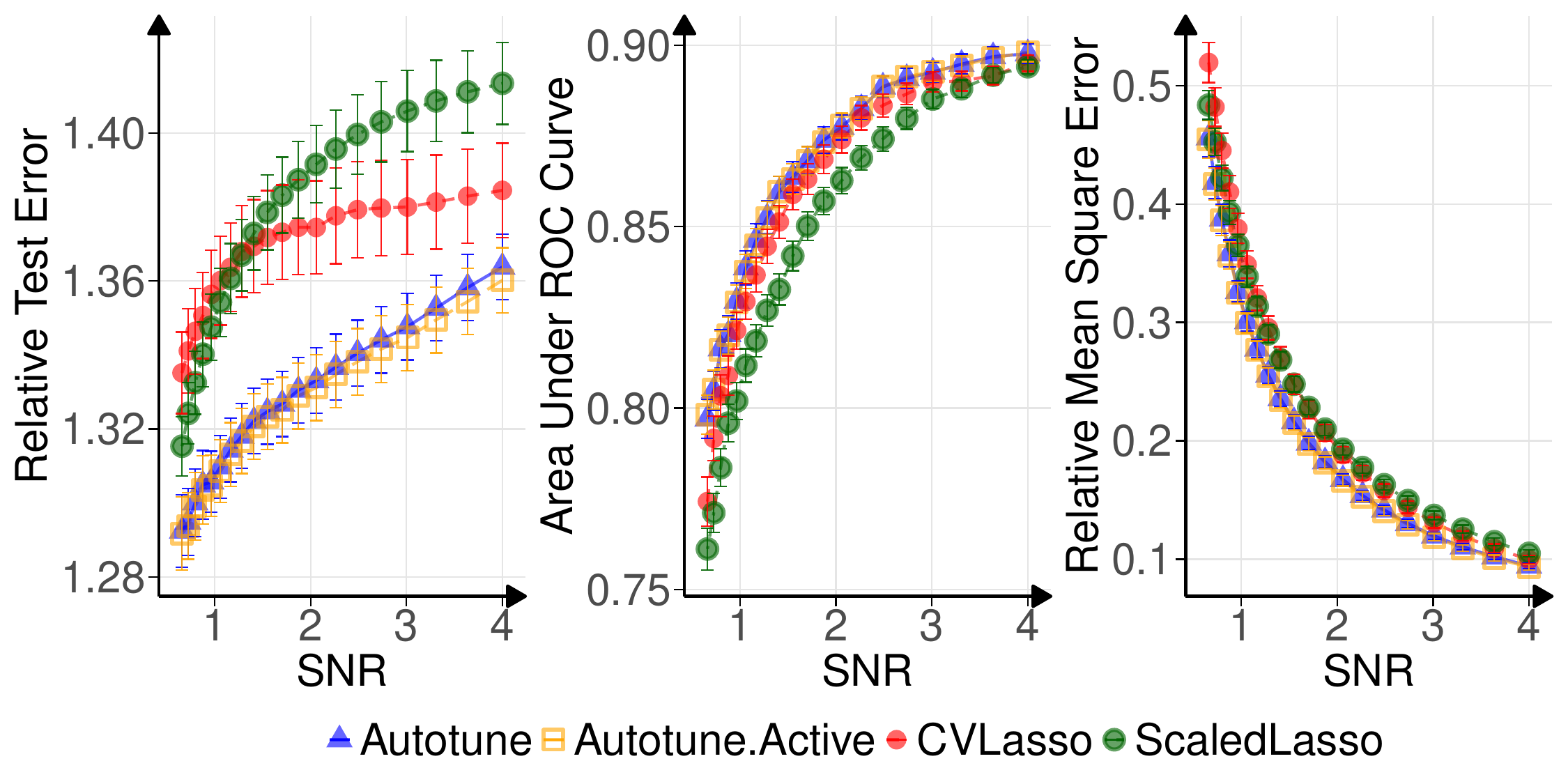}
    \end{tabular}
    \\
    \caption{RTE, AUROC and RMSE of $\autotune$, $\msf{autotune.active}$, CV, and Scaled Lasso plotted as a function of SNR for moderate-dimensional setup.}
    \label{fig: appen moderate dim accuracy plot btype 5}
\end{figure}
\clearpage
\begin{figure}[H]
    \ContinuedFloat
    \centering
    \begin{tabular}{c}
        \textbf{Moderate Dimensional setup, Beta-type 5 (continued):} \textbf{Bottom Row:} $\rho = 0.7$\\
        \includegraphics[trim=0in 0in 0in 0in, clip, width=0.92\textwidth]{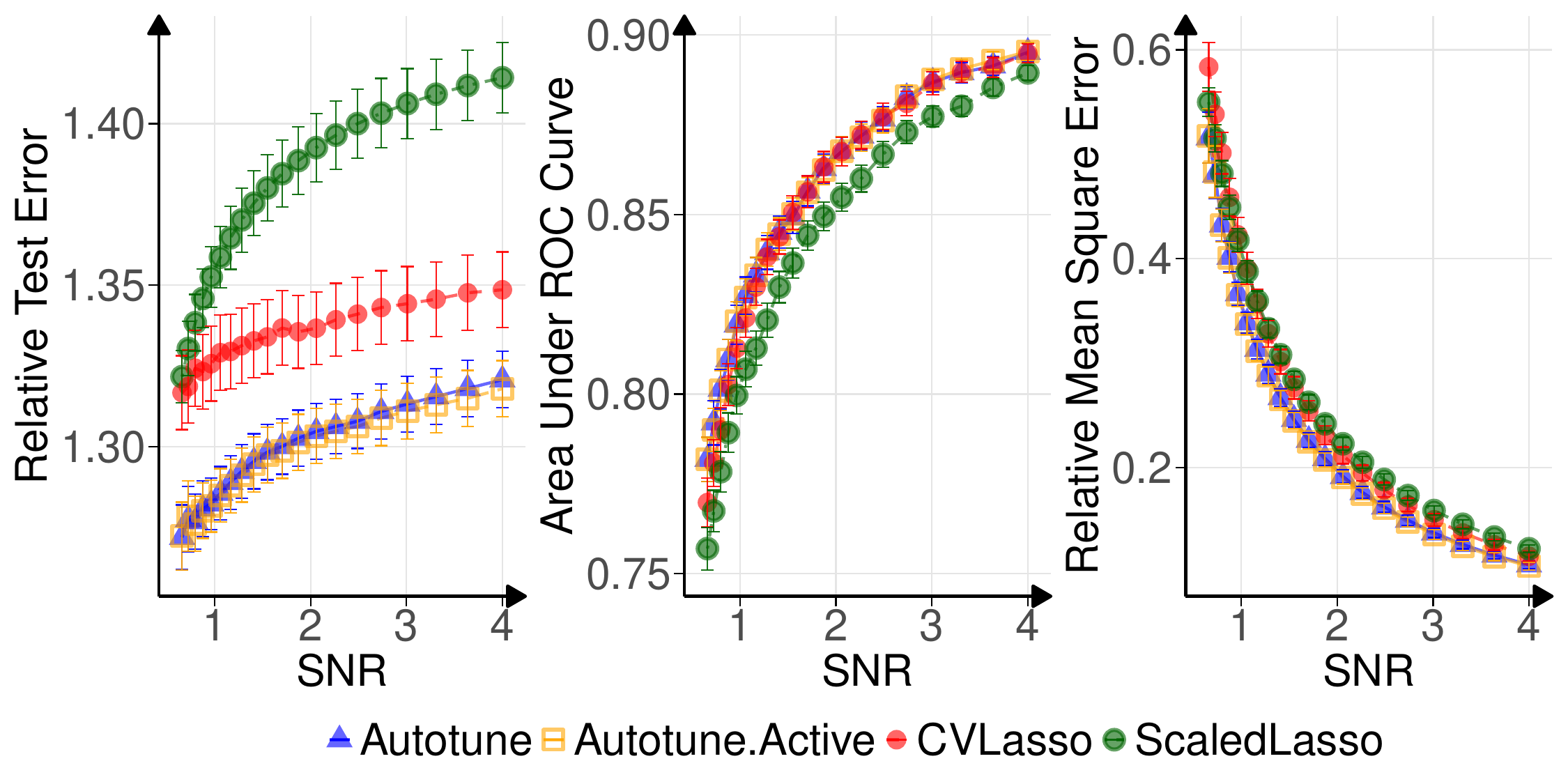}
    \end{tabular}
    \caption[]{Continued: bottom row with $\rho = 0.7$.}
\end{figure}

\subsection{Low Dimensional Setting: n = 80, p = 50, s = 5}
\leavevmode\par

\begin{figure}[H]
    \centering
    \begin{tabular}{c}
        \textbf{Low Dimensional setup:} $n = 80, \hspace{0.1cm} p=50, \hspace{0.1cm} s=5$, \textbf{Beta-type: 1}, Varying levels  \\
        of correlation, \textbf{Top Row:} $\rho = 0$, \textbf{Middle Row:} $\rho = 0.35$ and \textbf{Bottom Row:} $\rho = 0.7$\\
        \includegraphics[trim=0in 0.8in 0in 0in, clip, width=0.85\textwidth]{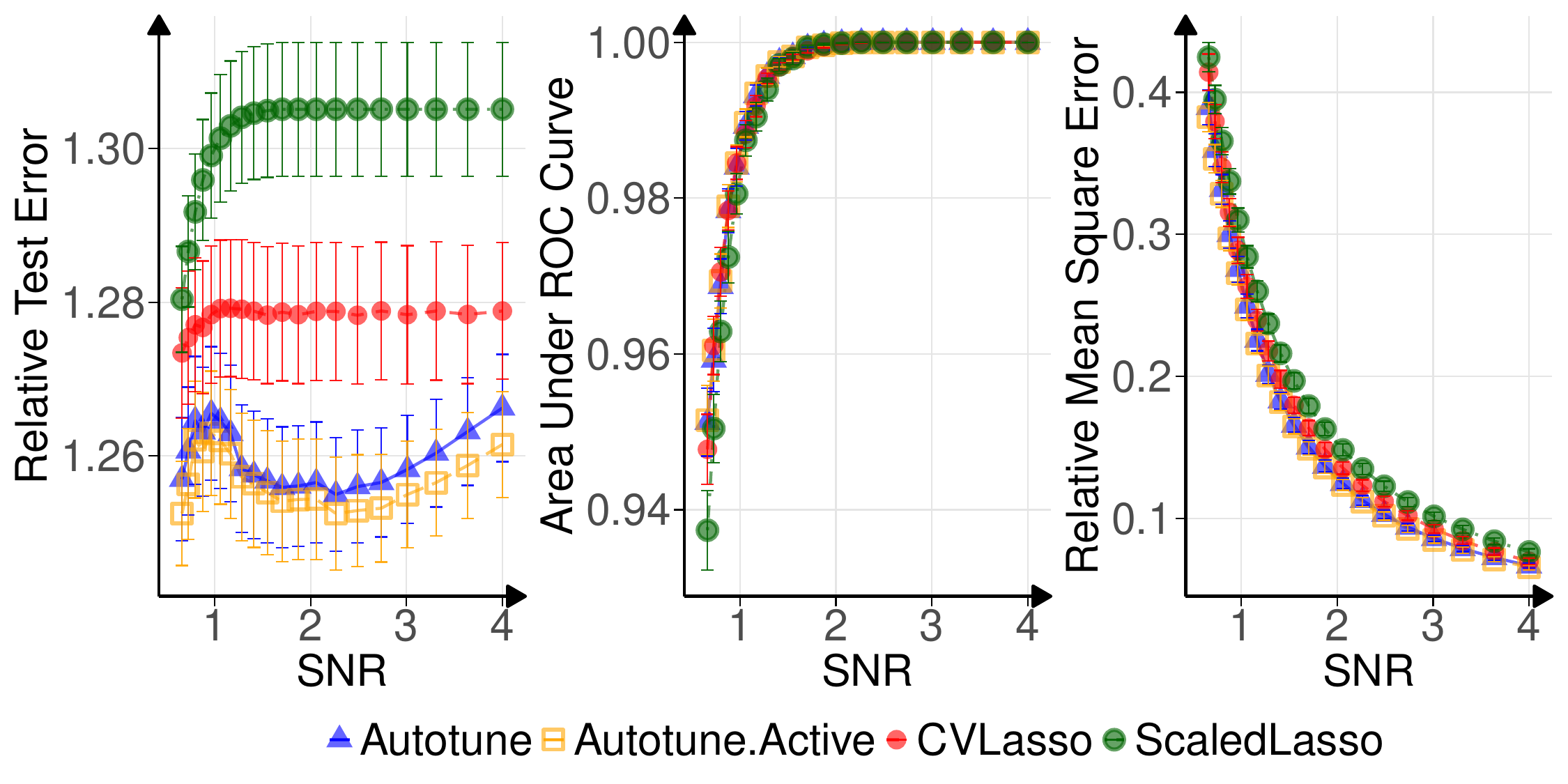}\\
        \includegraphics[trim=0in 0.8in 0in 0in, clip, width=0.85\textwidth]{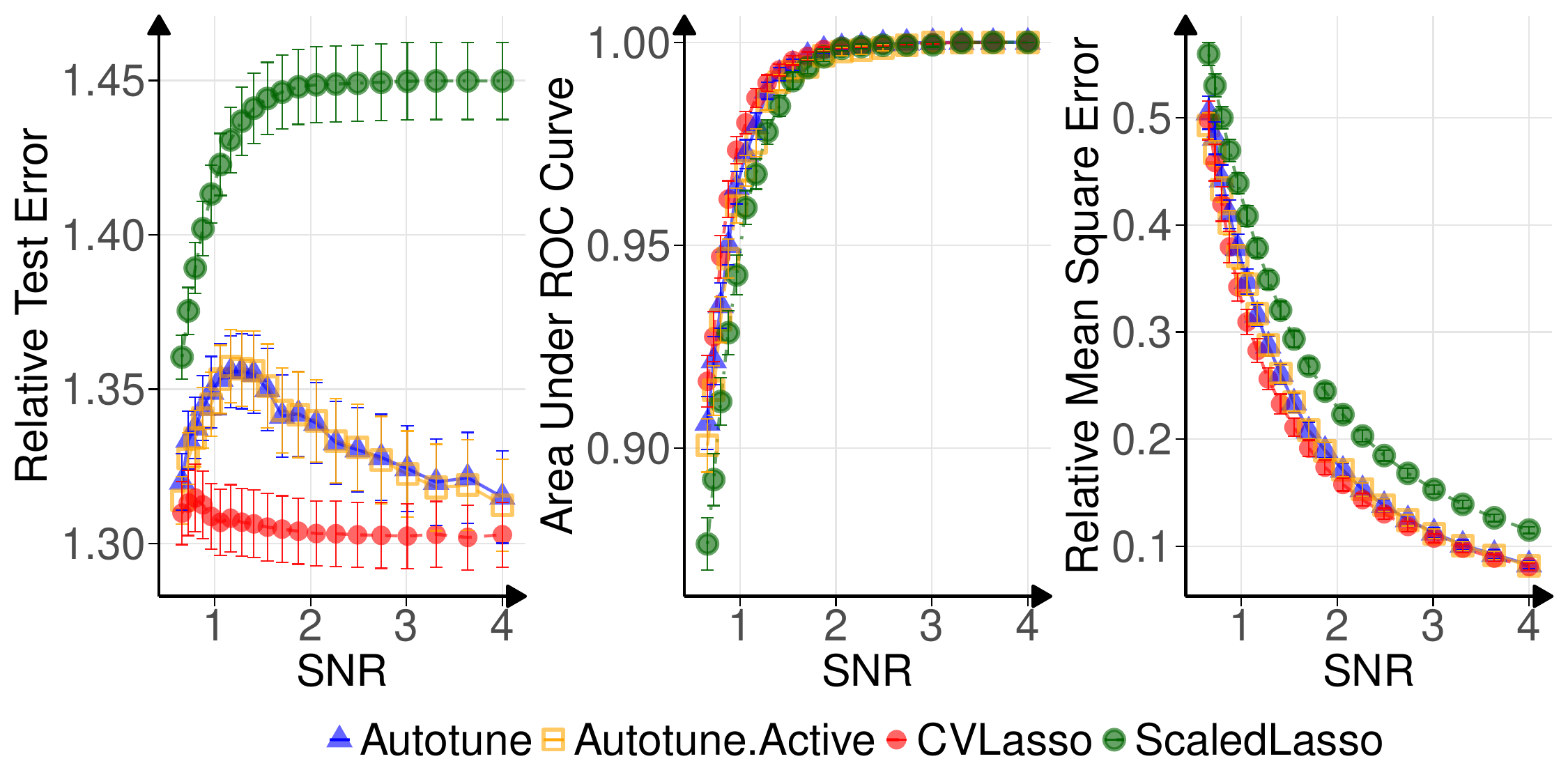}\\
        \includegraphics[trim=0in 0in 0in 0in, clip, width=0.85\textwidth]{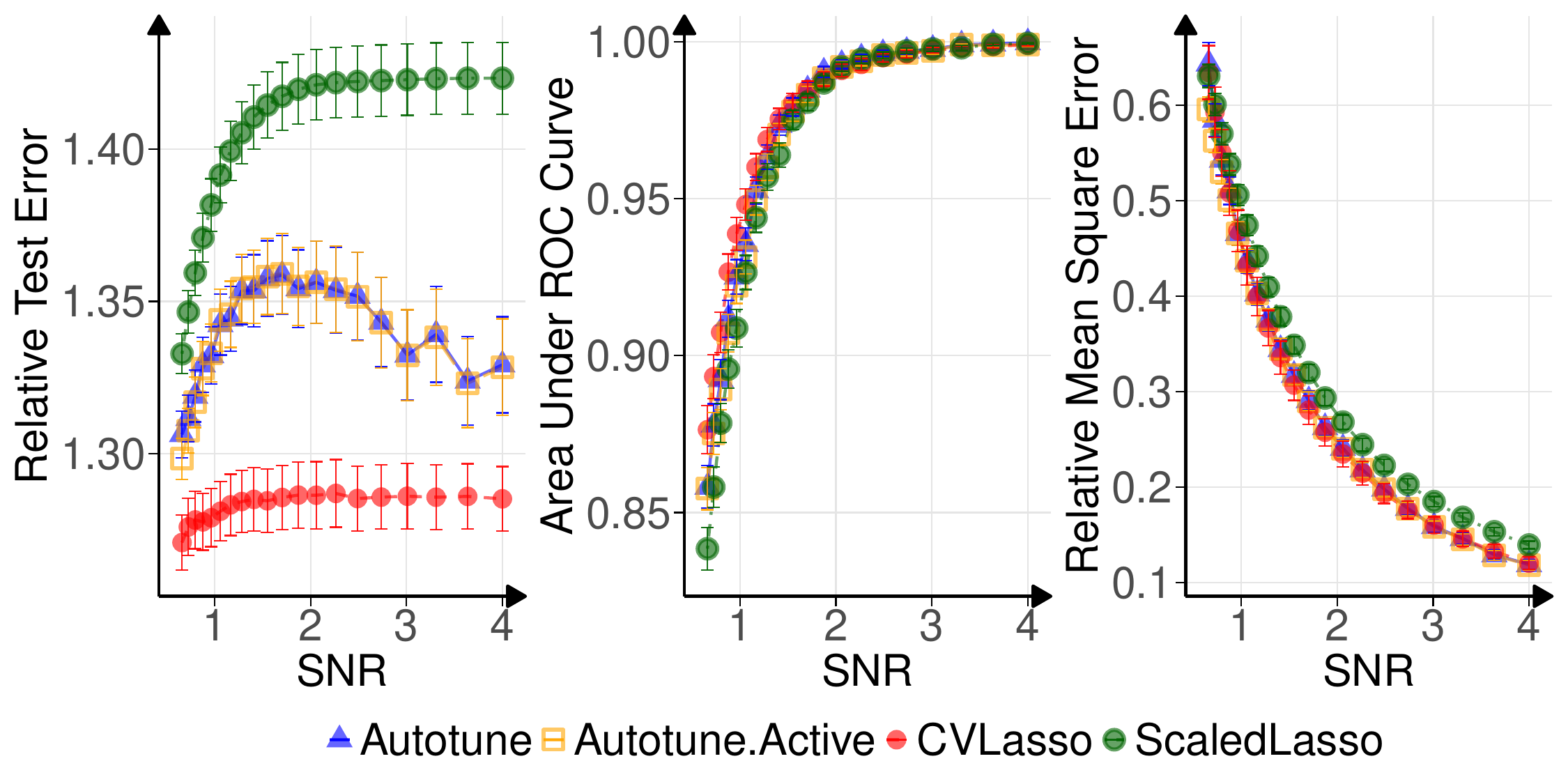}
    \end{tabular}
    \\
    \caption{RTE, AUROC and RMSE of $\autotune$, $\msf{autotune.active}$, CV, and Scaled Lasso plotted as a function of SNR for low-dimensional setup.}
    \label{fig: appen low dim accuracy plot btype 1}
\end{figure}

\begin{figure}[H]
    \centering
    \begin{tabular}{c}
        \textbf{Low Dimensional setup:} $n = 80, \hspace{0.1cm} p=50, \hspace{0.1cm} s=5$, \textbf{Beta-type: 2}, Varying levels  \\
        of correlation, \textbf{Top Row:} $\rho = 0$, \textbf{Middle Row:} $\rho = 0.35$ and \textbf{Bottom Row:} $\rho = 0.7$\\
        \includegraphics[trim=0in 0.8in 0in 0in, clip, width=0.92\textwidth]{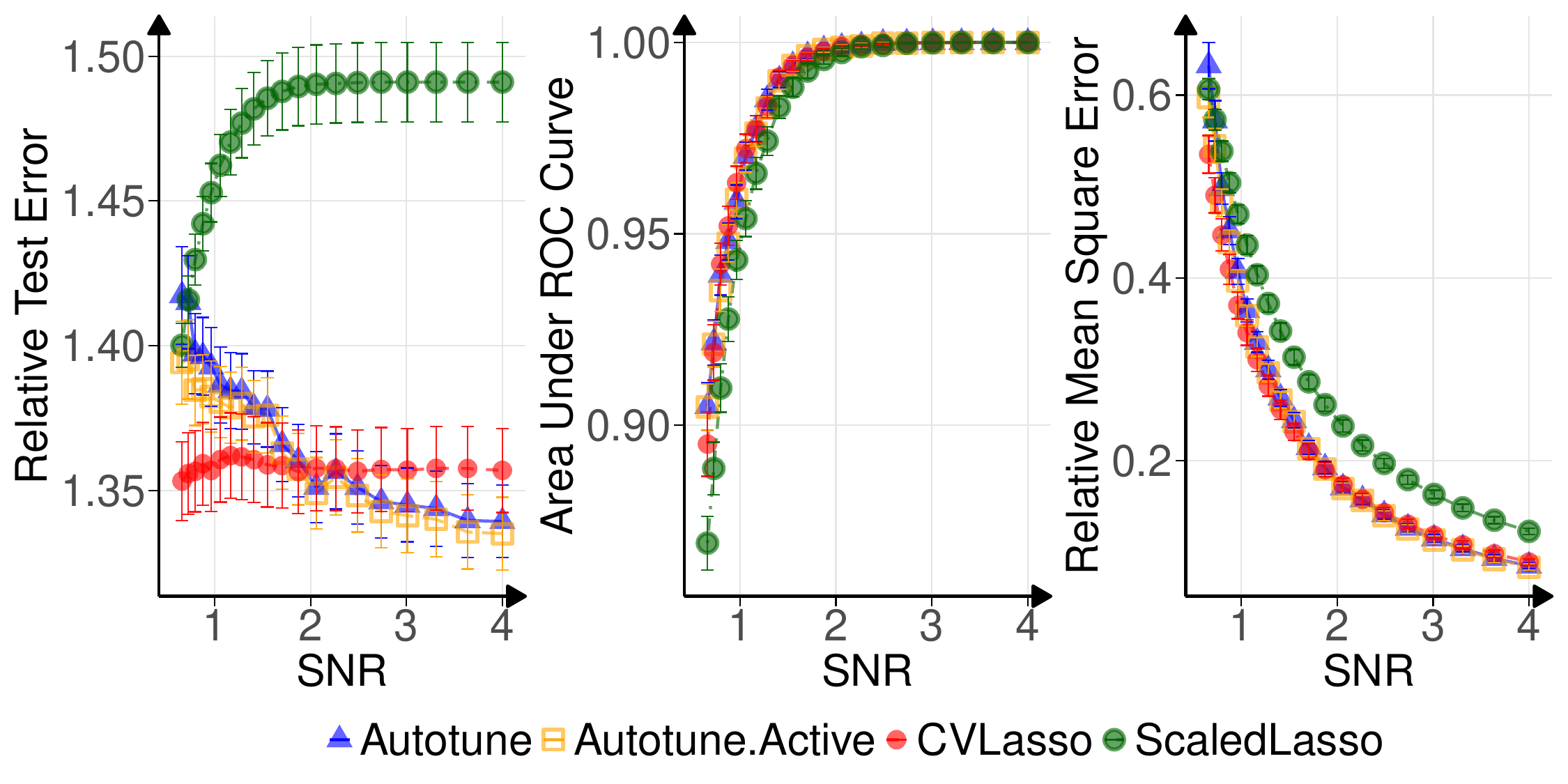}\\
        \includegraphics[trim=0in 0.8in 0in 0in, clip, width=0.92\textwidth]{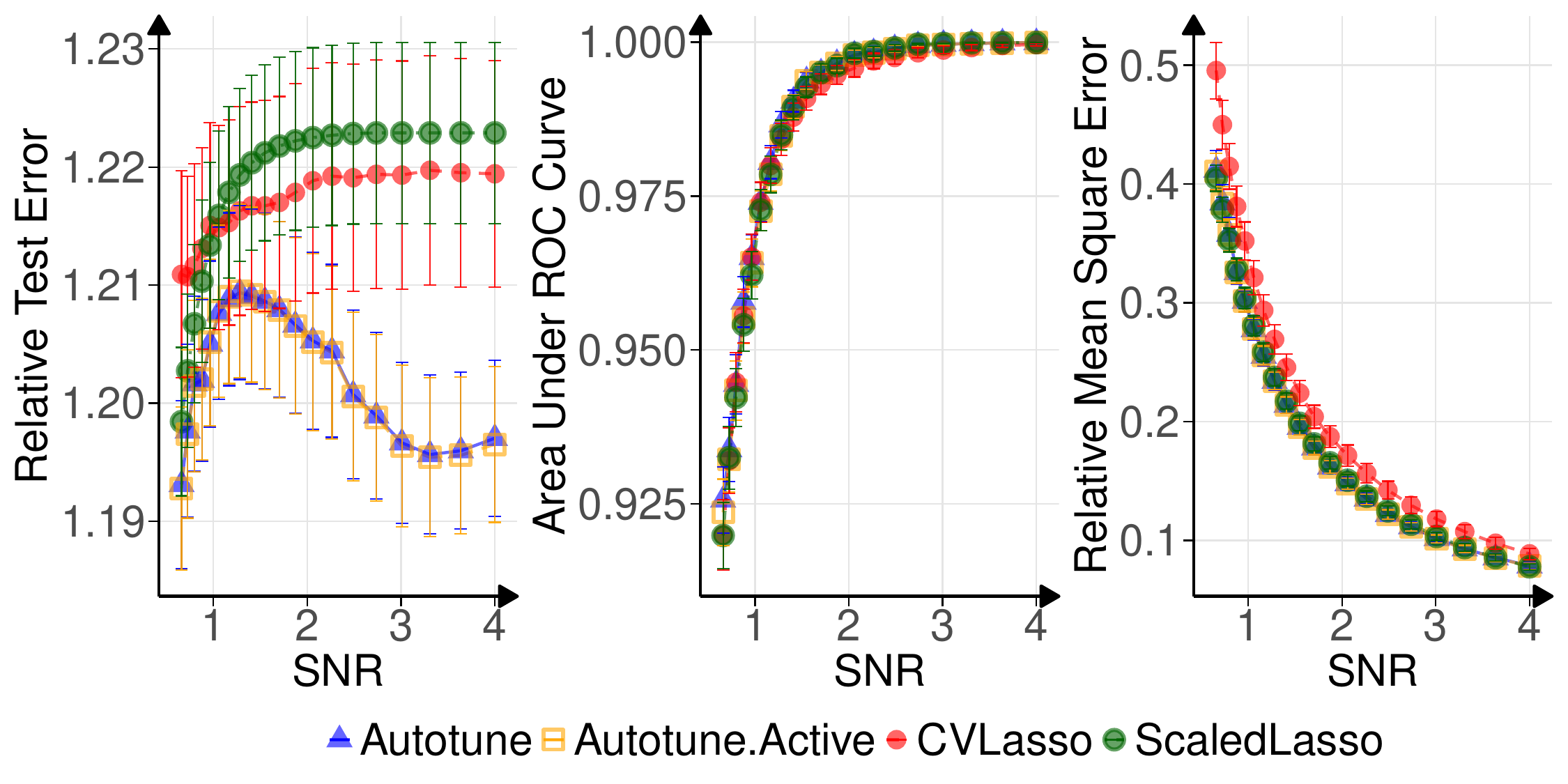}\\
        \includegraphics[trim=0in 0in 0in 0in, clip, width=0.92\textwidth]{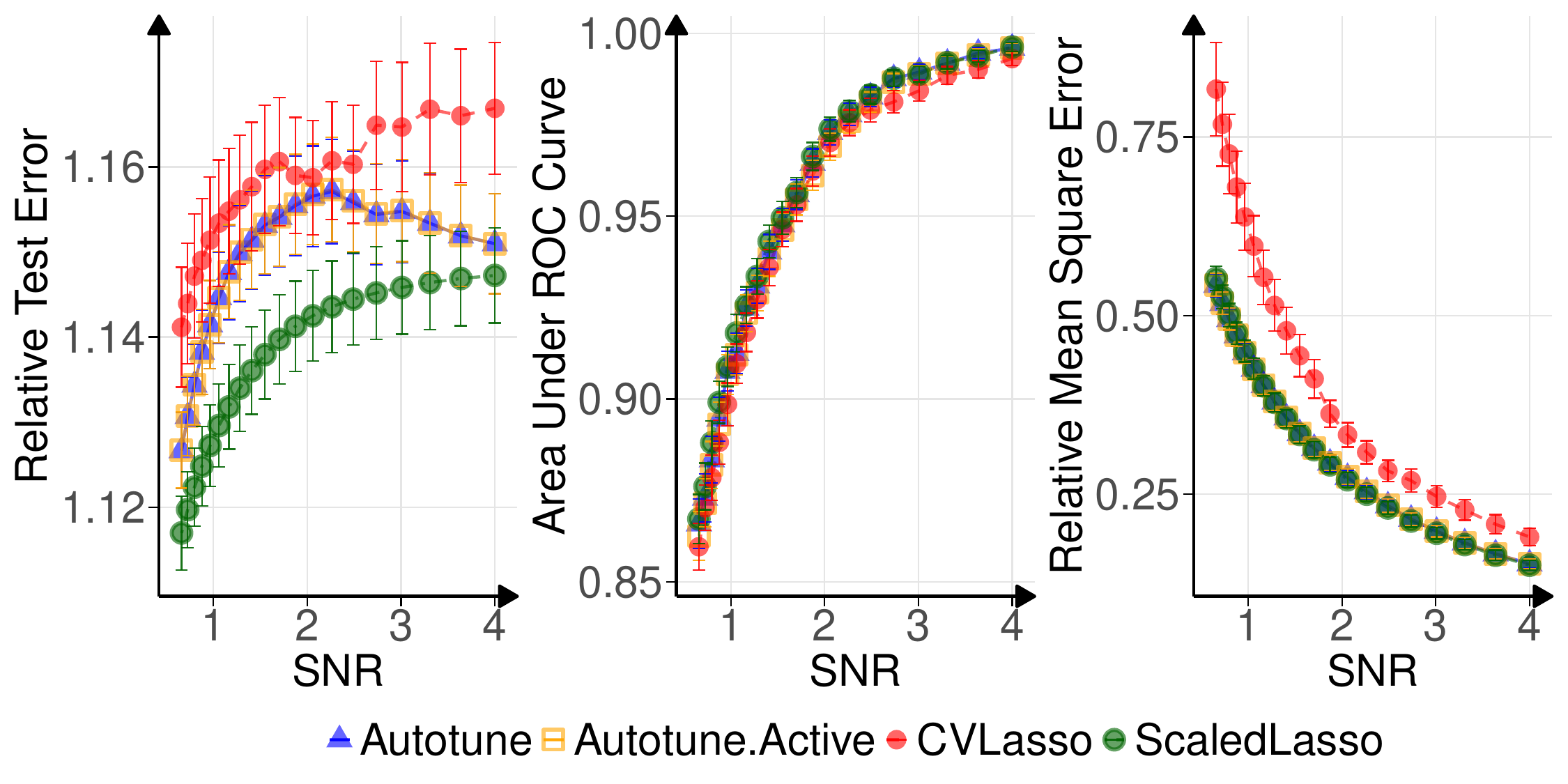}
    \end{tabular}
    \\
    \caption{RTE, AUROC and RMSE of $\autotune$, $\msf{autotune.active}$, CV, and Scaled Lasso plotted as a function of SNR for low-dimensional setup.}
    \label{fig: appen low dim accuracy plot btype 2}
\end{figure}

\begin{figure}[H]
    \centering
    \begin{tabular}{c}
        \textbf{Low Dimensional setup:} $n = 80, \hspace{0.1cm} p=50, \hspace{0.1cm} s=5$, \textbf{Beta-type: 3}, Varying levels  \\
        of correlation, \textbf{Top Row:} $\rho = 0$, \textbf{Middle Row:} $\rho = 0.35$ and \textbf{Bottom Row:} $\rho = 0.7$\\
        \includegraphics[trim=0in 0.8in 0in 0in, clip, width=0.92\textwidth]{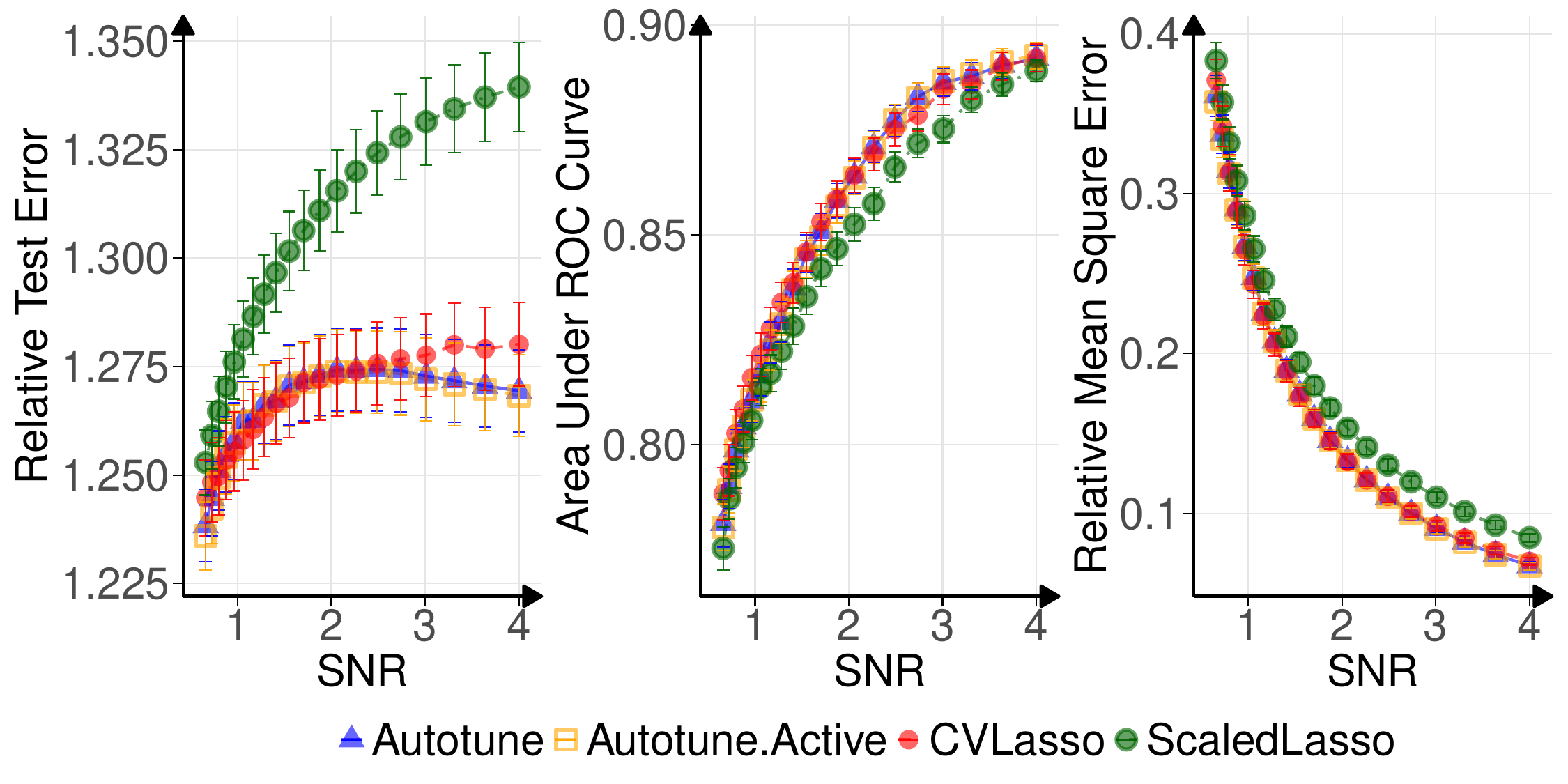}\\
        \includegraphics[trim=0in 0.8in 0in 0in, clip, width=0.92\textwidth]{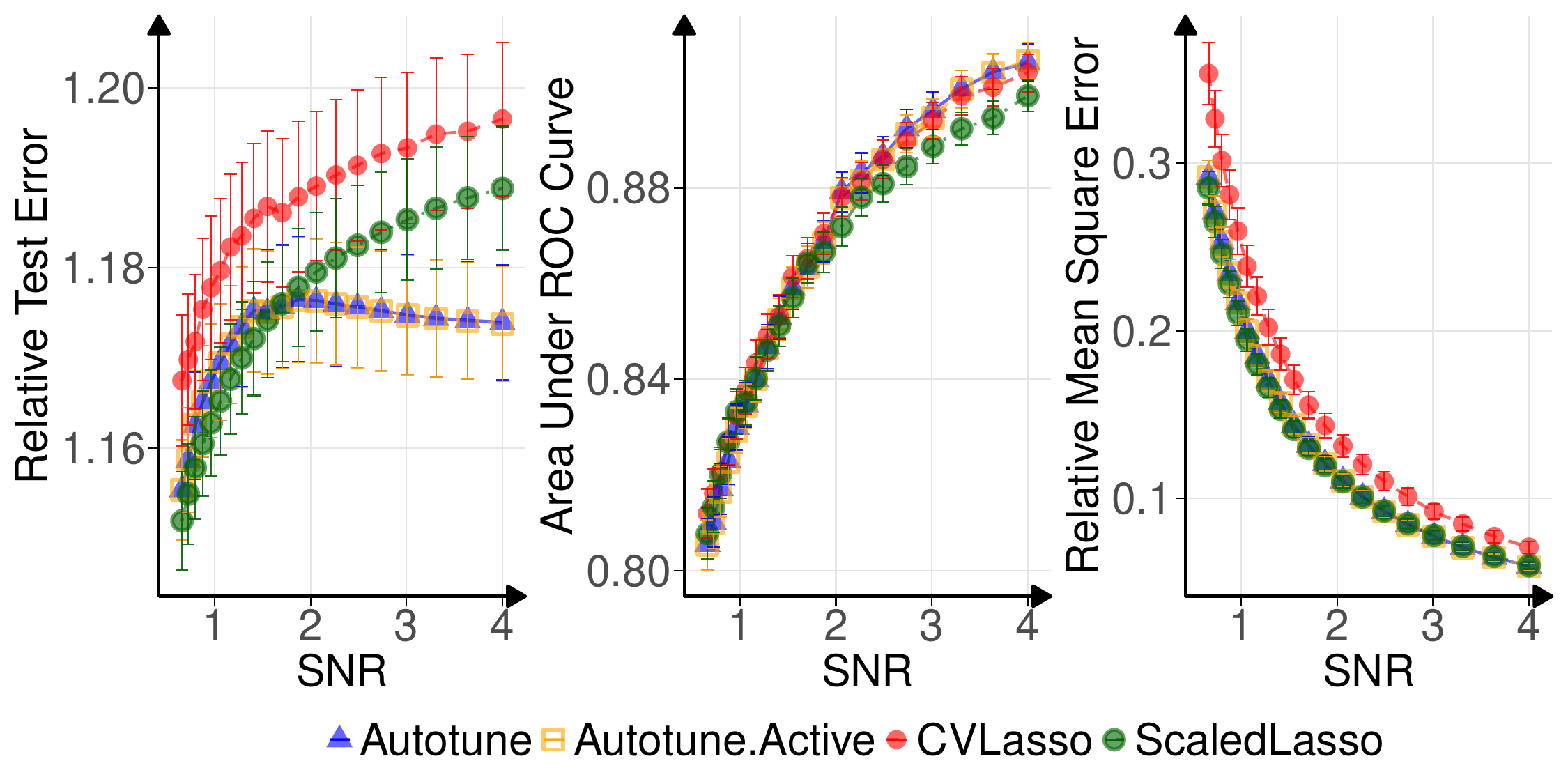}
    \end{tabular}
    \\
    \caption{RTE, AUROC and RMSE of $\autotune$, $\msf{autotune.active}$, CV, and Scaled Lasso plotted as a function of SNR for low-dimensional setup.}
    \label{fig: appen low dim accuracy plot btype 3}
\end{figure}
\clearpage
\begin{figure}[H]
    \ContinuedFloat
    \centering
    \begin{tabular}{c}
        \textbf{Low Dimensional setup, Beta-type 3 (continued):} \textbf{Bottom Row:} $\rho = 0.7$\\
        \includegraphics[trim=0in 0in 0in 0in, clip, width=0.92\textwidth]{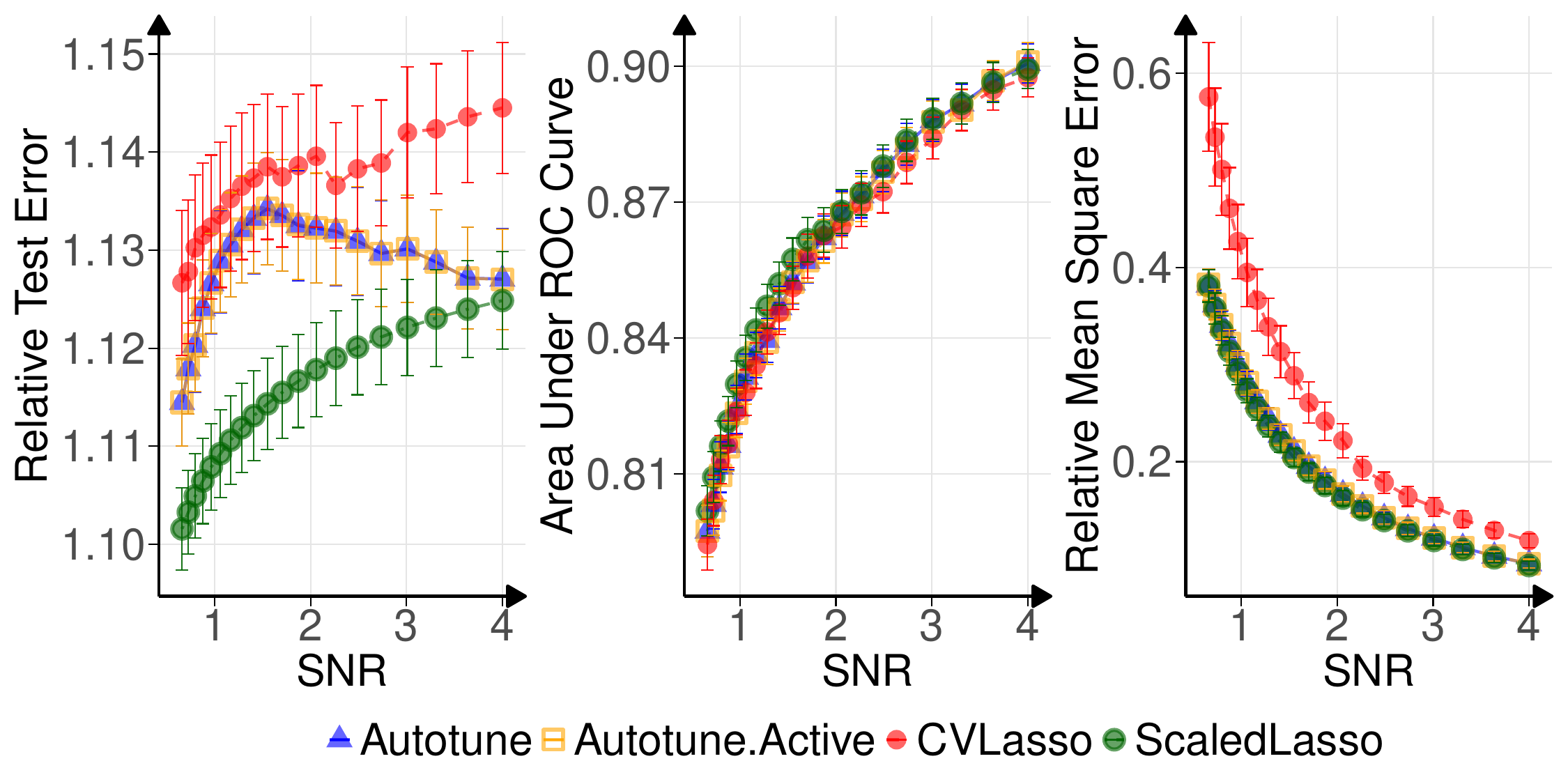}
    \end{tabular}
    \caption[]{Continued: bottom row with $\rho = 0.7$.}
\end{figure}

\begin{figure}[H]
    \centering
    \begin{tabular}{c}
        \textbf{Low Dimensional setup:} $n = 80, \hspace{0.1cm} p=50, \hspace{0.1cm} s=5$, \textbf{Beta-type: 5}, Varying levels  \\
        of correlation, \textbf{Top Row:} $\rho = 0$, \textbf{Middle Row:} $\rho = 0.35$ and \textbf{Bottom Row:} $\rho = 0.7$\\
        \includegraphics[trim=0in 0.8in 0in 0in, clip, width=0.92\textwidth]{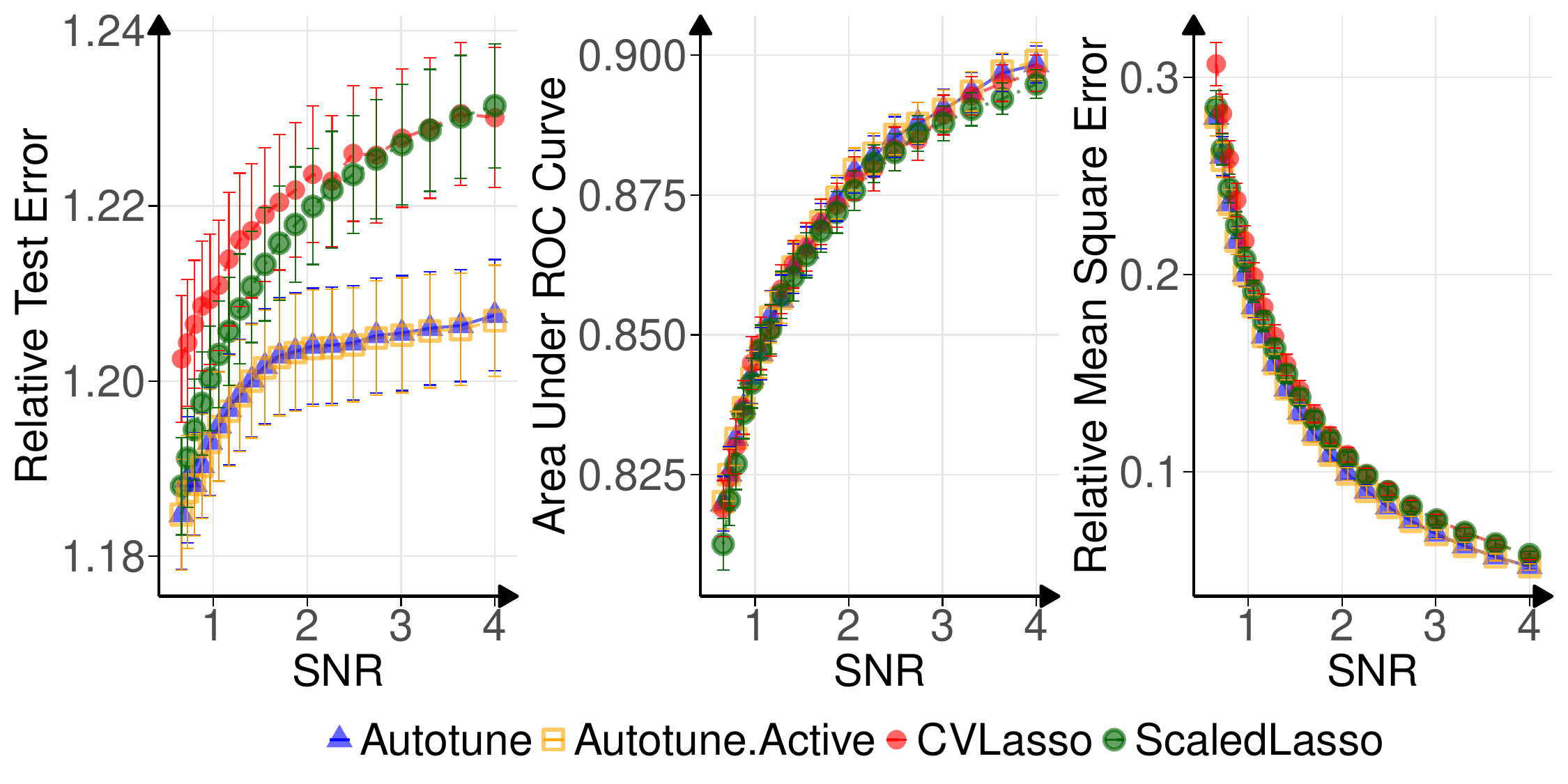}\\
        \includegraphics[trim=0in 0.8in 0in 0in, clip, width=0.92\textwidth]{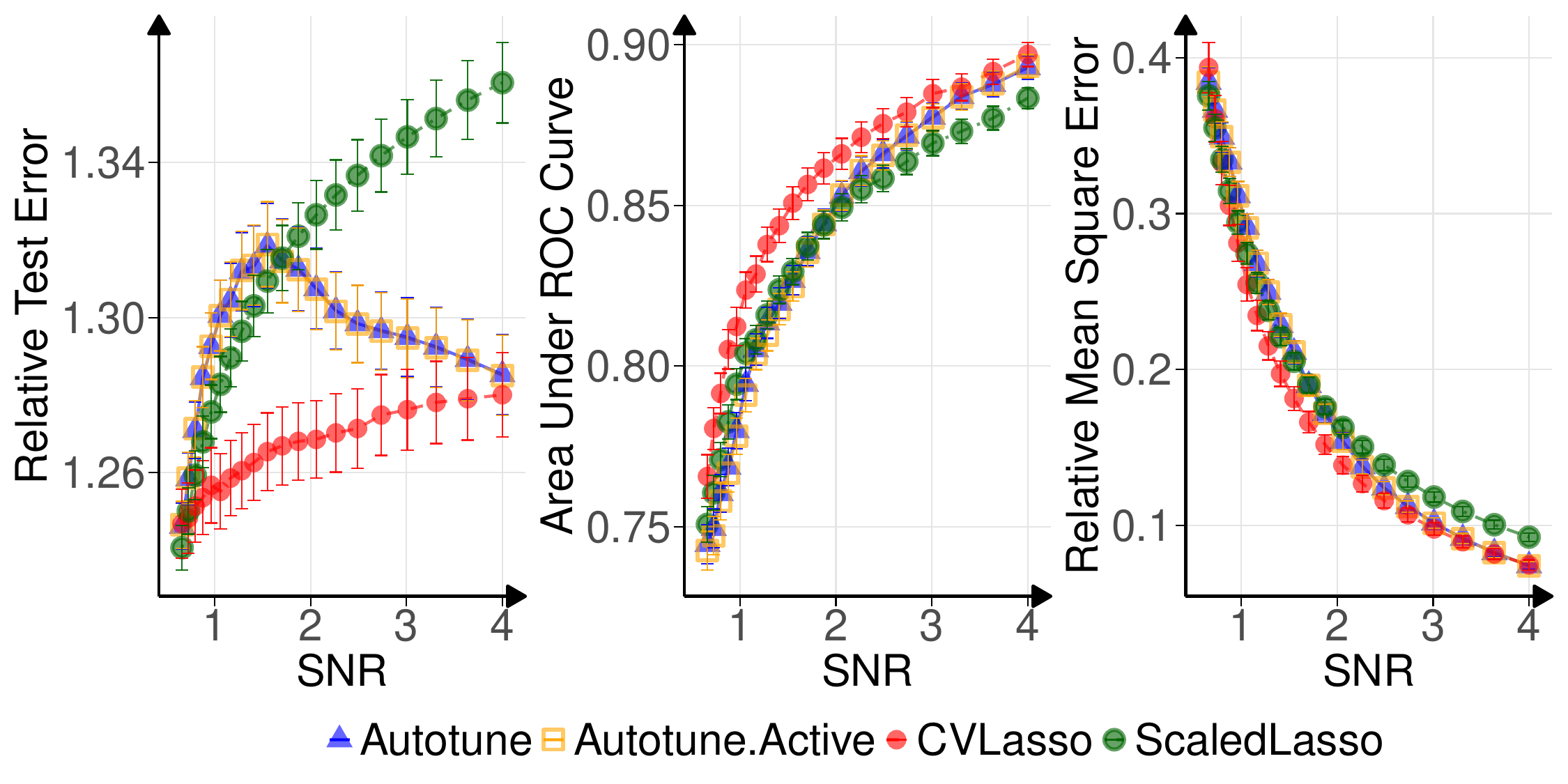}\\
        \includegraphics[trim=0in 0in 0in 0in, clip, width=0.92\textwidth]{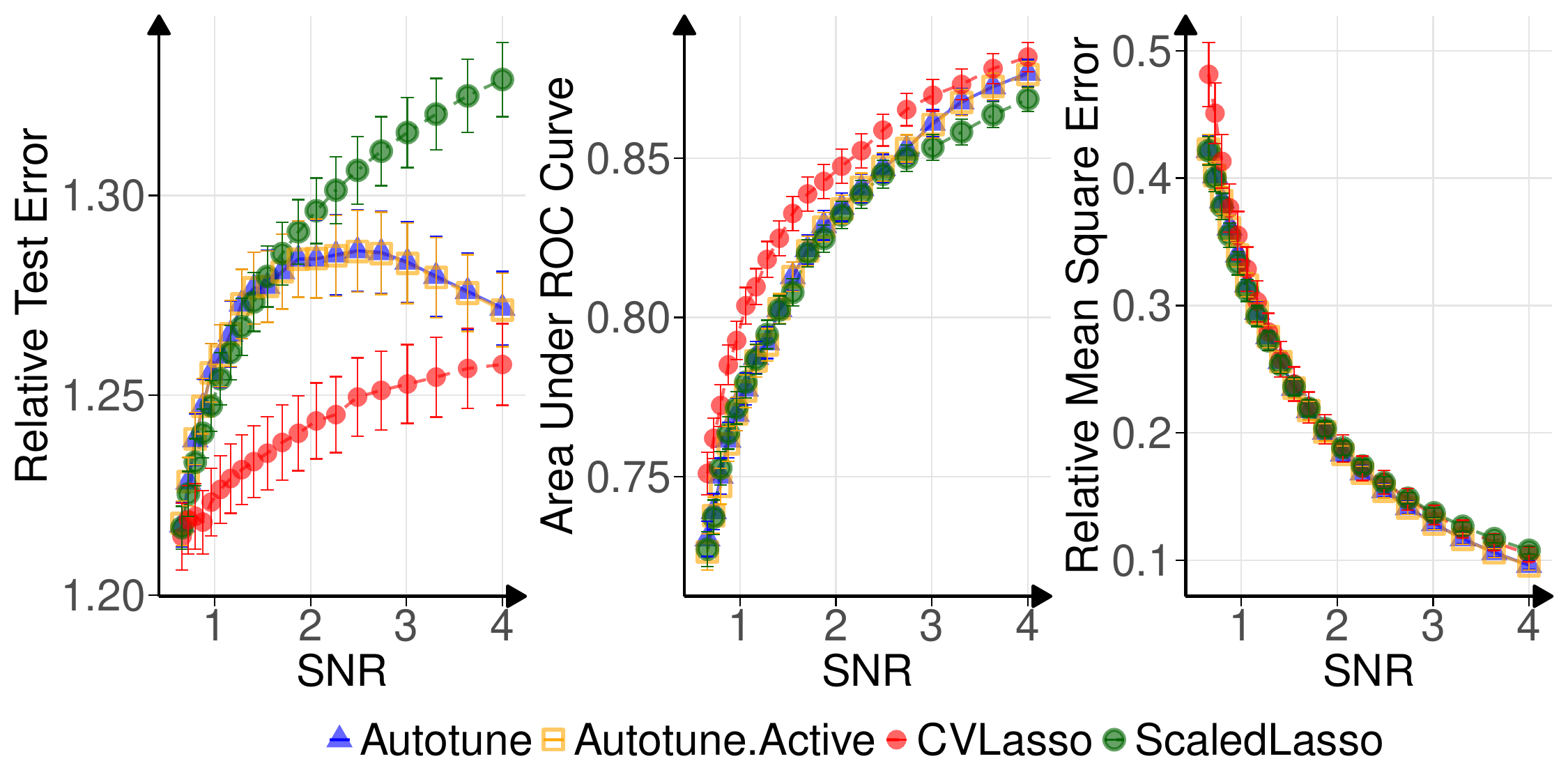}
    \end{tabular}
    \\
    \caption{RTE, AUROC and RMSE of $\autotune$, $\msf{autotune.active}$, CV, and Scaled Lasso plotted as a function of SNR for low-dimensional setup.}
    \label{fig: appen low dim accuracy plot btype 5}
\end{figure}

\subsection{Accuracy when only approximate sparsity holds}
\label{subsubsec: approx sparsity}
\leavevmode\par
\begin{figure}[H]
    \centering
    \begin{tabular}{c}
\textbf{High Dimensional setup:} $n = 80, \hspace{0.1cm} p=750, \hspace{0.1cm} s=5$, \textbf{Beta-type: 4}, Varying levels  \\
        of correlation, \textbf{Top Row:} $\rho = 0$, \textbf{Middle Row:} $\rho = 0.35$ and \textbf{Bottom Row:} $\rho = 0.7$\\
        \includegraphics[trim=0in 0.8in 0in 0in, clip, width=0.85\textwidth]{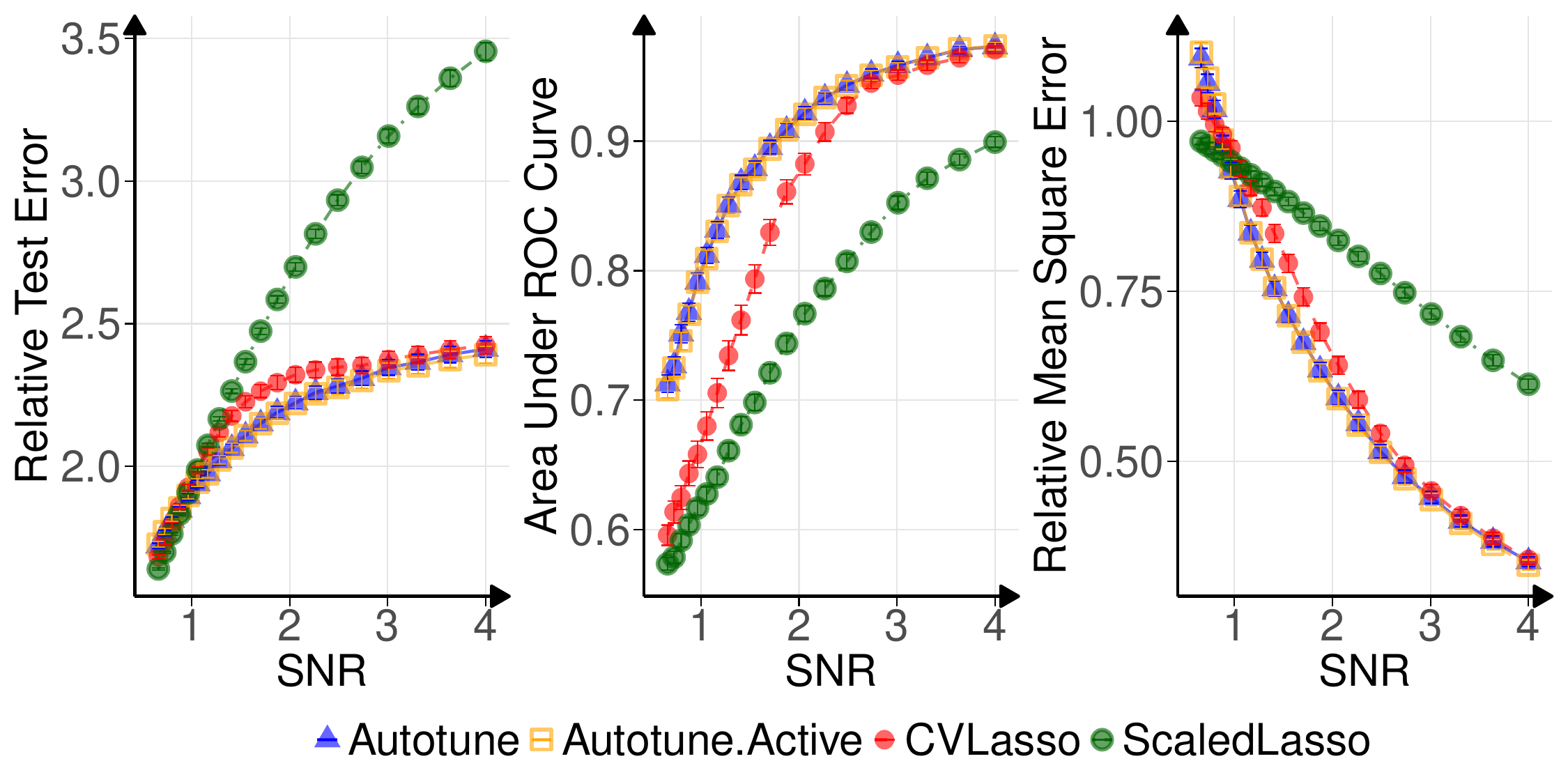}\\
        \includegraphics[trim=0in 0.8in 0in 0in, clip, width=0.85\textwidth]{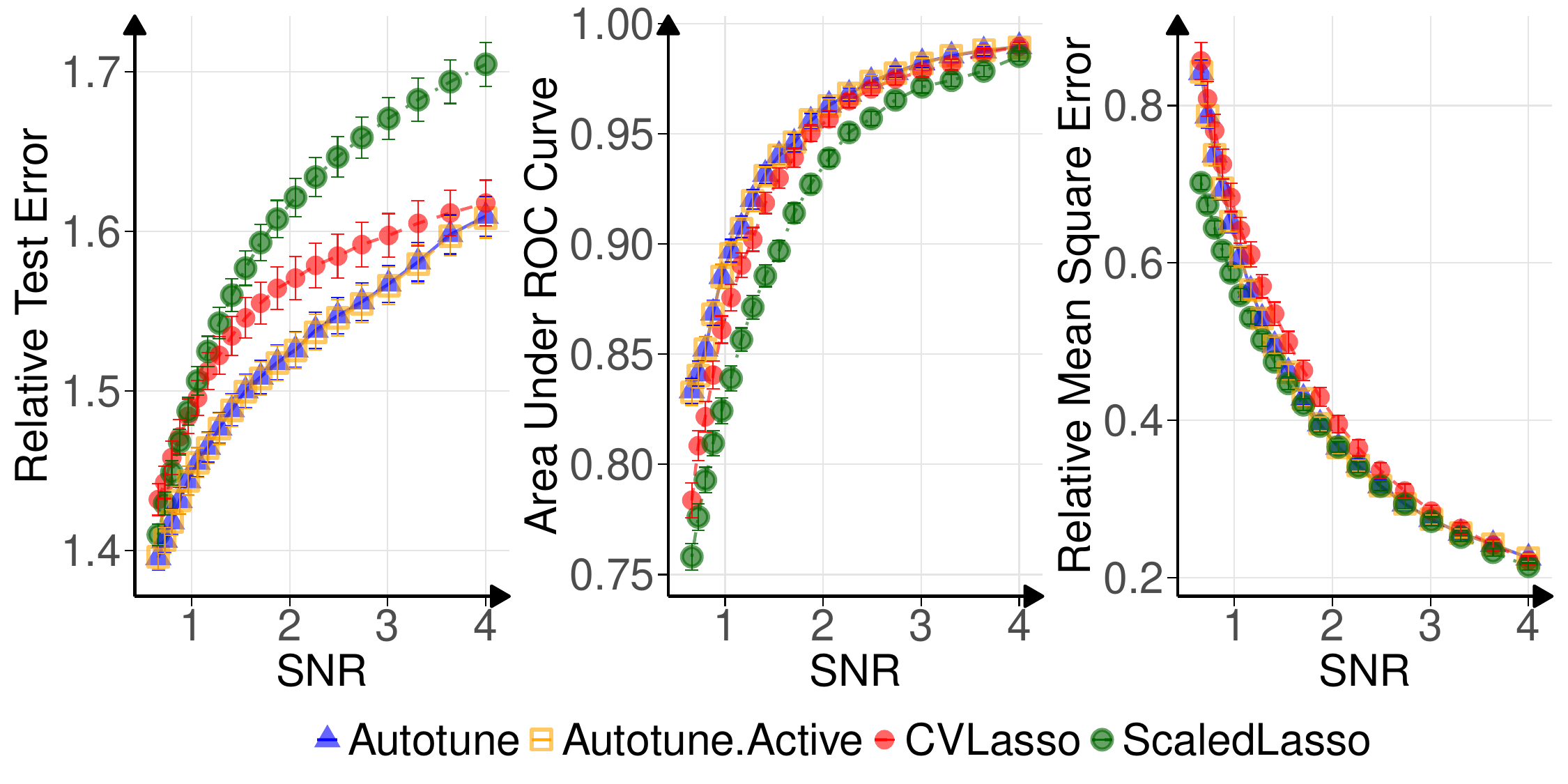}\\
        \includegraphics[trim=0in 0in 0in 0in, clip, width=0.85\textwidth]{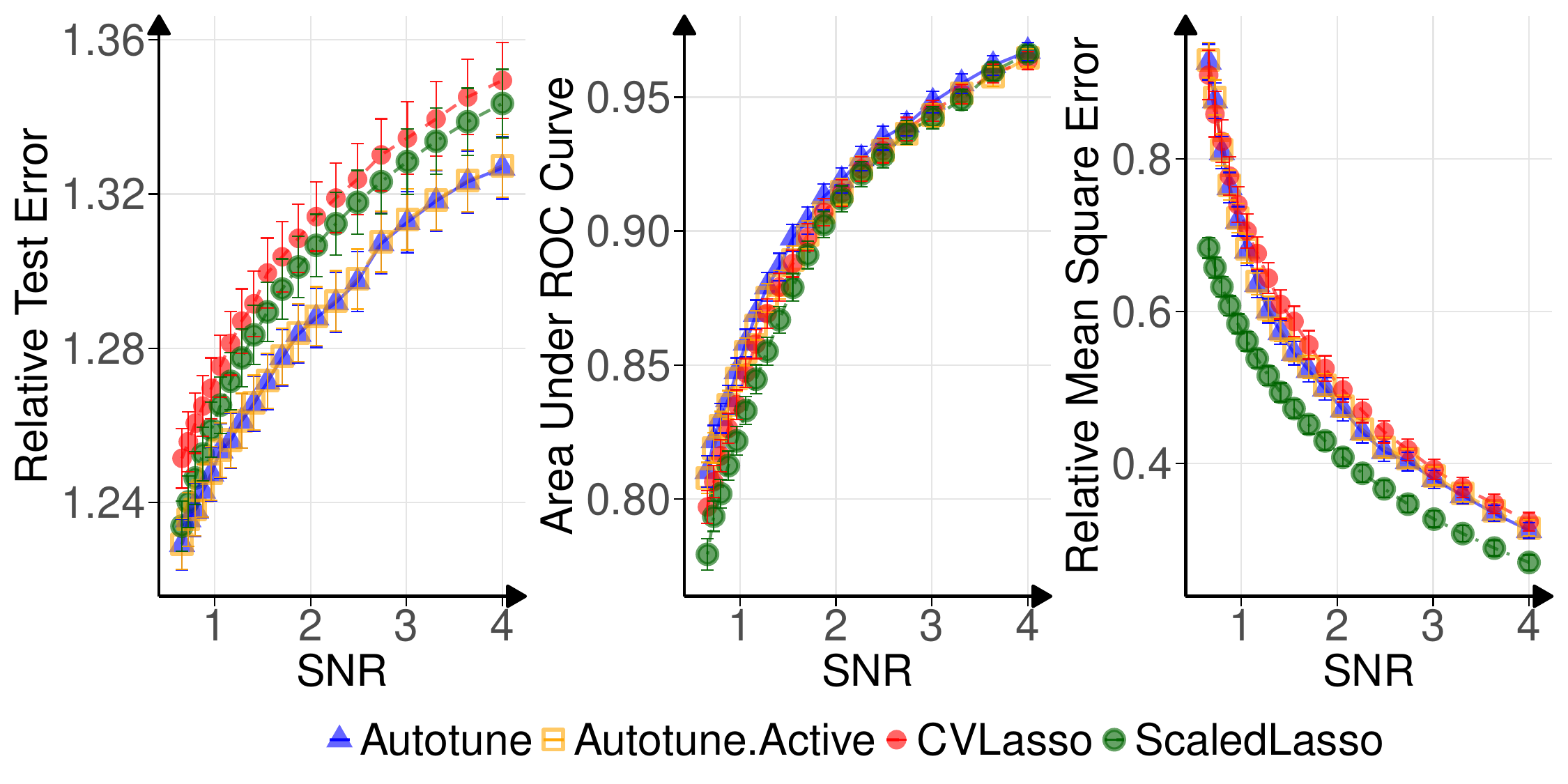}
    \end{tabular}
    \\
    \caption{RTE, AUROC and RMSE of $\autotune$, $\msf{autotune.active}$, CV, and Scaled Lasso plotted as a function of SNR for high-dimensional setup.}
    \label{fig: appen high dim accuracy plot btype 4}
\end{figure}

\begin{figure}[H]
    \centering
    \begin{tabular}{c}
         \textbf{Medium Dimensional setup:} $n = 80, \hspace{0.1cm} p=150, \hspace{0.1cm} s=5$, \textbf{Beta-type: 4}, Varying levels  \\
        of correlation, \textbf{Top Row:} $\rho = 0$, \textbf{Middle Row:} $\rho = 0.35$ and \textbf{Bottom Row:} $\rho = 0.7$\\
        \includegraphics[trim=0in 0.8in 0in 0in, clip, width=0.85\textwidth]{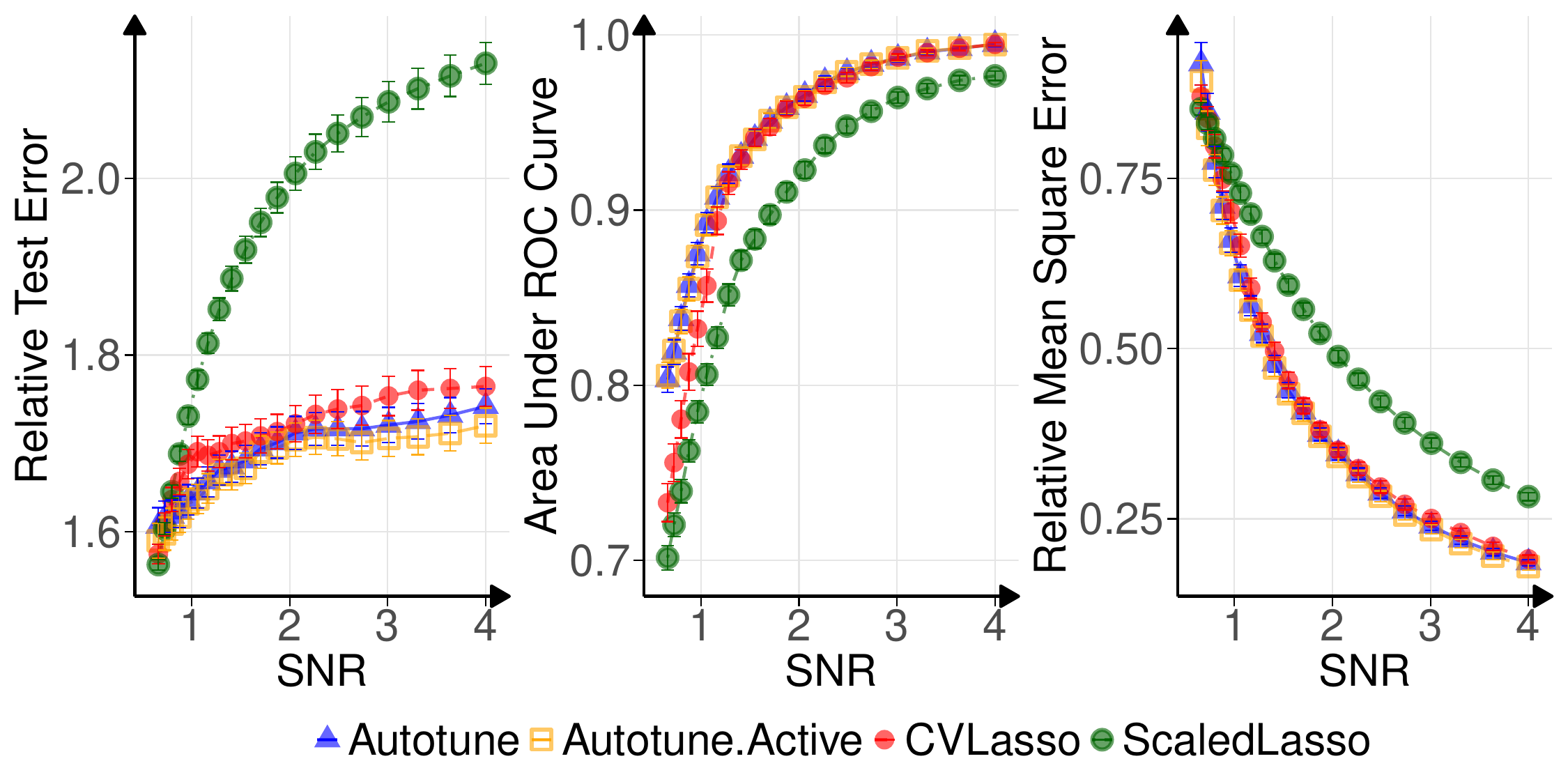}\\
        \includegraphics[trim=0in 0.8in 0in 0in, clip, width=0.85\textwidth]{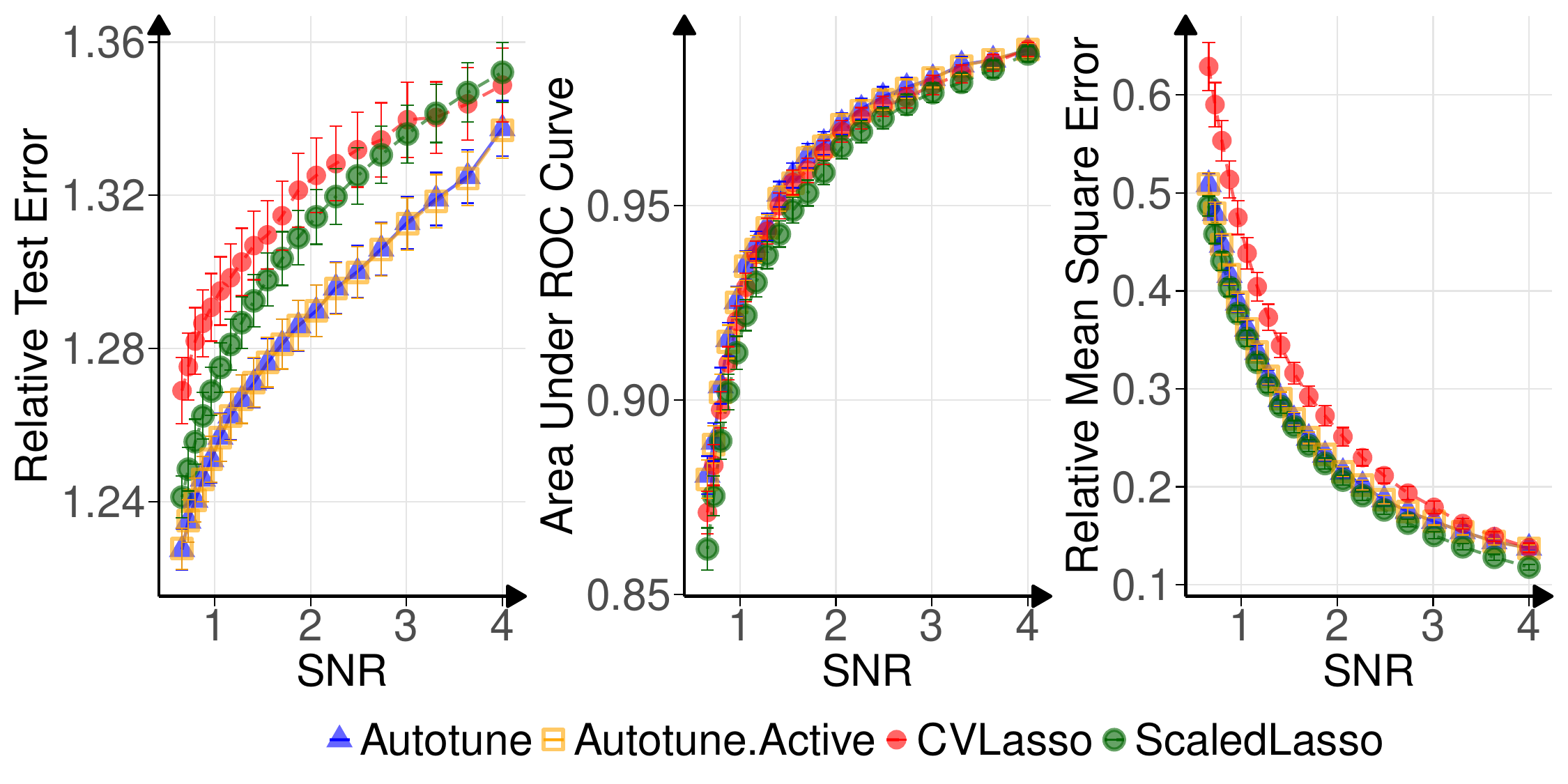}\\
        \includegraphics[trim=0in 0in 0in 0in, clip, width=0.85\textwidth]{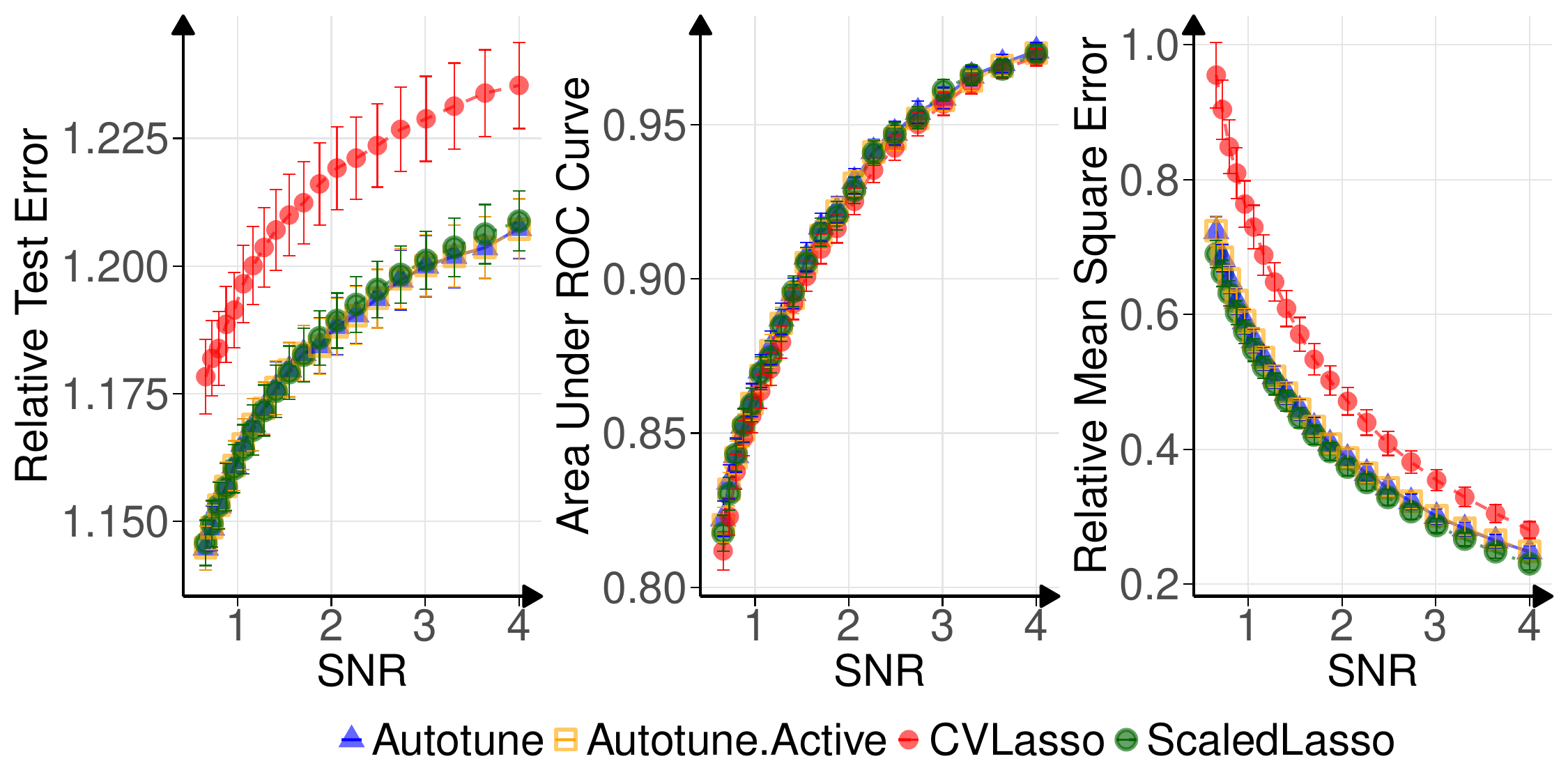}
    \end{tabular}
    \\
    \caption{RTE, AUROC and RMSE of $\autotune$, $\msf{autotune.active}$, CV, and Scaled Lasso plotted as a function of SNR for medium-dimensional setup.}
    \label{fig: appen medium dim accuracy plot btype 4}
\end{figure}

\begin{figure}[H]
    \centering
    \begin{tabular}{c}
         \textbf{Low Dimensional setup:} $n = 80, \hspace{0.1cm} p=150, \hspace{0.1cm} s=5$, \textbf{Beta-type: 4}, Varying levels  \\
        of correlation, \textbf{Top Row:} $\rho = 0$, \textbf{Middle Row:} $\rho = 0.35$ and \textbf{Bottom Row:} $\rho = 0.7$\\
        \includegraphics[trim=0in 0.8in 0in 0in, clip, width=0.85\textwidth]{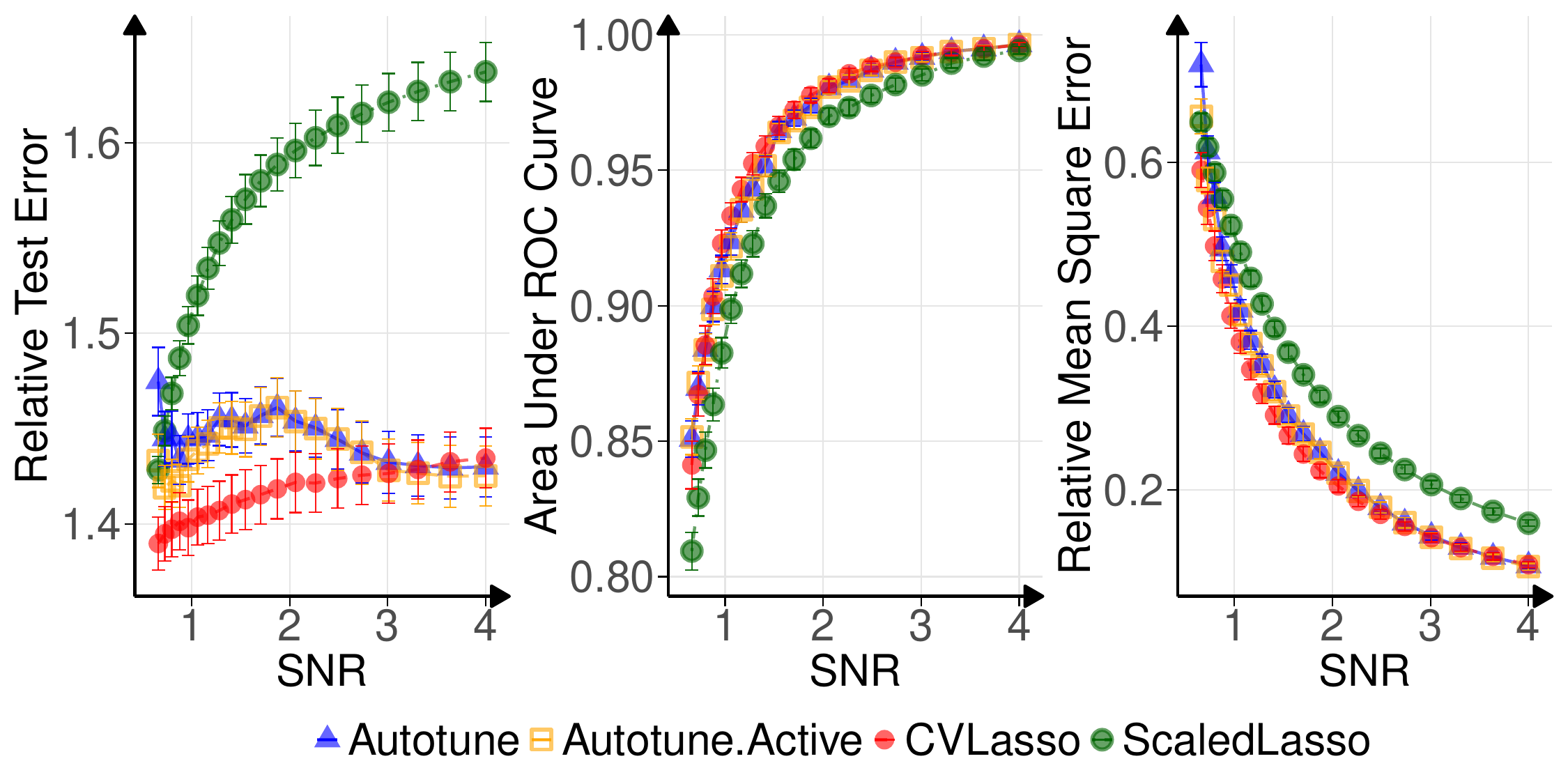}\\
        \includegraphics[trim=0in 0.8in 0in 0in, clip, width=0.85\textwidth]{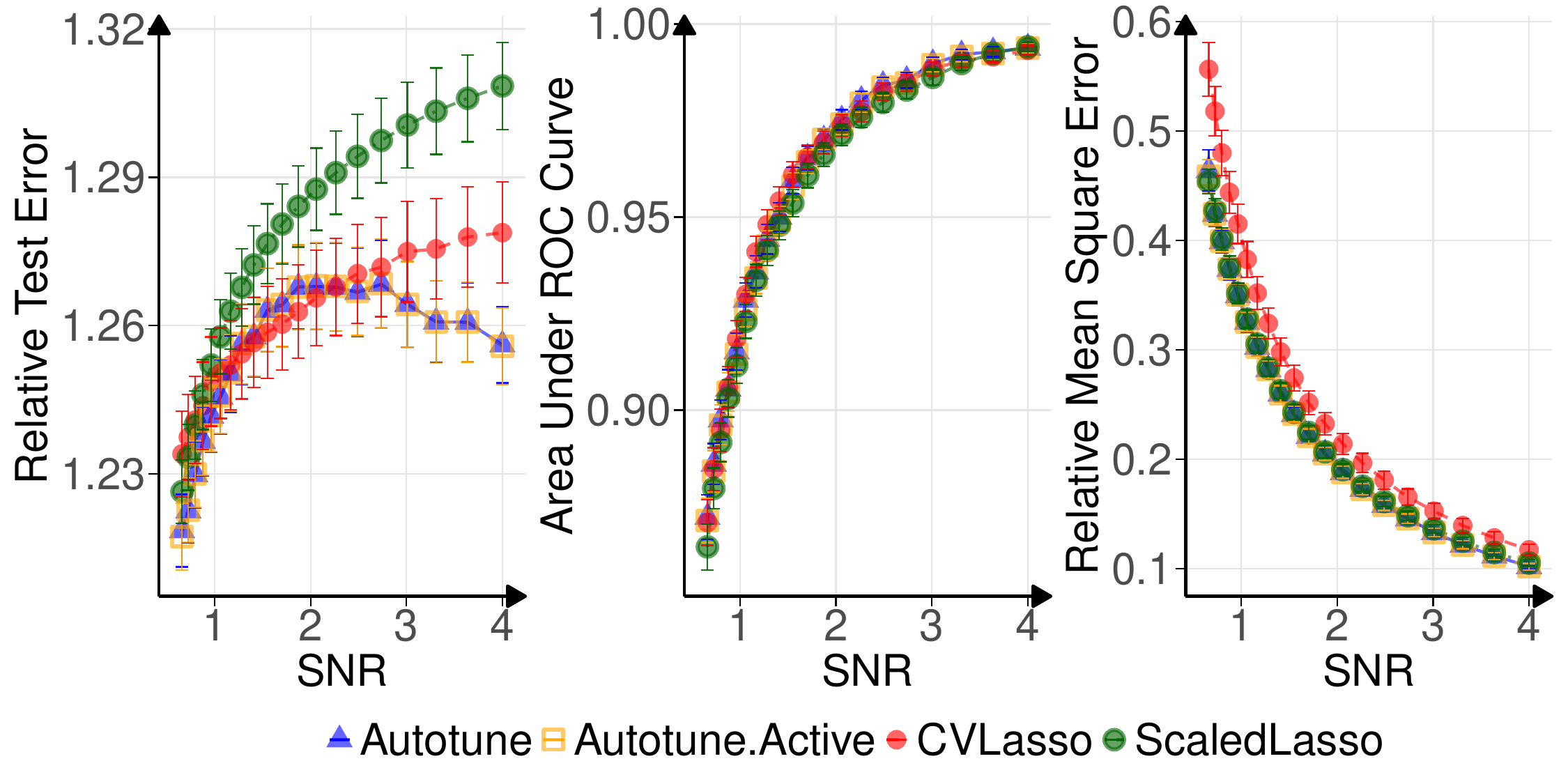}\\
        \includegraphics[trim=0in 0in 0in 0in, clip, width=0.85\textwidth]{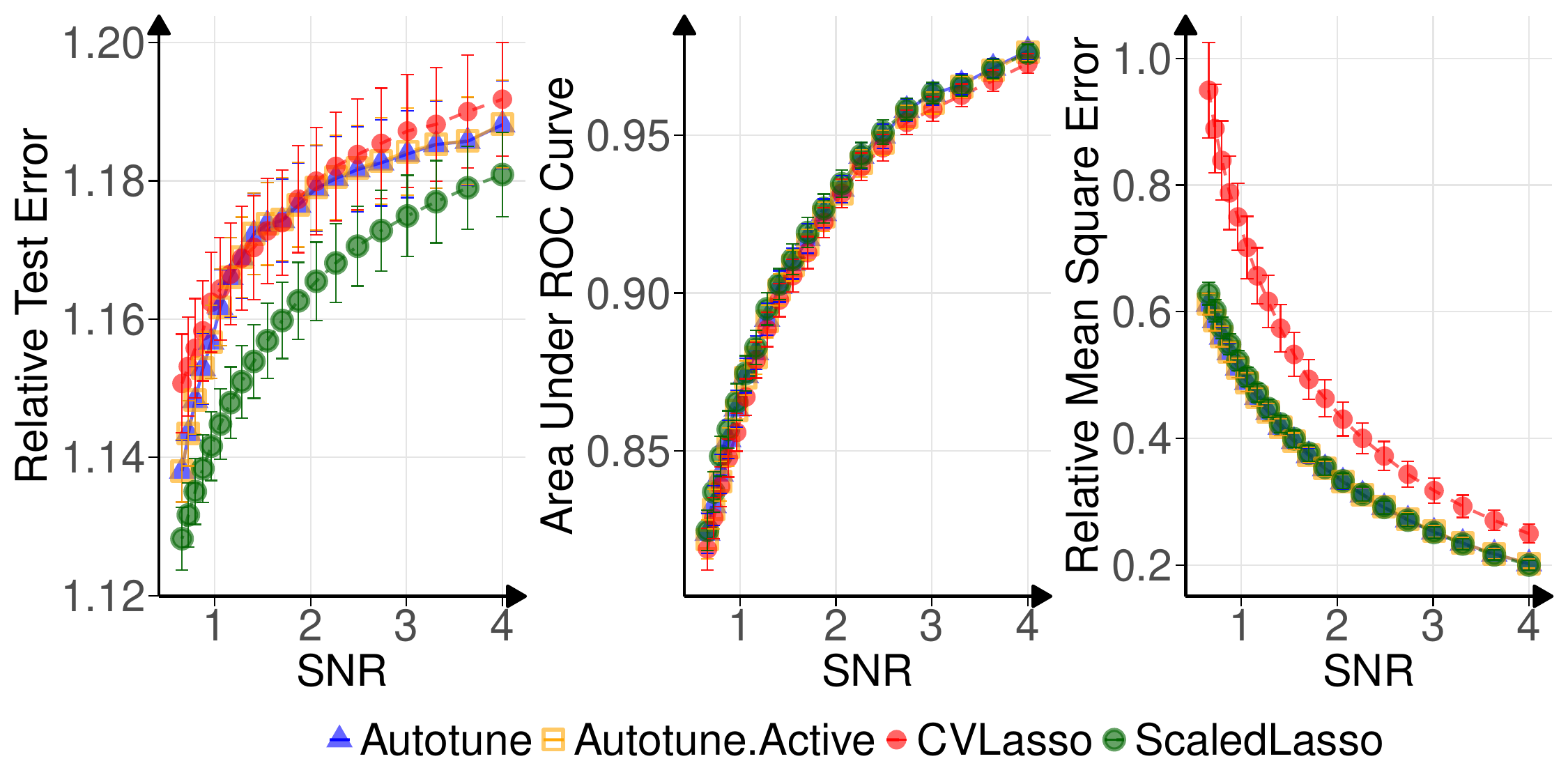}
    \end{tabular}
    \\
    \caption{RTE, AUROC and RMSE of $\autotune$, $\msf{autotune.active}$, CV, and Scaled Lasso plotted as a function of SNR for medium-dimensional setup.}
    \label{fig: appen low dim accuracy plot btype 4}
\end{figure}

\clearpage

\section{Deferred results of Section \ref{sec: simulation}}
\subsection{\texorpdfstring{Deferred tables comparing runtimes of $\autotune$ Lasso with CV and Scaled Lasso}{Deferred runtime tables for autotune, CV, and Scaled Lasso}}
\leavevmode\par

\begin{table}[H]
\caption{Table of mean runtimes (with SD in the brackets) for Figure \ref{fig: teaser into autotune}. $n = 80, s = 5,$ $\rho= 0.35, \mrm{SNR} = 1$, Beta-type 2 and $p$ varies from 50 to 750.}
\centering
\begin{tabular}[t]{ccccccc}
\toprule
Algorithm & 50 & 64 & 80 & 100 & 120 & 160\\
\midrule
$\autotune$ & 1.0 (0.1) & 1.0 (0.1) & 1.0 (0.1) & 1.2 (0.1) & 1.5 (0.1) & 1.9 (0.6)\\
CV Lasso & 85.1 (12.9) & 134.5 (11.5) & 122.3 (9.9) & 105.9 (9.4) & 126.4 (10.3) & 140.1 (10.7)\\
Scaled Lasso & 25.8 (3.6) & 28.7 (1.6) & 53.9 (4.6) & 50.1 (3.0) & 70.0 (4.9) & 72.8 (5.8)\\
\bottomrule
\end{tabular}

\centering
\begin{tabular}[t]{ccccccc}
\toprule
Algorithm & 200 & 240 & 320 & 400 & 560 & 750\\
\midrule
$\autotune$ & 2.4 (0.2) & 2.6 (0.5) & 3.9 (0.3) & 6.5 (0.9) & 9.9 (0.9) & 11.1 (1.4)\\
CVLasso & 136.3 (12.2) & 163.2 (16.1) & 180.8 (15.5) & 200.1 (17.8) & 152.4 (15.8) & 165.2 (14.9)\\
Scaled Lasso & 75.8 (6.1) & 94.8 (6.6) & 100.8 (8.3) & 105.9 (10.4) & 162.6 (19.7) & 200.4 (19.1)\\
\bottomrule
\end{tabular}
\label{tab: runtime table}
\end{table}

\begin{table}[H]
\caption{Table of mean runtimes (with SD in the brackets) for left figure of Figure \ref{fig: runtime plots}. $n = 200, s = 15,$ $\rho= 0.35, \mrm{SNR} = 1$, Beta-type 2 and $p$ varies from 100 to 1000.}
\centering
\begin{tabular}[t]{ccccccc}
\toprule
Algorithm & 100 & 160 & 180 & 200 & 250 & 300\\
\midrule
Autotune & 4.5 (1.9) & 4.6 (0.7) & 5.6 (1.6) & 7.9 (0.5) & 7.1 (0.5) & 7.5 (0.6)\\
CVLASSO & 140.0 (19.6) & 482.2 (45.9) & 1203.1 (76.4) & 521.5 (32.7) & 411.4 (26.8) & 434.8 (28.2)\\
Scaled Lasso & 79.9 (31.8) & 270.0 (93.6) & 332.1 (28.7) & 749.7 (51.5) & 613.4 (39.5) & 664.0 (46.0)\\
\bottomrule
\end{tabular}
\centering
\begin{tabular}[t]{cccccc}
\toprule
Algorithm & 400 & 500 & 600 & 800 & 1000\\
\midrule
Autotune & 13.2 (1.1) & 15.5 (1.9) & 20.5 (3.1) & 34.2 (13.0) & 34.9 (5.9)\\
CVLASSO & 513.6 (25.3) & 303.0 (40.5) & 346.4 (27.7) & 373.3 (26.3) & 422.9 (32.1)\\
Scaled Lasso & 775.6 (51.6) & 821.5 (63.7) & 962.8 (94.9) & 1272.5 (149.7) & 1383.9 (100.4)\\
\bottomrule
\end{tabular}
\label{tab: runtimes for fig b}
\end{table}

\newpage

\subsection{\texorpdfstring{Runtime comparison of $\Autotune$ vs $\msf{autotune.active}$ Lasso}{Runtime comparison of Autotune and autotune.active Lasso}}
\label{subsec: autotune_vs_active_runtime}
\leavevmode\par

\begin{figure}[H]
    \centering
    \begin{subfigure}[t]{0.495\textwidth}
        \centering
        \includegraphics[page = 1, trim=0in 0in 0in 0in, clip, width=\textwidth]{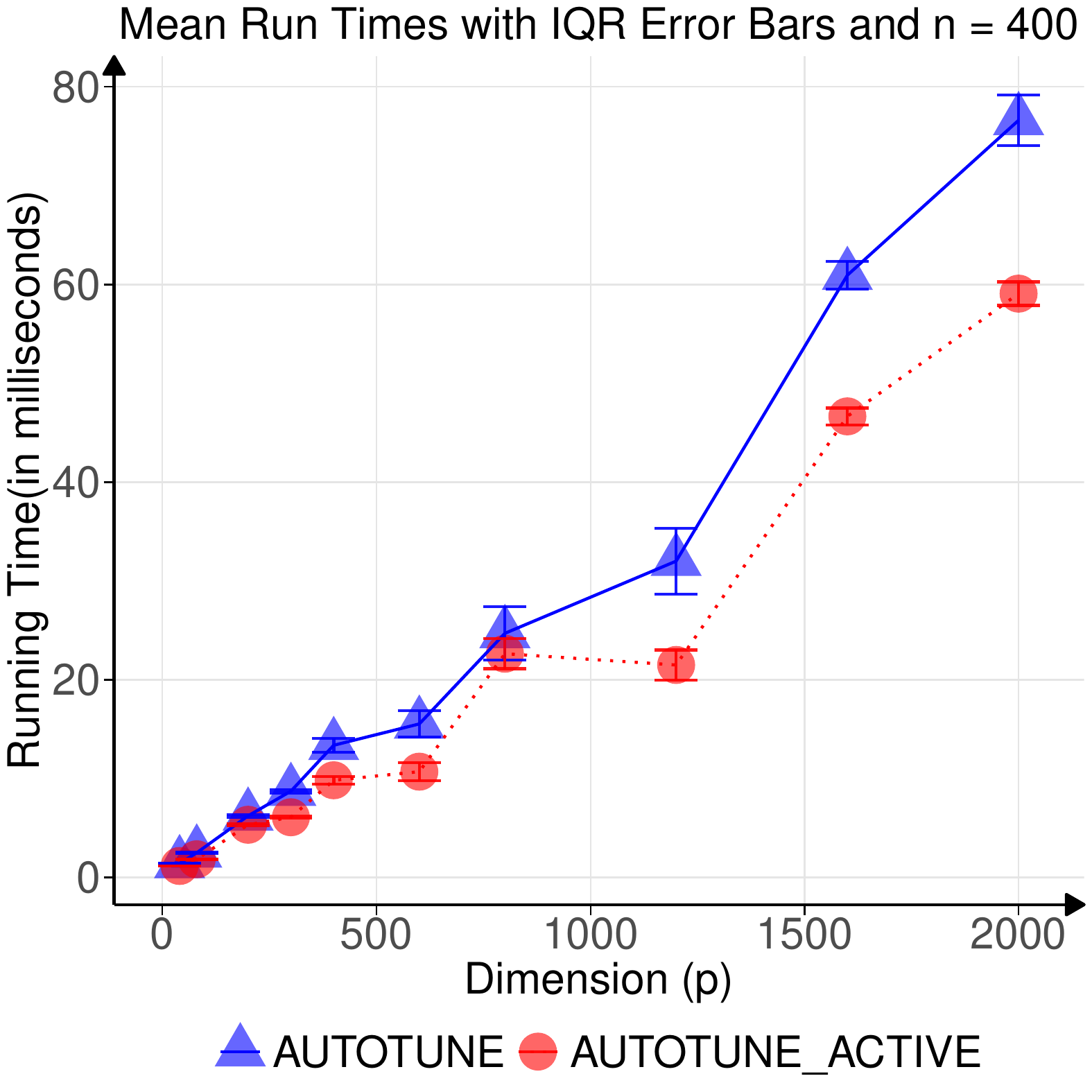}
        \caption{Beta-type 1, $\rho = 0$, SNR = 2}
        \label{fig: appen_runtime1}
    \end{subfigure}
    \begin{subfigure}[t]{0.495\textwidth}
        \centering
        \includegraphics[page = 1, trim=0in 0in 0in 0in, clip, width=\textwidth]{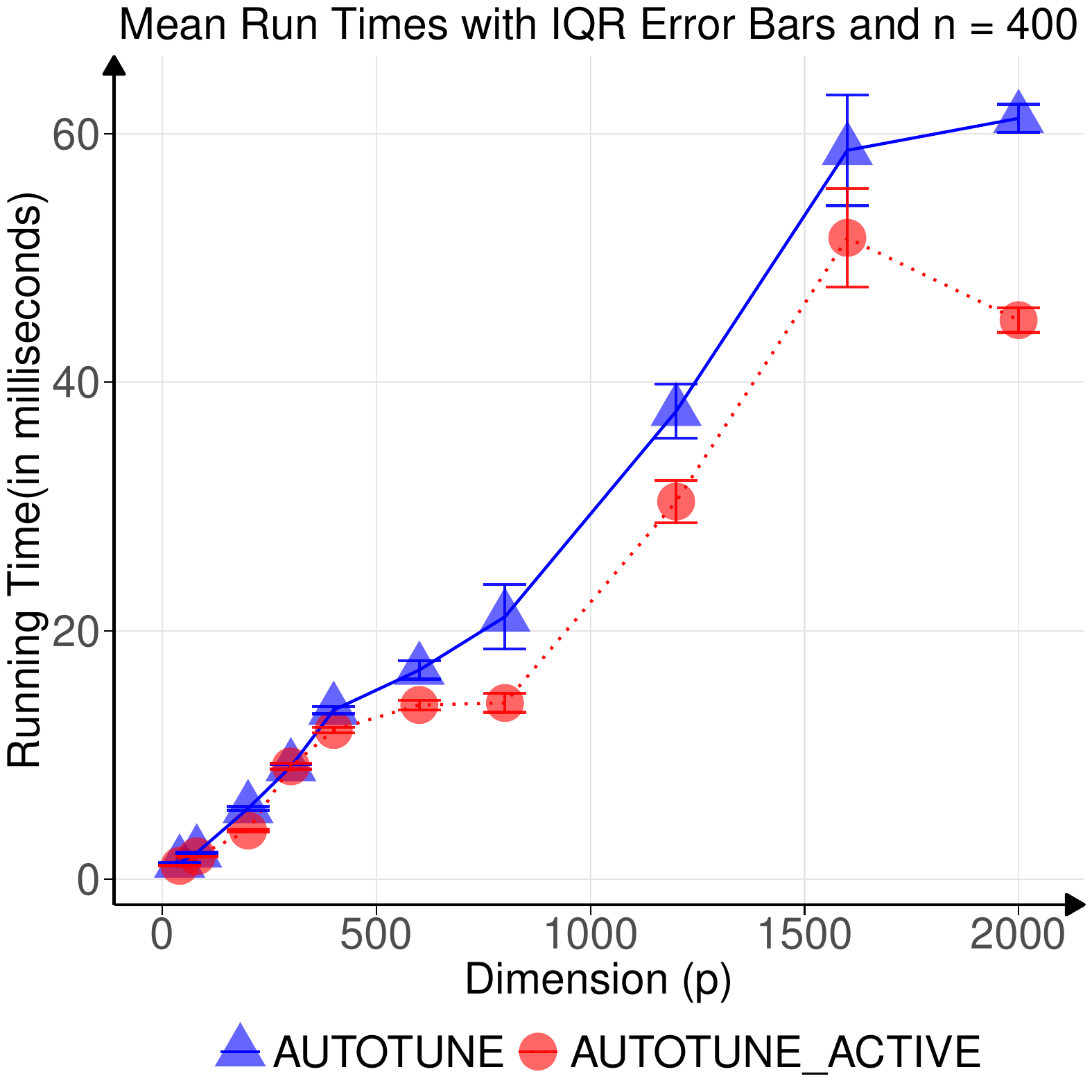}
        \caption{Beta-type 1, $\rho = 0.35$, SNR = 2}
        \label{fig: appen_runtime2}
    \end{subfigure}
    \vspace{0.5em}  
    \centering
    \begin{subfigure}[b]{0.495\textwidth}
        \centering
        \includegraphics[page = 1, trim=0in 0in 0in 0in, clip, width=\textwidth]{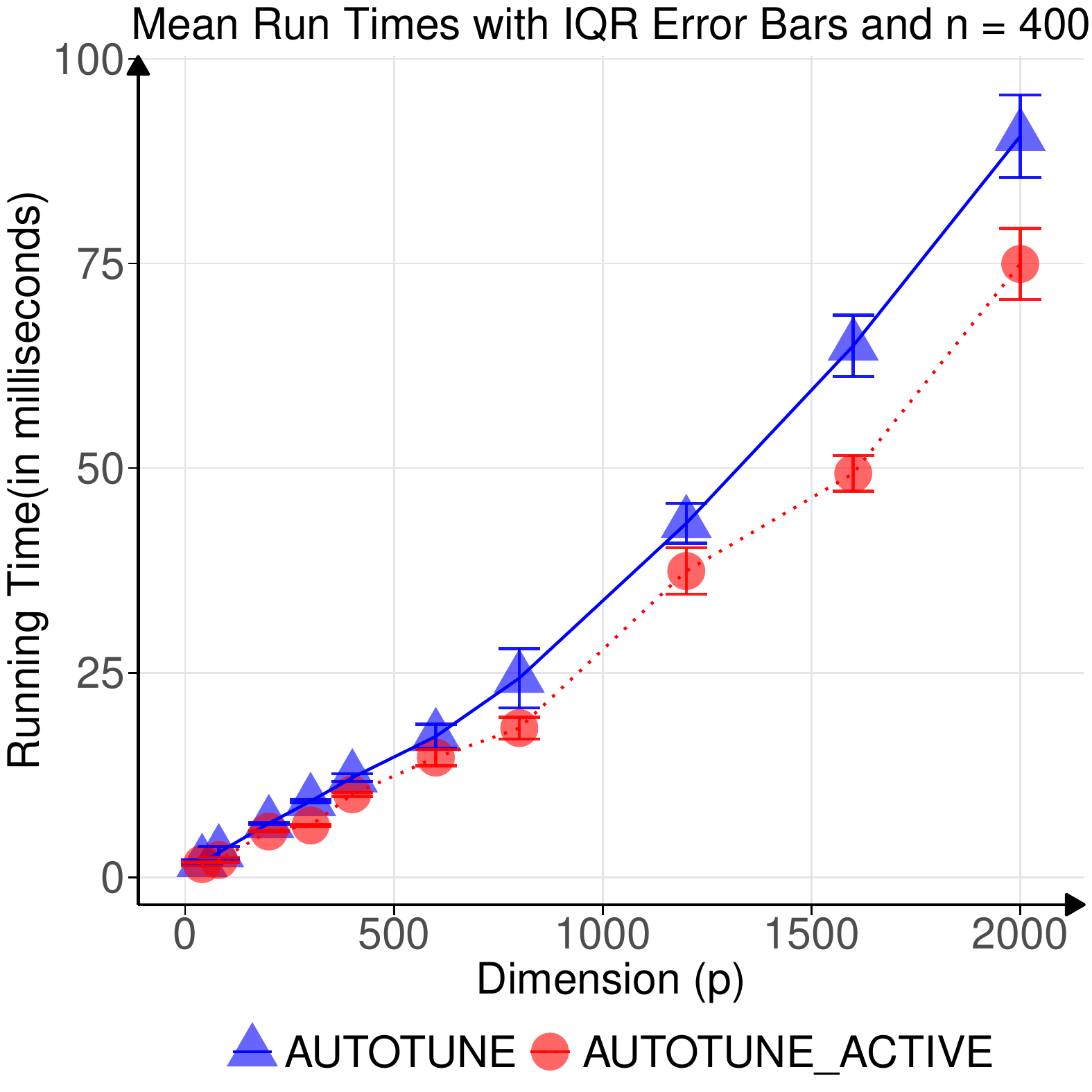}
        \caption{Beta-type 2, $\rho = 0$, SNR = 2}
        \label{fig: appen_runtime3}
    \end{subfigure}
    \begin{subfigure}[b]{0.495\textwidth}
        \centering
        \includegraphics[page = 1, trim=0in 0in 0in 0in, clip, width=\textwidth]{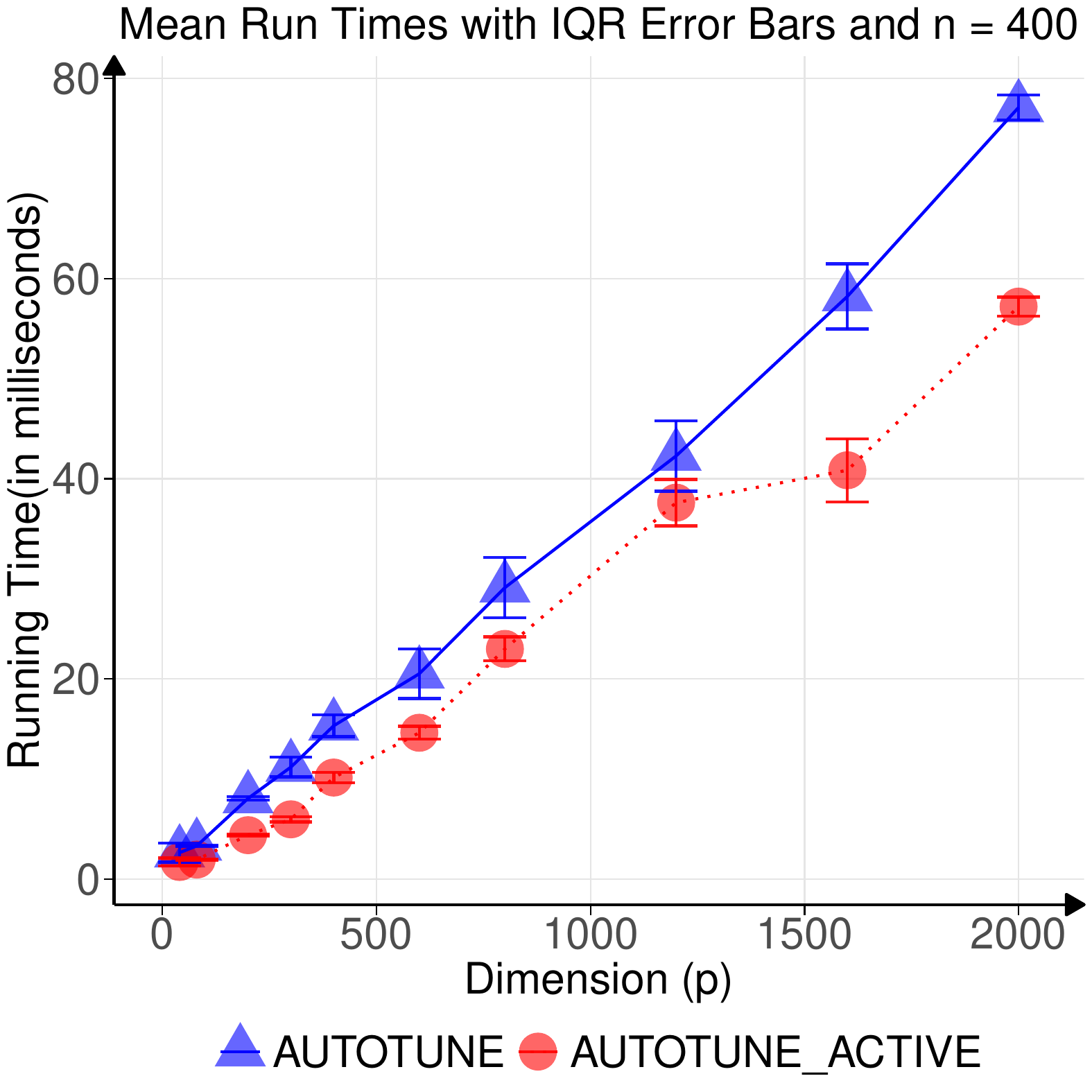}
        \caption{Beta-type 2, $\rho = 0.35$, SNR = 2}
        \label{fig: appen_runtime4}
    \end{subfigure}
    \caption{Runtime of $\autotune$ lasso with its active set version, $\msf{autotune.active}$ lasso as a function of increasing $p$ for $n = 400, s = 5$.}
    \label{fig: appen_runtime}
\end{figure}

\subsection{\texorpdfstring{Sensitivity of $\sigma$ estimation to $\alpha$}{Sensitivity of sigma estimation to alpha}}
\label{subsec: sigma alpha comp}
\leavevmode\par

\begin{figure}[H]
    \centering
    \begin{subfigure}[t]{0.495\textwidth}
        \centering
        \includegraphics[page = 1, trim=0in 0in 0in 0in, clip, width=\textwidth]{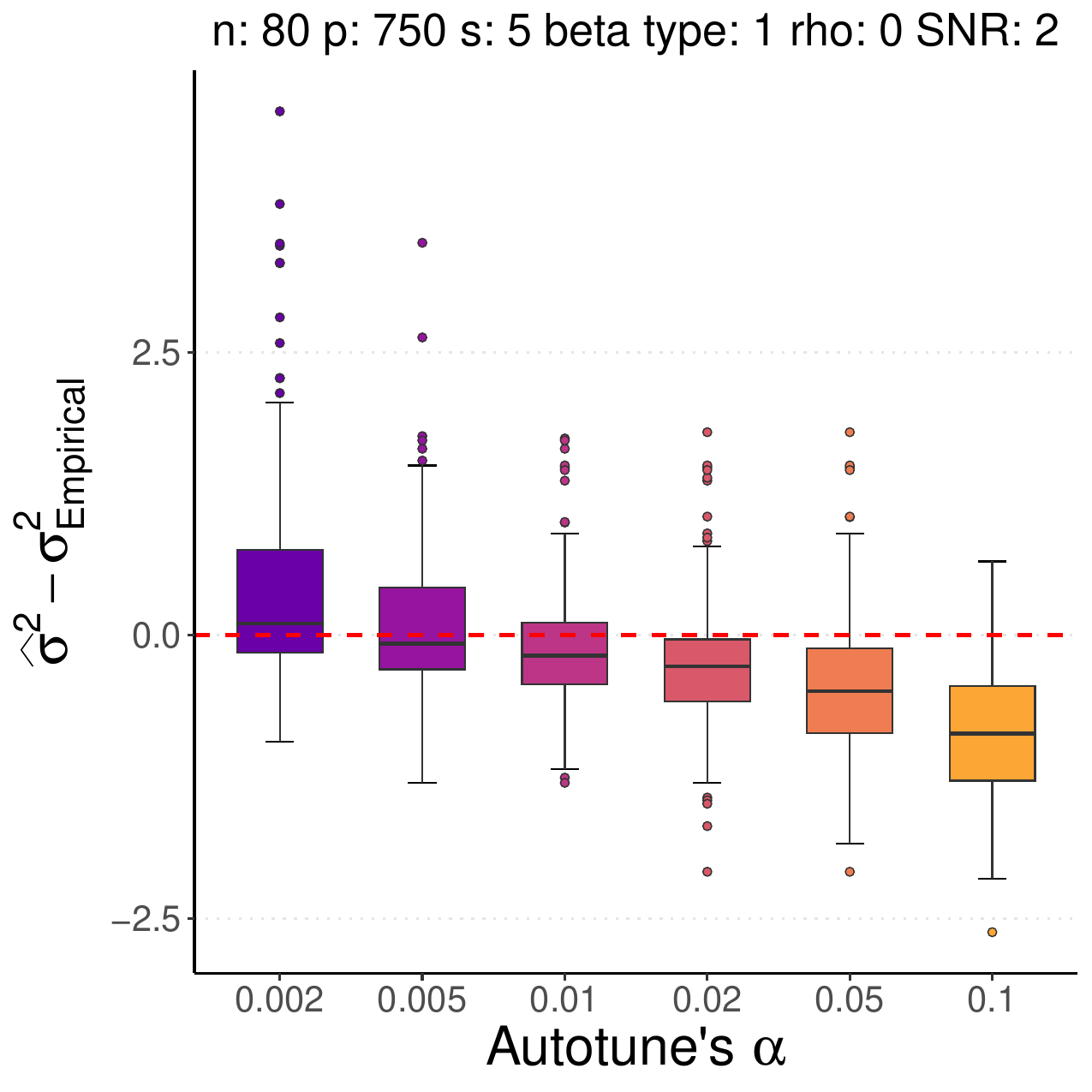}
        \label{fig: alpha_sensitivity1}
    \end{subfigure}
    \begin{subfigure}[t]{0.495\textwidth}
        \centering
        \includegraphics[page = 1, trim=0in 0in 0in 0in, clip, width=\textwidth]{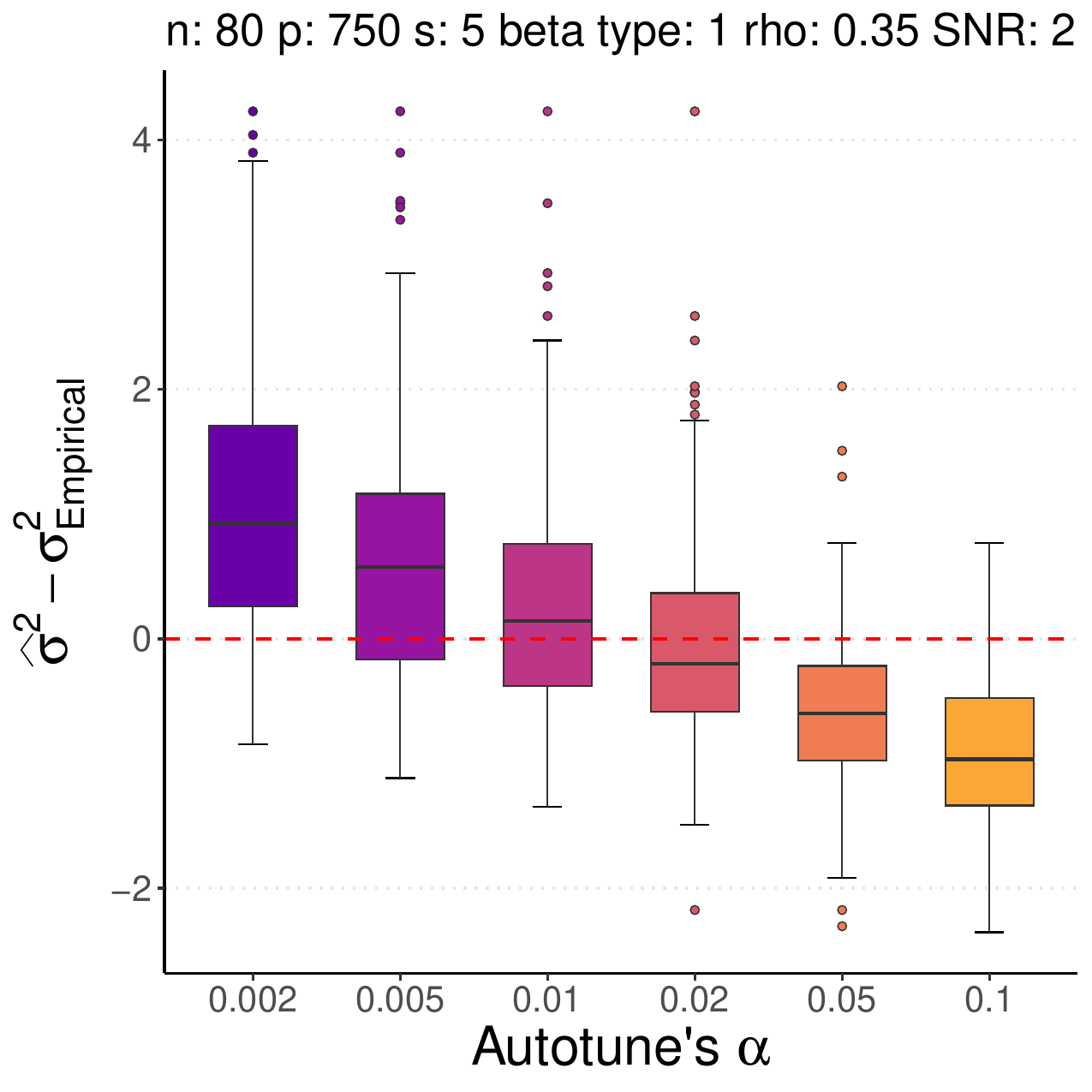}
        \label{fig: alpha_sensitivity2}
    \end{subfigure}
    \vspace{0.5em}  
    \centering
    \begin{subfigure}[t]{0.495\textwidth}
        \centering
        \includegraphics[page = 1, trim=0in 0in 0in 0in, clip, width=\textwidth]{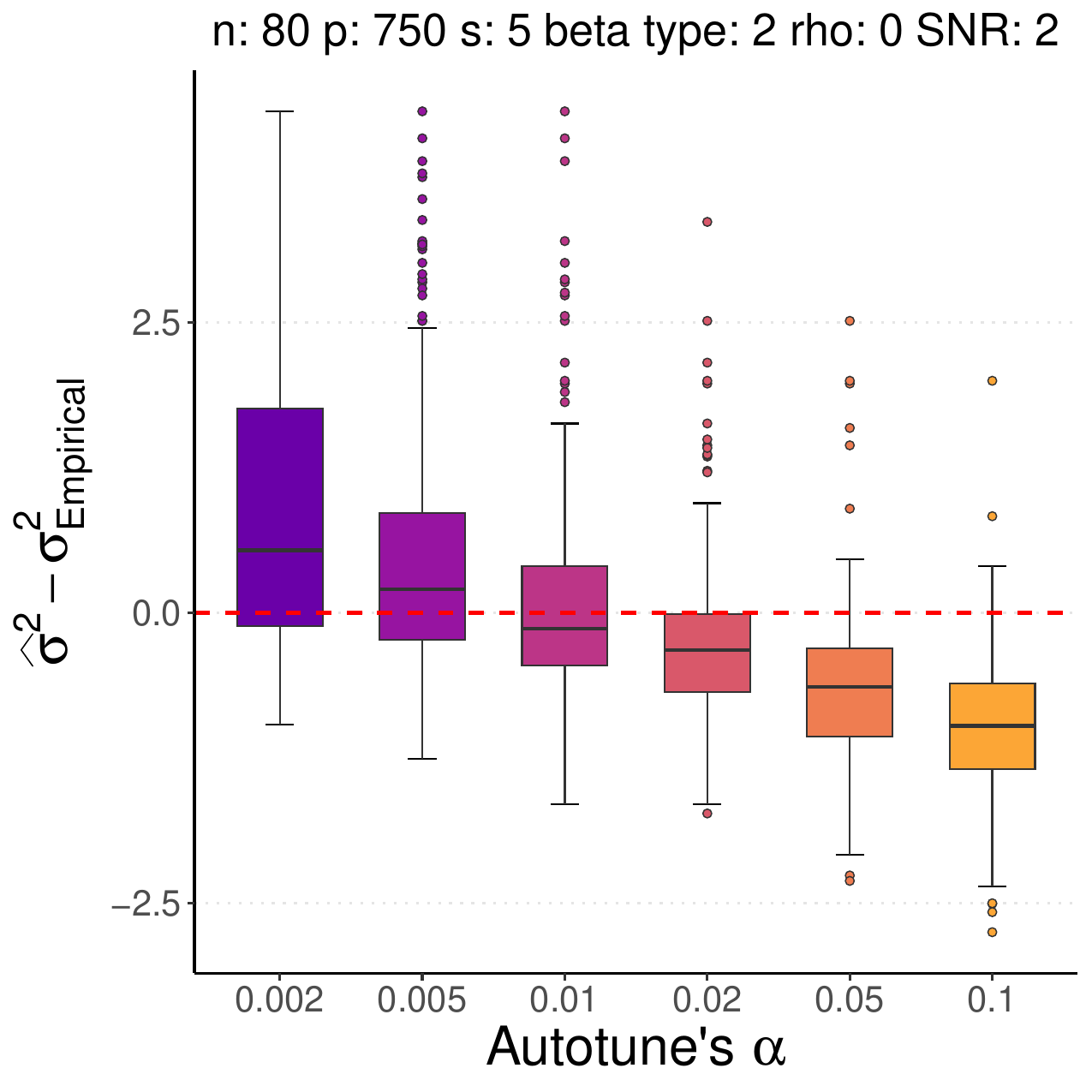}
        \label{fig: alpha_sensitivity3}
    \end{subfigure}
    \begin{subfigure}[t]{0.495\textwidth}
        \centering
        \includegraphics[page = 1, trim=0in 0in 0in 0in, clip, width=\textwidth]{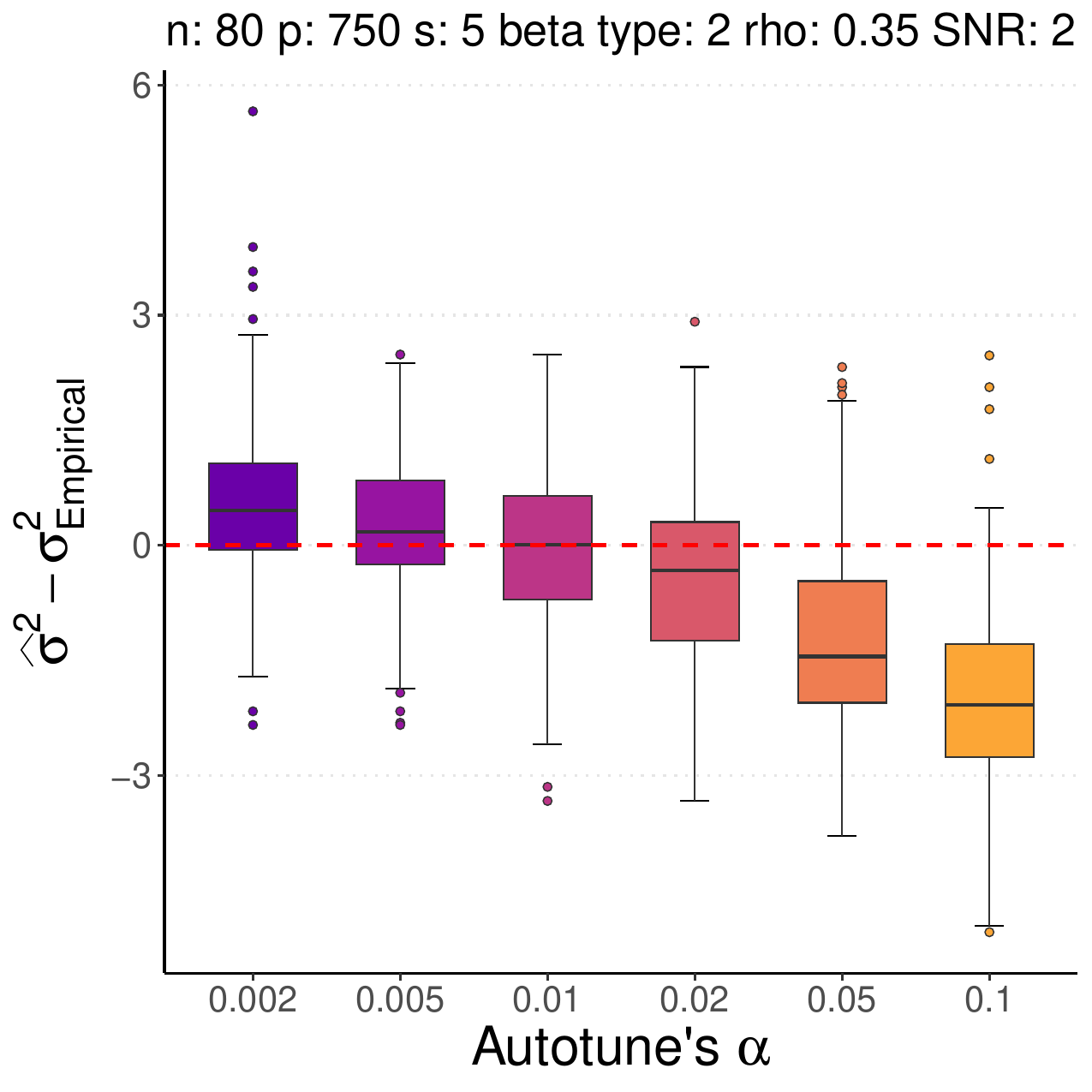}
        \label{fig: alpha_sensitivity4}
    \end{subfigure}
    \caption{Comparison of $\Autotune$ Lasso's noise variance estimation in the high-dimensional setup among different choices of $\alpha$ in the inner sequential F-test (see \SUlink).}
    \label{fig: alpha_sensitivity}
\end{figure}

\section{Deferred results for Data Analysis Section}

\subsection{Deferred experimental details of Data Analysis Section}
\label{app: experimental details of data analysis}

We plot the RTE of the solution path taken by the CV Lasso (orange) and $\autotune$ (blue, with endpoint denoted by a blue star) on the test data along with normalized CV errors on the training dataset (red) and demarcate the finally chosen $\lambda$ with labeled vertical lines in Figure \ref{fig: sp500} and \ref{fig: appen sp500 plots}. 
We also generate $100$ bootstrap samples of our training set, fit all the methods on each sample, and then evaluate the O.O.S. $\text{R}^2$ of the trained models. We report these using side-by-side boxplots in Figure \ref{fig: sp500} and \ref{fig: boot sp500 appen}. Results of AIC and BIC Lasso were omitted as they were not competitive.

\subsection{Deferred tables for Data Analysis Section}
\leavevmode\par
\begin{table}[!ht]
\centering
\caption{RTE of various algorithms for different training dataset sizes $n$. BIC results were dropped as they were same as the results of AIC.}
\begin{tabular}{cccccc}
\hline
$n$ & $\autotune$ & Scaled & CV(min) & CV(1se) & AIC  \\
\hline
150 & \textbf{0.749} & 0.845 & 0.832 & 0.819 & 0.832  \\
160 & \textbf{0.633} & 0.694 & 0.682 & 0.676 & 0.682 \\
170 & \textbf{0.515} & 0.567 & 0.561 & 0.555 & 0.561  \\
180 & \textbf{0.343} & 0.437 & 0.437 & 0.427 & 0.437  \\
190 & \textbf{0.250} & 0.341 & 0.336 & 0.331 & 0.336 \\
200 & \textbf{0.255} & 0.340 & 0.335 & 0.325 & 0.335 \\
210 & \textbf{0.332} & 0.403 & 0.397 & 0.379 & 0.397  \\
\hline
\end{tabular}
\label{tab: sp500 test mse}
\end{table}

\begin{table}[!ht]
\centering
\caption{Number of nonzero estimated regression coefficients by various algorithms while modeling DJIA. AIC results were dropped for space reasons.
}
\begin{tabular}{cccccc}
  \hline
 $n$ & $\Autotune$ & Scaled & CV(min) & CV(1se) & BIC \\ 
  \hline
150 & \textbf{48} & 65 & 67 & 59 & 62 \\ 
160 & \textbf{57} & 64 & 61 & 58 & 61 \\
170 & 57 & 68 & 60 & 58 & \textbf{54} \\ 
180 & 62 & 63 & 59 & \textbf{55} & 58 \\ 
190 & \textbf{47} & 64 & 60 & 57 & 60 \\ 
200 & \textbf{41} & 64 & 61 & 58 & 61 \\ 
210 & \textbf{57} & 70 & 62 & 61 & 59 \\ 
  \hline
\end{tabular}

\label{tab: sp500 model selection}
\end{table}

\newpage
\subsection{Supplementary plots for Data Analysis Section}
\label{subsec: additional plots of sp500}
\leavevmode\par

\begin{figure}[H]
    \centering
    \begin{subfigure}[t]{0.32\textwidth}
        \centering
        \includegraphics[ width=\textwidth]{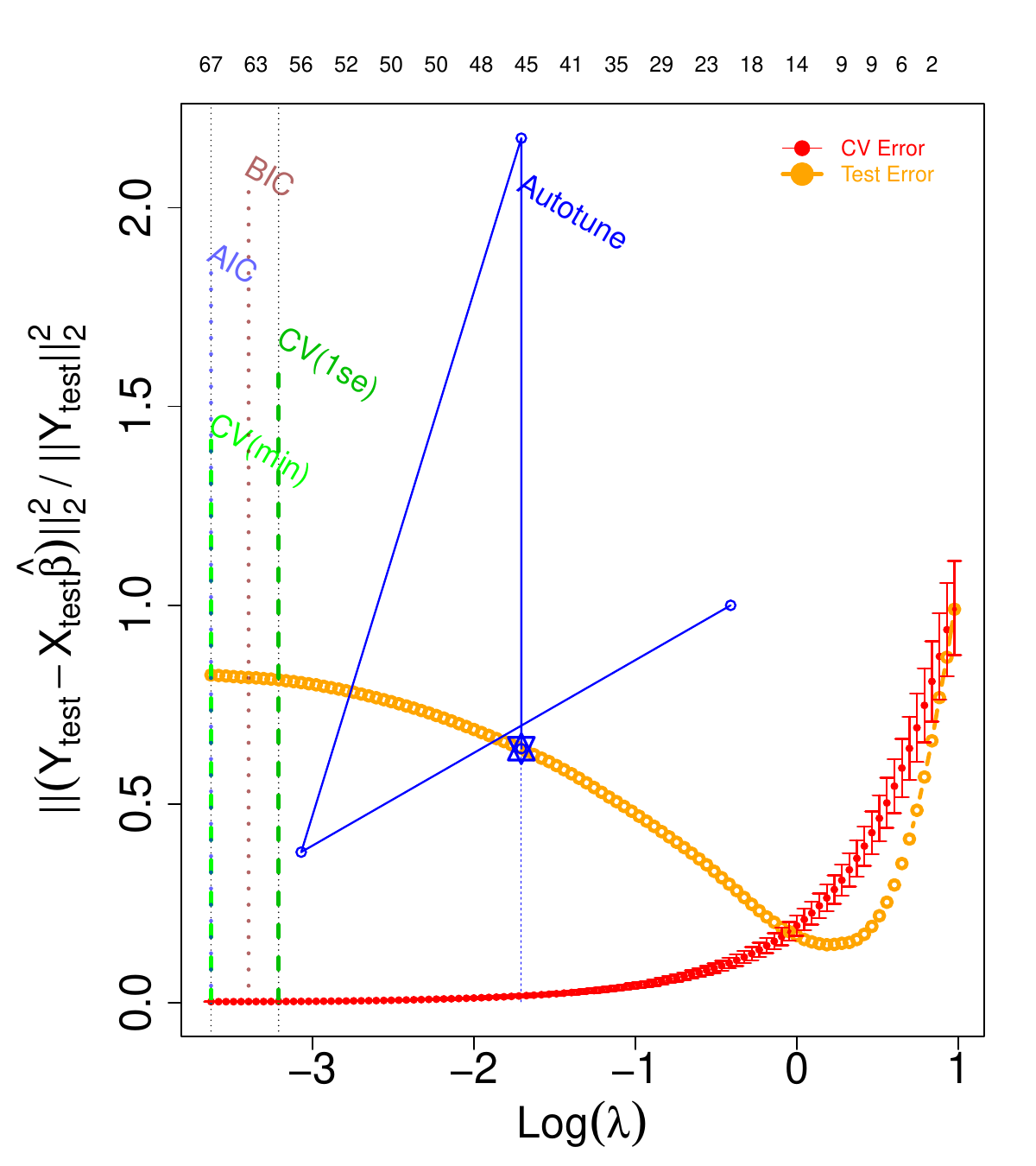}
        \caption{train set size $n = 150$, test set size $252 - n=102$.}
        \label{fig: sp500 n150}
    \end{subfigure}
    \begin{subfigure}[t]{0.32\textwidth}
        \centering
        \includegraphics[ width=\textwidth]{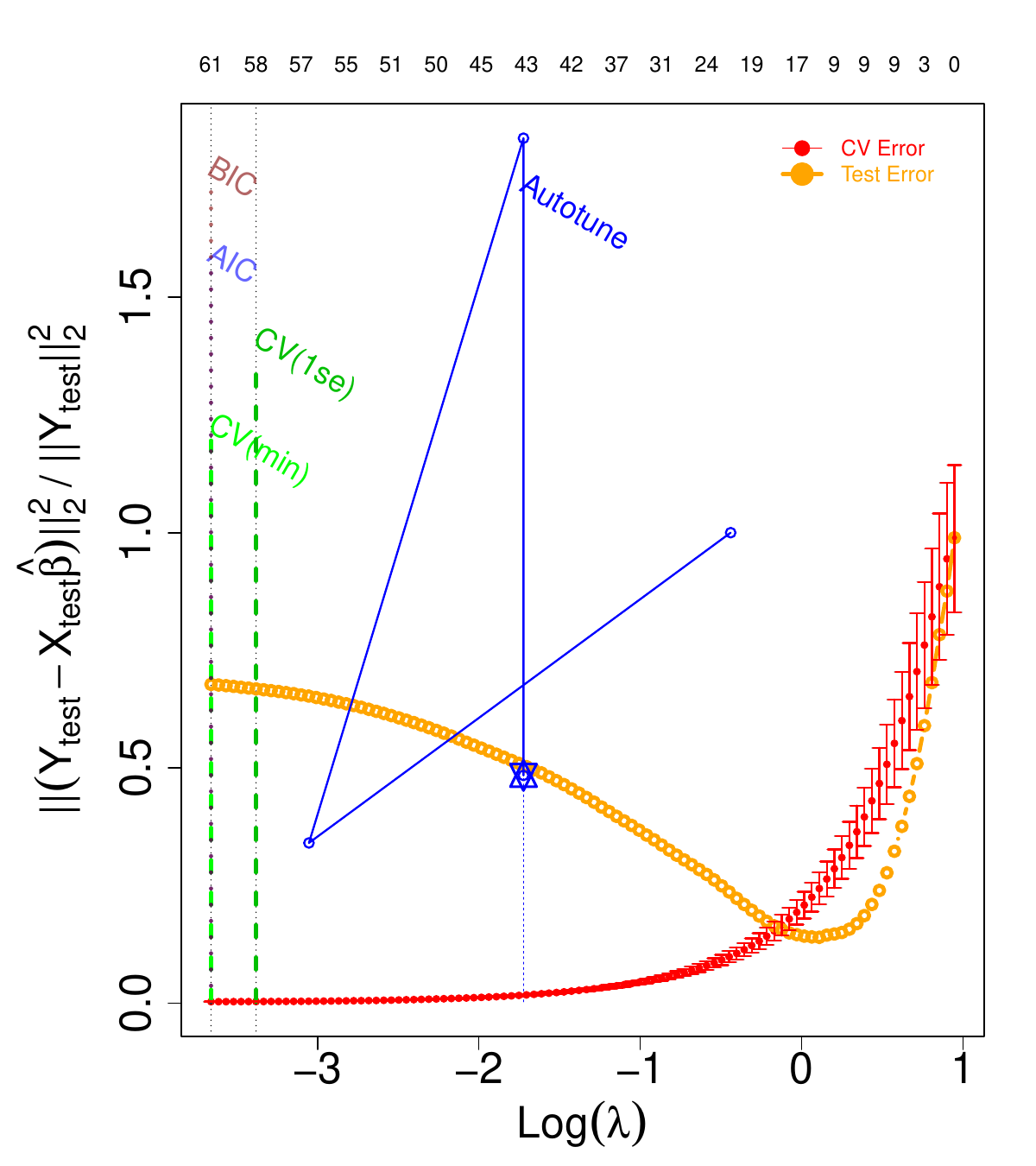}
        \caption{train set size $n = 160$, test set size $252 - n=92$.}
        \label{fig: sp500 n160}
    \end{subfigure}
    \begin{subfigure}[t]{0.32\textwidth}
        \centering
        \includegraphics[ width=\textwidth]{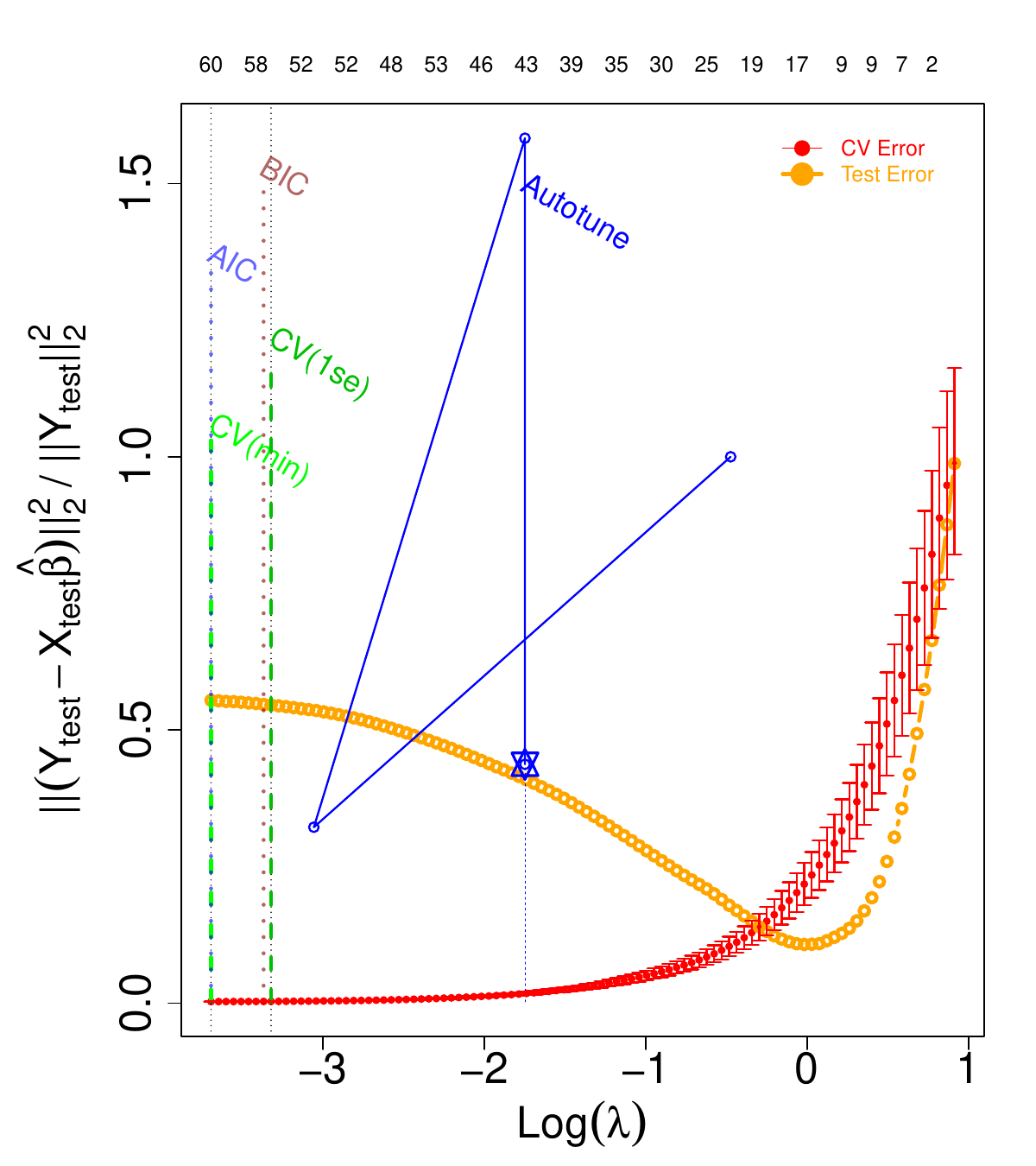}
        \caption{train set size $n = 170$, test set size $252 - n=82$.}
        \label{fig: sp500 n170}
    \end{subfigure}
    \vspace{0.5em}  
    \centering
    \begin{subfigure}[t]{0.32\textwidth}
        \centering
        \includegraphics[ width=\textwidth]{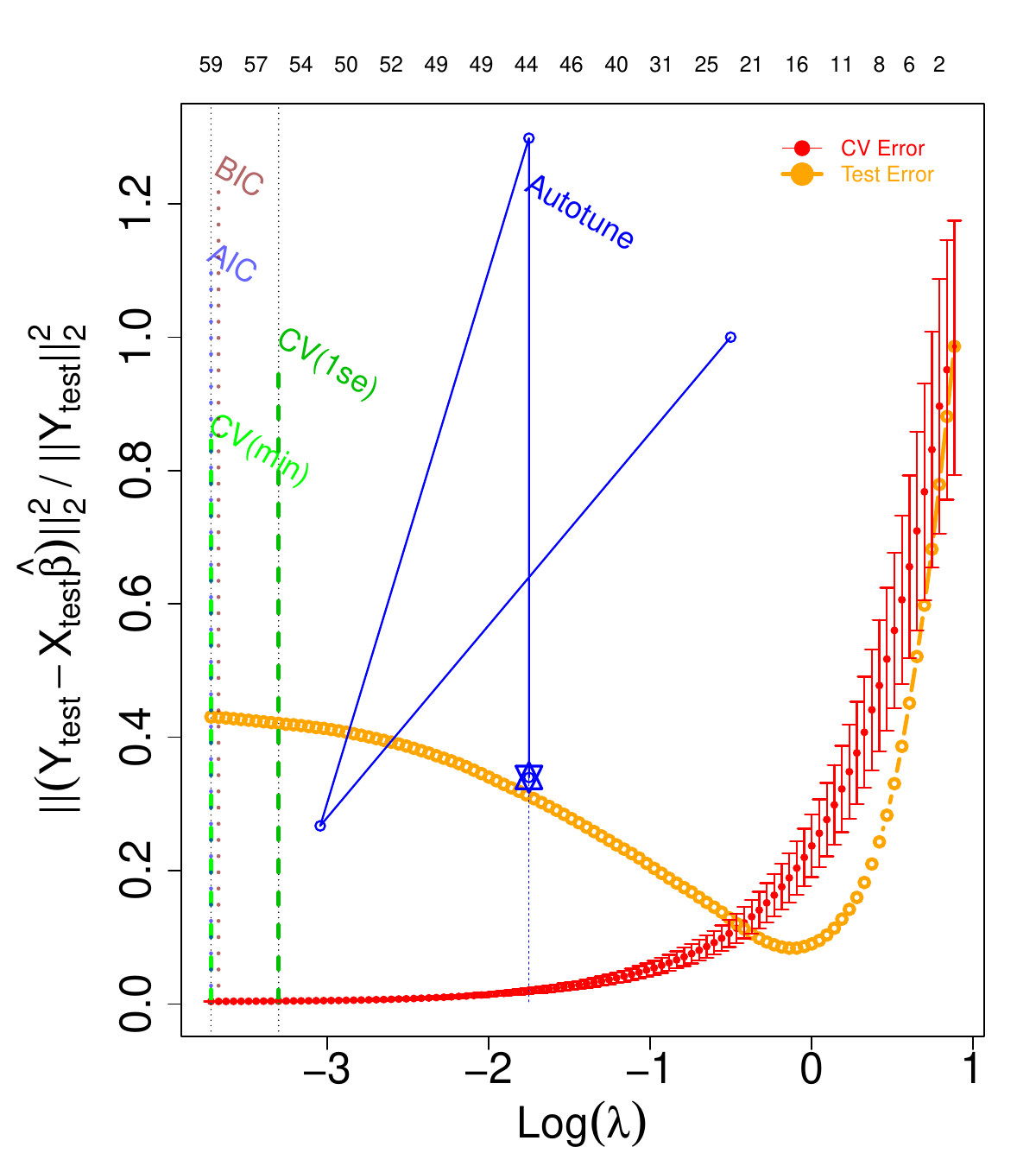}
        \caption{train set size $n = 180$, test set size $252 - n=72$.}
        \label{fig: sp500 n180}
    \end{subfigure}
    \begin{subfigure}[t]{0.325\textwidth}
        \centering
        \includegraphics[ width=\textwidth]{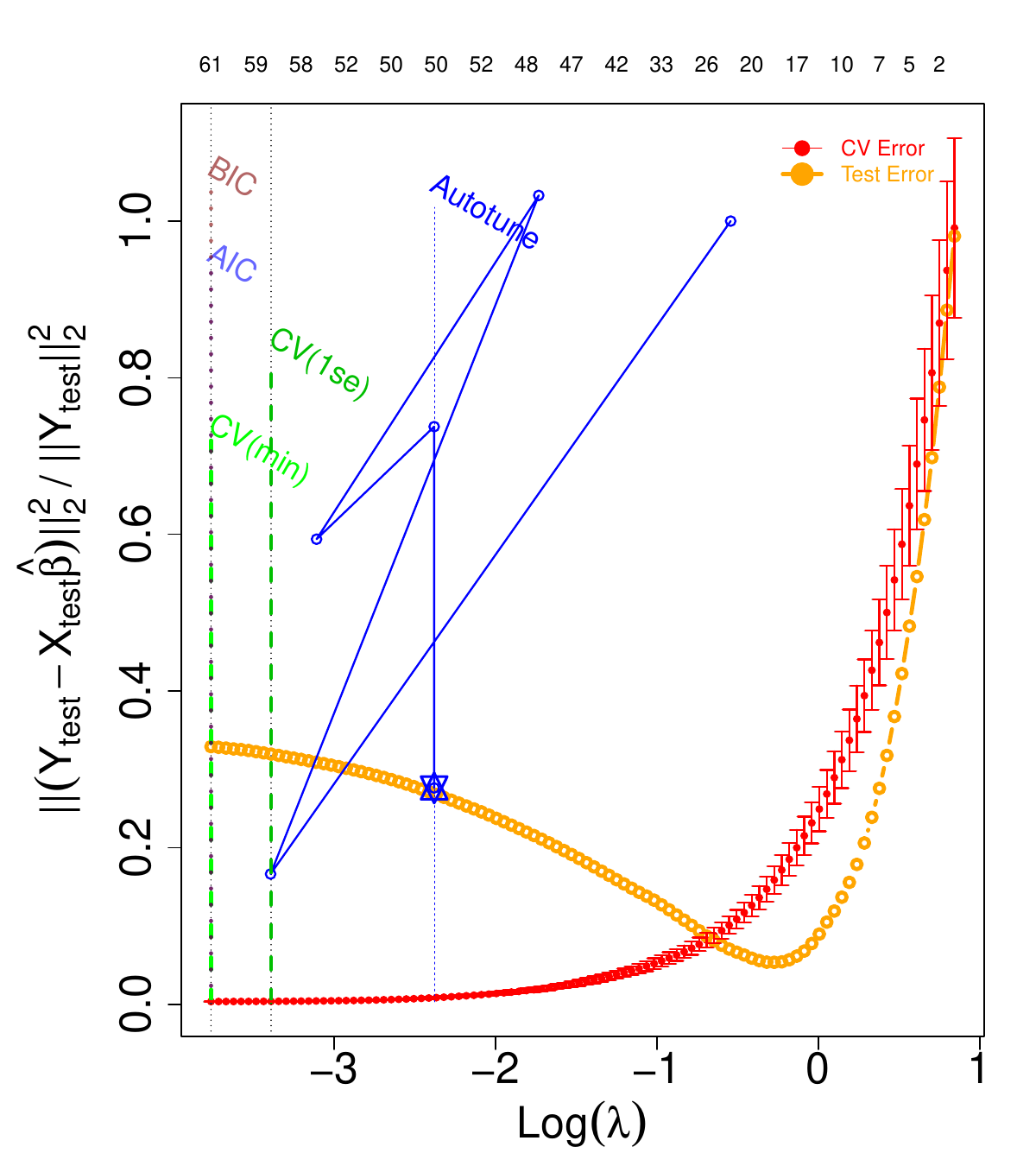}
        \caption{train set size $n = 200$, test set size $252 - n=52$.}
        \label{fig: sp500 n200}
    \end{subfigure}
    \begin{subfigure}[t]{0.325\textwidth}
        \centering
        \includegraphics[ width=\textwidth]{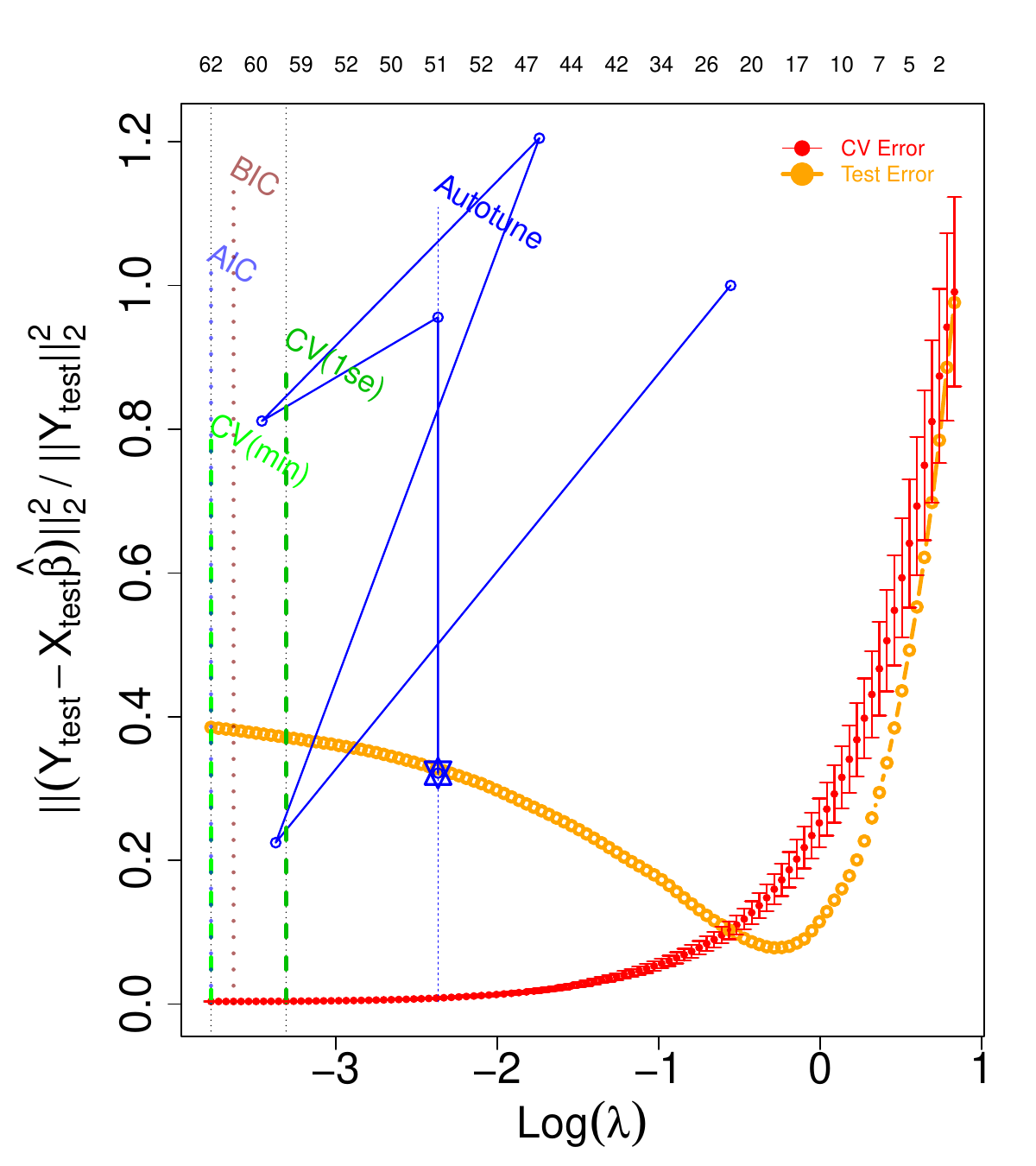}
        \caption{train set size $n = 210$, test set size $252 - n =42$}
        \label{fig: sp500 n210}
    \end{subfigure}
    \caption{Solution paths of CV Lasso in orange and $\autotune$ in blue (with endpoint denoted by blue star) for multiple train/test splits. Finally, selected $\lambda$ of different algorithms are indicated by labeled vertical lines.}
    \label{fig: appen sp500 plots}
\end{figure}

\begin{figure}[H]
    \centering
    \begin{subfigure}[t]{0.32\textwidth}
        \centering
        \includegraphics[ width=\textwidth]{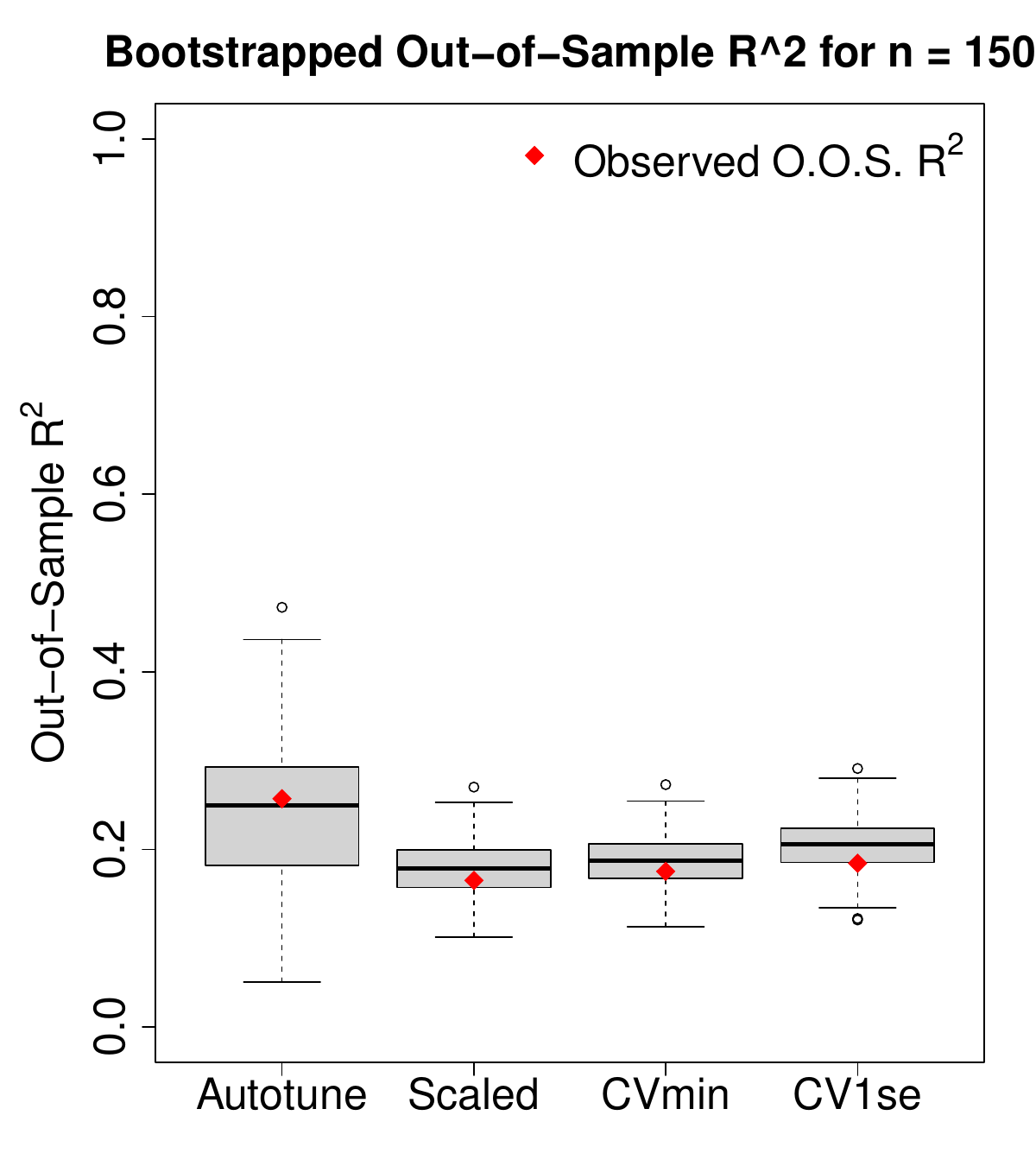}
        \caption{train set size $n = 150$, test set size $252 - n=102$.}
        \label{fig: sp500 n150 boot}
    \end{subfigure}
    \begin{subfigure}[t]{0.32\textwidth}
        \centering
        \includegraphics[ width=\textwidth]{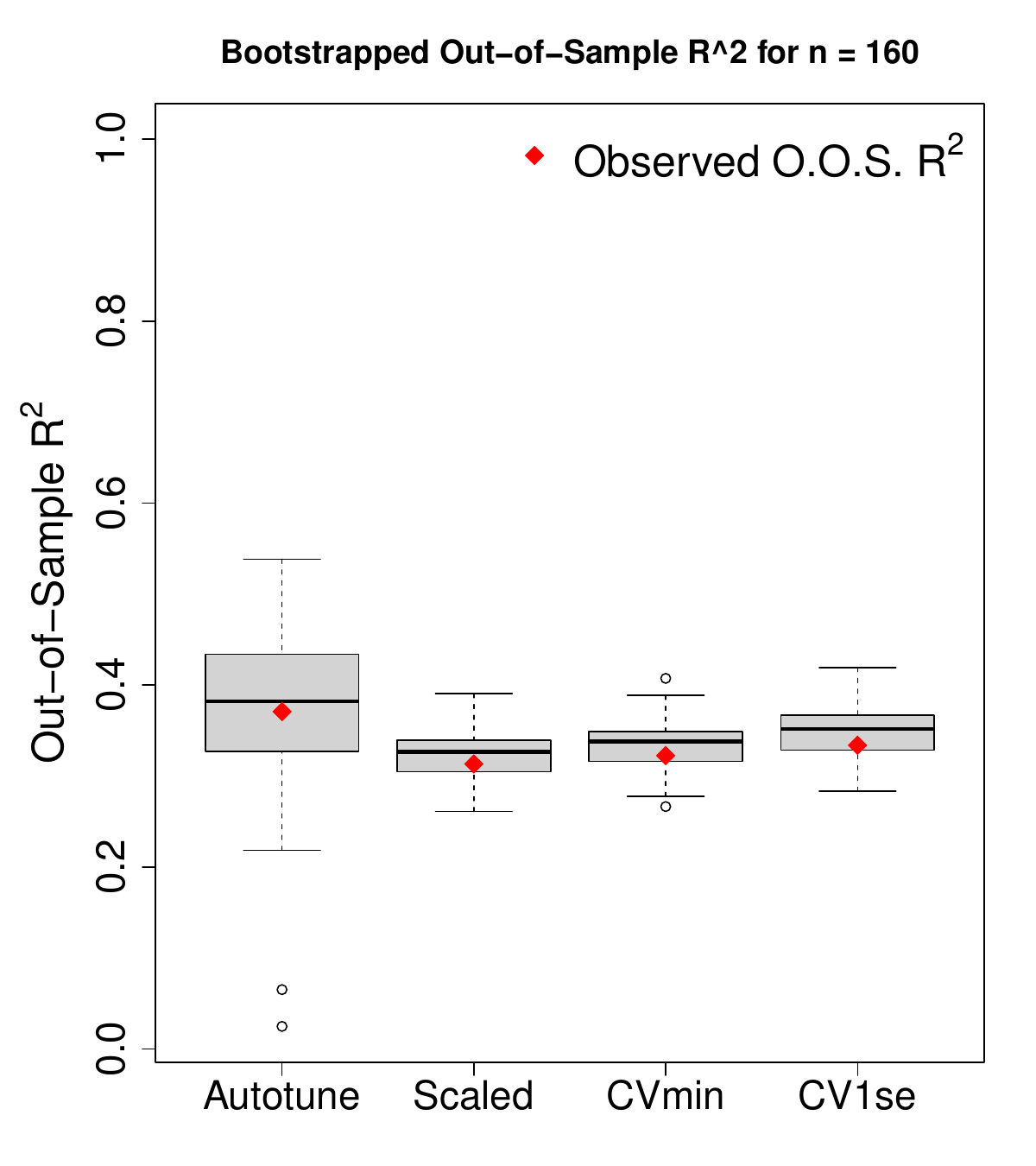}
        \caption{train set size $n = 160$, test set size $252 - n=92$.}
        \label{fig: sp500 n160 boot}
    \end{subfigure}
    \begin{subfigure}[t]{0.32\textwidth}
        \centering
        \includegraphics[ width=\textwidth]{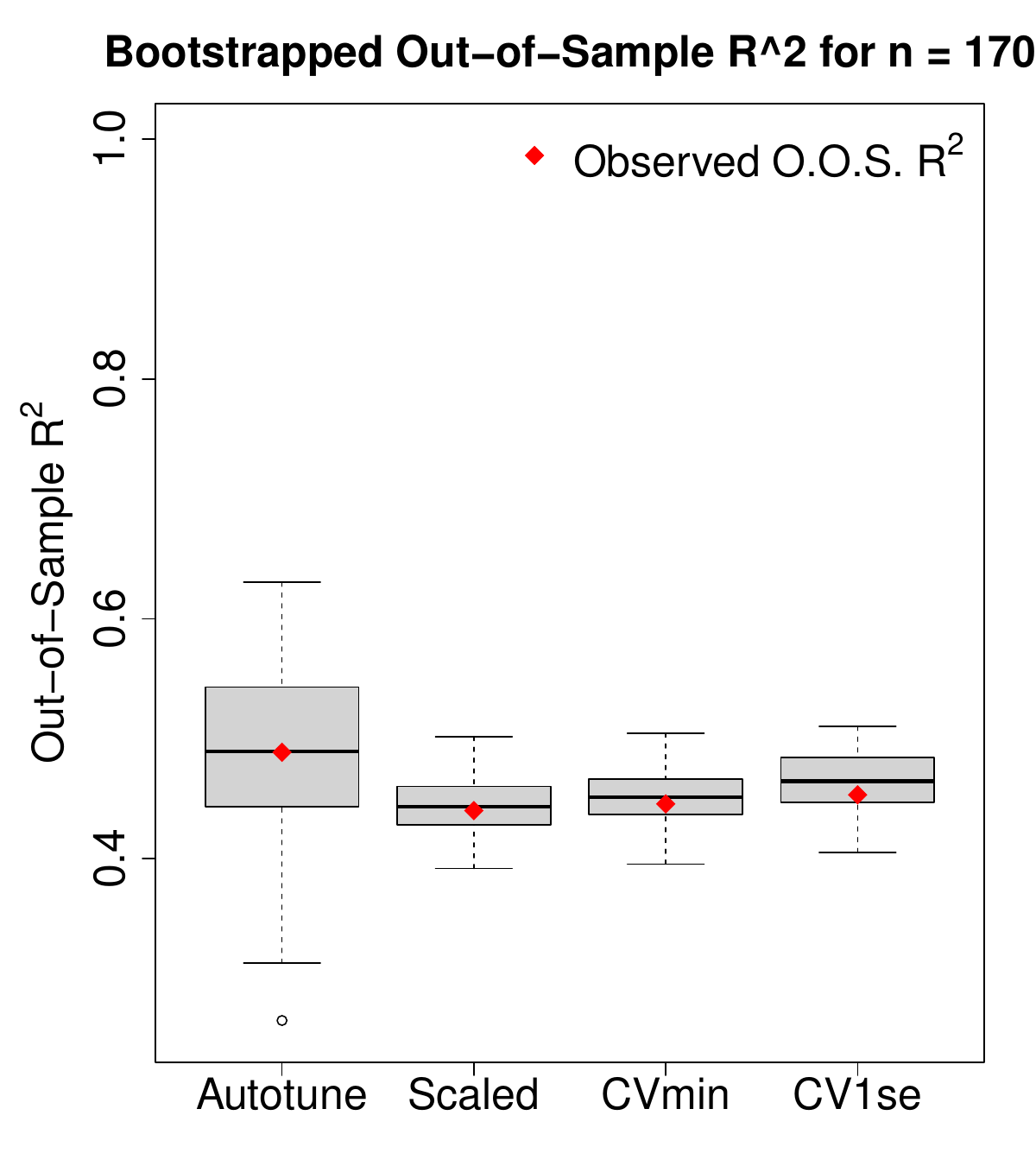}
        \caption{train set size $n = 170$, test set size $252 - n=82$.}
        \label{fig: sp500 n170 boot}
    \end{subfigure}
    \vspace{0.5em}  
    \centering
    \begin{subfigure}[t]{0.32\textwidth}
        \centering
        \includegraphics[ width=\textwidth]{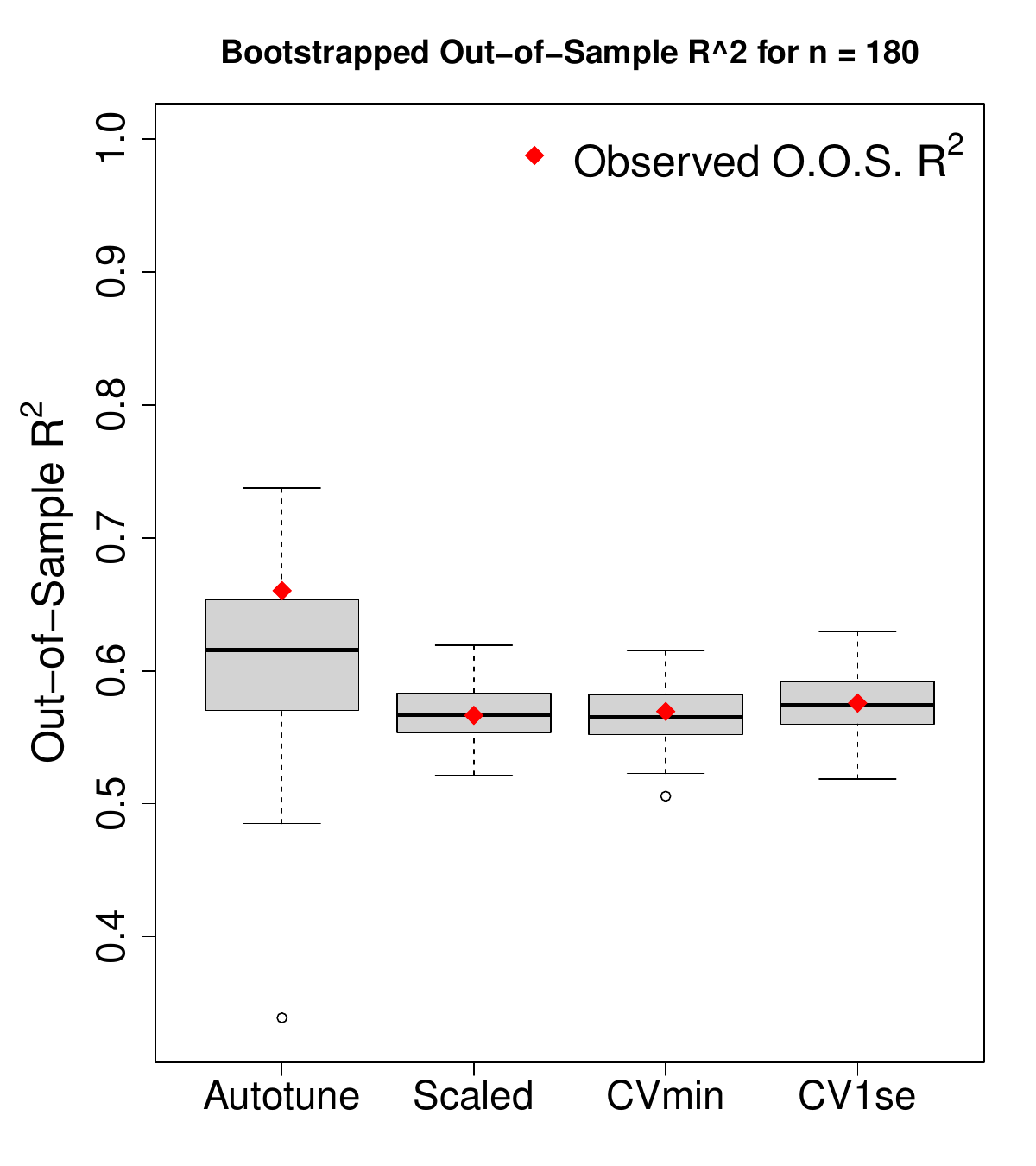}
        \caption{train set size $n = 180$, test set size $252 - n = 72$.}
        \label{fig: sp500 n180 boot}
    \end{subfigure}
    \begin{subfigure}[t]{0.32\textwidth}
        \centering
        \includegraphics[ width=\textwidth]{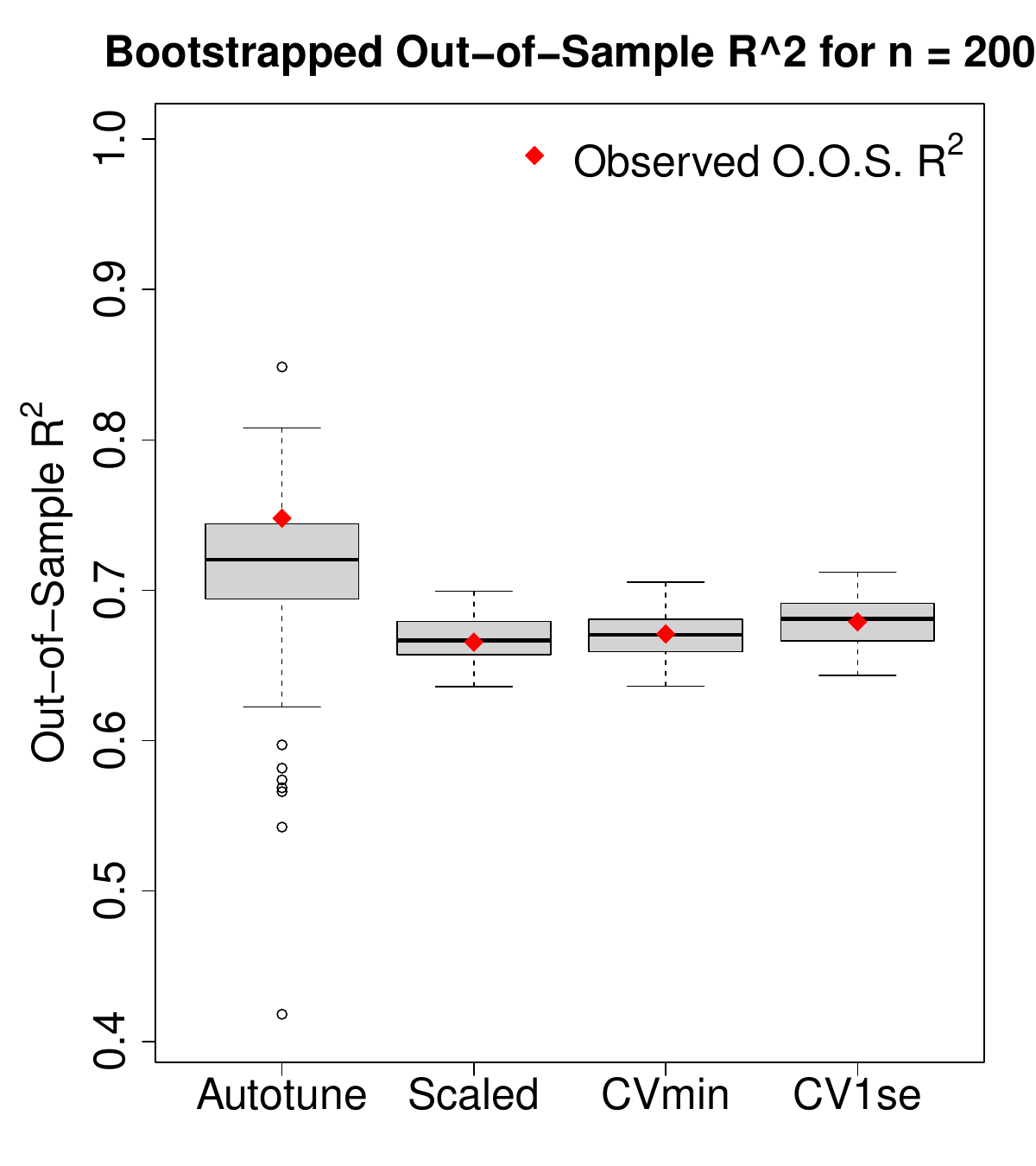}
        \caption{train set size $n = 200$, test set size $252 - n=52$.}
        \label{fig: sp500 n200 boot}
    \end{subfigure}
    \begin{subfigure}[t]{0.32\textwidth}
        \centering
        \includegraphics[ width=\textwidth]{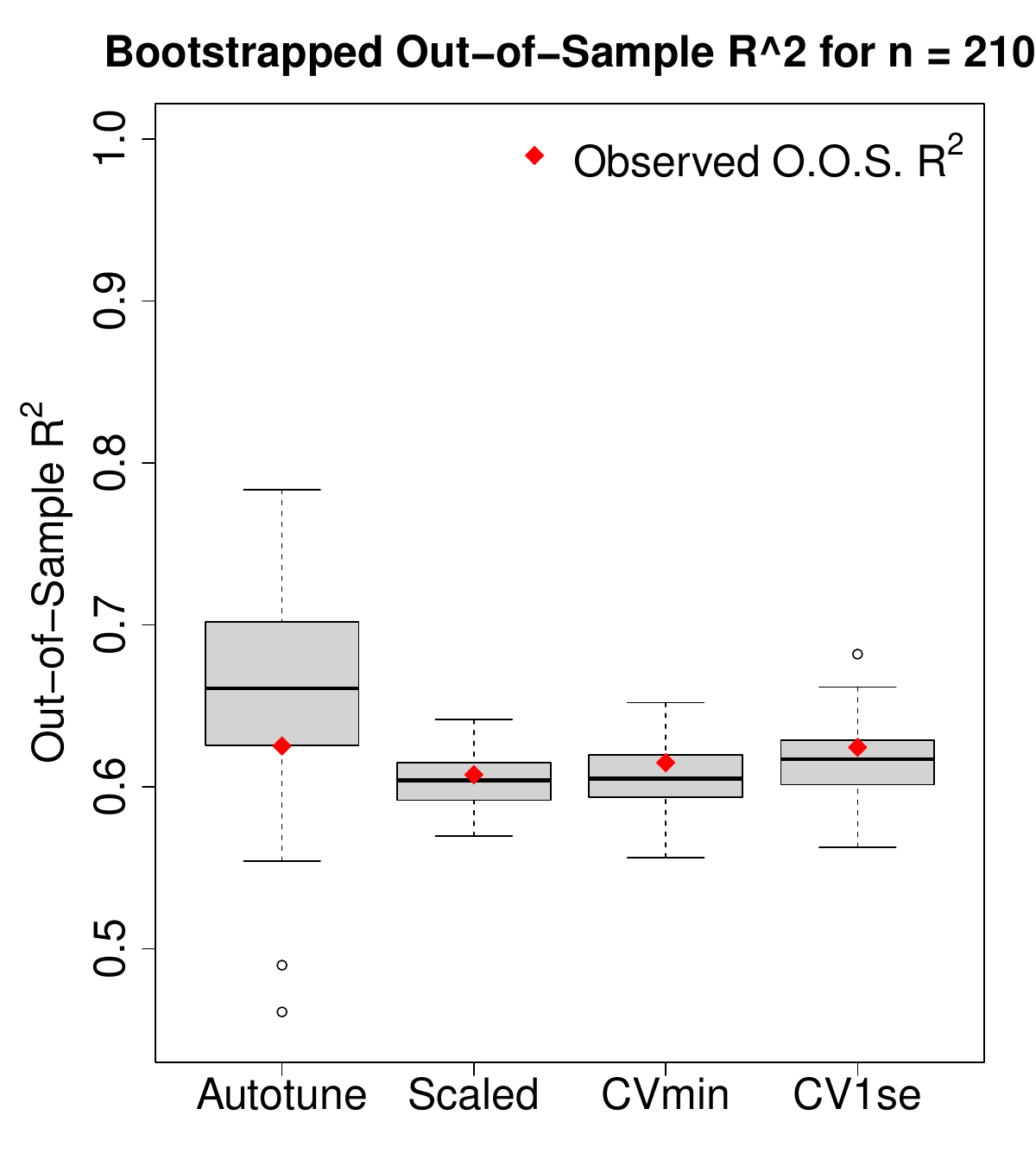}
        \caption{train set size $n = 210$, test set size $252 - n =42$}
        \label{fig: sp500 n210 boot}
    \end{subfigure}
    \caption{Boxplot of 100 bootstraped out-of-sample R$^2$ for different tuning strategies.}
    \label{fig: boot sp500 appen}
\end{figure}

\clearpage
\section{Empirical illustration on additional real life datasets}
\label{app: more real life datasets}

We benchmark $\autotune$ against CV Lasso of $\glmnet$, sparse-ho and skglm in three more datasets: ames\_housing \cite{AmesHousingR}, a RNA-seq dataset about the mammalian eyes, scheetz \href{https://iowabiostat.github.io/data-sets/Scheetz2006/Scheetz2006.html}{(https://iowabiostat.github.io/data-sets/Scheetz2006/Scheetz2006.html)} and a financial dataset e2006\_tfidf \cite{kogan-etal-2009-predicting}. ames\_housing originally had 79 features, we do one-hot-encoding of the categorical variables to increase $p$ to 237 and make it a high-dimensional dataset. e2006\_tfidf originally had 150360 features, we took 27500 predictors most correlated with the response due to computational limitations. Details of the dimensions of the table are given in \cref{tab: dataset_summary}.

In \cref{tab: oos_r2_combined}, $\autotune$ gave the best Out-of-Sample R$^2$ in all but one dataset. $\autotune$ is uniformly the fastest algorithm and in e2006 and ames dataset, $\autotune$ is atleast 7 times faster than the runner-up (\cref{tab: runtime_combined}), which was CV lasso of $\glmnet$ in all the datasets.
{

\begin{table}[ht]
\centering
\begin{tabular}{lcccc}
\toprule
Dataset & Training sample size & Number of features & Test sample size \\
\midrule
ames\_housing & 73 & 237 & 1387 \\
E2006-tfidf & 16087 & 27500 & 3308 \\
Scheetz2006 & 84 & 18975 & 36 \\
\bottomrule
\end{tabular}
\caption{Summary of datasets used in experiments}
\label{tab: dataset_summary}
\end{table}

\begin{table}[ht]
\centering
\begin{tabular}{lccc}
\toprule
Method & \multicolumn{3}{c}{Dataset} \\
\cmidrule(lr){2-4}
 & e2006-tfidf & Scheetz2006 & ames\_housing \\
\midrule
autotune   & \textbf{0.5128} & 0.4789 & \textbf{0.7588} \\
cv.glmnet  & 0.4922 & 0.4636 & 0.7040 \\
skglm      & 0.4909 & \textbf{0.5543} & 0.6893 \\
sparse-ho   & 0.4908 & 0.4444 & 0.6800 \\
\bottomrule
\end{tabular}
\caption{Out-of-sample R$^2$ comparison across datasets}
\label{tab: oos_r2_combined}
\end{table}

\begin{table}[ht]
\centering
\begin{tabular}{lccc}
\toprule
Method & \multicolumn{3}{c}{Dataset} \\
\cmidrule(lr){2-4}
 & e2006-tfidf & Scheetz2006 & ames\_housing \\
\midrule
autotune   & \textbf{45.8850} & \textbf{0.5320} & \textbf{0.0130} \\
cv.glmnet  & 332.8920 & 0.6770 & 0.0920 \\
skglm      & 696.9188 & 1.2791 & 0.9029 \\
sparse-ho   & 271.6784 & 1.4643 & 0.0976 \\
\bottomrule
\end{tabular}
\caption{Runtime (in seconds) comparison across datasets}
\label{tab: runtime_combined}
\end{table}

}
\newpage

}{%
  
}

\end{document}